\documentclass[cpp,11pt]{wmthesis}
\usepackage[linesnumbered, ruled, vlined]{algorithm2e}
\usepackage[pdftex]{graphicx}
\usepackage{graphicx}
\usepackage{balance}
\usepackage{csquotes}
\usepackage{caption}
\usepackage{makecell}
\usepackage{moresize}
\usepackage{pbox}
\usepackage{mathtools}
\usepackage[TABBOTCAP]{subfigure}
\usepackage{paralist}
\usepackage{hyperref}
\usepackage[T1]{fontenc}
\usepackage{balance}
\usepackage[dvipsnames,table,xcdraw]{xcolor}
\usepackage{multirow}
\usepackage{multicol}
\usepackage{amsmath}
\usepackage{listings}
\usepackage{lscape}
\usepackage{setspace}
\usepackage{verbatim}
\usepackage[all]{nowidow}
\usepackage{float}
\usepackage{xspace}
\usepackage{amssymb}
\usepackage{ifthen}
\usepackage{pifont}
\usepackage{textcomp}
\usepackage{pict2e,picture}
\usepackage{url}
\usepackage{epsfig}
\usepackage[subfigure]{tocloft}
\usepackage{amsfonts}
\usepackage{amssymb}
\usepackage{latexsym}
\usepackage{booktabs}
\usepackage{tabularx}
\usepackage{rotating}
\usepackage[numbers]{natbib}
\usepackage{mdwlist}
\usepackage{colortbl}
\usepackage{enumitem}
\usepackage{wrapfig}
\usepackage{verbatim}
\usepackage{courier}
\usepackage{paralist}
\usepackage{setspace}
\usepackage{xspace}
\usepackage[most]{tcolorbox}
\usepackage[normalem]{ulem}
\usepackage{float}
\usepackage{array}
\newcommand{\PreserveBackslash}[1]{\let\temp=\\#1\let\\=\temp}
\newcolumntype{C}[1]{>{\PreserveBackslash\centering}p{#1}}
\newcolumntype{R}[1]{>{\PreserveBackslash\raggedleft}p{#1}}
\newcolumntype{L}[1]{>{\PreserveBackslash\raggedright}p{#1}}
\usepackage{tikz}
\newcommand*\circled[1]{\tikz[baseline=(char.base)]{
            \node[shape=circle,draw,inner sep=0.5pt] (char) {#1};}}

\input{macro}

\def\BEGINITEMIZE{\begin{itemize}}
\def\ENDITEMIZE{\end{itemize}}

\def\defaultpenalty{1000} \clubpenalty=\defaultpenalty
\widowpenalty=\defaultpenalty

\newcommand{\inputfig}[2][\empty]{ %
   \begin{center} %
   	\ifx\empty#1 \includegraphics{../figs/#2}
	\else\scalebox{#1}{\includegraphics{../figs/#2}}\fi
   \end{center}}

\renewcommand\listfigurename{\begin{center}\Large\normalfont LIST OF FIGURES\vspace{-.35in}\end{center}}
\renewcommand\listtablename{\begin{center}\Large\normalfont LIST OF TABLES\vspace{-.35in}\end{center}}

\begin{document}
\doublespacing

\thesisTitle{Intelligent Software Tooling for Improving Software Development}
\thesisAuthor[Nathan Allen Cooper]{Nathan Cooper}
\thesisMonth{May}
\thesisYear{2023}
\thesisAdvisor{Professor Denys Poshyvanyk}
\thesisLocation{Encinitas, California, USA}
\thesisDegreeOne{Bachelor of Science, University of West Florida, 2018}
\thesisCommittee[Computer Science]{Professor Denys Poshyvanyk}
\thesisCommittee[Computer Science]{Assistant Professor Oscar Chaparro}
\thesisCommittee[Computer Science]{Assistant Professor Adwait Nadkarni}
\thesisCommittee[Computer Science]{Assistant Professor Huajie Shao}
\thesisCommittee[Computer Science]{Associate Professor Robert Michael Lewis}
\thesisCommittee[Computer Science]{Professor Andrian Marcus}

\thesisDedication{To my mom, without whom I would not be here and to whom I owe everything. From you I have learned the kindness, respect, and love that the world needs.\\I love you mom.}

\thesisAcknowledge{acknowledge}
\thesisAbstract{abstract} 

\makeThesisProlog

\chapter{Introduction}
\label{sec:intro}
\begin{displayquote}
We're making this analogy that AI is the new electricity. Electricity transformed industries: agriculture, transportation, communication, manufacturing. \signed{Andrew Ng (Baidu)}
\end{displayquote}

The 21st century has seen the fruition of the computation revolution in the 20th through the ubiquity of computers in the world from large datacenters down to wearables. The heart of this fast adoption across all sectors of life I believe to be due to the software that has unlocked the usefulness of these computers. With many of the top businesses in the world being software first companies \footnote{\url{https://www.forbes.com/lists/global2000}}, \ie companies whose main product is software, and the projected growth of software developer jobs to grow over the 2020s by 25\% \footnote{\url{https://www.bls.gov/ooh/computer-and-information-technology/software-developers.htm}}, it is safe to say that improving the construction, reliability, maintenance, and security of software can have a huge impact on the world.

The computation revolution has also brought the storage and access to large amounts of data that is extremely diverse. Such access has lead to a huge success of one specific branch of Artificial Intelligence called Deep Learning, where statistical methods inspired by the human brain are used to learn a model on data for a specific task such as classifying a movie review as positive or negative. With the growing access to software artifacts on sites such as GitHub and GitLab and the processing of such data in the field of Mining Software Repositories (MSR), Software Development tools have begun to explore using Deep Learning to improve or automate tasks. These tasks include code search \cite{husain2019codesearchnet}, program repair \cite{Tufano2018}, code generation \cite{hindle2016naturalness, tu2014localness, hellendoorn2019code, raychev2014code, bruch2009learning, jin2018hidden, izadi2022codefill, aye2021learning, Hellendoorn2017, hou2010towards, nguyen2019combining, robbes2008program, robbes2010improving, svyatkovskiy2021fast, svyatkovskiy2019pythia, yang2017language, chen2021evaluating, ciniselli2021empirical}, bug detection \cite{Deshmukh2017, 10.1145/3276517, 10.1145/3360588}, and GUI sketch generation \cite{beltramelli2018pix2code, Moran2018} to name a few. \new{With this in mind, this dissertation focuses on exploring the following research question: \textit{In what ways can the software development process be improved through leveraging deep learning techniques on the vast amounts of unstructured software engineering artifacts?} This research question gives rise to the following hypothesis:}

\begin{displayquote}
\new{Deep learning allows for learning novel automations from software engineering artifacts that can help facilitate the software development process.}
\end{displayquote}

\new{To answer this question and verify this hypothesis we both expanded on the set of tasks as well as improving the performance of existing tasks in the research field of deep learning for software engineering (DL4SE)}. 

\section{Contributions}
\label{intro:contributions}



Chapter \ref{sec:dlslr} presents our survey that we performed which canvases the application of deep learning for software engineering tasks. It gives insights into the progress that has been made in the community and where future work is needed and is instrumental in guiding this dissertation's work. The survey collected and analyzed a total of \includedpapers~papers across a 10 year period, which included \includedsetasks~software engineering tasks.

Chapter \ref{sec:tango} discusses our work on leveraging advances in computer vision that uses troves of Android screenshots to learn high level representations that are useful in determining if videos of bugs in mobile applications are duplicates of each other. This work uses a popular self-supervised learning approach, \ie no labeled data is required making it extremely scalable, called SimCLR \cite{Chen:SimCLR'20} to learn a representation of frames in videos that are then used to compute a global video representation of video-based bug reports that can then be used to compare against other video representations in a corpus of previously submitted bug reports to determine if a new video is a duplicate or not.

Chapter \ref{sec:athena} \new{investigates our work on improving upon information retrieval techniques in impact analysis (IA) by incorporating knowledge about a system through its call-graph. We introduce a novel IA approach, called \athena, that combines a software system's call graph information with a conceptual coupling approach that uses advances in deep representation learning of code without the need for change histories. Our approach is unsupervised and therefore does not require any labeled IA data. Previous IA benchmarks are small, containing less than ten software projects, and suffer from tangled commits, making it difficult to measure accurate results. Therefore, we constructed a large-scale IA benchmark, from 25 open source software projects, that utilizes fine-grained commit information from bug fixes.}

Finally, Chapter \ref{sec:completeformer} discusses our exploration on the generalizability of current state of the art code completion architectures. Specifically, a recent trend in NLP research has been the idea of generalizing to longer sequences than seen during training \cite{dai2019transformer, press2022alibi, wu2022memorizing}. Press \etal \cite{press2022alibi} introduced a simple and efficient change to the standard decoder-only Transformer's positional encoding scheme that showed an ability to generalize to longer sequences than trained on compared to other approaches such as Rotary embeddings \cite{su2021roformer} or T5 bias \cite{raffel2020exploring}. \new{We present a large empirical study evaluating this generalization property in the context of code completion of a total of four positional encoding schemes proposed in the literature, namely Sinusoidal, xPOS, ALiBi, and T5, as they have been at the heart of this generalization debate in NLP. We found that none of these solutions successfully generalize to unseen lengths and that the only safe solution is to ensure the representativeness in the training set of all lengths likely to be encountered at inference time.}


In addition to the works discussed in this dissertation, the author has worked and collaborated on a wide array of other works, which included the following topics: multi-task learning for software engineering \cite{mastropaolo2021studying}, support for interpreting neural code models \cite{palacio2023theory}, program repair \cite{connor2022can}, soundness in cryptographic misuse tools \cite{ami2022crypto}, and reproduction of mobile bugs \cite{Bernal-Cardenas:ICSE'20,havranek_v2s_2021,9839513}. Lastly, all work discussed in this dissertation was not done in isolation and is the product of a large collaborative effort among my talented colleagues, collaborators, mentees, and mentors.
\chapter{Background \& Related Work}
\label{sec:relwork}

In this section we will provide the necessary background on software development and maintenance in general and specific to mobile. Additionally, we will provide a landscape of the related work that exist in this area.



\section{Duplicate Bug Report Detection Tooling}

In this subsection, we outline work related to our research, \tango, that helps developers find duplicate video-based bug reports in an issue tracking system. This research is specifically related to work in near duplicate video retrieval, analysis of graphical software artifacts, and duplicate detection of textual bug reports.

\textbf{Near Duplicate Video Retrieval.} Extensive research has been done outside SE in near-duplicate video retrieval, which is the identification of similar videos to a query video (\eg exact copies \cite{Douze:TM'10,Kraaij:11,Jiang:16,Hao:TM'17} or similar  events \cite{Chen:ICDM'06,Jiang:IVR'07,Revaud:CVPR'13,Kordopatis-Zilos:TM'19}).

The closest work to ours with \tango is by Kordopatis-Zilos \etal~\cite{Kordopatis-Zilos:TM'19}, who addressed the problem of retrieving videos of incidents (\eg accidents). 
In their work, they explored using handcrafted-based \cite{Lowe:JCV'04,Bay:ECCV'06,Wu:MM'07,Zhao:PAMI'07,Jing:19,Huang:CVPR'97} (\eg SURF or HSV histograms) and DL-based \cite{Chechik:JMLR'10,Kordopatis-Zilos:ICCVW'17,Lecun:CNN'98,Kordopatis-Zilos:17,Tran:ICCV'15,Carreira:17} (\eg CNNs) visual feature extraction techniques and ways of combining the extracted visual features \cite{Jegou:CVPR'10,Revaud:CVPR'13,Sivic:CCV'03,Jiang:IVR'07,Cai:MM'11,Kordopatis-Zilos:17} (\eg VLAD).
While we do make use of the best performing model (CNN+ Bag of Visual Words (BoVW)) from this work~\cite{Kordopatis-Zilos:TM'19}, we did not use the proposed handcrafted approaches, as these were designed for scenes about real-world incidents, rather than for mobile bug reporting. We also further modified and extended this approach given our different domain, through the combination of visual and textual information modalities, and adjustments to the CCN+BoVW model, including the layer configuration and training objective. We discuss \tango in-depth in Chapter \ref{sec:tango}.

\textbf{Analysis of Graphical Software Artifacts.} The analysis of graphical software artifacts to support software engineering tasks has been common in recent years. Such tasks include mobile app testing \cite{Jones:2014,Moran:ICST16,Hu:FSE18,Bernal-Cardenas:ICSE'20,8094439,7965246,8094467,7181432,Fazzini2023}, developer/user behavior modelling \cite{Caetano:02,Bao:ICSE15,Frisson:CHI16,10.5555/2820518.2820534}, GUI reverse engineering and code generation \cite{Dixon:CHI11,Nguyen:ASE15,Beltramelli:EICS18,Chen2018,Moran2018,Chen:FSE20}, analysis of programming videos \cite{MacLeod:ICPC'15,Lasecki:ACM'15,Yadid:2016,Ponzanelliz:TSE'19,Alahmadi:EMSE20,Zhao:ICSE19}, and GUI understanding and verification \cite{Chang:UIST11,Zhao:ICSE20,moran_automated_2018,10.1145/3238147.3238203}. None of these works deal with finding duplicate video-based bug reports, which is our focus.

\textbf{Detection of Duplicate Textual Bug Reports.}
Many research projects have focused on detecting duplicate textual bug reports
\cite{11Bettenburg:MSR08,Borg2014,Chaparro2016a,Chaparro2019,He2020,Hindle2016,Hindle2018,Klein2014,Lazar2014,Lerch2013a,Liu2013,Nguyen2012,Panichella2019,Rakha2018,Rakha:TSE'18,Rodrigues2020,Runeson2007,Sun2010,Sun2011,Sureka2010,Thung2014,Tian2012,Wang2008,Wang2019,Wang2020,21Zhou:CIKM12}. Similar to \tango, most of the proposed techniques return a ranked list of duplicate candidates \cite{Kang2017,Chaparro2016a}.  The work most closely related to \tango is by Wang \etal \cite{Wang2019}, who leveraged attached mobile app images to better detect duplicate textual reports. Visual features were extracted from the images (\eg representative colors), using computer vision, which were combined with textual features extracted from the text to obtain an aggregate similarity score. While this is similar to our work, \tango is intended to be applied to videos rather than single images and focuses on video-based bug reports alone, without any extra information such as bug descriptions.

\section{Impact Analysis}

\new{The task of IA is concerned with determining the rippling effects a change to a software system may incur. It is composed of two parts, the first is the change entities of which there are several types, including feature requests, bug reports, or refactorings. The second being the impact set that contains all the entities that would be impacted by the initial change \cite{lehnert2011taxonomy}. In this dissertation, we structure our dataset such that, for each IA task, there is a single change entity which is the initial method modified by a given bug fix and the impact set is the set of affected methods.}

\new{Lehnert \cite{lehnert2011taxonomy} constructed a taxonomy of different IA approaches illustrating the different underlying techniques that have been utilized for IA, ranging from program slicing~\cite{tip1994survey, gallagher1990using, zhao2002change, korpi2007supporting, vidacs2007macro, santelices2010probabilistic} and call graph construction~\cite{ren2005chianti, ryder2001change, xia2004change, badri2005supporting} to information retrieval~\cite{antoniol2000identifying, canfora2005impact, canfora2006fine, vaucher2008discovering, poshyvanyk2009using, wang2018integrated} and history mining \cite{gall2003cvs, xing2004data, xing2004understanding, ying2004predicting, zimmermann2005mining,revelle_exploratory_2009,5306298}. Most of these approaches can be classified as coupling approaches, where a coupling metric is used as a proxy for how likely a change to one entity will impact another. For example, Briand \etal \cite{briand1999using} introduced structural coupling metrics such as Coupling Between Objects (CBOs) where usage of methods or attributes between classes classified them as coupled or information-flow-based coupling where the number of method invocations (weighted by their parameter counts) between classes are used for IA. Dynamic coupling metrics take into account dynamic analysis, such as dynamic binding and polymorphism \cite{arisholm2004dynamic, hassoun2004dynamic}. Evolutionary and logical coupling \cite{gall2003cvs, zimmermann2005mining} use co-change occurrences to determine coupling. Poshyvanyk \etal \cite{poshyvanyk2009using} introduced the use of information retrieval on source code to determine a conceptual coupling metric using Latent Semantic Indexing (LSI) \cite{deerwester1990indexing} for IA. Huang \etal~\cite{huang2014probabilistic} are among the only to leverage Neural Networks to tackle the problem of IA. However, unlike our proposed technique, they rely on handcrafted features that are then used to train a neural network.}

\new{The most similar work to our own is work by Wang \etal~\cite{wang2018integrated} wherein they combine an LSI model with doc2vec \cite{le2014distributed}, a canonical neural network that takes a bag of words and generates a rich vector representation. However, they focus on using change requests to find impacted methods whereas we strictly use method information for both the change and the impact set. Additionally, we leverage information from a software system's call graph to improve the performance of our technique.}

\new{As for the evaluation, most previous IA techniques for estimating impact sets have been evaluated on fewer than ten open-source projects. This dissertation constructed an evaluation benchmark based on well-curated and accurate impact sets sourced via manual labeling of changed lines in bug fixing commits from 25 open-source projects~\cite{herbold2022fine}.}

\section{Neural Code Representation}


\new{Neural representation of text has been common place in natural language processing since the advent of word2vec \cite{mikolov2013distributed}, with recent Transformer \cite{vaswani2017attention} based methods such as BERT \cite{devlin2018bert} achieving state-of-the-art results across a multitude of tasks. Unlike natural language, programming languages can be represented in multiple ways. For instance, one useful representation that has been commonly used is the Abstract Syntax Tree (AST). White \etal \cite{White2016} introduced a DL method that utilized source code and ASTs to perform code clone detection. Similarly, Alon \etal \cite{alon2019code2vec} introduced code2vec, which learned a dense vector representation of code snippets that were useful for tasks such as method renaming and retrieval. Code2vec relied on AST representations to generate these dense vector representations. Feng \etal \cite{feng2020codebert} adapted the successful BERT model to function with bimodal code information, namely, code and the code's corresponding documentation. Using masked token prediction \cite{devlin2018bert} and replaced token detection \cite{clark2020electra} pretraining tasks followed by a downstream finetuning scheme, the authors were able to achieve high performance on many code-related benchmarks including code search. Similar to White~\etal and Alon~\etal, Guo~\etal~\cite{guo2020graphcodebert} integrated code specific information, namely Data Flow Graphs (DFGs), into the training of a BERT-like model by introducing two additional pretraining objectives aside from the masked token prediction task. These two objectives involve predicting edge information of the DFGs and aligning the representations of the code and its DFG. Guo~\etal~\cite{guo2022unixcoder} introduced UnixCoder, a unified model for code understanding and generation leveraging different attention masks to accomplish several pretraining objectives, \ie masked language modeling, causal language modeling, and denoising. The authors also incorporated AST information, which through ablations, was illustrated to improve performance across multiple tasks including code search. Finally, they also define two tasks, including multi-modal contrastive learning and cross-modal generation, to utilize the multi-modal inputs and align code representations among programming languages based on code comments.}

\new{Although these transformer-based code representation models achieved state-of-the-art performance across multiple code-related tasks, they have not been applied to the task of IA. Our work is the first to evaluate three representative neural models, (\ie CodeBERT, GraphCodeBERT, and UniXCoder), on IA with and without incorporating additional contextual information related to the call graph of the entire software system. Next, we discuss the necessary background of these models to aid in the understanding of our approach description (Chapter ~\ref{athena:approach}), by focusing our discussion on the overarching architecture upon which they all rely -- the Transformer~\cite{vaswani2017attention}.}


\section{Transformer Architecture}

Originally, the Transformer architecture was proposed for the task of machine translation leveraging an encoder-decoder architecture~\cite{vaswani2017attention}. However, the main advantage of the Transformer architecture is the usage of attention~\cite{bahdanau2014neural} without any recurrent mechanism, as was used in previous architectures for machine translation, such as Long-Short Term Memory networks~\cite{hochreiter1997long}. More specifically, the Transformer's attention mechanism uses three vectors that represent keys, values, and queries, which are different token representations akin to an information retrieval setting. 

In this setting, the query vector represents a word that is fed into the model, and the key and value vectors represent the "memory" of the model, \ie all the words that have been processed or generated previously. Key vectors that have a high dot product with the query vector have a stronger "memory" signal that focuses the attention of "head" with the keys' matching value vectors. Transformers leverage multiple types of attention, referred to as multi-head attention, wherein each head can specialize its attention for different things (\eg co-referencing, parts of speech, \etc). In essence, the attention mechanism can be thought of as a fuzzy dictionary look up with an input word being transformed into a query vector, and searches for keys in the memory (\ie other words in the sequence) and then a softmax function is used to "select" which previous word values to "pay attention" to and how much to weight them. This can be represented mathematically by:
\begin{equation}
    A_{head} = softmax(\alpha Q_{head} \cdot K_{head}) \cdot V
\end{equation}

\noindent where $A_{head}$ is the attention head, and $K$, $Q$ and $V$ are the query, key and value vectors. The main benefit of this formulation is that it is completely differentiable allowing for gradient optimization to be applied. The information that is routed between the layers is represented by the $V$ vector in the above equation. For CodeBERT and GraphCodeBERT, only the encoder portion of the Transformer is used whereas for UnixCoder, both encoder and decoder portions are used. However, during the embedding phase, only the encoder is used in UnixCoder.

\section{Code Completion}

We discuss the literature related to (i) code completion, (ii) techniques aimed at improving the generalizability of Transformers, and (iii) NLP studies aimed at investigating the extent to which Transformers can generalize to instances different from those seen during training.

\noindent\textbf{Code Completion:} Code completion has been studied extensively for several years in SE \cite{hindle2016naturalness, tu2014localness, hellendoorn2019code, raychev2014code, bruch2009learning, jin2018hidden, izadi2022codefill, aye2021learning, Hellendoorn2017, hou2010towards, nguyen2019combining, robbes2008program, robbes2010improving, svyatkovskiy2021fast, svyatkovskiy2019pythia, yang2017language, chen2021evaluating, ciniselli2021empirical, austin2021program, fried2022incoder, nijkamp2022codegen, allal2023santacoder}. It has seen many iterations going from techniques able to generate simple predictions such as method calls \cite{mandelin2005jungloid} to recent DL models able to predict multiple code statements \cite{chen2021evaluating, ciniselli2021empirical, fried2022incoder, nijkamp2023codegen, allal2023santacoder}.

Our work can be thought of as a research thrust continuation to that of Ciniselli \etal \cite{ciniselli2021empirical,9463129}. In their work, they explored the applicability of Transformer models for the task of code completion, especially as the number of tokens to complete grows. They found a T5 architecture to perform fairly well on this task, reporting however a performance degradation as the number of tokens to predict grew. While Ciniselli \etal \cite{ciniselli2021empirical} study how a T5 model trained on a specific dataset can work with prediction tasks of different complexity, we study how models trained on different datasets featuring instances characterized by different lengths generalize to unseen lengths. 

Hendrycks \etal \cite{hendrycks2021measuring} and Chen \etal \cite{chen2021evaluating} proposed a systematic evaluation for code generation tools using functional-correctness. However, their focus was on generating complete programs rather than completing existing code. 

Fried \etal \cite{fried2022incoder} investigated a novel infilling pretraining scheme for decoder-only Transformer architectures that allow them to use bidirectional context to complete code. They found this infilling scheme allows for models that achieve a higher code completion rate than left context only models for single and multi-line code completions. Our study is complementary to these since we focus on encoder-decoder models and on the generalization to code completion length, rather than general performance.

\noindent\textbf{Methods to Improve Length Generalization:} There has been a plethora of literature on different techniques to improve generalization of inference length of Transformer models. Neishi and Yoshinaga \cite{neishi2019relation} demonstrated that replacing the positional encoding scheme in Transformers with a Recurrent Neural Network (RNN) improves machine translation performance on sentences longer than those seen during training. In a similar vein, Dai \etal \cite{dai2019transformer} take inspiration from RNNs by adding a segment-level recurrence mechanism to improve performance on long sequences. Dubois \etal \cite{dubois2020location} introduced a separate location and content based attention to improve generalization to longer sequences than those seen during training. Newman \etal \cite{newman2020eos} showed that training models to predict the end of sequence (EOS) token significantly degraded a model's ability to generalize to longer sequences than those seen during training. Specifically, they found that the hidden states of models trained on predicting EOS tokens lead to stratification of the hidden state manifold by length and get clustered into \textit{length attractors}, which are areas where the EOS token is given the highest-probability prediction. This causes a failure to generalize to longer sequences that are not present when omitting the prediction of the EOS token. Lastly, several works \cite{kiyono2021shape, likhomanenko2021cape, press2022alibi, su2021roformer, sun2022length} introduced various modifications to the positional encoding schemes of Transformers to improve generalization to longer sequences not seen during training. Among those, we considered in our study the four described in Section \ref{completeformer:back} due their good representativeness of the different types of encoding schemas, namely Absolute Positional Encoding (APE) and Relative Positional Encoding (RPE). In addition, we considered the T5 model since Press \etal \cite{press2022alibi} also showed it had an ability to slightly generalize to longer sequence lengths than it had seen during training.

\noindent\textbf{Evaluations of Length Generalization:} The most similar work to ours (from the NLP field) is Rosendahl \etal \cite{rosendahl2019analysis}'s study on analyzing a variety of positional encoding schemes and their ability to generalize to longer machine translation sentences than those seen during training. Similar to other works \cite{neishi2019relation, press2022alibi}, they found that relative positional encodings are superior to absolute positional encodings for generalizing to longer sequences. Lake and Baroni \cite{lake2018generalization} and Hupkes \etal \cite{hupkes2020compositionality}'s work focused on measuring the composability of language models. One type of composition was specific to generalization of sequence length and they both found current language models to be unable to perform well on such tasks. 

\chapter{A Systematic Literature Review on the Use of Deep Learning in Software Engineering Research}
\label{sec:dlslr}

Software engineering (SE) research investigates questions pertaining to the design, development, maintenance, testing, and evolution of software systems. As software continues to pervade a wide range of industries, both open- and closed-source code repositories have grown to become unprecedentedly large and complex. This has resulted in an increase of unstructured, unlabeled, yet important data including requirements, design documents, source code files, test cases, and defect reports. Previously, the software engineering community has applied canonical machine learning (ML) techniques to identify patterns and unique relationships within this data to automate or enhance many tasks typically performed manually by developers. Unfortunately, the process of implementing ML techniques can be a tedious exercise in careful feature engineering, wherein researchers experiment with identifying salient attributes of data that can be leveraged to help solve a given problem or automate a given task.

However, with recent improvements in computational power and the amount of memory available in modern computer architectures, an advancement to traditional ML approaches has arisen called Deep Learning. Deep learning represents a fundamental shift in the manner by which machines learn patterns from data by \textit{automatically} extracting salient features for a given computational task as opposed to relying upon human intuition. Deep Learning approaches are characterized by architectures comprised of several layers that perform mathematical transformations on data passing through them. These transformations are controlled by sets of learnable parameters that are adjusted using a variety of learning and optimization algorithms. These computational layers and parameters form models that can be trained for specific tasks by updating the parameters according to a model's performance on a set of training data. Given the immense amount of structured and unstructured data in software repositories that are likely to contain hidden patterns, DL techniques have ushered in advancements across a range of tasks in software engineering research including automatic program repair~\cite{Tufano2018}, code suggestion~\cite{Gu2018}, defect prediction~\cite{Wang2016}, malware detection \cite{Li2018}, feature location~\cite{Corley2015}, among many others~\cite{Ma2018, Wan2018, Liu2018, White2016, Xu2016, Guo2017, Tian2018, Liu2017,8919098}. A recent report from the 2019 NSF Workshop on Deep Leaning \& Software Engineering has referred to this area of work as Deep Learning for Software Engineering (DL4SE)~\cite{dlse19-report}. 

The applications of DL to improve and automate SE tasks points to a clear synergy between ongoing research in SE and DL \cite{devanbu_deep_2020}. However, in order to effectively chart the most impactful path forward for research at the intersection of these two fields, researchers need a clear map of what has been done, what has been successful, and what can be improved.
In an effort to map and guide research at the intersection of DL and SE, we conducted a systematic literature review (SLR) to identify and systematically enumerate the synergies between the two research fields. As a result of the analysis performed in our SLR, we synthesize a detailed \textit{research roadmap} of past work on DL techniques applied to SE tasks\footnote{It should be noted that another area, known as Software Engineering for Deep Learning (SE4DL), which explores improvements to engineering processes for DL-based systems, was also identified at the 2019 NSF workshop. However, the number of papers we identified on this topic was small, and mostly centered around emerging testing techniques for DL models. Therefore, we reserve a survey on this line of research for future work.} (\ie DL4SE), complete with identified open challenges and best practices for applying DL techniques to SE-related tasks and data. Additionally, we analyzed the impacts of these DL-based approaches and discuss some observed concerns related to the potential reproducibility and replicability of our studied body of literature.

We organize our work around five major Research Questions (RQs) that are fundamentally centered upon the \textit{components of learning}. That is, we used the various components of the machine learning process as enumerated by Abu-Mostafa~\cite{abu-mastafa}, to aid in grounding the creation of our research roadmap and exploration of the DL4SE topic. Our overarching interest is to identify best practices and promising research directions for applying DL frameworks to SE contexts. 
Clarity in these respective areas will provide researchers with the tools necessary to effectively apply DL models to SE tasks. 
To answer our RQs, we created a taxonomy of our selected research papers that highlights important concepts and methodologies characterized by the types of software artifacts analyzed, the learning models implemented, and the evaluation of these approaches. We discovered that while DL in SE has been successfully applied to many SE tasks, there are common pitfalls and details that are critical to the components of learning that are often omitted. Therefore, in addition to our taxonomy that describes how the components of learning have been addressed, we provide insight into components that are often omitted, alongside strategies for avoiding such omissions. As a result, this paper provides the SE community with important guidelines for applying DL models that address issues such as sampling bias, data snooping, and over- and under-fitting of models. Finally, we provide an online appendix with all of our data and results to facilitate reproducability and encourage contributions from the community to continue to taxonomize DL4SE research\footnote{\url{http://wm-semeru.github.io/dl4se/}}~\cite{watson_palacio_cooper_moran_poshyvanyk}.
\newpage

\section{Research Question Synthesis}
\label{dlslr:rq_synthesis}

\begin{wrapfigure}{l}{0.5\textwidth} 
	\centering
	\includegraphics[width=0.5\columnwidth]{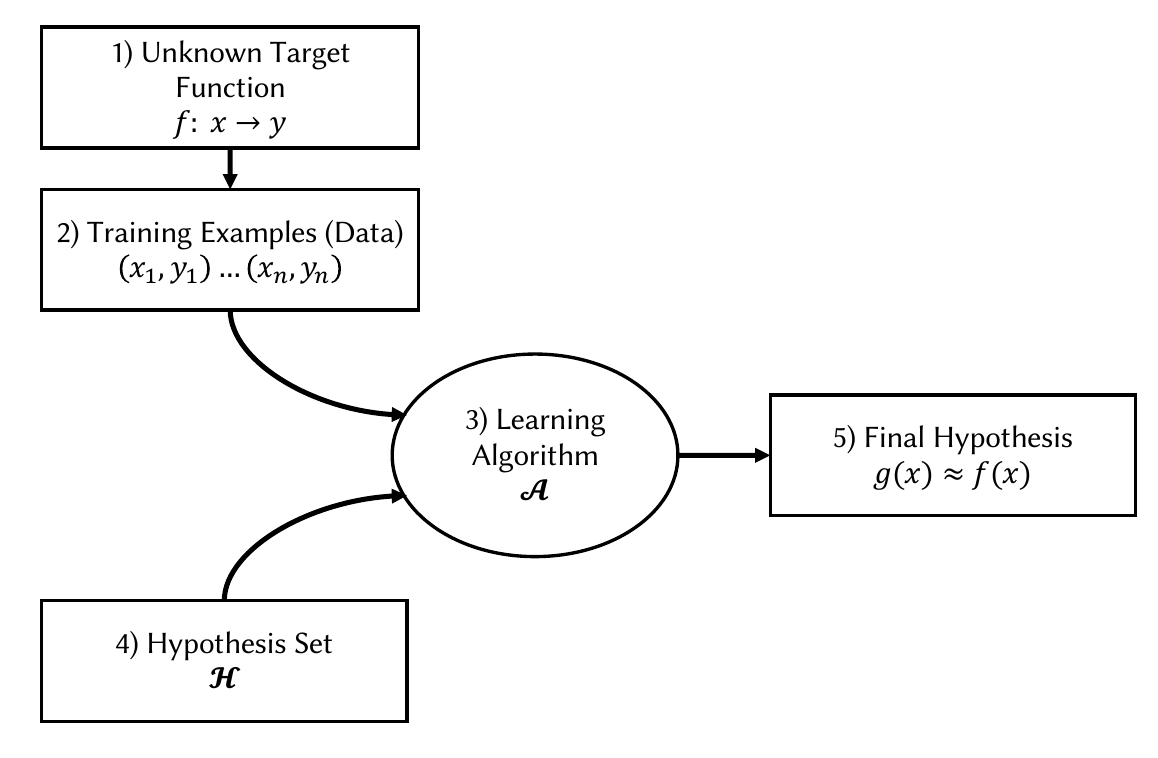}
	\caption{The Components of Learning}	
	\label{fig:components}
\end{wrapfigure}

The synthesis and generation of research questions (RQs) is an essential step to any systematic literature review (SLR). In order to study the intersection of DL \& SE, our intention was to formulate RQs that would naturally result in the derivation of a taxonomy of the surveyed research, establish coherent guidelines for applying DL to SE tasks, and address common pitfalls when implementing these complex models. Therefore, in order to properly accomplish these tasks and frame our review, we centered the synthesis of our RQs on the \textit{components of learning}~\cite{abu-mastafa}, which are illustrated in Figure~\ref{fig:components}. The components of learning are a formalization introduced by Abu-Mostafa~\cite{abu-mastafa} in an effort to enumerate the conditions for computational learning. By framing our top-level research questions according to these components, we can ensure that that analysis component of our literature review effectively captures the essential elements that any research project applying a \textit{deep} learning-based solution should discuss, allowing for a thorough taxonomic inspection. Given that these components represent essential elements that should be described in any application of computational learning, framing our research questions in this manner allows for the extrapolation of observed trends related to those elements that are commonly included or omitted from the surveyed literature. This, in turn, allows us to make informed recommendations to the community related to the reproducibility of our surveyed work. In the remainder of this section, we detail how each of our top-level research questions were derived from the elements of learning. Note that, in order to perform our analysis to a sufficient level of detail, in addition to our top-level RQs, we also define several Sub-RQs that allow for a deeper analysis of some of the more complex elements of learning. We provide the full list of all the research questions at the end of this section.

\subsection{The First Element of Learning: The Target Function}

The first component of learning is an unknown \textit{target function} ($f:x\rightarrow y$), which represents the relationship between two observed phenomenon $x$ and $y$. The target function is typically tightly coupled to the task to which a learning algorithm is applied. By analyzing the target function to be learned, one can infer the input and output of the model, the type of learning, hypothetical features to be extracted from the data and potential applicable architectures. To capture the essence of this component of learning we formulated the following research question:

\begin{tcolorbox}[enhanced,skin=enhancedmiddle,borderline={1mm}{0mm}{MidnightBlue}]
	\textit{\textbf{RQ$_1$:} What types of SE tasks have been addressed by DL-based approaches?}
\end{tcolorbox}  

In understanding what SE tasks have been analyzed, we are able to naturally present a taxonomy of what tasks have yet to be explored using a DL-based approach. We were also able to infer why certain SE tasks may present unique challenges for DL models as well as the target users of these DL approaches, given the SE task they address.

\subsection{The Second Element of Learning: The (Training) Data}

The second component of learning is defined by the \textit{data} that is presented to a given learning algorithm in order to learn this unknown target function. Here, we primarily focused on studying the input and output training examples and the techniques used in DL approaches to prepare the data for the model. An understanding of the training examples presents greater insight into the target function while also providing further intuition about the potential features and applicable DL architectures that can be used to extract those features. Thus, in capturing this component of learning, we aimed to derive a taxonomy of the data used, how it was extracted and preprocessed, and how these relate to different DL architectures and SE tasks. This taxonomy captures relationships between data and the other elements of learning, illustrating effective (and ineffective) combinations for various SE-related applications. Our intention is that this information can inform researchers of effective combinations and potentially unexplored combinations of data/models/tasks to guide future work. Thus, our second RQ is formulated as follows:

\begin{tcolorbox}[enhanced,skin=enhancedmiddle,borderline={1mm}{0mm}{MidnightBlue}]
	\textit{\textbf{RQ$_2$:} How are software artifacts being extracted, prepared, and used in DL-based approaches for SE tasks?}
\end{tcolorbox}  

\noindent Given the multi-faceted nature of selecting, creating, and preprocessing data, we specifically examine three sub-research questions that explore the use of SE data in DL approaches in depth: 

\begin{itemize}\setlength\itemsep{0.4em}
	\item{\textit{\textbf{RQ$_{2a}$:} What types of SE data are being used?}}
	\item{\textit{\textbf{RQ$_{2b}$:} How is this data being extracted and pre-processed into formats that are consumable by DL approaches?}}
	\item{\textit{\textbf{RQ$_{2c}$:} What type of exploratory data analysis is conducted to help inform model design and training?}}
\end{itemize}

\noindent \textit{RQ$_{2a}$} explores the different types of data that are being used in DL-based approaches. Given the plethora of different software artifacts currently stored in online repositories, it is important to know which of those artifacts are being analyzed and modeled. \textit{RQ$_{2b}$} examines how data is being extracted and pre-processed into a format that a DL model can appropriately consume. The results of this RQ will enumerate the potential tools and techniques to mine different types of data for various DL applications within SE. Additionally, the representation of data is often dependent on the DL architecture and its ability to extract features from that representation, which lends importance to the discussion of the relationship between DL architectures and the data they process. \textit{RQ$_{2c}$} investigates what type of exploratory data analysis is conducted to help inform model design and training. In order to perform effectively, DL models typically require large-scale datasets, and the quality of the learned hypothesis is a product of the quality of data from which the model learns. Therefore, since the quality of a given DL model is often directly associated with its data, we examined how research performed (or didn't perform) various analyses to avoid common data- related pitfalls recognized by the ML/DL community, including sampling bias and data snooping.

\subsection{\hspace{-0.25cm}The Third \& Fourth Elements of Learning: The Learning Algorithm \& Hypothesis Set}

Next, we jointly examine both the third and fourth components of learning, the \textit{learning algorithm} and \textit{hypothesis set}, in a single research question due to their highly interconnected nature. The learning algorithm is a mechanism that navigates the hypothesis set in order to best fit a given model to the data. The learning algorithm typically consists of a numeric process that uses a probability distribution over the input data to appropriately approximate the optimal hypothesis from the hypothesis set. The hypothesis set is a set of all hypotheses, based on the learning algorithm, to which the input can be mapped. This set changes because it is a function of the possible outputs given the input space, and is dependent on the learning algorithm's ability to model those possible outputs. Taken together the learning algorithm and the hypothesis set are referred to as the learning model, thus, our third RQ is formulated as follows:

\begin{tcolorbox}[enhanced,skin=enhancedmiddle,borderline={1mm}{0mm}{MidnightBlue}]
	\textit{\textbf{RQ$_3$:} What deep learning models are used to support SE tasks?}
\end{tcolorbox} 

\noindent Given the various types of DL model architectures and optimization techniques that may be applicable to various SE tasks, we examine \textit{RQ$_3$} through the lens of three sub-RQs, which address the aforementioned attributes of the learning model individually. 

\begin{itemize}\setlength\itemsep{0.4em}
	\item{\textit{\textbf{RQ$_{3a}$:} What types of model architectures are used to perform automated feature engineering of the data related to various SE tasks?}}
	\item{\textit{\textbf{RQ$_{3b}$:} What learning algorithms and training processes are used in order to optimize the models?}}
	\item{\textit{\textbf{RQ$_{3c}$:} What methods are employed to combat over- and under-fitting of the models?}}
\end{itemize}

\noindent Firstly, \textit{RQ$_{3a}$} explores the different types of model architectures that are used to perform automated feature engineering of different SE artifacts for various SE tasks. As part of the analysis of this RQ we also examine how the type of architecture chosen to model a particular target function relates to the types of features that are being extracted from the data. Secondly, \textit{RQ$_{3b}$} examines the different learning algorithms and training processes that are used to optimize various DL models. As part of this analysis, we explore a variety of different learning algorithms whose responsibility is to properly capture the hypothesis set for the given input space. The different optimization algorithms and training processes used to tune the weights of the model are an important step for finding the target hypothesis that best represents the data. Lastly, \textit{RQ$_{3c}$} analyses the methods used to combat over- and under-fitting. Our intention with this RQ is to understand the specific methods (or lack thereof) used in SE research to combat over- or under-fitting, and the successes and shortcomings of such techniques.

\subsection{The Fifth Element of Learning: The Final Hypothesis}

Our fourth RQ addresses the component of learning known as the \textit{final hypothesis}, which is the target function learned by the model that is used to predict aspects of previously unseen data points. In essence, in order to investigate this component of learning in the context of SE applications, we examine the \textit{effectiveness} of the learned hypothesis as reported according to a variety of metrics across different SE tasks. Our intention with this analysis is to provide an indication of the advantages of certain data selection and processing pipelines, DL architectures, and training processes that have been successful for certain SE tasks in the past. Thus, our fourth RQ is formulated as follows:

\begin{tcolorbox}[enhanced,skin=enhancedmiddle,borderline={1mm}{0mm}{MidnightBlue}]
	\textit{\textbf{RQ$_4$:} How well do DL tasks perform in supporting various SE tasks?}
\end{tcolorbox}

\noindent Analyzing the effectiveness of DL applied to a wide range of SE tasks can be a difficult undertaking due to the variety of different metrics and evolving evaluation settings used in different contexts. Thus we examined two primary aspects of the literature as sub-RQs in order to provide a holistic illustration of DL effectiveness in SE research:

\begin{itemize}\setlength\itemsep{0.4em}
	\item{\textit{\textbf{RQ$_{4a}$:} What ``baseline'' techniques are used to evaluate DL models and what benchmarks are used for these comparisons?}}	
	\item{\textit{\textbf{RQ$_{4b}$:} How is the impact or automatization of DL approaches measured and in what way do these models promote generalizability?}}
\end{itemize}

\noindent Understanding the metrics used to quantify the comparison between DL approaches is important for informing future work regarding methods for best measuring the efficacy of newly proposed techniques. Thus, \textit{RQ$_{4a}$} explores trade-offs related to model complexity and accuracy. In essence, we examine applications of DL architectures through the lens of the \textit{Occam's Razor Principal}, which states that ``the least complex model that is able to learn the target function is the one that should be implemented''~\cite{NIPS2000_1925}.
We attempted to answer this overarching RQ by first delineating the baseline techniques that are used to evaluate new DL models and identifying what metrics are used in those comparisons. An evaluation that contains a comparison with a baseline approach, or even non-learning based solution, is important for determining the increased effectiveness of applying a new DL framework. \textit{RQ$_{4b}$} examines how DL-based approaches are impacting the automatization of SE tasks through measures of their effectiveness and in what ways these models generalize to practical scenarios, as generalizability of DL approaches in SE is vital for their usability. For instance, if a state-of-the-art DL approach is only applicable within a narrowly defined set of circumstances, then there may still be room for improvement.

\subsection{Analyzing Trends Across RQs}

Our last RQ encompasses all of the components of learning by examining the extent to which our analyzed body of literature properly accounts for and describes each element of learning. In essence, such an analysis explores the potential \textit{reproducibility} \& \textit{replicability} (or lack thereof) of DL applied to solve or automate various SE tasks. Therefore, our final RQ is formulated as follows:

\begin{tcolorbox}[enhanced,skin=enhancedmiddle,borderline={1mm}{0mm}{MidnightBlue}]
	\textit{\textbf{RQ$_5$:} What common factors contribute to the difficulty when reproducing or replicating DL4SE studies?}
\end{tcolorbox}

Our goal with this RQ is to identify common DL components which may be absent or underdescribed in our surveyed literature. In particular, we examined both the \textit{reproducibility} and \textit{replicability} of our primary studies as they relate to the sufficient presence or absence of descriptions of the elements of computational learning. Reproducibility is defined as the ability to take the exact same model with the exact same dataset from a primary study and produce the same results~\cite{acm-artifact-review}. Conversely, replicability is defined as the process of following the methodology described in the primary study such that a similar implementation can be generated and applied in the same or different contexts~\cite{acm-artifact-review}. The results of this RQ will assist the SE community in understanding what factors are being insufficiently described or omitted from approach descriptions, leading to difficulty in reproducing or replicating a given approach.

Lastly, given the analysis we perform as part of \textit{RQ$_5$} we derive a set of guidelines that both enumerate methods for properly applying DL techniques to SE tasks, and advocate for clear descriptions of the various different elements of learning. These guidelines start with the identification of the SE task to be studied and provide a step by step process through evaluating the new DL approach. Due to the high variability of DL approaches and the SE tasks they are applied to, we synthesized these steps to be flexible and generalizable. In addition, we provide checkpoints throughout this process that address common pitfalls or mistakes that future SE researchers can avoid when implementing these complex models. Our hope is that adhering to these guidelines will lead to future DL approaches in SE with an increased amount of clarity and replicability/reproducibility.

\subsection{Research Questions At-a-Glance}

We provide our full set of research questions below:
{\small
\begin{itemize}\setlength\itemsep{0.4em}
	
\item{\textit{\textbf{RQ$_1$: What types of SE tasks have been addressed by DL-based approaches?}}}

\item{\textit{\textbf{RQ$_2$: How are software artifacts being extracted, prepared, and used in DL-based approaches for SE tasks?}}}
	\begin{itemize}\setlength\itemsep{0.4em}
		\item{\textit{\textbf{RQ$_{2a}$:} What types of SE data are being used?}}
		\item{\textit{\textbf{RQ$_{2b}$:} How is this data being extracted and pre-processed into formats that are consumable by DL approaches?}}
		\item{\textit{\textbf{RQ$_{2c}$:} What type of exploratory data analysis is conducted to help inform model design and training?}}
	\end{itemize}

\item{\textit{\textbf{RQ$_3$: What deep learning models are used to support SE tasks?}}}
	\begin{itemize}\setlength\itemsep{0.4em}
		\item{\textit{\textbf{RQ$_{3a}$:} What types of model architectures are used to perform automated feature engineering of the data related to various SE tasks?}}
		\item{\textit{\textbf{RQ$_{3b}$:} What learning algorithms and training processes are used in order to optimize the models?}}
		\item{\textit{\textbf{RQ$_{3c}$:} What methods are employed to combat over- and under-fitting of the models?}}
	\end{itemize}

\item{\textit{\textbf{RQ$_4$: How well do DL tasks perform in supporting various SE tasks?}}}
	\begin{itemize}\setlength\itemsep{0.4em}
		\item{\textit{\textbf{RQ$_{4a}$:} What ``baseline'' techniques are used to evaluate DL models and what benchmarks are used for these comparisons?}}		
		\item{\textit{\textbf{RQ$_{4b}$:} How is the impact or automatization of DL approaches measured and in what way do these models promote generalizability?}}

	\end{itemize}

\item{\textit{\textbf{RQ$_5$: What common factors contribute to the difficulty when reproducing or replicating DL studies in SE?}}}

\end{itemize}}

\textbf{Note:} For this dissertation, I have decided to only include the results from the research questions that apply to this dissertation. Specifically, only RQ$_1$, RQ$_3$, and RQ$_4$.

\section{RQ$_1$: What types of SE tasks have been addressed by DL-based approaches?}
\label{dlslr:rq1}

\begin{figure*}[t]
	\centering
	\includegraphics[width=1\textwidth]{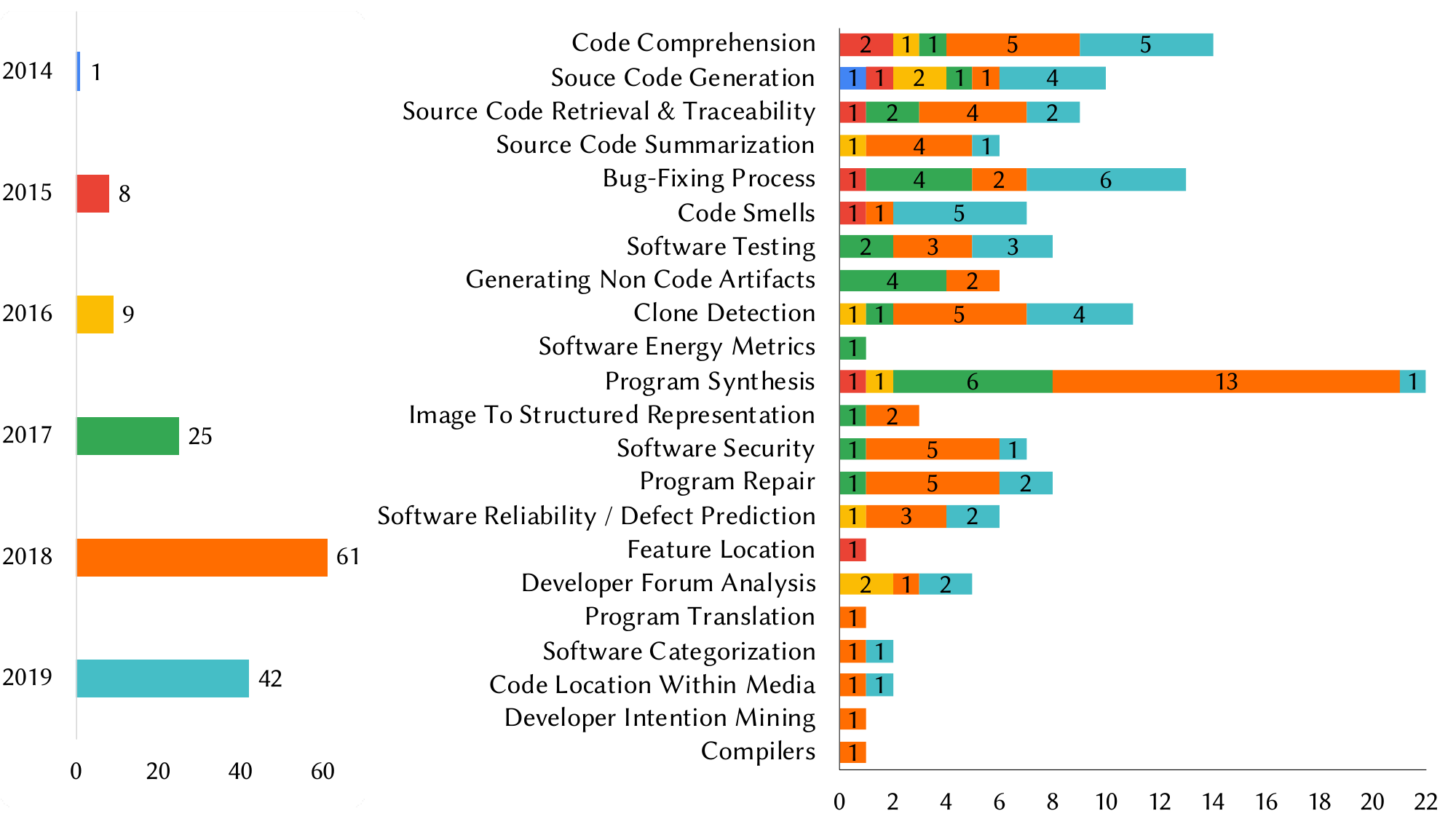}
	\caption{Papers published per year according to SE task. Note that a single paper can be associated with multiple SE Tasks.}	
	\label{fig:sebreakdown}
\end{figure*}

This RQ explores and quantifies the different applications of DL approaches to help improve or automate various SE tasks. Out of the \includedpapers papers we analyzed for this SLR, we identified \includedsetasks ~separate SE tasks where a DL-based approach had been applied. Figure \ref{fig:sebreakdown} provides a visual breakdown of how many SE tasks we found within these \includedpapers primary studies across a 10 year period. 
Unsurprisingly, there was very little work done between the years of 2009 and 2014. However, even after the popularization of DL techniques brought about by results achieved by approaches such as  AlexNet~\cite{NIPS2012_4824}, it took the SE community nearly $\approx3$ years to begin exploring the use of DL techniques for SE tasks. This also coincides with the offering and popularization of DL frameworks such as PyTorch and TensorFlow. The first SE tasks to use a DL technique were those of  \textit{Source Code Generation}, \textit{Code Comprehension}, \textit{Source Code Retrieval \& Traceability}, \textit{Bug-Fixing Processes}, and \textit{Feature Location}. Each of these tasks uses source code as their primary form of data. Source code served as a natural starting point for applications of DL techniques given the interest in large scale mining of open source software repositories in the research community, and relative availability of large-scale code datasets to researchers. Access to a large amount of data and a well-defined task is important for DL4SE, since in order for DL to have an effective application two main components are needed: i) a large-scale dataset of data to support the training of multi-parameter models capable of extracting complex representations and ii) a task that can be addressed with some type of predictable target. One of the major benefits of DL implementations is the ability for automatic feature extraction. However, this requires data associated with the predicted target variable.

It was not until 2017 that DL was used extensively in solving SE tasks as shown in Figure \ref{fig:sebreakdown}, with a large increase in the number of papers, more than doubling over the previous year from 9 to 25. During this period, the set of target SE tasks also grew to become more diverse, including tasks such as \textit{Code Smell Detection}, \textit{Software Security}, and \textit{Software Categorization}. However, there are three main SE tasks that have remained the most active across the years: \textit{Code Comprehension}, \textit{Source Code Retrieval \& Traceability}, and \textit{Program Synthesis}.
The most popular of the three being Program Synthesis, composing a total of 22 papers out of the \includedpapers we collected. We suspect that a variety of reasons contribute to the multiple applications of DL in program synthesis. First and foremost, is that the accessibility to data is more prevalent. Program synthesis is trained using a set of input-output examples. This makes for accessible, high quality training data, since one can train the DL model to generate a program, given a set of existing or curated specifications. The second largest reason is the clear mapping between training examples and a target programs. Given that it has proven difficult to engineer effective features that are capable to predict or infer programs, DL techniques are able to take advantage of the structured nature of this problem and extracting effective hierarchical representations. We display the full taxonomy in Table \ref{tab:setask}, which associates the cited primary study paired with its respective SE task.

\begin{table}[t]
\centering
\caption{SE Task Taxonomy}
\vspace{-0.3cm}
\label{tab:setask}
\footnotesize
\begin{tabular}{|c|c|}
\hline
\multicolumn{1}{|c|}{\textbf{SE Task}}                      	& \multicolumn{1}{c|}{\textbf{Papers}} \\ \hline
\multicolumn{1}{|c|}{Code Comprehension}                    	& \multicolumn{1}{c|}{\cite{Levy2017, BenNun2018, Piech2015, Hellendoorn2018, Le2018a, allamanis2018learning, Allamanis2015, codereviewlearn, 10.5555/3015812.3016002, DBLP:conf/iclr/YinNABG19, 10.1109/ICSE.2019.00084, zhang2019novel, 10.1145/3196321.3196334, guo:saner19}} \\ \hline
\multicolumn{1}{|c|}{Souce Code Generation}                 	& \multicolumn{1}{c|}{\cite{Karampatsis2019, Hellendoorn2017, White2015a, Gu2016, Chen2018, sun2019, 10.1145/2594291.2594321, gao:saner19, nguyen:saner19, cvitkovic:icml19}} \\ \hline
\multicolumn{1}{|c|}{Source Code Retrieval \& Traceability} 	& \multicolumn{1}{c|}{\cite{Chen2018a, Guo2017, Gu2018, Deshmukh2017, Lam2015, Chen2019, allamanis2018learning, 10.1145/3196398.3196408, xie:saner19}} \\ \hline
\multicolumn{1}{|c|}{Source Code Summarization}					& \multicolumn{1}{c|}{\cite{Chen2018a, Wan2018, Allamanis2016, tufano2019learning, 10.1145/3196398.3196408, 10.1145/3196321.3196334}} \\ \hline
\multicolumn{1}{|c|}{Bug-Fixing Process} 						& \multicolumn{1}{c|}{\cite{Lee2017, Murali2017, Deshmukh2017, Lam2015, gupta2019, gupta2017deepfix, DBLP:conf/iclr/YinNABG19, Liu2018d, tufano2019learning, 10.1145/3276517, 10.1145/3360588, huo:tse19, zhang:saner19}} \\ \hline
\multicolumn{1}{|c|}{Code Smells} 								& \multicolumn{1}{c|}{\cite{Liu2018, Allamanis2015, 8812134, tufano2019learning, 8811893, thaller:saner19, fakhoury:saner19}} \\ \hline
\multicolumn{1}{|c|}{Software Testing} 							& \multicolumn{1}{c|}{\cite{Liu2017, Godefroid2017, Cummins2018, Si2018, DBLP:conf/aaai/LiuLPW19, Zhang2018a, 8730177}} \\ \hline
\multicolumn{1}{|c|}{Non Code Related Software Artifacts} 		& \multicolumn{1}{c|}{\cite{Jiang2017, Schroeder2017, Choetkiertikul2018, Choetkiertikul2017, Choetkiertikul2019, Huang2018, 10.1145/3213846.3213876}} \\ \hline
\multicolumn{1}{|c|}{Clone Detection} 							& \multicolumn{1}{c|}{\cite{Li2017, White2016, Saini2018, Gao2018, Liu2018c, Tufano2018a, zhao2018deepsim, yu2019neural, zhang2019novel, perez:msr19, buch:saner19}} \\ \hline
\multicolumn{1}{|c|}{Software Energy Metrics} 					& \multicolumn{1}{c|}{\cite{Romansky2017}} \\ \hline
\multicolumn{1}{|c|}{Program Synthesis} 						& \multicolumn{1}{p{5.5cm}|}{\cite{Gaunt2017, Cai2017, Zhang2018, Devlin2017, Sun2018, DBLP:conf/iclr/MuraliQCJ18, DBLP:journals/corr/ReedF15, Liu2016, Balog2016, DBLP:conf/icml/DevlinUBSMK17, Bunel2018, Vijayakumar2018, DBLP:conf/iclr/ChenLS18, Hellendoorn2018a, Arabshahi2018, Zohar2018, Shin2018, Ellis2018, Ellis2018a, Liang2018, DBLP:conf/iclr/ParisottoMS0ZK17, Bavishi2019}} \\ \hline
\multicolumn{1}{|c|}{Image To Structured Representation} 		& \multicolumn{1}{c|}{\cite{10.5555/3305381.3305483, Chen2018, Moran2018}} \\ \hline
\multicolumn{1}{|c|}{Software Security} 						& \multicolumn{1}{c|}{\cite{Han2017, Zhao2018, Chen2018d, Gao2018, Harer2018, Dam2018, 8812083}} \\ \hline
\multicolumn{1}{|c|}{Program Repair} 							& \multicolumn{1}{c|}{\cite{Bhatia2018, Wang2017, Harer2018, tufano2019empirical, Liu2018d, gupta2019, gupta2017deepfix, white:saner19}} \\ \hline
\multicolumn{1}{|c|}{Software Reliability / Defect Prediction} 	& \multicolumn{1}{c|}{\cite{Wang2016, Wen2018, 8502853, dam:msr19, hoang:msr19, liu:saner19}} \\ \hline
\multicolumn{1}{|c|}{Feature Location} 							& \multicolumn{1}{c|}{\cite{Corley2015}} \\ \hline
\multicolumn{1}{|c|}{Developer Forum Analysis} 					& \multicolumn{1}{c|}{\cite{Chen2016, Xu2016, Lin2018, wang:msr19, guo:saner19}} \\ \hline
\multicolumn{1}{|c|}{Program Translation} 						& \multicolumn{1}{c|}{\cite{Chen2018e}} \\ \hline
\multicolumn{1}{|c|}{Software Categorization} 					& \multicolumn{1}{c|}{\cite{DBLP:conf/aaai/BuiJY18, bui:saner19}} \\ \hline
\multicolumn{1}{|c|}{Compilers} 				            	& \multicolumn{1}{c|}{\cite{katz:saner19}} \\ \hline
\multicolumn{1}{|c|}{Code Location Within Media} 				& \multicolumn{1}{c|}{\cite{Ott2018, Zhao:ICSE19}} \\ \hline
\multicolumn{1}{|c|}{Developer Intention Mining} 				& \multicolumn{1}{c|}{\cite{Huang2018}} \\ \hline
\multicolumn{1}{|c|}{Software Resource Control} 				& \multicolumn{1}{c|}{\cite{8811988, 10.1145/3213846.3213876}} \\ \hline
\end{tabular}%
\end{table}

\subsection{Results of Exploratory Data Analysis}

In performing our exploratory data analysis, we derived two primary findings. \revision{First, it is clear that SE researchers apply DL techniques to a diverse set of tasks, as 70\% of our derived SE task distribution was comprised of distinct topics that were evenly distributed ($\approx$ 3-5\%). Our second finding is that the SE task was the most informative feature we extracted ($\approx 4.04B$), meaning that it provides the highest level of discriminatory power in predicting the other features (\eg elements of learning) related to a given study. In particular, we found that SE tasks had strong correlations to data (1.51B), the loss function used (1.14B) and the architecture employed (1.11B). This suggests that there are DL framework components that are better suited to address specific SE tasks, as authors clearly chose to implement certain combinations of DL techniques associated with different SE tasks.}
For example, we found that SE tasks such as program synthesis, source code generation and program repair were highly correlated with the preprocessing technique of tokenization. Additionally, we discovered that the SE tasks of source code retrieval and source code traceability were highly correlated with the preprocessing technique of neural embeddings. When we analyzed the type of architecture employed, we found that code comprehension, prediction of software repository metadata, and program repair were highly correlated with both recurrent neural networks and encoder-decoder models. When discussing some of the less popular architectures we found that clone detection was highly correlated with siamese deep learning models and security related tasks were highly correlated with deep reinforcement learning models.
Throughout the remaining RQs, we look to expand upon the associations we find to better assist software engineers in choosing the most appropriate DL components to build their approach. 

\subsection{Opportunities for Future Work}

Although the applications of DL-based approaches to SE related tasks is apparent, there are many research areas of interest in the SE community as shown in ICSE'20's topics of interest\footnote{https://conf.researchr.org/track/icse-2020/icse-2020-papers\#Call-for-Papers} that DL has not been used for. Many of these topics have no \textit{readily apparent} applicability for a DL-based solution. Still, some potentially interesting topics that seem well suited or positioned to benefit from DL-based techniques have yet to be explored by the research community or are underrepresented. Topics of this unexplored nature include software performance, program analysis, cloud computing, human aspects of SE, parallel programming,  feature location, defect prediction and many others. Some possible reasons certain SE tasks have yet to gain traction in DL-related research is likely due to the following:

\begin{itemize}
    \item There is a lack of available, ``clean'' data in order to support such DL techniques;
    \item The problem itself is not well-defined, such that a DL model would struggle to be effectively trained and optimized;
    \item No current architectures are adequately fit to be used on the available data.
\end{itemize}

We believe that one possible research interest could be in the application of new DL models toward commonly explored SE tasks. For example, a DL model that is gaining popularity is the use of transformers, such as BERT, to represent sequential data \cite{devlin2018bert}. It is possible that models such as this could be applied to topics related to clone detection and program repair. There is sufficient exploration in the use of DL within these topics to determine if these new architectures would be able to create a more meaningful representation of the data when compared to their predecessors.

\mybox{\textbf{Summary of Results for RQ$_{1}$}:}{gray!60}{gray!20}{Researchers have applied DL techniques to a diverse set of tasks, wherein \textit{program synthesis}, \textit{code comprehension}, and \textit{source code generation} are the most prevalent. The SE task targeted by a given study is typically a strong indicator of the other details regarding the other components of learning, suggesting that certain SE tasks are better suited to certain combinations of these components. Our associative rule learning analysis showed a strong correlation amongst SE task, data type, preprocessing techniques, loss function used and DL architecture implemented, indicating that the SE task is a strong signifier of what other details about the approach are present. While there has been a recent wealth of work on DL4SE, there are still underrepresented topics that should be considered by the research community, including different topics in software testing and program analysis.}

\section{RQ$_3$: What Deep Learning Models are Used to Support SE Tasks?}
\label{dlslr:rq3}

In RQ$_2$ we investigated how different types of SE data were used, preprocessed, and analyzed for use in DL techniques. In this section, we shift our focus to the two key components of DL models: the \textit{architecture} and the \textit{learning algorithm}. The type of architecture selected for use in a DL application reveals key aspects of the types of features that researchers hope to model for a given SE task. Thus, we aim to empirically determine if certain architectures pair with specific SE tasks. Additionally, we aim to explore the diversity of the types of architectures used across different SE tasks and whether or not idiosyncrasies between architectures might be important when considering the specific SE task at hand. We also examined how various architectures are used in conjunction with different learning or optimization algorithms. Specifically, we aimed to create a taxonomy of different learning algorithms and determine if there was a correlation between the DL architectures, the learning algorithms and the SE tasks. 

\subsection{\textit{RQ$_{3A}$}: What types of model architectures are used to perform automated feature engineering of the data related to various SE tasks?} 
\label{rq3a}

 \begin{figure*}[t]
	\centering
	\includegraphics[width=\columnwidth]{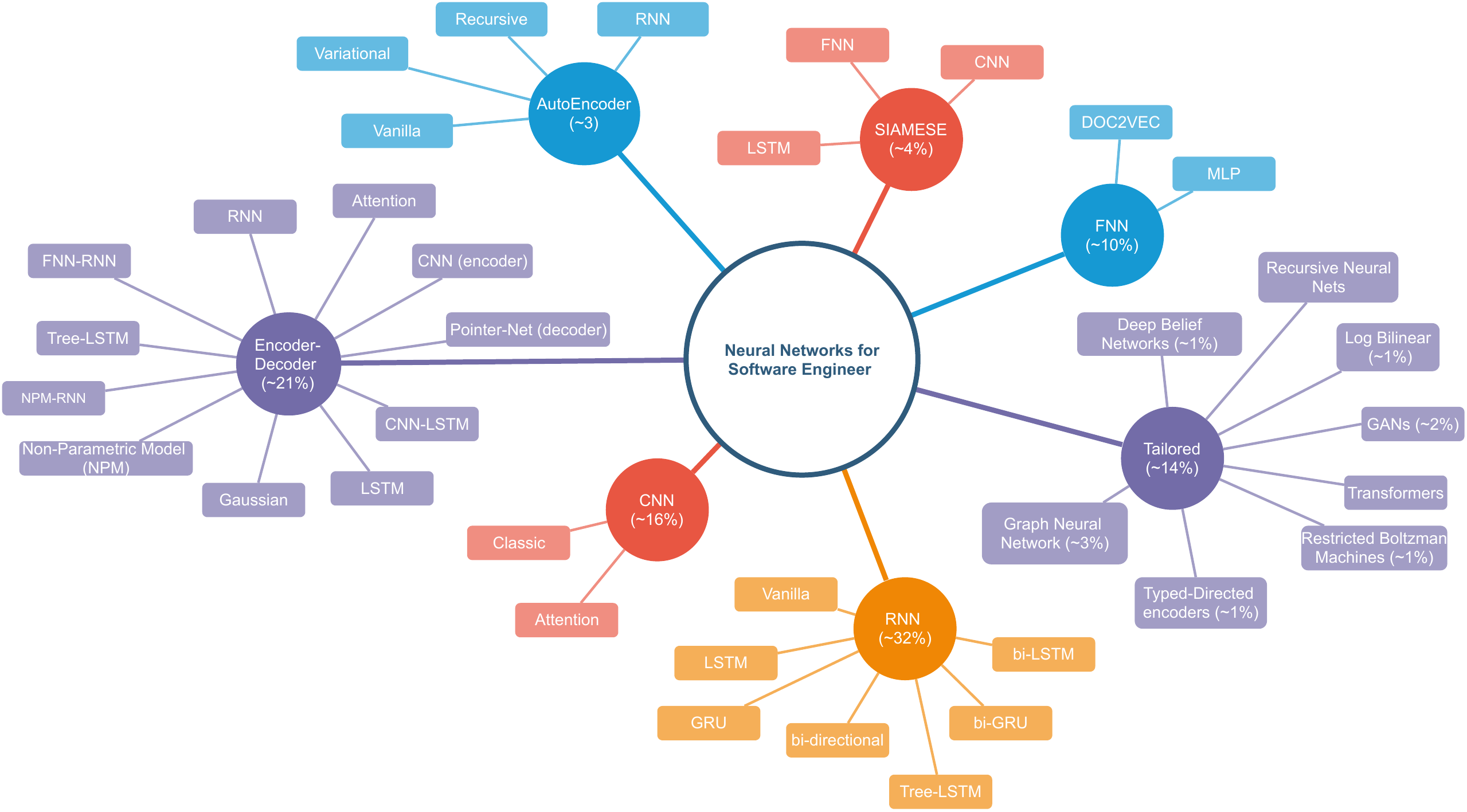}
	\caption{DL Model Taxonomy \& Type Distribution}
	\label{fig:dlmodel}
\end{figure*} 

In this section, we discuss the different types of DL models software engineers are using to address SE tasks. \revision{Figure~\ref{fig:dlmodel} illustrates the various different DL architecture types that we extracted from our selected studies. We observe seven major architecture types: \textit{Recurrent Neural Networks} (\rnncount), \textit{Encoder-Decoder Models} (\encdeccount), \textit{Convolutional Neural Networks} (\cnncount), \textit{Feed-Forward Neural Networks (FNNs)} (\fnncount), \textit{AutoEncoders} (\autocount), \textit{Siamese Neural Networks} (\siamcount), as well as a subset of other custom, highly tailored architectures.} We observe an additional level of diversity within each of these different types of architectures with Encoder-Decoder models illustrating the most diversity, followed by RNNs and the tailored techniques. The diversity of Encoder-Decoder models is expected, as this type of model is, in essence, a combination of two distinct model types, and is therefore extensible to a range of different combinations and hence architectural variations. The variance in RNNs is also logical. RNNs excel in modeling sequential data since the architecture is formulated such that a weight matrix is responsible for representing the features between the sequence of inputs \cite{Goodfellow2016}, making them suitable to source code. Given that one of the most popular SE data types is source code which is inherently sequential data, the varied application of RNNS is expected. We also observe a number of architectures, such as Graph Neural Networks, that are specifically tailored for given SE tasks. For instances, graph-based neural networks have been adapted to better model the complex \textit{structural} relationships between code entities.

\begin{figure*}[t]
	\centering
	\includegraphics[width=\columnwidth]{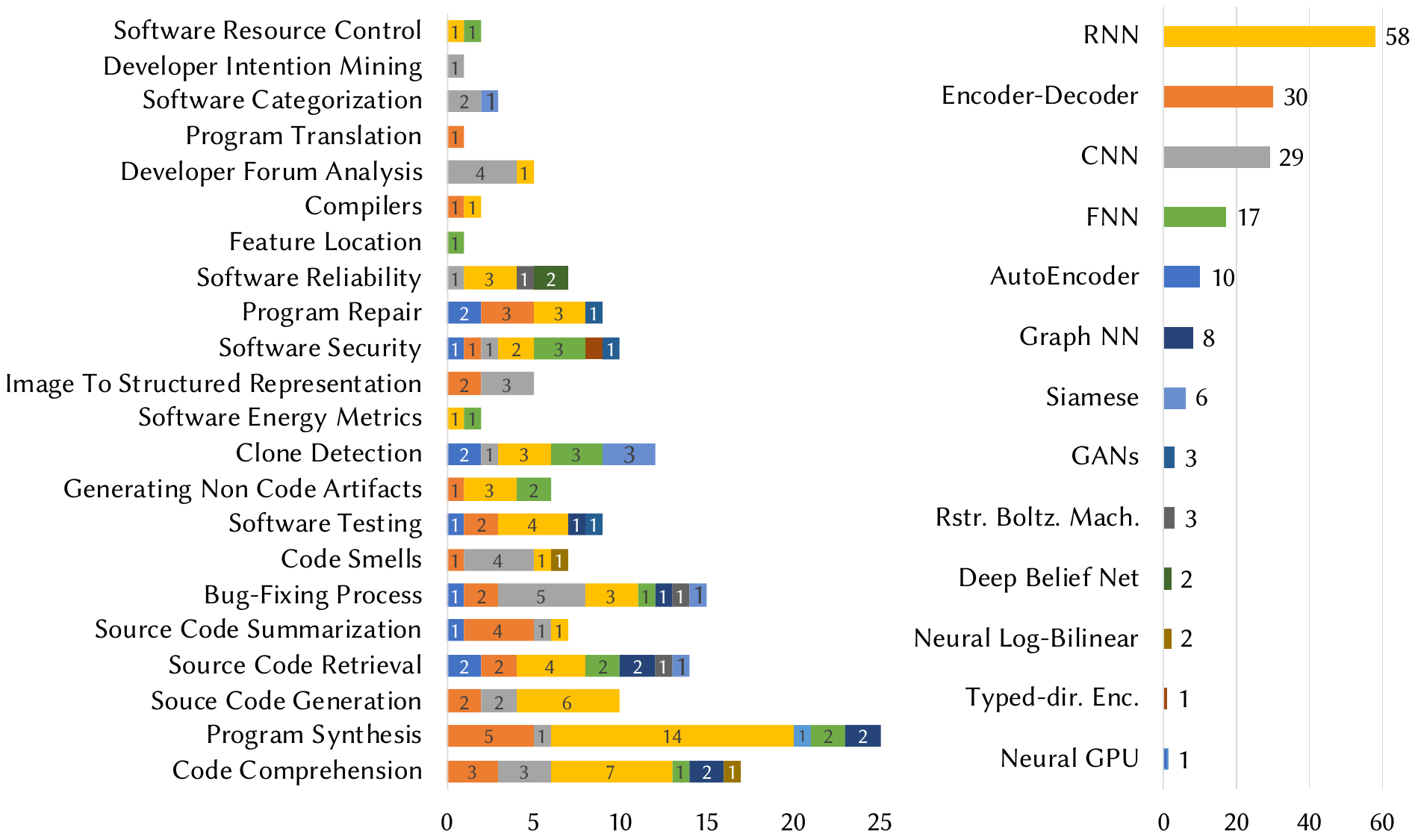}
	\caption{DL Architectures by the Task}	
	\label{dlslr:fig:architecture}
\end{figure*} 

Figure~\ref{dlslr:fig:architecture} delineates the prevalence of various different types of architectures according to the SE tasks to which they are applied. The popularity of our identified techniques closely mirrors their diversity. Examining this data, we find that RNNs are the most prevalent architectures, followed by Encoder-Decoder models, CNNs, and FNNs. The prevalence of RNNs is not surprising given the prevalence of source code as a utilized data type, as discussed above.  The flexibility of Encoder-Decoder models is also expected as they excel at understanding and ``translating'' between parallel sets of sequential data, which is a common scenario in SE data (\eg code and natural language). 
The encoder's responsibility is to translate the raw data into a latent representation that the decoder is capable of understanding and decoding into the target. 
Therefore, since neural embeddings were such a popular preprocessing technique for data formatting and preparation, it aligns with the high prevalence of the encoder-decoder DL architecture. CNNs serve as the most popular architectures for processing visual data, such as images, and hence are popular for visual SE data.

In addition to the prevalence, we observed certain trends between the DL architecture utilized and the corresponding SE task, as illustrated in Figure~\ref{dlslr:fig:architecture}. 
As expected, most of the SE tasks having to do with source code generation, analysis, synthesis, traceability, and repair make use of RNNs and encoder-decoder models.  
Likewise, SE tasks involving the use of images or media data have CNNs commonly applied to them.

We also observed some pertinent trends related to some of the less popular types of DL architectures, including: siamese networks, deep belief networks, Graph Neural Networks (GNNs) and auto-encoders. While these architectures have only been applied to a few taszks it is important to note that they have only recently gained prominence and become accessible outside of ML/DL research communities. 
It is possible that such architectures can highlight orthogonal features of SE data that other architectures may struggle to observe. For example, the use of GNNs may better capture the structure or control flow of code or possibly the transition to different mobile screens within a mobile application. There may also be an opportunity for the use of Siamese networks in software categorization, as they have been shown to classify data into unique classes accurately based only on a few examples \cite{Saini2018}.
One notable absence from our identified architecture types is \textit{deep reinforcement learning}, signaling its relative lack of adoption within the SE community. Deep reinforcement learning excels at modeling decision-making tasks. One could argue that deep reinforcement learning is highly applicable to a range of SE tasks that can be modeled as decisions frequently made by developers. This is a fairly open area of DL in SE that has not been sufficiently explored. The only type of SE task that had an application of Reinforcement Learning was related to program verification. In this paper the authors propose an approach that constructs the structural external memory representation of a program. They then train the approach to make multi-step decisions with an autoregressive model, querying the external memory using an attention mechanism. Then, the decision at each step generates subparts of the loop invariant~\cite{Si2018}. 

In addition to the discussion around the DL architectures and their relations to particular SE tasks, it is also important to understand trends related to the \textit{explicit} and \textit{implicit} features extracted from these different architectures. As we discuss in Section for (RQ$_{2B}$)\footnote{Section omitted for dissertation}, it is common for data to be fed into DL models only after being subjected to certain preprocessing steps. However, in supervised learning, once that data has been preprocessed, the DL model automatically extracts implicit features from the preprocessed data in order to associate those features with a label or classification. In unsupervised learning, the model extracts implicit features from the preprocessed data and groups similar datum together as a form of classification. We refer to the preprocessing steps as highlighting certain explicit features, since these steps frequently perform dimensionality reduction while maintaining important features. In our analysis we found the most common techniques for highlighting explicit features to be tokenization, abstraction, neural embeddings and vectorizing latent representations. These techniques attempt to highlight explicit features that are uniquely tailored to the data being analyzed. Once the data is fed into the model itself, the model is responsible for extracting implicit features to learn a relationship between the input data and target function. 
The extraction of explicit and implicit features dramatically impacts a DL model's ability to represent a target function, which can be used to predict unobserved data points. 

 \begin{figure*}[t]
	\centering
	\includegraphics[width=\columnwidth]{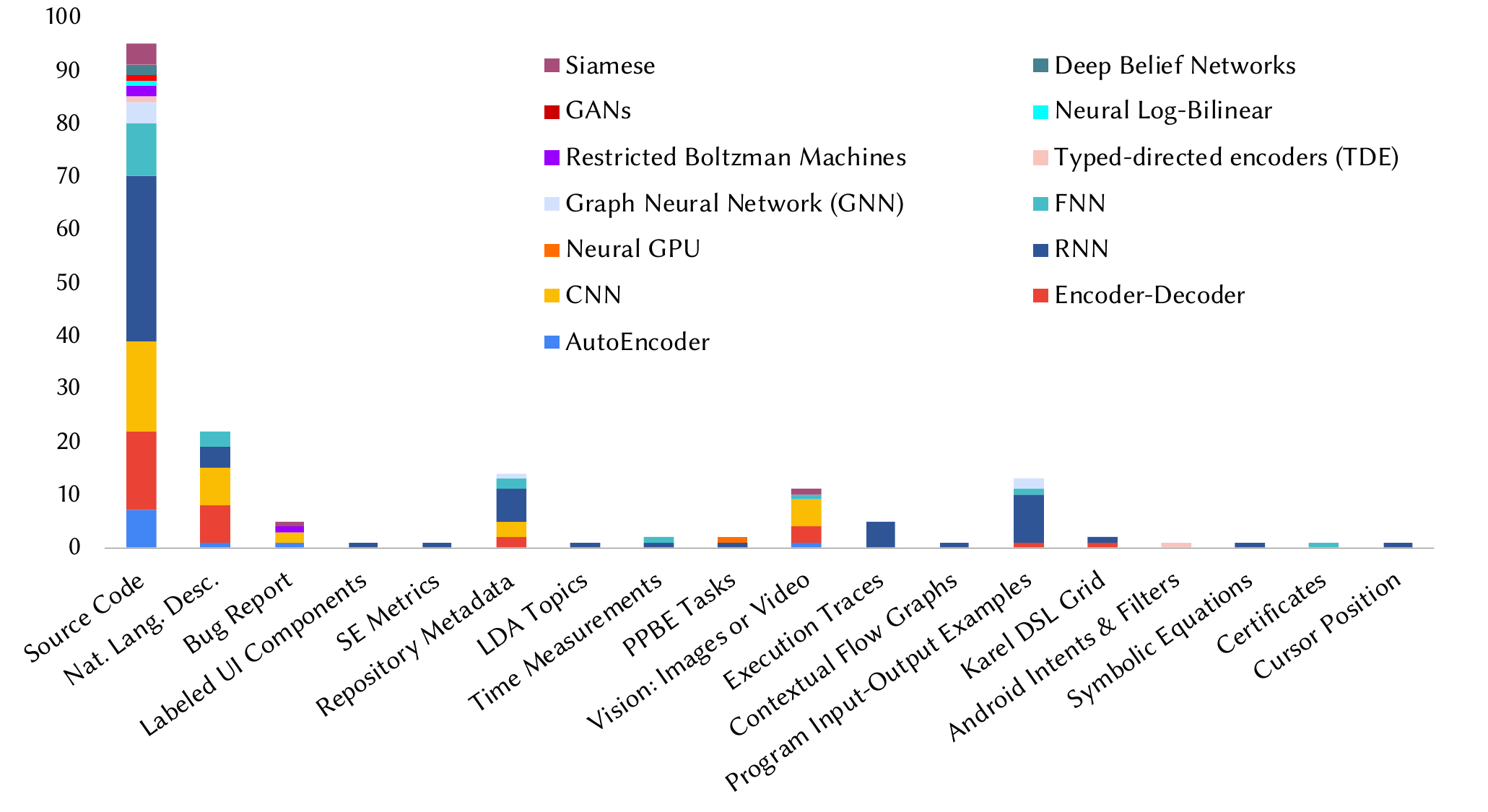}
	\caption{DL Architectures by Data Type}	
	\label{fig:archData}
\end{figure*}

Figure \ref{fig:archData} shows a breakdown of DL architectures by the type of data to which they are applied. This relationship between data and architecture is important since the architecture is partially responsible for the type of implicit features being extracted. For example, images and other visual data are commonly represented with a CNN. This is because CNNs are particularly proficient at modeling the spatial relationships of pixel-based data. We also discovered a strong correlation between RNNs and sequential data such as source code, natural language and program input-output examples. This correlation is expected due to RNNs capturing implicit features relating to the sequential nature of data. The models are able to capture temporal dependencies between text and source code tokens. Another correlation we observed was the use of CNNs for visual data or data which requires dimensionality reduction. This included the data types of images, videos, and even natural language and source code. CNNs have the ability to reduce features within long sequential data which makes them useful for tasks involving sentiment analysis or summarization. We also observed less popular combinations such as the use of deep belief networks (DBNs) for defect prediction~\cite{Wang2016}. Here, a DBN is used to learn semantic features of token vectors from a program's AST graph to better predict defects. A DBN can be a useful architecture in this situation due to its ability to extract the necessary semantic features from a tokenized vector of source code tokens. Those features are then used within their prediction models to drastically increase performance. 

\subsubsection{Results of Exploratory Data Analysis}

In our exploratory data analysis, we found that SE tasks greatly influence the architecture adopted in an approach. \revision{The mutual information value between the features of a SE task and a DL architecture is $1.11B$.} We also note that the SE research landscape has primarily focused on SE tasks that consist primarily of text-based data, including source code. \revision{This helps to explain why RNNs are used in \rnncount~of the papers analyzed in this SLR. The encoder-decoder architecture was also seen frequently (\encdeccount~of papers), which generally makes use of RNNs.}

\subsubsection{Opportunities for Future Work} We were able to correlate different DL architectures with particular SE tasks and data types, primarily due to the fact that a given architecture is typically suited for a specific type of implicit feature engineering. However, there exists a fundamental problem in the ability of current research to validate and quantify these implicit features the model is extracting. This leads to decreased transparency in DL models, which in turn, can impact their practical applicability and deployment for real problems. Thus, there exists an open research problem related to being able to explain how a given model was capable of predicting the target function, \textit{specifically} as it relates to SE data \cite{yuan2021explainability,yang2019enhancing,ANGELOV2020185,gilpin2019explaining,xie2020explainable}. While interpretability is a broader issue for the DL community, insights into implicit feature engineering specifically for SE data would be beneficial for DL4SE work. It is necessary for developers to understand what complex hierarchical features are used by the model for this prediction. This could demystify their ability to correctly predict the output for a given input datum.

The ability to increase the interpretability of DL4SE solution also contributes toward the novel field of SE4DL, where SE methods are applied to the creation, sustainability and maintenance of DL software. The ability to interpret DL based solutions could help to create more complete testing suites for DL based software. This paradigm becomes even more important as new and innovative DL architectures are being developed. The SE community could take inspiration from the recent success in the NLP community on developing benchmarks for explainability~\cite{ribeiro:2020}. Peeling back the "black box" nature of DL models should allow for an analysis on the integrity of the learning algorithms and an ability to better understand and build usable tools around their predictions. 

\mybox{\textbf{Summary of Results for RQ$_{3A}$}:}{gray!60}{gray!20}{\revision{Our analysis revealed seven major types of DL architectures that have been used in work on DL4SE including: \textit{Recurrent Neural Networks} (\rnncount), \textit{Encoder-Decoder Models} (\encdeccount), \textit{Convolutional Neural Networks} (\cnncount), \textit{Feed-Forward Neural Networks} (\fnncount), \textit{AutoEncoders} (\autocount), \textit{Siamese Neural Networks} (\siamcount), as well as a subset of other custom, highly tailored architectures.} RNNs and Encoder-Decoder models were both the most prevalent architecture used in our surveyed studies and the most diverse in terms of their varying configurations. We also discovered strong correlations between particular DL architectures to data types. For example, we found that architectures capable of capturing temporal differences within sequential data are used to study source code, natural language, repository metadata and program input-output examples. Likewise, architectures capable of capturing spatial and structural features from data have been used to study images, bug reports and program structures (ASTs, CFGs, etc.). }

\subsection{\textit{RQ$_{3B}$}: What learning algorithms and training processes are used in order to optimize the models?}
\label{rq3b}

In addition to the variety of DL models that can be used within a DL-based approach, the way in which the model is trained can also vary. To answer RQ$_{3B}$ we aimed to analyze the learning algorithms used in three primary ways: according to (i) the manner in which the weights of the model are updated, (ii) the overall error calculation, and (iii) by the optimization algorithm, which governs the parameters of the learning algorithm as training progresses. Learning algorithms that have been defined in ML/DL research are typically used in an ``off-the-shelf'' manner, without any alteration or adjustment, in the context of SE research. This is likely a result of researchers in SE being primarily interested in DL applications, rather than the intricacies of learning algorithms. 

In terms of the process for adjusting weights, the most prevalent technique employed among our analyzed studies was the incorporation of the gradient descent algorithm. The breakdown of learning algorithms throughout our SLR are as follows: \revision{We found $\approx76\%$ of the primary studies used some version of gradient descent to train their DL model. The remaining studies used gradient ascent $\approx2\%$, or policy based learning $\approx2\%$. Other studies did not explicitly specify their learning algorithm in the paper $\approx18\%$. Our exploratory data analysis revealed that papers published in recent years (2018 and 2019) have begun to employ learning algorithms that differ from gradient descent, such as reward policies or gradient ascent.} 

Our analysis reveled that there are a variety of ways that DL-based implementations calculate error. \revision{However, we did find that a majority of the papers we analyzed used cross entropy as their loss function \entropycount, which was most commonly paired with gradient descent algorithms. Other common loss functions that were used with gradient descent algorithms were negative log likelihood (\neglogcount), maximum log likelihood (\maxlogcount), and cosine loss (\cosinelog). There were a number of papers which did not provide any indication about the loss function within their learning algorithm ($\approx42\%$).} We did find that when the primary study was not using gradient descent as a way to adjust the weights associated with the DL model, the error functions used became a lot more diverse. For example, the work done by Ellis \etal learned to infer graphical programs from deep learning hand-drawn images. They used gradient ascent rather than descent as their learning algorithm and also used surrogate likelihood function as a way to calculate the error of the model \cite{Ellis2018a}. We found that approaches that implement reinforcement algorithms are based on a developed policy, which calculates the error associated with the action taken by the model and adjusts the weights.

Lastly, we examined the use of optimization algorithms to determine if there were any relevant patterns. We discovered that the choice of optimization algorithm is somewhat agnostic to the model, the weight adjustment algorithm and the error function. In many cases, the optimization algorithm was not reported within the primary study ($\approx53\%$ of the time). 
\revision{However, we did analyze the papers that provided this information and identified four major optimization algorithms: Adagrad (\adagradcount) , AdaDelta (\adadeltacount), RMSprop (\rmspropcount), and Adam (\adamcount).} Below, we briefly address each optimization algorithm in order to point out potential situations in which they should be used.

\emph{Adagrad} is an algorithm that adapts the learning rate based on the impact that the parameters have on classification. When a particular parameter is frequently involved in classification across multiple inputs, the amount of adjustment to those parameters is lower. Likewise, when the parameter is only associated with infrequent features, then the adjustment to that parameter is relatively high~\cite{Duchi2011}. A benefit of AdaGrad is that it removes the need for manual adjustment of the learning rates. However, the technique that AdaGrad calculates the degree by which it should adjust the parameters is using an accumulation the sum of the squared gradients. This can lead to summations of the gradient that are too large, often requiring an extremely small learning rate. 

\emph{AdaDelta} was formulated out of AdaGrad in order to combat the gradient size problem. Rather than consider all the sums of the past squared gradients, AdaDelta only considers the sum of the past squared gradients limited to a fixed size. Additionally, this optimization algorithm does not require a default learning rate as it is defined by an exponentially decaying average of the calculated squared gradients up to a fixed size \cite{Zeiler2012}. 

\emph{RMSprop} is the next optimization algorithm, however, this algorithm has not been published or subjected to peer review. This algorithm was developed by Hinton \etal and follows the similar logic of AdaDelta. The way in which RMSprop battles the diminishing learning rates that AdaGrad generates is by dividing the learning rate by the recent average of the squared gradients. The only difference is that AdaDelta uses the root means squared error in the numerator as a factor that contributes to the adjustment of the learning rate where RMSprop does not.

\emph{Adam}, the last of our optimization algorithms discussed, also calculates and uses the exponentially decaying average of past squared gradients similar to AdaDelta and RMSprop. However, the optimization algorithm also calculates the exponentially decaying average of the past gradients. Keeping this average dependent on gradients rather than just the squared gradients allows Adam to introduce a term which mimics the momentum of how the learning rate is moving. It can increase the rate at which the learning rate is optimized \cite{Kingma2014}. 

\subsubsection{Results of Exploratory Data Analysis}

\revision{We found that the loss function is correlated to the chosen technique to combat overfitting with a mutual dependence of $1.00B$. However, the SE community omits reporting the loss function in $\approx33\%$ of the papers we analyzed. Additionally, the loss function is correlated to SE task with a mutual dependence of $1.14B$}

\subsubsection{Opportunities for Future Work} A consequential highlight of our analysis of employed learning algorithms was the lack of data available from the primary studies. However, we did find a strong correlation between certain loss functions paired to specific learning algorithms. One aspect we believe could provide vital insight into the DL process is an analysis regarding how learning algorithms affect the parameters of the model for certain types of data. It would not only be important to study the type of data that learning algorithms and loss functions are associated with, but also what preprocessing techniques influence the learning algorithms and loss functions chosen. It is possible that some loss functions and learning algorithms are more efficient when applied to data that has been subjected to a particular preprocessing technique. Finding the optimal pairing of loss function and learning algorithm for an architecture/data pair remains an open problem. 

\mybox{\textbf{Summary of Results for RQ$_{3B}$}:}{gray!60}{gray!20}{Our analysis revealed four different techniques for updating the weights of the DL models, with the large majority making use of gradient descent. \revision{We found four major techniques that were utilized for calculating error, including cross entropy \entropycount, negative log likelihood \neglogcount, maximum log likelihood \maxlogcount, and cosine loss \cosinelog -- with cross entropy being the most prevalent. Finally, we observed the use of four major optimization algorithms, including Adagrad (\adagradcount) , AdaDelta (\adadeltacount), RMSprop (\rmspropcount), and Adam (\adamcount).}}

\subsection{\textit{RQ$_{3C}$}: What methods are employed to combat over- and under-fitting?}
\label{rq3c}

Two potential problems associated with the use of any type of learning based approach, whether that be canonical machine learning or deep learning, are \textit{overfitting} and \textit{underfitting}. Both of these issues are related to the notion of generalization, \ie how well does a trained ML/DL model perform on unseen data. Overfitting is the process of a model learning to fit the training data extremely well, yet not being able to generalize to unseen data, and hence is a poor approximation of the actual target function to be learned~\cite{Tetko1995NeuralNS}. 
Underfitting is typically described as the scenario in which a given model incurs a high error rate on a training set. This can occur when the model lacks the necessary complexity, is overly constrained, or has not had the sufficient training iterations to appropriately approximate the target function.   For RQ$_{3C}$, we are primarily interested in the specific methods employed by researchers to combat these two problems in the context of SE tasks.

\begin{figure*}[t]
	\centering
	\includegraphics[width=0.75\columnwidth]{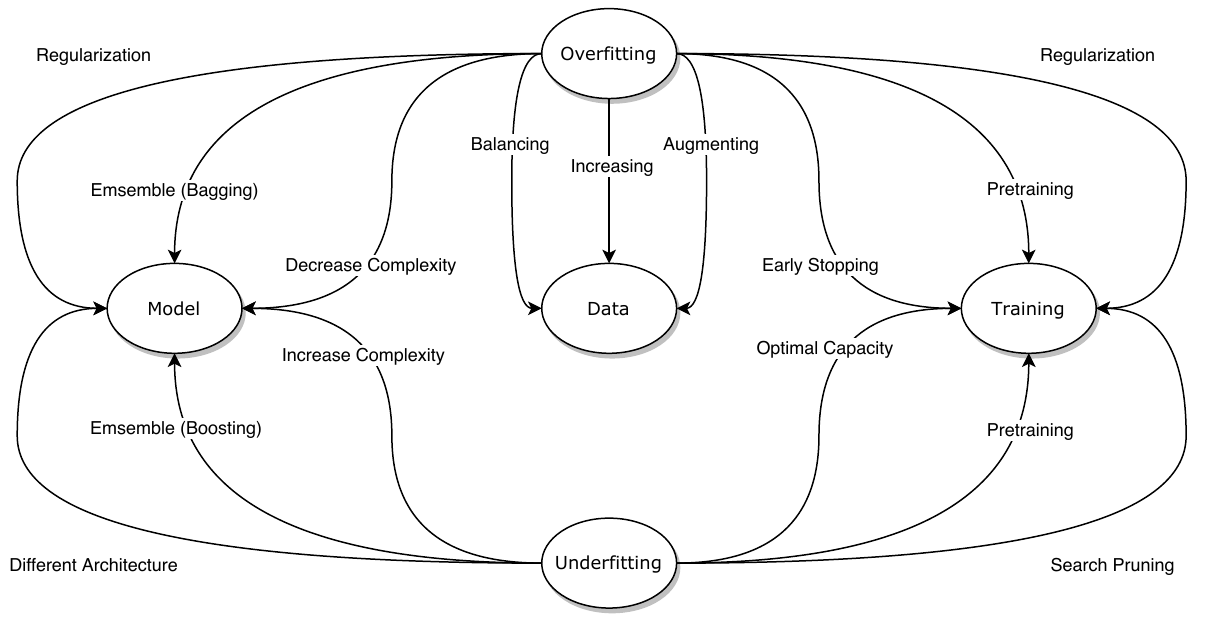}
	\caption{Overfitting and Underfitting Overview}	
	\label{fig:over_under_fit_overview}
\end{figure*}

Figure \ref{fig:over_under_fit_overview} provides an overview of some general methods used to combat overfitting and underfitting\footnote{Generated through an analysis of the following sources: \\ \tiny{\url{https://elitedatascience.com/overfitting-in-machine-learning}}, \\ \tiny{\url{https://hackernoon.com/memorizing-is-not-learning-6-tricks-to-prevent-overfitting-in-machine-learning-820b091dc42}}, \\ \tiny{\url{https://towardsdatascience.com/dont-overfit-how-to-prevent-overfitting-in-your-deep-learning-models-63274e552323}}, \\ \tiny{\url{https://elitedatascience.com/bias-variance-tradeoff}}}. The figure also addresses what parts of an ML/DL approach are affected by these techniques. As illustrated, there are three main types of regularization. The first regularizes the model, which includes things such as adding Dropout layers \cite{JMLR:v15:srivastava14a} or Batch Normalization \cite{DBLP:journals/corr/IoffeS15}. The second regularizes the data itself, either through adding more data or cleaning the data already extracted. The third type of regularization is applied to the training process, which modifies the loss function with L1 regularization, L2 regularization or incorporates early stop training.

As outlined in \cite{abu-mastafa}, the use of a validation set is a commonly used method for detecting if a model is overfitting or underfitting to the data, which is why it is very common to split data into training, validation and evaluation sets. The splitting of data helps to ensure that the model is capable of classifying unseen data points. This can be done in parallel with a training procedure, to ensure that overfitting is not occurring. We see cross-validation in $\approx11\%$ papers we analyzed. However, other potentially more effective techniques were seen less frequently. 

\begin{figure*}[t]
	\centering
	\includegraphics[width=0.90\textwidth]{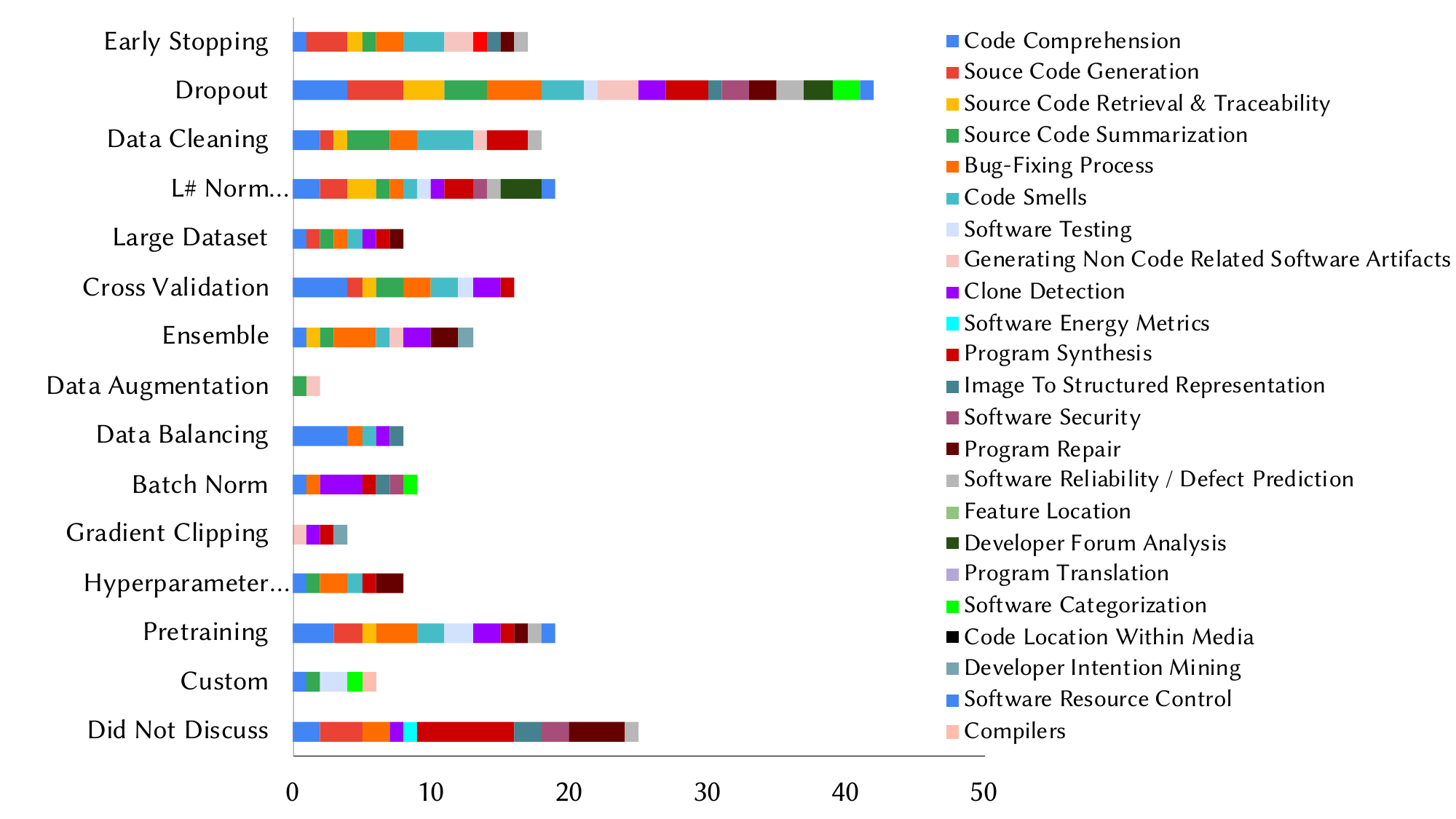}
	\caption{Overfitting Techniques per Task Type}	
	\label{fig:overfit_task}
\end{figure*}

We aimed to determine if a given SE task had any relationship with the methods employed to prevent over/under-fitting. Figure \ref{fig:overfit_task} analyzes the relationship between DL approaches and the techniques that combat overfitting. This figure shows that there are some techniques that are much more commonly applied to SE tasks than others. \revision{For example, \textit{dropout} (\dropoutcount) was the most commonly used regularization technique and is used in a variety of DL approaches that address different SE tasks, followed by \textit{data cleaning} (\cleancount), \textit{L1/L2 regularization} (\regcount), and \textit{early stopping} (\earlycount).} 
Dropout is one of the most popular regularization techniques because of its effectiveness and ease of implementation. Dropout randomly blocks signals from a node within a layer of the neural network with a certain probability determined by the researcher. This ensures that a single node doesn't overwhelmingly determine the classification of a given data point. We also observed a number of custom methods that were employed. These methods are configured to address the specific neural network architecture or data type being used. For example, in Sun et al. \cite{Sun2018}, they encourage diversity in the behavior of generated programs by giving a higher sampling rate to the perception primitives that have higher entropy over $K$ different initial states. In Delvin et al. \cite{Devlin2017} they perform multiple techniques to combat overfitting which include the even sampling of the dataset during training and ensuring that each I/O grid of every example is unique.  
In addition to the aforementioned techniques, we found a subset of more unique approaches including the use of deep reinforcement learning instead of supervised learning \cite{Wan2018}, gradient clipping, lifelong learning \cite{Gaunt2017}, modification of the loss function \cite{Bunel2018}, pretraining \cite{Wan2018, Si2018}, and ensemble modeling~\cite{Jiang2017}.

We also analyzed the relationships between techniques to combat over/under-fitting, and the underlying data type that a given model operated upon. We observed similar patterns in that there are a variety of techniques to combat overfitting regardless of the data type. The only exception to this pattern was seen when analyzing natural language, where L1/L2 regularization was predominately used.  Figure \ref{fig:overfit_task} illustrates that the techniques used to combat overfitting do not have a strong association with the SE task. Therefore, we observe that a range of different techniques are applicable across many different contexts. 

\revision{One of the more concerning trends that we observed is the number of papers categorized into the \textit{Did Not Discuss} (\naoverfitcount) category.} Given the importance of combating overfitting when applying a DL approach, it is troublesome that so many primary studies did not mention these techniques. We hope that our observation of this trend signals the importance of recording such information.

Combating underfitting is a more complex process, as there aren't a well-defined set of standard techniques that are typically applied. 
One method that can be used to combat underfitting is searching for the optimal capacity of a given model. The optimal capacity is the inflection point where the model starts to overfit to the training data and performs worse on the unseen validation set. One technique for achieving this optimal capacity include maximizing training time while monitoring performance on validation data. 
Other techniques include the use of a more complex model or a model better suited for the target function, which can be determined by varying the number of neurons, varying the number of layers, using different DL architectures, pretraining the model, and pruning the search space. From our SLR, the most commonly used underfitting techniques applied were pruning the search space of the model \cite{Vijayakumar2018, DBLP:conf/iclr/ChenLS18}, curriculum training \cite{Zhang2018, DBLP:journals/corr/ReedF15, DBLP:conf/iclr/ChenLS18} and pretraining \cite{Wan2018, Si2018}. We found that only 6/\includedpapers primary studies explicitly stated the implementation of an underfitting technique. This is a stark contrast to the number of studies implementing an overfitting technique, 97/\includedpapers.

Surprisingly, more than \naoverfitcount \hspace{.001cm} of our studied papers did not discuss any techniques used to combat overfitting or underfitting. Combating this issue is a delicate balancing act, as attempting to prevent one can begin to cause the other if the processes are not carefully considered. For example, having a heavily regularized learning model to prevent overfitting to a noisy dataset can lead to an inability to learn the target function, thus causing underfitting of the model. This is also possible while attempting to prevent underfitting. An increase in the number of parameters within the architecture to increase the complexity of the model can cause the model to learn a target function that is too specific to the noise of the training data. Therefore, the incorporation of techniques to address over- and under-fitting is crucial to the generalizability of the DL approach. 

\subsubsection{Opportunities for Future Research}

Given the relative lack of discussion of techniques to combat the over- and under-fitting observed in our studies, it is clear that additional work is needed in order to better understand different mitigation techniques in the context of SE tasks and datasets, culminating in a set of shared guidelines for the DL4SE community. In addition, more work needs to be done to analyze and understand specialized techniques for SE tasks, data types, and architectures. Similar to preprocessing data, the implementation of over- and underfitting techniques are subject to a set of variables or parameters that define how they work. An in-depth analysis on how these details and parameters change depending on the type of SE task, architecture or data, is beyond the scope of this review. However, it would be useful to the SE community to provide some intuition about what combination of over- and underfitting techniques to apply and what parameters inherent to those techniques will likely lead to beneficial results.

\mybox{\textbf{Summary of Results for RQ$_{3C}$}:}{gray!60}{gray!20}{\revision{Our analysis shows that \textit{dropout} (\dropoutcount) was the most commonly used method to combat over/under-fitting, followed by \textit{data cleaning} (\cleancount), \textit{L1/L2 regularization} (\regcount), and \textit{early stopping} (\earlycount). Nearly 1/4 of papers did not discuss such techniques.}}

\section{\textbf{RQ4: How well do DL tasks perform in supporting various SE tasks?}}
\label{dlslr:rq4}

In this RQ, we aim to explore the impact that DL4SE research has had through an examination of the effectiveness of the techniques proposed in our selected studies. we primarily analyze metrics on a per task basis and summarize the current state of benchmarks and baselines in DL4SE research. 

\subsection{\textbf{\textit{RQ$_{4A}$}: What ``baseline'' techniques are used to evaluate DL models and what benchmarks are used for these comparisons?}}
\label{rq4a}

For RQ$_{4A}$, we examine the baseline techniques and evaluation metrics used for comparison in DL4SE work. 
In general, while we did observe the presence of some common benchmarks for specific SE tasks, we also found that a majority of papers self-generated their own benchmarks. 
We observed that baseline approaches are extremely individualized, even within the same SE task. Some DL4SE papers do not compare against any baseline approaches while others compare against 3-4 different models. Therefore, we included the listing of baselines that each paper compared against in our supplemental material~\cite{watson_palacio_cooper_moran_poshyvanyk}. We found that many of the baseline approaches were canonical machine learning models or very simple neural networks. We suspect the reason for this is in part due to DL4SE being a relatively new field, meaning that there were not many available DL-based approaches to compare against. 
As the field of DL4SE begins to mature, we expect to see a transition to evaluations that include comparisons against previously implemented DL approaches.

One somewhat troubling pattern that we observed is that many model implementations do not include a publicly available implementation of a DL approach. This, in part, explains why there are so many highly individualized, baseline approaches. Since researchers do not have access to common baselines used for comparison, they are forced to implement their own version of a baseline. 
The robustness of the results of such papers may suffer from the fact that many papers did not include any information about the baselines themselves. Additionally, a unique implementation of the same baselines could lead to confounding results when attempting to examine purported improvements. While we expect that the set of existing, publicly available baselines will continue to improve over time, we also acknowledge the need for well-documented and publicly available baselines, and guidelines that dictate their proper dissemination. 

 \begin{figure*}[t]
	\centering
	\includegraphics[width=0.95\columnwidth]{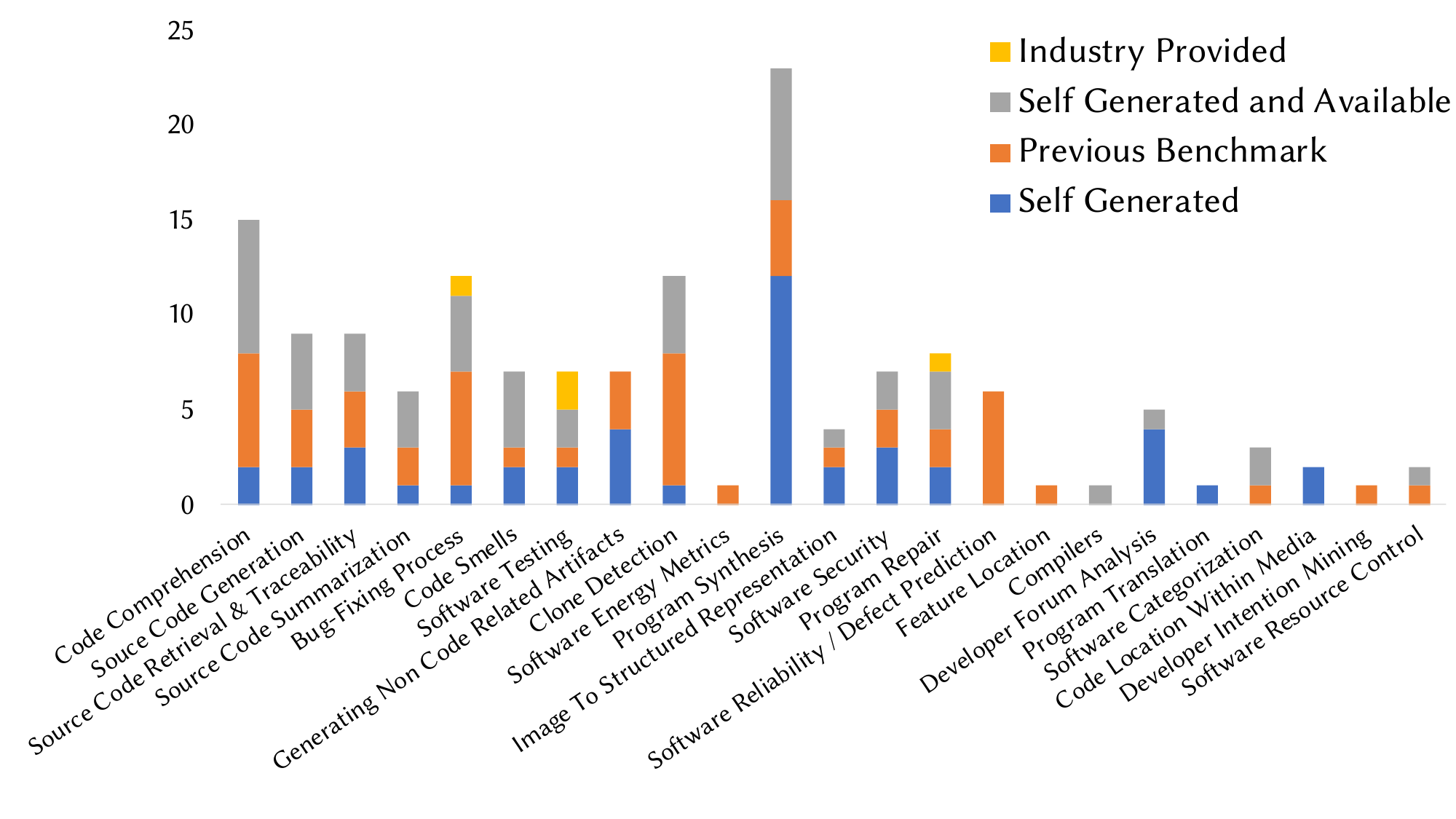}
	\caption{Benchmark Usage DL in SE}	
	\label{dlslr:fig:benchmarks}
\end{figure*}

Our online appendix~\cite{watson_palacio_cooper_moran_poshyvanyk} includes a list of all the benchmarks and baselines used for each paper within our SLR. The diversity and size of this list of benchmarks prohibited its inclusion to the text of this manuscript.  
However, we recorded the number of primary studies that used a previously curated benchmark as opposed to ones that curated their own benchmark. We noted that there is an overwhelming number of self-generated benchmarks. Additionally, we classified self-generated benchmarks into those that are publicly available and those that are not. Unfortunately, we found a majority of self-generated benchmarks may not be available for public use. The full breakdown of benchmarks used in the primary studies can be seen in Figure \ref{dlslr:fig:benchmarks}. This trend within DL4SE is worrying as there are few instances where DL approaches can appropriately compare against one another with available benchmarks. We hope that our online repository aids researchers by providing them with an understanding about which benchmarks are available for an evaluation of their approach within a specific SE task. Additionally, we urge future researchers to make self-generated benchmarks publicly available, which will provide a much needed resource not only for comparisons between approaches, but also for available data applicable to DL techniques.

Although the use of previously established benchmarks was not common among our studies, we did observe a subset of benchmarks that were used multiple times within our primary studies. For the SE task of clone detection, we found that the dataset BigCloneBench~\cite{6976121} was used frequently to test the quality of the DL frameworks. Also, for the task of defect prediction, we saw uses of the PROMISE dataset~\cite{Sayyad-Shirabad+Menzies:2005} as a way to compare previous DL approaches that addressed defect prediction in software. 

\subsubsection{Opportunities for Future Research} The use of baselines and benchmarks in DL4SE studies, for the purpose of evaluation, is developing into a standard practice. However, there exists a need for replicable, standardized, baseline approaches that can be used for comparison when applying a new DL approach to a given SE task. The baseline implementations should be optimized for the benchmark used as data for a non-biased evaluation. This requires a thorough and detailed analysis of each published benchmark, within a specific SE task, for high quality data that does not suffer from sampling bias, class imbalance, etc. Many of the primary studies used a comparative approach for their evaluation, however, with a lack of standardized baselines the evaluations are dependent on how optimized the baseline is for a particular dataset. This can lead to confusing or conflicting  results across SE tasks. We have started to see recent progress in the derivation and sharing of large-scale datasets with efforts such as the CodeXGlue dataset from Microsoft~\cite{lu2021codexglue}. 

\mybox{\textbf{Summary of Results for RQ$_{4A}$}:}{gray!60}{gray!20}{Our analysis revealed a general lack of well-documented, reusable baselines or benchmarks for work on DL4SE. A majority of the baseline techniques utilized in the evaluations of our studied techniques were self-generated, and many are not publicly available or reusable. While a subset of benchmark datasets do exist for certain tasks, there is a need for well-documented and vetted benchmarks.}

\subsection{\textbf{\textit{RQ$_{4B}$}: How is the impact or automatization of DL approaches measured and in what way do these models promote generalizability?}}
\label{rq4b}

Table \ref{tab:metrics} describes the distribution of metrics found in this SLR. In our analysis of utilized metrics within work on DL4SE, we observed that the metrics chosen are often related to the type of learning. \revision{Therefore, many of the supervised learning methods have metrics that analyze the resulting hypothesis, such as the accuracy (\accpercent),  precision (\precpercent), recall (\recallpercent), or F1 measure (\fpercent). In fact, classification metrics are reported in $\approx74\%$ of the papers.} These metrics are used to compare the supervised learning algorithms with the outputs representing the target hypothesis. Intuitively, the type of metric chosen to evaluate the DL-based approach is dependent upon the data type and architecture employed by the approach. The ``other'' category illustrated in Figure \ref{tab:metrics} is comprised of less popular metrics including: \textit{likert scale}, \textit{screen coverage}, \textit{total energy consumption}, \textit{coverage of edges}, \textit{ROUGE}, \textit{Jaccard similarity}, \textit{minimum absolute difference}, \textit{cross entropy}, \textit{F-rank}, \textit{top-k generalization}, \textit{top-k model-guided search accuracy}, \textit{Spearman's rank correlation coefficient}, and \textit{confusion matrices}. In addition to the use of these metrics, we found a limited number of statistical tests to support the comparison between two approaches. These statistical tests included: \textit{Kruskal's $\gamma$}, \textit{macro-averaged mean cost-error}, \textit{Matthew's correlation coefficient}, and \textit{median absolute error}. \revision{Surprisingly, only approximately 5\% of papers made use of statistical tests.} 

\revision{
\begin{landscape}
\begin{table}[t]
\centering
\caption{Metrics Used for Evaluation}
\vspace{-0.35cm}
\label{tab:metrics}
\resizebox{1.2\columnwidth}{!}{%
{\begin{tabular}{lll}
\toprule
Measurement Type                            & Metrics                                           & Studies \\ \midrule
\hspace{3mm}Alignment Scores                
                                            & \hspace{3mm}Rouge-L                               & \hspace{3mm}\cite{Wan2018} \\ 
                                            & \hspace{3mm}BLEU Score                            & \hspace{3mm}\cite{Chen2018a, Jiang2017, Wan2018, Gu2016, Chen2018, Harer2018, 10.1145/3196321.3196334, sun2019, 10.1109/ICSE.2019.00087, gao:saner19} \\ 
                                            & \hspace{3mm}METEOR Score                          & \hspace{3mm}\cite{Chen2018a, Wan2018} \\ \midrule
\hspace{3mm}Classification Measures         
                                            & \hspace{3mm}Precision                             & \hspace{3mm}\begin{tabular}[c]{@{}l@{}}\makecell[l]{\cite{Tufano2018, Liu2018, Li2017, Han2017, Lam2015, Allamanis2015, White2016, Xu2016, Wang2016, Guo2017, Deshmukh2017, Saini2018, Hellendoorn2018, Choetkiertikul2018, Choetkiertikul2017, Arabshahi2018, Gu2018, 10.1145/3213846.3213876}\\\cite{Lin2018, Dam2018, Huang2018, Wen2018, Tufano2018a, Ott2018, Moran2018, 8812134, 8811893, Zhao:ICSE19, DBLP:conf/aaai/BuiJY18, 10.1145/3196398.3196408, 10.1145/3360588, zhao2018deepsim, yu2019neural, zhang2019novel, 10.1109/ICSE.2019.00084, thaller:saner19, wang:msr19, perez:msr19, huo:tse19, xie:saner19, guo:saner19, fakhoury:saner19}}\end{tabular} \\\\
                                            & \hspace{3mm}Recall                                & \hspace{3mm}\begin{tabular}[c]{@{}l@{}}\makecell[l]{\cite{Liu2018, Li2017, Han2017, Allamanis2015, Xu2016, Wang2016, Guo2017, Deshmukh2017, Saini2018, Hellendoorn2018, Choetkiertikul2018, Choetkiertikul2017, Arabshahi2018, 10.1145/3213846.3213876, Lin2018, Liu2018c, Dam2018, Huang2018, Wen2018}\\\cite{Chen2019, Tufano2018a, Ott2018, 8812134, 8811893, Zhao:ICSE19, DBLP:conf/aaai/BuiJY18, 10.1145/3196398.3196408, 10.1145/3276517, 10.1145/3360588, zhao2018deepsim, DBLP:conf/iclr/YinNABG19, yu2019neural, zhang2019novel, 10.1109/ICSE.2019.00084, thaller:saner19, wang:msr19, perez:msr19, guo:saner19, xie:saner19, guo:saner19}}\end{tabular} \\\\
                                            & \hspace{3mm}Confusion Matrix                      & \hspace{3mm}\cite{Moran2018, 10.1145/3276517} \\\\
                                            & \hspace{3mm}Accuracy                              & \hspace{3mm}\begin{tabular}[c]{@{}l@{}}\makecell[l]{\cite{Lee2017, Murali2017, Hellendoorn2017, Li2017, Gaunt2017, 10.5555/3305381.3305483, Levy2017, Han2017, Cai2017, BenNun2018, Zhang2018, Devlin2017, Sun2018, DBLP:conf/iclr/MuraliQCJ18, Bhatia2018, DBLP:journals/corr/ReedF15, Piech2015, Chen2016, Xu2016, Liu2016, DBLP:conf/icml/DevlinUBSMK17, Deshmukh2017, Bunel2018, Vijayakumar2018, DBLP:conf/iclr/ChenLS18}\\\cite{ Hellendoorn2018a, Saini2018, Hellendoorn2018, Arabshahi2018, Chen2018, 10.1145/3213846.3213876, Lin2018, Chen2018e, Gao2018, Wang2017, allamanis2018learning, Zohar2018, Shin2018, Ellis2018a, Liang2018, Choetkiertikul2019, Huang2018, Ott2018, Harer2018, 8811988, tufano2019learning, 10.1145/3276517, sun2019, 10.5555/3015812.3016002, gupta2017deepfix, DBLP:conf/iclr/YinNABG19, bui:saner19, liu:saner19, white:saner19, nguyen:saner19, cvitkovic:icml19}}\end{tabular} \\\\
                                            & \hspace{3mm}ROC/AUC                               & \hspace{3mm}\cite{Zhao2018, Saini2018, Wen2018, Choetkiertikul2018, Choetkiertikul2017, Gao2018, Dam2018, Wen2018, 10.1145/3196398.3196408, codereviewlearn, dam:msr19, hoang:msr19, liu:saner19, buch:saner19} \\
                                            & \hspace{3mm}F-Score                               & \hspace{3mm}\cite{Liu2018, Han2017, Zhao2018, Allamanis2015, Xu2016, Allamanis2016, Wang2016, Choetkiertikul2018, Choetkiertikul2017, Le2018a, 10.1145/3213846.3213876, Dam2018, Huang2018, Wen2018, Tufano2018a, 8812134, 8811893, Zhao:ICSE19, DBLP:conf/aaai/BuiJY18, 10.1145/3360588, codereviewlearn, zhao2018deepsim, yu2019neural, zhang2019novel, thaller:saner19, dam:msr19, wang:msr19, perez:msr19, guo:saner19, xie:saner19, fakhoury:saner19} \\
                                            & \hspace{3mm}Matthews Correlation                  & \hspace{3mm}\cite{Choetkiertikul2018, thaller:saner19} \\
                                            & \hspace{3mm}Scott-Knott Test                      & \hspace{3mm}\cite{liu:saner19} \\
                                            & \hspace{3mm}Exam-Metric                           & \hspace{3mm}\cite{zhang:saner19} \\
                                            & \hspace{3mm}Clustering-Based                      & \hspace{3mm}\cite{8812083} \\ \midrule
\hspace{3mm}Coverage \& Proportions
                                            & \hspace{3mm}Rate or Percentages                   & \hspace{3mm}\cite{Gu2018, Cummins2018, Chen2018d, gupta2019, DBLP:conf/aaai/LiuLPW19, DBLP:conf/iclr/YinNABG19, 8730177, 8502853, fakhoury:saner19} \\
                                            & \hspace{3mm}Coverage-Based                        & \hspace{3mm}\cite{Liu2017, Godefroid2017, DBLP:conf/aaai/LiuLPW19, DBLP:conf/iclr/ParisottoMS0ZK17, Zhang2018a, wang:msr19} \\
                                            & \hspace{3mm}Solved Tasks                          & \hspace{3mm}\cite{Si2018, Ellis2018, Wen2018, Bavishi2019, gupta2019} \\
                                            & \hspace{3mm}Cost-Effectiveness                     & \hspace{3mm}\cite{liu:saner19, white:saner19} \\
                                            & \hspace{3mm}Total Energy or Memory Consumption    & \hspace{3mm}\cite{Romansky2017} \\ \midrule
\hspace{3mm}Distance Based
                                            & \hspace{3mm}CIDER                                 & \hspace{3mm}\cite{Wan2018, Zohar2018} \\
                                            & \hspace{3mm}Cross Entropy                         & \hspace{3mm}\cite{Hellendoorn2018a} \\
                                            & \hspace{3mm}Jaccard Distance                       & \hspace{3mm}\cite{DBLP:conf/iclr/MuraliQCJ18} \\
                                            & \hspace{3mm}Model Perplexity                      & \hspace{3mm}\cite{White2015a, katz:saner19, dam:msr19} \\
                                            & \hspace{3mm}Edit Distance                         & \hspace{3mm}\cite{katz:saner19, gao:saner19} \\
                                            & \hspace{3mm}Exact Match                           & \hspace{3mm}\cite{gao:saner19} \\
                                            & \hspace{3mm}Likert Scale                          & \hspace{3mm}\cite{Jiang2017} \\ \midrule
\hspace{3mm}Approximation Error
                                            & \hspace{3mm}Mean Absolute Error                   & \hspace{3mm}\cite{Choetkiertikul2018, Choetkiertikul2019} \\
                                            & \hspace{3mm}Minimum Absolute Difference           & \hspace{3mm}\cite{DBLP:conf/iclr/MuraliQCJ18} \\
                                            & \hspace{3mm}Macro-averaged Mean Absolute Error    & \hspace{3mm}\cite{Choetkiertikul2018, Choetkiertikul2017} \\
                                            & \hspace{3mm}Root Mean Squared Error               & \hspace{3mm}\cite{Schroeder2017} \\
                                            & \hspace{3mm}Median Absolute Error                 & \hspace{3mm}\cite{Choetkiertikul2019} \\
                                            & \hspace{3mm}Macro-averaged Mean Cost Error        & \hspace{3mm}\cite{Choetkiertikul2017} \\ \midrule
\hspace{3mm}Ranking
                                            & \hspace{3mm}F-Rank                                & \hspace{3mm}\cite{Gu2018} \\
                                            & \hspace{3mm}Top K - Based                         & \hspace{3mm}\cite{Shin2018, Moran2018, Liu2018d, 10.1145/2594291.2594321, huo:tse19} \\
                                            & \hspace{3mm}Spearmans Rank                        & \hspace{3mm}\cite{Tufano2018a} \\
                                            & \hspace{3mm}MRR                                   & \hspace{3mm}\cite{Karampatsis2019, Chen2018a, Hellendoorn2017, Lam2015, Corley2015, Gu2018, Chen2019, huo:tse19} \\
                                            & \hspace{3mm}Kruskal's $\gamma$                      & \hspace{3mm}\cite{Zhao2018} \\ \midrule
Timing                                      & \hspace{3mm}Time                                  & \hspace{3mm}\cite{White2016, Balog2016, Ellis2018, Ellis2018a, 8811988} \\ \midrule
\bottomrule
\end{tabular}}
}
\end{table}
\end{landscape}
}

 \begin{figure*}[t]
	\centering
	\includegraphics[width=0.85\linewidth]{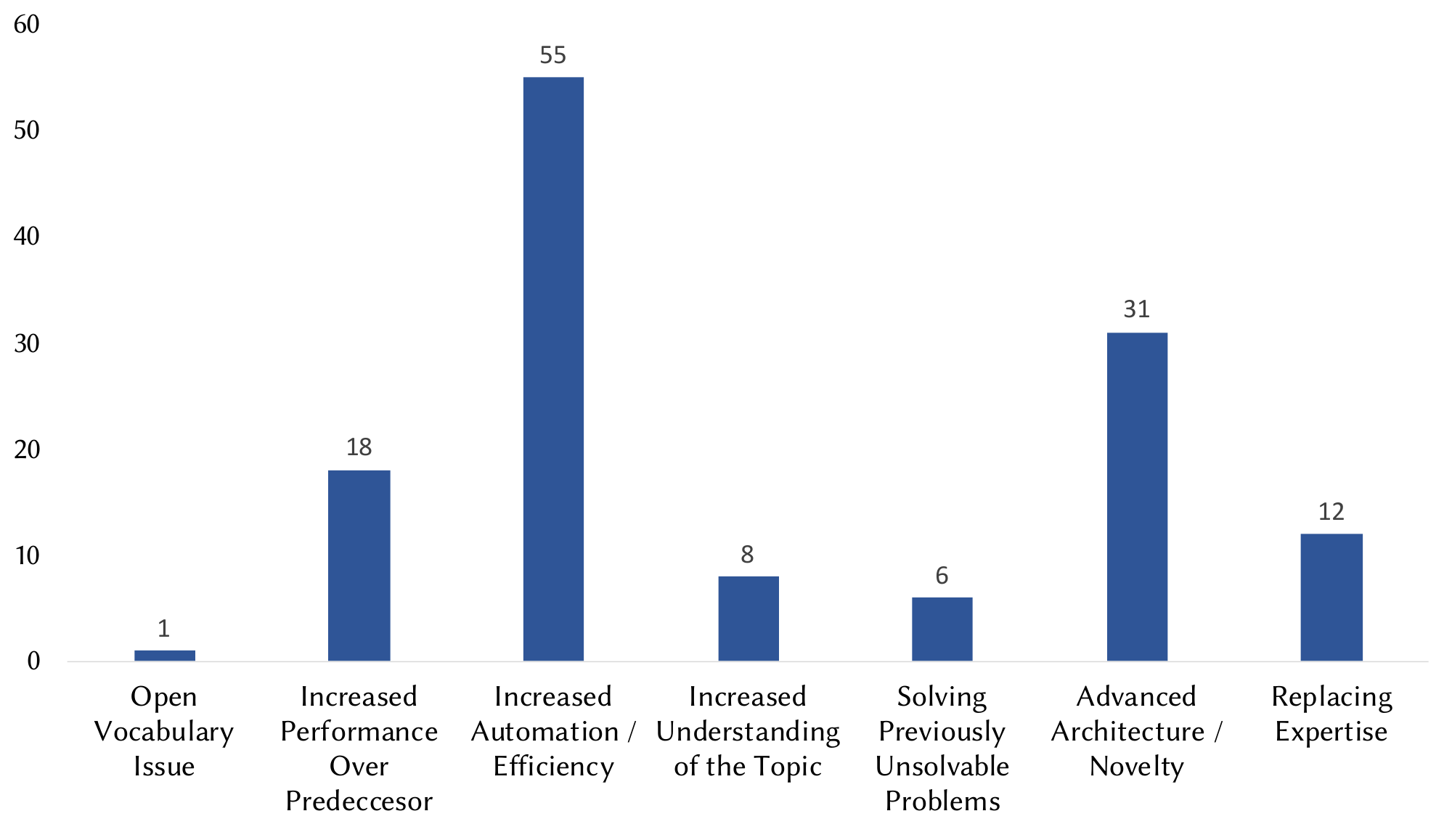}
	\caption{Impact of DL4SE}
	\label{dlslr:fig:impact}
\end{figure*} 

We classified each primary study into seven categories, which represents the major contribution of the work. The result of this inquiry can be seen in Figure \ref{dlslr:fig:impact}. 
\revision{We found three primary objectives that the implementation of a DL model is meant to address: (i) in $\approx43\%$ of papers observed, a DL approach was implemented with the main goal of increasing automation efficiency; (ii) in $\approx24\%$ of the papers observed, a DL approach was implemented with the main goal of advancing or introducing a novel architecture; (iii) in $\approx14\%$ of the papers observed, a DL approach was implemented with the main goal of increasing performance over a prior technique.}

\begin{wrapfigure}{l}{2.6in}
 \centering
	\includegraphics[width=2.4in]{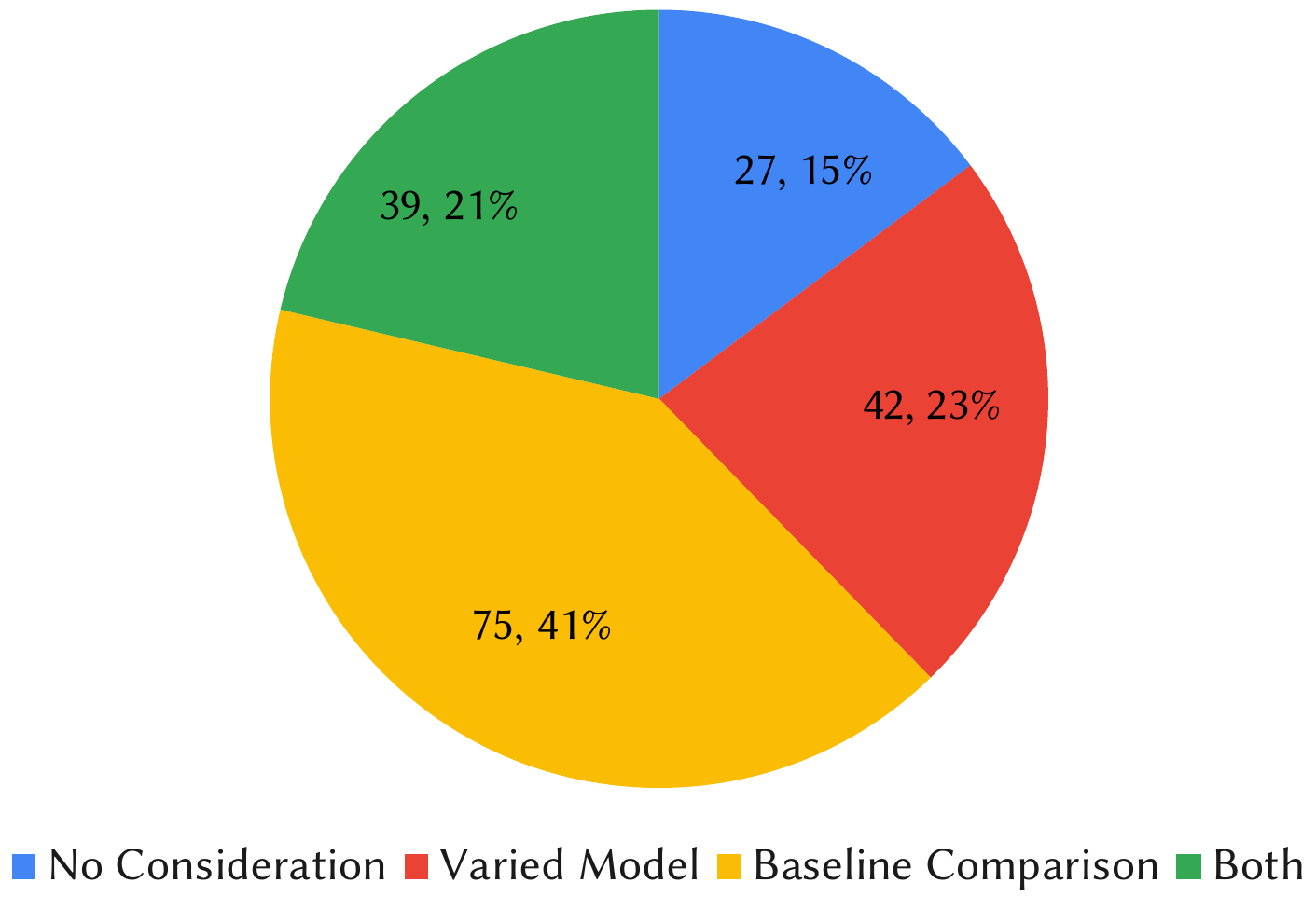}
	\caption{Evidence of Occam's Razor}
	\label{fig:occams}
\end{wrapfigure}

In addition to the primary studies major objectives, we also observed that many papers did not analyze the complexity or generalizability of their implemented models. Thus to examine this further, we analyzed our primary studies through the lends of Occam's Razor and model efficiency. A valid question for many proposed DL techniques applied to SE tasks is whether the complexity of the model is worth the gains in effectiveness or automation for a given task, as recent research has illustrated~\cite{Fu:FSE17}. This concept is captured in a notion known as \textit{Occam's Razor}. Occam's Razor is defined by two major viewpoints: 1) "Given two models with the same generalization error, the simpler one should be preferred because simplicity is desirable"~\cite{10.5555/3000292.3000299}, 2) "Given two models with the same training-set error, the simpler one should be preferred because it is likely to have lower generalization error" \cite{10.5555/3000292.3000299}. %
In the context of our SLR, we aimed to investigate the concept of Occam's Razor through analyzing whether authors considered technically ``simpler'' baseline techniques in evaluating their approaches.  In Figure \ref{fig:occams} we break the primary studies into four groups: 1) those that compare against less complex models and analyze the results; 2) those that manipulate the complexity of their own model by testing a variety of layers or nodes per layer; 3) those that perform both; 4) those that did not have any Occam's Razor consideration. Note that these are overlapping groupings and so the sum of papers exceeds the number of papers in our SLR.

Although a majority of the primary studies do consider Occam's Razor, there are still $\approx16\%$ of DL4SE studies that do not consider the principle. 
Without a consideration of Occam's Razor, it is possible that a canonical machine learning model or a simple statistical based approach could yield an optimal solution. This idea coincides with the findings mentioned by Fu et al. \cite{Fu:FSE17}, who discovered that by applying a simpler optimizer to fine tune an SVM they were able to outperform a DL model applied to the same task. Fu~\etal warn against the blind use of DL models without a thorough evaluation regarding whether the DL technology is a necessary fit for the problem~\cite{Fu:FSE17}. Interestingly, in $\approx23\%$ of the primary studies, the author's considered Occam's Razor by adjusting the complexity of the model being evaluated. This is done by varying the number of layers, the number of nodes, the size of embeddings, etc. The downside to this method is that there is no way to determine if the extraction of complex hierarchical features is more complex than what is necessary to address the SE task. The only way to properly answer this question is to compare against baseline approaches that are not as complex. In our DL4SE studies, this often took the form of a comparison to a canonical ML technique. 

\subsubsection{Results of Exploratory Data Analysis}

\revision{Our exploratory data analysis revealed  papers that combat overfitting, excluding data augmentation, omit ROC or AUC evaluations with a confidence level of 
$\approx0.95$. This metric is a common means by which comparisons to baseline approaches can be performed. Our exploratory data analysis of this RQ revealed that the automation impact is correlated to the SE task deduced from a mutual information of $0.71B$. This means that there is a subtle association between the SE task and the claimed automation impact of the approach.}

\subsubsection{Opportunities for Future Research} Throughout our analysis regarding the evaluation of DL4SE studies, it became apparent that there is a troubling lack of consistency of analysis, even within a given application to an SE task. Thus, there is an opportunity to develop guidelines and supporting evaluation infrastructure for metrics and approach comparisons. Such work would allow for clearer and more concise evaluations of new approaches, solidifying claims made from the results of a given evaluation.  DL models are evaluated on their ability to be generalizable, this is normally accomplished through the use of a testing set, which the model has not been trained on. However, these testing sets can suffer from under representing certain class of data that can be found in the real world. More work is needed on evaluating the quality of testing sets and determining how representative they are when being used for evaluation. Having the limitations of DL approaches well document will create a greater opportunity for these DL solutions to be applied in the context of industry and real software development scenarios.  
Lastly, it would be advantageous for the research community to develop a methodology that could demonstrate the \textit{need} for the complexity that DL offers when addressing a particular problem.

\mybox{\textbf{Summary of Results for RQ$_{4b}$}:}{gray!60}{gray!20}{\revision{Our analysis illustrates that a variety of metrics have been used to evaluate DL4SE techniques, with \textit{accuracy} (\accpercent), \textit{precision} (\precpercent), \textit{recall} (\recallpercent), and \textit{F1-measure} (\fpercent) being the most prominent.} In terms of claimed impact of our primary studies, the most claimed was \textit{increased automation or efficiency}, followed by \textit{advancing a DL architecture}, and \textit{replacing human expertise.} We also found that most studies did consider the concept of Occam's Razor and offered a comparison to a conceptually simpler learning model.}

\section{Threats to Validity}
\label{dlslr:threats}

Our systematic literature review was conducted according to the guidelines set forth by Kitchenham et al.~\cite{Kitchenham2007}. However, as with any SLR our review does exhibit certain limitations primarily related to our search methodology and our data extraction process employed to build our paper taxonomy.

\subsubsection{External Validity}

Issues related to external validity typically concern the generalization of the conclusions drawn by a given study. A potential threat to the external validity to our systematic literature review is the search string and filtering process used to identify meaningful DL4SE studies. It is possible that our search string missed studies that should have been included in our review. This could be due to a missed term or combination of terms that may have returned more significant results. We mitigated this threat by testing a variety of DL and SE terms such as:

\begin{enumerate}
    \item (“Deep Learning” OR “Neural Network”)
    \item (“Learning”) AND (“Neural” OR “Automatic” OR “Autoencoder” OR “Represent”)
    \item (“Learning”) AND (“Supervised” OR “Reinforcement” OR “Unsupervised” OR “Semi-supervised”)
    \item (“Learning” OR “Deep” OR “Neural” OR “Network”)
    \item (“Learning” OR “Deep” OR “Neural”)
    \item (“Artificial Intelligence” OR “Learning” OR “Representational” OR “Neural” OR “Network”)
\end{enumerate}

We evaluated these potential search strings through an iterative process as outlined by Kitchenham et al. The utilized search string "Deep" \texttt{OR} "Learning" \texttt{OR} "Neural" returned the greatest number of DL4SE studies. This search string was also chosen to limit selection bias since it ``cast the widest net'' in order to bolster completeness and limit potential biases introduced by more restrictive search strings. However, the trade-off was that it required a much more substantial effort to remove studies that were not applicable to DL4SE.

We also face potential selection bias of the studies to be included into our SLR. We attempt to mitigate this threat through the use of inclusion and exclusion criteria, which is predefined before the filtering process begins, and which we have listed in our online appendix~\cite{watson_palacio_cooper_moran_poshyvanyk}. This criteria is also helpful in reducing the manual effort of filtering papers given our broad search string. We also perform snowballing as a means to mitigate selection bias. In this method, we collect all the references from the primary studies that passed our inclusion and exclusion criteria and determine if any of those references should be considered for the SLR.

Additionally, to further illustrate the generalizability of our paper sampling methodology, we perform a probability sampling to determine if we capture a significant proportion of DL4SE papers. We found that our expert sampling strategy captures a statistically significant number of studies, such that we are confident in our taxonomy's representation. Therefore, we feel that the trends highlighted in this review can be generalized to the entire body of DL4SE work.

Another potential threat to our systematic literature review consists of the venues chosen for consideration. For our review, we included the top SE, PL, and AI related conferences and journals. We included venues with at least a C CORE ranking~\cite{core}, which helped us to determine venue impact. Although it is possible that not considering other conferences and journals caused us to miss some pertinent studies, we wanted to capture trends as seen in top SE, PL, and AI venues. Furthermore, we provide our current taxonomy and list of venues on our website, and welcome contributions from the broader research community. We intend for our online appendix to serve as a "living" document that continues to survey and categorize DL4SE research.

\subsubsection{Internal Validity}

A major contribution of this dissertation lies in our derived taxonomy that characterizes the field of DL4SE. To mitigate any mistakes in our taxonomy, we followed a process inspired by open coding in constructivist grounded theory~\cite{Charmaz:groundedtheory} where each attribute classification of a primary study within our SLR was reviewed by at least three authors. However, while multiple evaluators limit the potential bias of this process, the classifications are still potentially affected by the collective views and opinions of the authors. Therefore, in effort to provide transparency into our process and bolster the integrity of our taxonomy, we have released all data extraction and classifications in our online repository \cite{watson_palacio_cooper_moran_poshyvanyk}. In releasing this information, authors of the works included in the SLR can review our classifications.

\subsubsection{Construct Validity}

One point of construct validity is the conclusions we draw at the end of each research question. In order to draw these conclusions, we performed an exploratory data analysis using rule association mining. In this analysis, we mine associations between attributes of DL solutions to SE tasks, which provides inspiration to further research why certain attributes show a strong or weak correlation.

Another threat to construct validity is our methodology for data synthesis and taxonomy derivation. To mitigate this threat we followed a systematic and reproducible process for analyzing the primary studies and creating a resulting taxonomy. To reduce the potential bias of data extraction, the authors developed and agreed upon a data extraction form to apply to each study. 
For our taxonomy, primary studies were categorized by three authors and refined by one additional authors. Through this process, we limit the number of individual mistakes in extracting the data and synthesizing the taxonomy.



\section{Bibliographical Notes}

The paper supporting the content described in this Chapter was written in collaboration with other members of the SEMERU group at William \& Mary and a researcher from George Mason University. I have received permission from the publisher and co-authors to reprint sections of this work:

Watson, C., \textbf{Cooper, N.}, Palacio, D. N., Moran, K., \& Poshyvanyk, D. (2020). A Systematic Literature Review on the Use of Deep Learning in Software Engineering Research. ACM Transactions on Software Engineering and Methodology (TOSEM), 31(2), 1-58.
\chapter{Combining Visual and Textual Information for Detecting Duplicate Video-Based Bug Reports}
\label{sec:tango}

Many modern mobile applications (apps) allow users to report bugs in a graphical form, given the GUI-based nature of mobile apps. For instance, Android and iOS apps can include built-in screen-recording capabilities in order to simplify the reporting of bugs by end-users and crowd-testers~\cite{bugclipper,testfairy,bugsee}. The reporting of visual data is also supported by many crowd-testing and bug reporting services for mobile apps~\cite{bugclipper,testfairy,ubertesters,Instabug,bugsee,snaffu,bugreplay,bugsquasher,birdeatsbugs,outklip,10.1145/3540250.3549131}, which intend to aid developers in collecting, processing, and understanding the reported bugs~\cite{3Bettenburg:FSE08,Mao:ASE17,PALOMBA2018143,7332475}.

The proliferation of sharing images to convey additional context for understanding bugs, \eg in Stack Overflow Q\&As \cite{SO2004}, has been steadily increasing over the last few years \cite{Nayebi2020}. Given this and the increased integration of screen capture technology into mobile apps, developers are likely to face a growing set of challenges related to processing and managing app screen-recordings in order to triage and resolve bugs --- and hence maintain the quality of their apps. 

One important challenge that developers will likely face in relation to video-related artifacts is determining whether two videos depict and report the same bug (\ie detecting duplicate video-based bug reports), as it is currently done for textual bug reports~\cite{Rakha:TSE'18,Rakha2018,Bettenburg:ICSM08}. When video-based bug reports are collected at scale, either via a crowdsourced testing  service~\cite{bugclipper,testfairy,ubertesters,Instabug,bugsee,snaffu,bugreplay,bugsquasher,birdeatsbugs,outklip} or by popular apps, the sizable corpus of collected reports will likely lead to significant developer effort dedicated to determining if a new bug report depicts a previously-reported fault, which is necessary to avoid redundant effort in the bug triaging and resolution process~\cite{Bettenburg:ICSM08,Li2019,3Bettenburg:FSE08,Mao:ASE17}. In a user study which we detail later in this chapter (Sec.~\ref{tango:user_study}), we investigated the effort required for experienced programmers to identify duplicate video-based bug reports and found that participants reported a range of difficulties for the task (\eg a single change of one step can result in two very similar looking videos showing entirely different bugs), and spent about $20$ seconds of comprehension effort on average per video viewed.
If this effort is extrapolated to the large influx of bug reports that could be collected on a daily basis \cite{Rakha:TSE'18,Rakha2018,Bettenburg:ICSM08,Chaparro2016a}, it illustrates the potential for the excessive effort associated with video-based duplicate bug detection. This is further supported by the plans of a large company that supports crowd-sourced bug reporting (name omitted for anonymity), 
which we contacted as part of eliciting the design goals for this research, who stated that they anticipate increasing developer effort in managing video-based reports and that they are planning to build a feature in their framework to support this process.

To aid developers in determining whether video-based bug reports depict the same bug, this work introduces \tango, a novel approach that analyzes both visual and textual information present in mobile screen-recordings using tailored computer vision (CV) and text retrieval (TR) techniques, with the goal of generating a list of candidate videos (from an issue tracker) similar to a target video-based report. In practice, \tango is triggered upon the submission of a new video-based report to an issue tracker. A developer would then use \tango to retrieve the video-based reports that are most similar (\eg top-5) to the incoming report for inspection. If duplicate videos are found in the ranked results, the new bug report can be marked as a duplicate in the issue tracker. Otherwise, the developer can continue to inspect the ranked results until she has enough confidence that the newly reported bug was not reported before (\ie it is not a duplicate).

\tango operates \textit{purely} upon the graphical information in videos in order to offer flexibility and practicality. These videos may show the \textit{unexpected behavior} of a mobile app (\ie a crash or other misbehavior) and the \textit{steps to reproduce} such behavior. Two videos are considered to be \textit{duplicates} if they show the same unexpected behavior (\aka a bug) regardless of the steps to reproduce the bug.
Given the nature of screen-recordings, video-based bug reports are likely to depict unexpected behavior towards the end of the video.
\tango attempts to account for this by leveraging the temporal nature of video frames and weighting the importance of frames towards the end of videos more heavily than other segments.

We conducted two empirical studies to measure: (i) the \textit{effectiveness} of different configurations of \tango by examining the benefit of combining visual and textual information from videos, as opposed to using only a single information source; and~(ii) \tangos ability to \textit{save developer effort} in identifying duplicate video-based bug reports. To carry out these studies, we collected a set of 180 video-bug reports from six popular apps used in prior research~\cite{Bernal-Cardenas:ICSE'20,Chaparro:FSE'19,Moran:FSE15,Moran:ICST16}, and defined 4,860 duplicate detection tasks that resemble those that developers currently face for textual bug reports -- wherein a corpus of potential duplicates must be ranked according to their similarity to an incoming bug report. 

The results of these studies illustrate that \tangos most effective configuration, which selectively combines visual and textual information, achieves 79.8\% mRR and 73.2\% mAP, an average rank of 1.7, a {\sc Hit}@1 of 67.7\%, and a {\sc Hit}@2 of 83\%. This means that \tango is able to suggest correct duplicate reports in the top-2 of the ranked candidates for 83\% of duplicate detection tasks.
The results of the user study we conducted with experienced programmers demonstrate that
on a subset of the tasks, \tango can reduce the time they spend in finding duplicate video-based bug reports by $\approx65\%$.

In summary, the main contributions of this work are the following:
\begin{enumerate}
    \item{\tango, a duplicate detection approach for video-based bug reports of mobile apps which is able to accurately suggest duplicate reports;}
    \item{The results of a comprehensive empirical evaluation that measures the \textit{effectiveness} of \tango in terms of suggesting candidate duplicate reports;}
    \item{The results of a user study with experienced programmers that illustrates \tango's practical applicability by measuring its potential for \textit{saving developer effort}; and}
    \item{A benchmark (included in our online appendix \cite{tango_appendix}) that enables (i) future research on video-based duplicate detection, bug replication, and mobile app testing, and (ii) the replicability of this work. The benchmark contains 180 video-based bug reports with duplicates, source code, trained models,  duplicate detection tasks, \tango's output, and detailed evaluation results.}
\end{enumerate}

\section{Tango's Approach}
\label{tango:approach}

\tango (\tangolong) is an automated approach based on CV and TR techniques, which leverages visual and textual information to detect duplicate video-based bug reports.

\subsection{\tango Overview}

\begin{figure*}[t]
	\centering
	\includegraphics[width=0.98\textwidth]{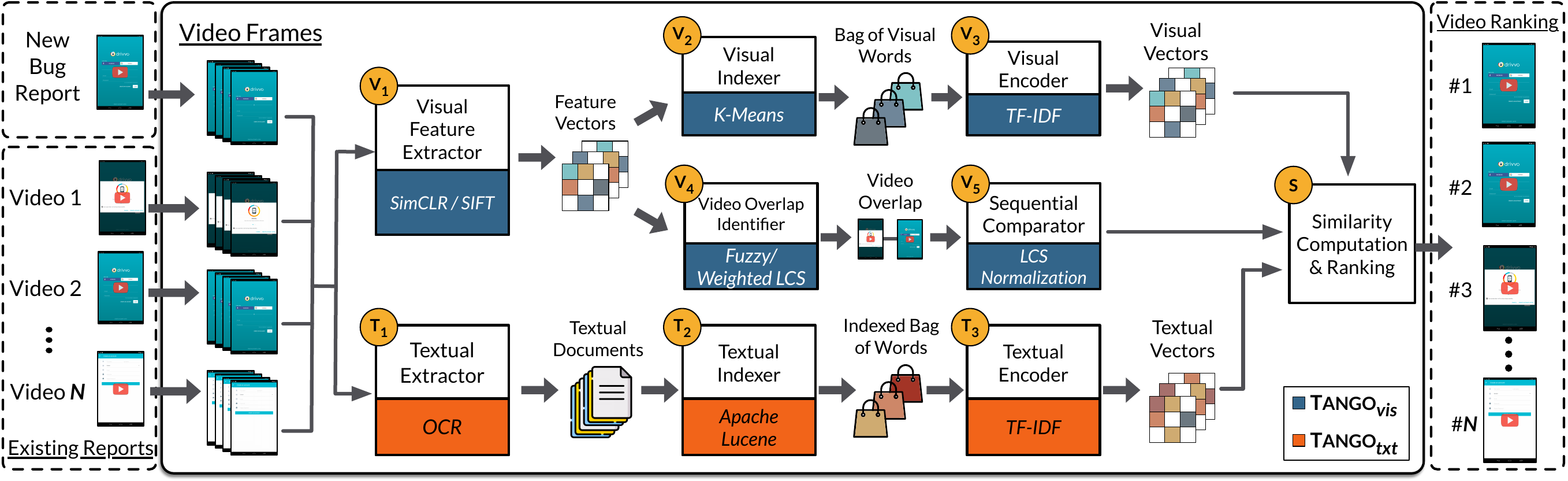}
	\caption{\normalsize{The \tango approach for detecting duplicate video-based bug reports.}}
	\label{tango:fig:approach}
\end{figure*}

\tango models duplicate bug report detection as an information retrieval problem. Given a new video-based bug report, \tango computes a similarity score between the new video and videos previously submitted by app users in a bug tracking system. The new video represents the query and the set of existing videos represent the corpus. \tango sorts the corpus of videos in decreasing order by similarity score and returns a ranked list of candidate videos. In the list, those videos which are more likely to show the same bug as the new video are ranked higher than those that show a different bug.

\tango has two major components, which we refer to as \tangov and \tangot (Fig.~\ref{tango:fig:approach}), that compute video similarity scores independently. \tangov computes the \textit{visual similarity} and \tangot computes the \textit{textual similarity} between videos. The resulting similarity scores are linearly combined to obtain a final score that indicates the likelihood of two videos being duplicates. In designing \tangov, we incorporated support for two methods of computing visual similarity --- one of which is sensitive to the \textit{sequential order} of visual data, and the other one that is not --- and we evaluate the effectiveness of these two techniques in experiments described in Sec.~\ref{tango:eval}-\ref{tango:results}.

The first step in \tango's processing pipeline (Fig.~\ref{tango:fig:approach}) is to decompose the query video, and videos from the existing corpus, into their constituent frames using a given sampling rate (\ie 1 and 5 frames per second - fps). Then, the \tangov and \tangot components of the approach are executed in parallel. The \textit{un-ordered} \tangov pipeline is shown at the top of Fig.~\ref{tango:fig:approach}, comprising steps \circled{\textbf{\scriptsize V$_\textnormal{\textbf{1}}$}}-\circled{\textbf{\scriptsize V$_\textnormal{\textbf{3}}$}}; the \textit{ordered} \tangov pipeline is illustrated in the middle of Fig.~\ref{tango:fig:approach}, comprising steps \circled{\textbf{\scriptsize V$_\textnormal{\textbf{1}}$}}, \circled{\textbf{\scriptsize V$_\textnormal{\textbf{4}}$}}, and \circled{\textbf{\scriptsize V$_\textnormal{\textbf{5}}$}}; and finally, the \tangot pipeline is illustrated at the bottom of Fig.~\ref{tango:fig:approach} through steps \circled{\textbf{\scriptsize T$_\textnormal{\textbf{1}}$}}-\circled{\textbf{\scriptsize T$_\textnormal{\textbf{3}}$}}. Any of these three pipelines can be used to compute the video ranking independently or in combination (\ie combining the two \tangov together, one \tangov pipeline with \tangot, which we call \tangoc, or all three -- see Sec.~\ref{tango:configurations}). Next, we discuss these three pipelines in detail.

\subsection{T{\small ANGO}$_{vis}$: Measuring Unordered Visual Video Similarity}

The \textit{unordered} version of \tangov computes the visual similarity ($S_{vis}$) of video-based bug reports by extracting visual features from video frames and converting these features into a vector-based representation for a video using a Bag-of-Visual-Words (BoVW) approach \cite{Sivic:CCV'03,Jiang:IVR'07}. This process is depicted in the top of Fig. \ref{tango:fig:approach}. The visual features are extracted by the \textit{visual feature extractor} model (\circled{\textbf{\scriptsize V$_\textnormal{\textbf{1}}$}} in Fig.~\ref{tango:fig:approach}). Then, the \textit{visual indexer} \circled{\textbf{\scriptsize V$_\textnormal{\textbf{2}}$}} assigns to each frame feature vector a visual word from a visual codebook and produces a BoVW for a video. The \textit{visual encoder} \circled{\textbf{\scriptsize V$_\textnormal{\textbf{5}}$}}, based on the video BoVW, encodes the videos using a TF-IDF representation that can be used for similarity computation.

\subsubsection{Visual Feature Extraction}

 The \textit{visual feature extractor}~\circled{\textbf{\scriptsize V$_\textnormal{\textbf{1}}$}} can either use the SIFT~\cite{Lowe:JCV'04} algorithm to extract features, or SimCLR~\cite{Chen:SimCLR'20}, a recently proposed Deep Learning model capable of learning visual representations in an unsupervised, contrastive manner. \tango's implementation of SimCLR is adapted to extract visual features from app videos. 
 
The first method by which \tango can extract visual features is using the Scale-Invariant Feature Transform (SIFT)~\cite{Lowe:JCV'04} algorithm. SIFT is a state-of-the-art model for extracting local features from images that are invariant to scale and orientation. These features can be matched across images for detecting similar objects. This matching ability makes SIFT promising for generating features that can help locate duplicate images (in our case, duplicate video frames) by aggregating the extracted features. \tango's implementation of SIFT does not resize images and uses the top-10 features that are the most invariant to changes and are based on the local contrast of neighboring pixels, with higher contrast usually meaning more invariant. This is done to reduce the number of SIFT features, which could reach at least three orders of magnitude for a single frame, and make the \textit{visual indexing} \circled{\textbf{\scriptsize V$_\textnormal{\textbf{2}}$}} (through $k$-Means -- see Sec. \ref{tango:visual_indexing}) computationally feasible.
 
The other technique that \tango can use to extract features is SimCLR. In essence, the goal of this technique is to generate robust visual features that cluster similar images together while maximizing the distance between dissimilar images in an abstract feature space. This is accomplished by (i) generating sets of image pairs (containing one original image and one augmented image) and applying a variety of random augmentations (\ie image cropping, horizontal flipping, color jittering, and gray-scaling); (ii) encoding this set of image pairs using a base encoder, typically a variation of a convolutional neural network; and (iii) training a multi-layer-perceptron (MLP) to produce feature vectors that increase the cosine similarity between each pair of image variants and decrease the cosine similarity between negative examples, where negative examples for a given image pair are represented as all other images not in that pair, for a given training batch. %
 \tango's implementation of SimCLR employs the ResNet50~\cite{He:CVPR'16} CNN architecture as the base encoder, where this architecture has been shown to be effective~\cite{Chen:SimCLR'20}.
 
To ensure that \tango's \textit{visual feature extractor} is tailored to the domain of mobile app screenshots, we trained this component on the entire RICO dataset~\cite{Deka:UIST'17}, which contains over 66k Android screenshots from over 9k of the most popular apps on Google Play.  %
Our implementation of  SimCLR was trained using a batch size of $1,792$ and $100$ epochs, the same hyperparameters (\eg learning rate, weight decay, \etc) recommended by Chen \etal \cite{Chen:SimCLR'20} in the original SimCLR paper, and resized images to 224$\times$224 to ensure consistency with our base ResNet50 architecture. The training process was carried out on an Ubuntu 20.04 server with three NVIDIA T4 Tesla 16GB GPUs.

The output of the \textit{feature extractor} for SimCLR is a feature vector (of size 64) for each frame of a given video.

\subsubsection{Visual Indexing}
\label{tango:visual_indexing}

While the SimCLR or SIFT feature vectors generated by \tango's \textit{visual feature extractor} \circled{\textbf{\scriptsize V$_\textnormal{\textbf{1}}$}} could be used to directly compute the similarity between video frames, recent work has suggested that a BoVW approach combined with a TF-IDF similarity measure is more adept to the task of video retrieval~\cite{Kordopatis-Zilos:TM'19}. Therefore, to transform the SimCLR or SIFT feature vectors into a BoVW representation, \tango uses a \textit{visual indexing process} \circled{\textbf{\scriptsize V$_\textnormal{\textbf{2}}$}}.

This process produces an artifact known as a Codebook that maps SimCLR or SIFT feature vectors to ``visual words'' --- which are discrete representations of a given image, and have been shown to be suitable for image and video recognition tasks~\cite{Kordopatis-Zilos:TM'19}. %
The Codebook derives these visual words by clustering feature vectors and computing the centroids of these clusters, wherein each centroid corresponds to a different visual word. %

The Codebook makes use of the $k$-Means clustering algorithm, where the $k$ represents the diversity of the visual words, and thus can affect the representative power of this indexing process. \tango's implementation of this process is configurable to 1k, 5k, or 10k for the $k$ number of clusters (\ie the number of visual words - VW). 1k VW and 10k VW were selected as recommended by Kordopatis-Zilos \etal \cite{Kordopatis-Zilos:TM'19} and we included 5k VW as a ``middle ground'' to better understand how the number of visual words impacts \tango's performance. A Codebook is generated only once for a given $k$, however, it must be trained before it can be applied to convert an input feature vector to its corresponding visual word(s). Once trained, a Codebook can then be used to map visual words from frame feature vectors without any further modification. Thus, we trained \tango's six Codebooks, three for SIFT and three for SimCLR, using features extracted from $15,000$ randomly selected images from the RICO dataset~\cite{Deka:UIST'17}. We did not use the entire RICO dataset due to computational constraints of the $k$-means algorithm. %

After the feature vector for a video frame is passed through the \textit{visual indexing} process, it is mapped to its BoVW representation by a trained Codebook. To do this, the Codebook selects the closest centroid to each visual feature vector, based on Euclidean distance. For SIFT, this process may generate more than one feature vector for a single frame, due to the presence of multiple SIFT feature descriptors. In this case, \tango assigns multiple visual words to each frame. For SimCLR, \tango assigns one visual word to each video frame, as SimCLR generates only one vector per frame.

\subsubsection{Visual Encoding} After the video is represented as a BoVW, the \textit{visual encoder} \circled{\textbf{\scriptsize V$_\textnormal{\textbf{3}}$}} computes the final vector representation of the video through a TF-IDF-based approach~\cite{Salton:TFIDF86}. The term frequency (TF) is computed as the number of visual words occurrences in the BoVW representation of the video, and the inverse document frequency (IDF) is computed  as the number of occurrences of a given visual word in the BoVW representations built from the RICO dataset. Since RICO does not provide videos but individual app screenshots, we consider each RICO image as one document. We opted to use RICO to compute our IDF representation for two reasons: (i) to combat the potentially small number of frames present in a given video recording, and (ii) to bolster the generalizability of our similarity measure across a diverse set of apps.

\subsubsection{Similarity Computation}

Given two videos, \tangov encodes them into their BoVW representations, and each video is represented as one visual TF-IDF vector. These vectors are compared 
using cosine similarity, which is taken as the \textit{visual similarity} \circled{\textbf{\scriptsize S}} of the videos ($S_{vis}=S_{BoVW}$).

\subsection{T{\small ANGO}$_{vis}$: Measuring Ordered Visual Video Similarity}

The \textit{ordered} version of \tangov considers the sequence of video frames when comparing two videos and is capable of giving more weight to common frames nearer the end of the videos, as this is likely where buggy behavior manifests. 
To accomplish this, the \textit{feature vector extractor} \circled{\textbf{\scriptsize V$_\textnormal{\textbf{1}}$}} is used to derive descriptive vectors from each video frame using either SimCLR or SIFT. \tango determines how much the two videos overlap using an adapted longest common substring (LCS) algorithm \circled{\textbf{\scriptsize V$_\textnormal{\textbf{4}}$}}. Finally, during the \textit{sequential comparison process} \circled{\textbf{\scriptsize V$_\textnormal{\textbf{5}}$}}, \tango calculates the similarity score by normalizing the computed LCS score.
\subsubsection{Video Overlap Identification} 

In order to account for the sequential ordering of videos, \tango employs two different versions of the longest common substring (LCS) algorithm. The first version, which we call fuzzy-LCS (f-LCS), modifies the comparison operator of the LCS algorithm to perform fuzzy matching instead of exact matching between frames in two videos. This fuzzy matching is done differently for SimCLR and SIFT-derived features. For SimCLR, given that each frame is associated with only a single visual word, the standard BoVW vector would be too sparse for a meaningful comparison. Therefore, we compare the feature vectors that SimCLR extracts from the two frames \textit{directly} using cosine similarity. For SIFT, we utilize the BoVW vectors derived by the \textit{visual encoder} \circled{\textbf{\scriptsize V$_\textnormal{\textbf{3}}$}}, but at a per-frame level.

The second LCS version, which we call weighted-LCS~(w-LCS), uses the same fuzzy matching that f-LCS performs. However, the similarity produced in this matching is then weighted depending on where the two frames from each video appeared. Frames that appear later in the video are weighted more heavily, since that is where the buggy behavior is typically occurring in a video-based bug report, and thus should be given more weight for duplicate detection. The exact weighting scheme used is $\frac{i}{m} \times \frac{j}{m}$, where $i$ is the $ith$ frame of video A, $m$ is the \# of frames in video A, $j$ is the $jth$ frame of video B, and $n$ is the \# of frames in video B.

\subsubsection{Sequential Comparison}

In order to incorporate the LCS overlap measurements into \tango's overall video similarity calculation, the overlap scores must be normalized between zero and one ($[0,1]$). To accomplish this, we consider the case where two videos overlap perfectly to be the upper bound of the possible LCS score between two videos, and use this to perform the normalization. For f-LCS, this is done by simply dividing by the \# of frames in the smaller video since the $max$ possible overlap that could occur is when the smaller video is a subsection in the bigger video, calculated as $overlap/min$ where $overlap$ denotes the amount the two videos share in terms of their frames and $min$ denotes the \# of frames in the smaller of the two videos. For w-LCS, if the videos are different lengths, we align the end of the shorter video to the \textit{end} of the longer video and consider this the upper bound on the LCS score, which is normalized as follows:

\begin{equation}
\label{eq:wlcs}
   S_{w-LCS} = \frac{overlap}{\sum_{i=min}^{1}\frac{i}{min}\times\frac{max - i}{max}}
\end{equation}

\noindent where $S_{w-LCS}$ is the normalized similarity value produced by w-LCS, $overlap$ and $min$ are similar to the f-LCS calculation and $max$ denotes the \# of frames in the longer of the two videos. The denominator in Eq. \ref{eq:wlcs} calculates the maximum possible overlap that can occur if the videos were exact matches, summing across the similarity score of each frame pair. Our online appendix contains the detailed f/w-LCS algorithms with examples \cite{tango_appendix}.

\subsubsection{Similarity Computation} f-LCS and w-LCS output the \textit{visual similarity} \circled{\textbf{\scriptsize S}} score $S_{f-LCS}$ and $S_{w-LCS}$, respectively. This can be combined with $S_{BoVW}$ to obtain an aggregate visual similarity score:  $S_{vis} = (S_{BoVW} + S_{f-LCS})/2$ or $S_{vis} = (S_{BoVW} + S_{w-LCS})/2$. We denote these \tangov variations as \bflcs and \bwlcs, respectively.

\subsection{Determining the Textual Similarity between Videos}
\label{tango:approach_textual}

In order to determine the textual similarity between video-based bug reports, \tango leverages the textual information from labels, titles, messages, \etc found in the app GUI components and screens depicted in the videos.

\tangot adopts a standard text retrieval approach based on Lucene \cite{Hatcher2004} and Optical Character Recognition (OCR)~\cite{tesseract-ocr,pytesseract} to compute the textual similarity ($S_{txt}$) between video-based bug reports. First, a textual document is built from each video in the issue tracker (\circled{\textbf{\scriptsize T$_\textnormal{\textbf{1}}$}} in Fig. \ref{tango:fig:approach}) using OCR to extract text from the video frames. The textual documents are pre-processed using standard techniques to improve similarity computation, namely tokenization, lemmatization, and removal of punctuation, numbers, special characters, and one- and two-character words. The pre-processed documents are indexed for fast similarity computation \circled{\textbf{\scriptsize T$_\textnormal{\textbf{2}}$}}. Each document is then represented as a vector using TF-IDF and the index~\cite{Salton:TFIDF86} \circled{\textbf{\scriptsize T$_\textnormal{\textbf{3}}$}}. 

In order to build the textual documents from the videos, \tangot applies OCR on the video frames through the Tesseract engine \cite{tesseract-ocr,pytesseract} in the \textit{textual extractor} \circled{\textbf{\scriptsize T$_\textnormal{\textbf{1}}$}}. We experiment with three strategies to compose the textual documents using the extracted frame text. The first strategy (all-text) concatenates all the text extracted from the frames. The second strategy (unique-frames) concatenates all the text extracted from unique video frames, determined by applying exact text matching (before text pre-processing). The third strategy (unique-words) concatenates the unique words in the frames (after pre-processing).

\subsubsection{Similarity Computation} \tango computes the \textit{textual similarity} ($S_{txt}$) in \circled{\textbf{\scriptsize S}} using Lucene's scoring function~\cite{lucene-tfidfsimilarity} based on cosine similarity and document length normalization.

\subsection{Combining Visual and Textual Similarities}
\label{tango:approach_combination}
\tango combines both the visual ($S_{vis}$) and textual ($S_{txt}$) similarity scores produced by \tangov and \tangot, respectively (\circled{\textbf{\scriptsize S}} in Fig. \ref{tango:fig:approach}). \tango uses a linear combination approach to produce an aggregate similarity value:
\begin{equation}\label{eq:pythagoras}
S_{comb} = (1 - w) \times S_{vis} + w \times S_{txt}
\end{equation}
where $w$ is a weight for $S_{vis}$ and $S_{txt}$, and takes a value between zero (0) and one (1). Smaller $w$ values weight $S_{vis}$ more heavily, and larger values weight $S_{txt}$ more heavily. We denote this approach as \tangoc.

Based on the combined similarity, \tango generates a ranked list of the video-based bug reports found in the issue tracker. This list is then inspected by the developer to determine if a new video reports a previously reported bug.

\section{\tango's Empirical Evaluation Design}
\label{tango:eval}

We empirically evaluated \tango with two goals in mind: (i) determining how effective \tango is at detecting duplicate video-based bug reports, when considering different configurations of components and parameters, and (ii) estimating the effort that \tango can save developers during duplicate video bug detection. Based on these goals, we defined the following research questions (RQs):

\begin{enumerate}[label=\textbf{RQ$_\arabic*$:}, ref=\textbf{RQ$_\arabic*$}, wide, labelindent=5pt]\setlength{\itemsep}{0.2em}
    \item \label{tango:rq:individual_performance}{\textit{How effective is \tango when using either visual or textual information alone to retrieve duplicate videos?}}
    \item \label{tango:rq:lcs_performance}{\textit{What is the impact of combining frame sequence and visual information on \tango's detection performance?}}
    \item \label{tango:rq:combination_performance}{\textit{How effective is \tango when combining both visual and textual information for detecting duplicate videos?}}
    \item \label{tango:rq:time}{\textit{How much effort can \tango{} save developers in finding duplicate video-based bug reports?}}
\end{enumerate}

To answer our \textbf{RQs}, we first collected video-based bug reports for six Android apps (Sec.~\ref{tango:data_collection}), and based on them, defined a set of duplicate detection tasks (Sec.~\ref{tango:scenarios_and_tasks}). We instantiated different configurations of \tango by combining its components and parameters (Sec.~\ref{tango:configurations}), and executed these configurations on the defined tasks (Sec.~\ref{tango:evaluation_methodology}). Based on standard metrics, applied on the video rankings that \tango produces, we measured \tango's effectiveness (Sec.~\ref{tango:evaluation_methodology}). We answer \ref{tango:rq:individual_performance}, \ref{tango:rq:lcs_performance}, and \ref{tango:rq:combination_performance} based on the collected measurements. To answer \ref{tango:rq:time}  (Sec.~\ref{tango:user_study}), we conducted a user study where we measured the time humans take to find duplicates for a subset of the defined tasks, and estimated the time \tango can save for developers. We present and discuss the evaluation results in Sec.~\ref{tango:results}.

\subsection{Data Collection}
\label{tango:data_collection}

We collected video-based bug reports for six open-source Android apps, namely AntennaPod (APOD)~\cite{antennapod}, Time Tracker (TIME)~\cite{time-tracker}, Token (TOK)~\cite{token}, GNUCash (GNU)~\cite{gnucash}, GrowTracker (GROW)~\cite{growtracker}, and Droid Weight (DROID)~\cite{droidweight}. We selected these apps because they have been used in previous studies \cite{Bernal-Cardenas:ICSE'20,Chaparro:FSE'19,Moran:FSE15,Moran:ICST16}, support different app categories (finance, productivity, \etc), and provide features that involve a variety of GUI interactions (taps, long taps, swipes, \etc). Additionally, none of these apps are included as part of the RICO dataset used to train \tango's SimCLR model and Codebooks, preventing the possibility of data snooping. Since video-based bug reports are not readily available in these apps' issue trackers, we designed and carried out a systematic procedure for collecting them. 

In total, we collected 180 videos that display 60 distinct bugs -- 10 bugs for each app and three videos per bug (\ie three duplicate videos per bug). From the 60 bugs, five bugs (one bug per app except for DROID) are reported in the apps' issue trackers. These five bugs were selected
because they were the only ones in the issue trackers that we were able to reproduce based on the provided bug descriptions. During the reproduction process, we discovered five \textit{additional} new bugs in the apps not reported in the issue trackers (one bug each for APOD, GNU, and TOK, and two bugs for TIME) for a total of 10 confirmed real bugs.

The remaining 50 bugs were introduced in the apps through mutation by executing MutAPK~\cite{Escobar-Velasquez:ASE'19}, a mutation testing tool that injects bugs (\ie mutants) into Android APK binaries via a set of 35 mutation operators that were derived from a large-scale empirical study on \textit{real} Android application faults. Given the empirically-derived nature of these operators, they were shown to accurately simulate real-world Android faults~\cite{Escobar-Velasquez:ASE'19,escobar-velasquez_enabling_2022}. We applied MutAPK to the APKs of all six apps. Then, from the mutant list produced by the tool, we randomly selected 7 to 10 bugs for each app, and ensured that they could be reproduced and manifested in the GUI. To diversify the bug pool, we selected the bugs from multiple mutant operators and ensured that they affected multiple app features/screens.

When selecting the 60 bugs, we ensured they manifest graphically and were reproducible by manually replicating them on a specific Android emulator configuration (virtual Nexus 5X with Android 7.0 configured via Android Studio). For all the bugs, we screen-recorded the bug and the reproduction scenario. We also generated a textual bug report (for bugs that did not have one) containing the description of the unexpected and expected app behavior and the steps to reproduce the bug.

To generate the remaining 120 video-based bug reports, we asked two professional software engineers and eight computer science (CS) Ph.D. students to replicate and record the bugs (using the same Android emulator), based only on the textual description of the unexpected and expected app behavior.
The participants have between 2 and 10 years of programming experience (median of 6 years). 

All the textual bug reports given to the study participants contained \textit{only} a brief description of the observed and expected app behavior, with \textit{no specific reproduction steps}.
We opted to perform the collection in this manner to ensure the robustness of our evaluation dataset by maximizing the diversity of video-based reproduction steps, and simulating a real-world scenario where users are aware of the observed (incorrect) and expected app behavior, and must produce the reproduction steps themselves.

We randomly assigned the bugs to the participants in such a way that each bug was reproduced and recorded by two participants, and no participant recorded the same bug twice. Before reproducing the bugs, the participants spent five minutes exploring the apps to become familiar with their functionality. Since some of the participants could not reproduce the bugs after multiple attempts (mainly due to bug misunderstandings) and some of the videos were incorrectly recorded (due to mistakes), we reassigned these bugs among the other participants, who were able to reproduce and record them successfully.

Our bug dataset consists of 35 crashes and 25 non-crashes, and include a total of 470 steps (397 taps, 12 long taps, 14 swipes, among other types), with an average of 7.8 steps per video. The average video length is $\approx28$ seconds.

\subsection{Duplicate Detection Tasks}
\label{tango:scenarios_and_tasks}

For each app, we defined a set of tasks that mimic a realistic scenario for duplicate detection.  
Each duplicate detection task is composed of a new video (\ie the new bug report, \aka the query) and a set of existing videos (\ie existing bug reports in the issue tracker, \aka the corpus). In practice, a developer would determine if the new video is a duplicate by inspecting the corpus of videos in the order given by \tango (or any other approach). For our task setup, the corpus contains both duplicate and non-duplicate videos. There are two different types of duplicate videos that exist in the corpus: (i) those videos that are a duplicate of the query (the \textit{Same Bug} group), and (ii) those videos which are duplicates of each other, but are not a duplicate of the query (the \textit{Different Bug} group). This second type of duplicate video is represented by bug reports marked as duplicates in the issue tracker and their corresponding master reports \cite{Rakha:TSE'18,Sun2011,Chaparro2016a}.
Each non-duplicate video reports a distinct bug.

We constructed the duplicate detection tasks on a \textit{per app} basis, using the 30 video reports collected for each app (\ie three video reports for each of the 10 bugs, for a total of 30 video reports per app). We first divided all the 30 videos for an app into three groups, each group containing 10 videos (one for each bug) created by one or more participants. Then, we randomly selected a video from one bug as the query and took the other two videos that reproduce the same bug as the \textit{Same Bug} duplicate group (\ie the ground truth). Then, we selected one of the remaining nine bugs and added its three videos to the \textit{Different Bug} duplicate group. Finally, we selected one video from the remaining eight bugs, and used these as the corpus' \textit{Non-Duplicate} group. This resulted in a total of 14 distinct bug reports per task (two in the \textit{Same Bug} group, three in the \textit{Different Bug} group, eight in the \textit{Non-Duplicate} group, and the query video). After creating tasks based on all the combinations of query and corpus, we generated a total of $810$ duplicate detection tasks per app or $4,860$ aggregating across all apps.

We designed the duplicate detection setting described above to mimic a scenario likely to be encountered in crowd-sourced app testing, where duplicates of the query, other duplicates not associated with the query, and other videos reporting unique bugs, exist in a shared corpus for a given app. While there are different potential task settings, we opted not to vary this experimental variable in order to make for a feasible analysis that allowed us to explore more thoroughly the different \tango configurations.

\subsection{\tango Configurations}
\label{tango:configurations}
 
We designed \tangov and \tangot to have different configurations. \tangov's configurations are based on different visual feature extractors (SIFT or SimCLR), video sampling rates (1 and 5 fps), \# of visual words (1k, 5k, and 10k VW), and approaches to compute video similarity (BoVW, f-LCS, w-LCS, \bflcs, and \bwlcs). \tangot's configurations are based on the same sampling rates (1 and 5 fps) and the approaches to extract the text from the videos (all-text, unique-frames, and unique-words). \tangoc combines \tangov and \tangot as described in Sec.~\ref{tango:approach_combination}.

\subsection{\tango's Execution and Effectiveness Measurement}
\label{tango:evaluation_methodology}

We executed each \tango configuration on the $4,860$ duplicate detection tasks and measured its effectiveness using standard metrics used in prior text-based duplicate bug detection research~\cite{Rakha:TSE'18,Sun2011,Chaparro2016a}. For each task, we compare the ranked list of videos produced by \tango and the expected duplicate videos from the ground truth.

We measured the \textit{rank} of the first duplicate video found in the ranked list, which serves as a proxy for how many videos the developer has to watch in order to find a duplicate video. A smaller \textit{rank} means higher duplicate detection effectiveness. Based on the \textit{rank}, we computed the \textit{reciprocal rank} metric: $1/rank$. We also computed the \textit{average precision} of \tango, which is the average of the precision values achieved at all the cutting points k of the ranked list (\ie precision@k). Precision@k is the proportion of the top-k returned videos that are duplicates according to the ground truth. We also computed \textit{HIT@k} (\aka Recall Rate@k \cite{Rakha:TSE'18,Sun2011,Chaparro2016a}), which is the proportion of tasks that are successful for the cut point k of the ranked list. A task is successful if at least one duplicate video is found in the top-k results returned by \tango. We report HIT@k for cut points k = 1-2 in this dissertation, and 1-10 in our online appendix~\cite{tango_appendix}.

Additionally, we computed the average of these metrics over sets of duplicate detection tasks: mean reciprocal rank (mRR), mean average precision (mAP), and mean rank ($\mu$ rank or $\mu$Rk) per app and across all apps. Higher mRR, mAP, and HIT@k values indicate higher duplicate detection effectiveness. These metrics measure the overall performance of a duplicate detector.

We focused on comparing mRR values to decide if one \tango configuration is more effective than another, as we consider that it
better reflects the usage scenario of \tango. In practice, the developer would likely stop inspecting the suggested duplicates (given by \tango) when she finds the first correct duplicate. This scenario is captured by mRR, through the \textit{rank} metric, which considers only the first correct duplicate video as opposed to the entire set of duplicate videos to the query (as mAP does).

\subsection{Investigating \tango's Effort Saving Capabilities}
\label{tango:user_study}

We conducted a user study in order to estimate the effort that developers would spend while manually finding video-based duplicates. This effort is then compared to the effort measurements of the best \tango configuration, based on $\mu$~\textit{rank} and \textit{HIT@k}. This study and the data collection procedure were conducted remotely due to COVID-19 constraints.

\subsubsection{Participants and Tasks}
One professional software engineer and four CS Ph.D. students from the data collection procedure described in Sec.~\ref{tango:data_collection} participated in this study. The study focused on APOD, the app that all the participants had in common from the data collection. We randomly selected 20 duplicate detection tasks, covering all 10 APOD bugs. 

\subsubsection{Methodology}

Each of the 20 tasks was completed by two participants. Each participant completed four tasks, each task's query video reporting a unique bug. The assignment of the tasks to the participants was done randomly. For each task, the participants had to watch the new video (the query) and then find the videos in the corpus that showed the same bug of the new video (\ie find the duplicate videos). All the videos were anonymized so that the duplicate videos were unknown to the participants. To do this, we named each video with a number that represents the video order and the suffix ``vid'' (\eg ``2\_vid.mp4'').

The corpus videos were given in random order and the participants could watch them in any order. To make the bug of the new video clearer to the participants, we provided them with the description of the unexpected and expected app behavior, taken from the textual bug reports that we generated for the bugs. We consider the randomization of the videos as a reasonable baseline given that other baselines (\eg video-based duplicate detectors) do not currently exist and the video-based bug reports in our dataset do not have timestamps (which can be used to give a different order to the videos). This is a threat to validity that we discuss in Sec. \ref{tango:limits}.

\subsubsection{Collected Measurements}

Through a survey, we asked each participant to provide the following information for each task: (i) the name of the first video they deemed a duplicate of the query, (ii) the time they spent to find this video, (iii) the number of videos they had to watch until finding the first duplicate (including the duplicate), (iv) the names of other videos they deemed duplicates, and (v) the time they spent to find these additional duplicates. We instructed the participants to perform the tasks without any interruptions in order to minimize inaccuracies in the time measurements.

\subsubsection{Comparing \tango and Manual Duplicate Detection}

The collected measurements from the participants were compared against the effectiveness obtained by executing the best \tango configuration on the 20 tasks, in terms of $\mu$~\textit{rank} and \textit{HIT@k}. We compared the avg. number of videos the participants watched to find one duplicate against the avg. number of videos they would have watched had they used~\tango.

\section{\tango's Evaluation Results}
\label{tango:results}

\subsection{RQ1: Using Only Visual or Textual Information}

\begin{table}[t]
\centering
\small
\caption{Effectiveness for the best  \tango configurations that use either visual (SimCLR/SIFT) or textual (OCR\&IR) information.}
\label{tab:all_best_indiv}
\begin{tabular}{c|c|ccc|cc}
\toprule
\textbf{App}             & \textbf{Config.} & \textbf{mRR} & \textbf{mAP} & \textbf{\footnotesize{$\mu$Rk}} & \textbf{\footnotesize{HIT@1}} & \textbf{\footnotesize{HIT@2}} \\ \midrule
\multirow{3}{*}{\footnotesize{APOD}}    & \footnotesize{SIFT}           & 64.6\%       & 51.1\%       & 3.0             & 47.7\%       & 71.7\%       \\
                         				& \footnotesize{SimCLR}         & 80.0\%       & 66.8\%       & 1.7             & \textbf{68.1\%}       & 82.6\%       \\
                         				& \footnotesize{OCR\&IR}        & \textbf{80.8\%}       & \textbf{75.3\%}       & \textbf{1.5}             & 65.7\%       & \textbf{88.6\%}      \\ \hline
\multirow{3}{*}{\footnotesize{DROID}}   & \footnotesize{SIFT}           & 66.3\%       & 55.0\%       & 2.5             & 49.1\%       & 69.5\%       \\
                        				& \footnotesize{SimCLR}         & 64.6\%       & 59.2\%       & 2.6             & 49.5\%       & 61.7\%       \\
                        				& \footnotesize{OCR\&IR}        & \textbf{67.9\%}       & \textbf{64.7\%}       & \textbf{2.3}             & \textbf{52.0\%}       & \textbf{69.8\%}       \\ \hline
\multirow{3}{*}{\footnotesize{GNU}}     & \footnotesize{SIFT}           & 66.1\%       & 57.2\%       & 2.2             & 47.4\%       & 68.4\%       \\
                         				& \footnotesize{SimCLR}         & 81.8\%       & 75.1\%       & 1.6             & 70.1\%       & 85.3\%       \\
                         				& \footnotesize{OCR\&IR}        & \textbf{84.5\%}       & \textbf{82.3\%}       & \textbf{1.4}             & \textbf{72.2\%}       & \textbf{92.0\%}       \\ \hline
\multirow{3}{*}{\footnotesize{GROW}}    & \footnotesize{SIFT}           & 56.0\%       & 49.9\%       & 3.0             & 36.5\%       & 54.3\%       \\
                         				& \footnotesize{SimCLR}         & 72.7\%       & 68.8\%       & 2.0             & 57.4\%       & 75.6\%       \\
                         				& \footnotesize{OCR\&IR}        & \textbf{76.8\%}      & \textbf{69.0\%}       & \textbf{1.9}             & \textbf{63.6\%}       & \textbf{80.1\%}       \\ \hline
\multirow{3}{*}{\footnotesize{TIME}}    & \footnotesize{SIFT}           & 49.2\%       & 40.7\%       & 3.3             & 26.7\%       & 46.4\%       \\
                         				& \footnotesize{SimCLR}         & \textbf{74.8\%}       & \textbf{67.6\%}       & \textbf{2.3}             & \textbf{63.7\%}       & \textbf{75.9\%}       \\
                         				& \footnotesize{OCR\&IR}        & 47.4\%       & 37.7\%       & 4.0             & 28.3\%       & 44.4\%       \\ \hline
\multirow{3}{*}{\footnotesize{TOK}}     & \footnotesize{SIFT}           & 39.0\%       & 32.1\%       & 4.4             & 17.0\%       & 33.7\%       \\
                         				& \footnotesize{SimCLR}         & \textbf{77.7\%}       & \textbf{69.3\%}       & \textbf{1.6}             & \textbf{60.6\%}       & \textbf{86.7\%}       \\
                         				& \footnotesize{OCR\&IR}        & 61.3\%       & 53.3\%       & 2.6             & 42.6\%       & 60.7\%       \\ \midrule
\multirow{3}{*}{\footnotesize{\hspace{-0.1cm}\textbf{Overall}}} & \footnotesize{SIFT}           & 56.9\%       & 47.7\%       & 3.1             & 37.4\%       & 57.3\%       \\
                                        & \footnotesize{SimCLR}         & \textbf{75.3\%}       & \textbf{67.8\%}       & \textbf{1.9}             & \textbf{61.6\%}       & \textbf{78.0\%}       \\
                         				& \footnotesize{OCR\&IR}        & 69.8\%       & 63.7\%       & 2.3             & 54.1\%       & 72.6\%       \\ \bottomrule
\end{tabular}
\end{table}

We analyzed the performance of \tango 
when using only visual or textual information exclusively. In this section, we present the results for \tango's best performing configurations. However, complete results can be found in our online appendix~\cite{tango_appendix}.
Table \ref{tab:all_best_indiv} shows the results for \tangov and \tangot when using SimCLR, SIFT, as the \textit{visual feature extractor}, and OCR as the \textit{textual extractor}. For simplicity, we use SimCLR, SIFT, and OCR\&IR to refer to SimCLR-based \tangov,  SIFT-based \tangov, and \tangot, respectively. The best results for each metric are illustrated in bold on a per app basis.
The results provided in Table \ref{tab:all_best_indiv} are those for the best parameters of the SimCLR, SIFT, and OCR\&IR feature extractors, which are (BoVW, 5 fps, 1k VW), (w-LCS, 1 fps, 10k VW), and (all-text, 5 fps), respectively.

Table \ref{tab:all_best_indiv} shows that \tangov is more effective when using SimCLR rather than SIFT across all the apps, achieving an overall mRR, mAP, avg. rank, HIT@1, and HIT@2 of 75.3\%, 67.8\%, 1.9, 61.6\%, and 78\%, respectively. SimCLR is also superior to OCR\&IR overall, whereas SIFT performs least effectively of the three approaches.
When analyzing the results per app, we observe that SimCLR is outperformed by OCR\&IR (by 0.7\% - 4\% difference in mRR) for APOD, DROID, GNU and GROW; with OCR\&IR being the most effective for these apps. SimCLR outperforms the other two approaches for TIME and TOK by more than 16\% difference in mRR. The differences explain the overall performance of SimCLR and OCR\&IR. SimCLR is more consistent in its performance compared to OCR\&IR and SIFT. Across apps, the mRR standard deviation of SimCLR is 6.2\%, which is lower than that for SIFT and OCR\&IR: 11.1\% and 13.9\%, respectively. The trend is similar for mAP and avg. rank. 

Since the least consistent approach across apps is \tangot in terms of effectiveness, we investigated the root causes for its lower performance on TIME and TOK.  After manually watching a subset of the videos for these apps, we found that their textual content was quite similar across bugs.
Based on this, we hypothesized that the amount of vocabulary shared between duplicate videos (from the same bugs) and non-duplicate videos (across different bugs) affected the discriminatory power of Lucene-based \tangot (see Sec.~\ref{tango:approach_textual}).

To verify this hypothesis, we measured the shared vocabulary of duplicate and non-duplicate video pairs, similarly to Chaparro \etal's analysis of textual bug reports \cite{Chaparro2016a}. We formed unique pairs of duplicate and non-duplicate videos from all the videos collected for all six apps. For each app, we formed 30 duplicate and 405 non-duplicate pairs, and we measured the avg. amount of shared vocabulary of all pairs, using the vocabulary agreement metric used by Chaparro \etal~\cite{Chaparro2016a}. Table \ref{tab:vocab_agreement} shows the vocabulary agreement of duplicate ($V_d$) and non-duplicate pairs ($V_{nd}$) as well as the mRR and mAP values of  \tangot  for each app. The table reveals that the vocabulary agreement of duplicates and non-duplicates is very similar for TIME and TOK, and dissimilar for the other apps. The absolute difference between these measurements (\ie $|V_d - V_{nd}|$) for TIME and TOK is 0.3\% and 8.6\%, while for the other apps it is above 16\%. We found 0.94 / 0.91 Pearson correlation~\cite{Freedman:07} between these differences and the mRR/mAP values. 

The results indicate that, for TIME and TOK, the similar vocabulary between duplicate and non-duplicate videos negatively affects the discriminatory power of \tangot, which suggests that for some apps, using only textual information may be sub-optimal for duplicate detection.

\begin{table}[t]
\centering
\small
\caption{Vocabulary agreement \& effectiveness for the best \tangot.} 
\label{tab:vocab_agreement}
\begin{tabular}{c|ccc|cc}
\toprule
\multirow{2}{*}{\textbf{App}} & \multicolumn{3}{c|}{\textbf{Vocabulary agreement}} & \multirow{2}{*}{\textbf{mRR}} & \multirow{2}{*}{\textbf{mAP}} \\ \cline{2-4}
        & \textbf{$V_d$} & \textbf{$V_{nd}$} & \textbf{$|V_d - V_{nd}|$} &        &        \\ 
\midrule
APOD    & 70.8\%         & 37.9\%            & 32.9\%                    & 80.8\% & 75.3\% \\
DROID   & 73.9\%         & 57.0\%            & 16.9\%                    & 67.9\% & 64.7\% \\
GNU     & 82.2\%         & 58.6\%            & 23.6\%                    & 84.5\% & 82.3\% \\
GROW    & 67.0\%         & 41.7\%            & 25.4\%                    & 76.8\% & 69.0\% \\
TIME    & 86.0\%         & 86.3\%            & 0.3\%                     & 47.4\% & 37.7\% \\
TOK     & 69.6\%         & 61.0\%            & 8.6\%                     & 61.3\% & 53.3\% \\ 
\midrule
\textbf{Overall} & 74.2\%         & 56.7\%            & 17.5\%                    & 69.8\% & 63.7\% \\ 
\bottomrule
\end{tabular}
\end{table}

\begin{tcolorbox}[enhanced,skin=enhancedmiddle,borderline={1mm}{0mm}{MidnightBlue}]
\textbf{Answer for \ref{tango:rq:individual_performance}}: SimCLR performs the best overall with an mRR and HIT@1 of 75.3\% and 61.6\%, respectively. For 4 of 6 apps, OCR\&IR outperforms SimCLR by a significant margin. However, due to issues with vocabulary overlap, it performs worse overall. SIFT is the worst-performing technique across all the apps.
\end{tcolorbox}

\subsection{RQ2: Combining Visual and Frame Sequence Information}

To answer \ref{tango:rq:lcs_performance}, we compared the effectiveness of the best configuration of \tango when using visual information alone (SimCLR, BoVW, 5fps, 1k VW) and when combining visual \& frame sequence information~(\ie \bflcs and~\bwlcs).

The results are shown in Table \ref{tab:all_best_lcs}. Overall, using \tango with BoVW alone is more effective than combining the approaches; \tango based on BoVW achieves 75.3\%, 67.8\%, 1.9, 61.6\%, and 78\% mRR, mAP, avg. rank, HIT@1, and HIT@2, respectively. Using BoVW and w-LCS combined is the least effective approach. BoVW alone and \bflcs are comparable in performance. However, BoVW is more consistent in its performance across apps: 6.2\% mRR std. deviation vs. 6.6\% and 9.2\% for  \bflcs and \bwlcs.

The per-app results reveal that \bwlcs consistently is the least effective approach for all apps except for GROW, for which \bwlcs performs best. After watching the videos for GROW, we found unnecessary steps in the beginning/middle of the duplicate videos, which led to their endings being weighted more heavily by w-LCS, where steps were similar. In contrast, BoVW and \bflcs give a lower weight to these cases thus reducing the overall video similarity. 

The lower performance of  \bflcs and \bwlcs, compared to BoVW, is partially explained by the fact that f-LCS and w-LCS are more restrictive by definition. Since they find the longest common sub-strings of frames between videos, small variations (\eg extra steps) in the reproduction steps of the bugs may lead to drastic changes in similarity measurement for these approaches. Also, these approaches only find one common substring (\ie the longest one), which may not be highly discriminative for duplicate detection. In the future, we plan to explore additional approaches for aligning the frames, for example, by using an approach based on longest common sub-sequence algorithms \cite{Gusfield:LCS'97} that can help align multiple portions between videos. Another potential reason for these results may lie in the manner that \tango combines visual and sequential similarity scores -- weighting both equally. In future work, we plan to explore additional combination techniques.

\begin{table}[t]
\centering
\small
\caption{Effectiveness for the best \tangov configuration using either visual information (BoVW) or a combination of visual and frame sequence information (\bflcs and \bwlcs).}
\vspace{-0.5em}
\label{tab:all_best_lcs}
\begin{tabular}{c|c|ccc|cc}
\toprule
\textbf{App}             & \textbf{Config.} & \textbf{mRR} & \textbf{mAP} & \textbf{\footnotesize{$\mu$Rk}} & \textbf{\footnotesize{HIT@1}} & \textbf{\footnotesize{HIT@2}} \\ \midrule
\multirow{3}{*}{\footnotesize{APOD}}    & \footnotesize{\bflcs}      & 79.3\%       & \textbf{67.8\%}      & \textbf{1.7}             & 66.2\%       & 82.3\%           \\
                         				& \footnotesize{\bwlcs}      & 77.2\%       & 65.5\%      & 1.9             & 64.2\%       & 80.1\%      \\
                         				& \footnotesize{BoVW}        & \textbf{80.0\%}       & 66.8\%      & \textbf{1.7}             & \textbf{68.1\%}       & \textbf{82.6\%}    \\ \hline
\multirow{3}{*}{\footnotesize{DROID}}   & \footnotesize{\bflcs}      & \textbf{64.8\%}       & \textbf{60.7\%}      & \textbf{2.6}             & \textbf{50.2\%}       & 61.6\%       \\
                         				& \footnotesize{\bwlcs}      & 63.7\%       & 54.8\%      & 2.7             & 48.9\%       & 62.3\%        \\
                         				& \footnotesize{BoVW}        & 64.6\%       & 59.2\%      & \textbf{2.6}             & 49.5\%       & \textbf{61.7\%}     \\ \hline
\multirow{3}{*}{\footnotesize{GNU}}     & \footnotesize{\bflcs}      & \textbf{83.3\%}       & \textbf{75.6\%}      & \textbf{1.6}             & \textbf{73.2\%}       & \textbf{85.6\%}       \\
                         				& \footnotesize{\bwlcs}      & 77.3\%       & 65.7\%      & 1.8             & 62.3\%       & 83.6\%          \\
                         				& \footnotesize{BoVW}        & 81.8\%       & 75.1\%      & \textbf{1.6}             & 70.1\%       & 85.3\%        \\ \hline
\multirow{3}{*}{\footnotesize{GROW}}    & \footnotesize{\bflcs}      & 76.0\%       & 70.2\%      & 2.0             & 64.2\%       & 75.2\%        \\
                         				& \footnotesize{\bwlcs}      & \textbf{81.3\%}       & \textbf{75.0\%}      & \textbf{1.7}             & \textbf{70.9\%}       & \textbf{82.8\%}         \\
                         				& \footnotesize{BoVW}        & 72.7\%       & 68.8\%      & 2.0             & 57.4\%       & 75.6\%       \\ \hline
\multirow{3}{*}{\footnotesize{TIME}}    & \footnotesize{\bflcs}      & 70.4\%       & 63.4\%      & \textbf{2.3}             & 54.4\%       & 74.3\%        \\
                         				& \footnotesize{\bwlcs}      & 63.8\%       & 58.5\%      & 2.8             & 48.0\%       & 64.9\%        \\
                         				& \footnotesize{BoVW}        & \textbf{74.8\%}       & \textbf{67.6\%}      & \textbf{2.3}             & \textbf{63.7\%}       & \textbf{75.9\%}        \\ \hline
\multirow{3}{*}{\footnotesize{TOK}}     & \footnotesize{\bflcs}      & 73.4\%       & 65.6\%      & 1.7             & 54.0\%       & 82.5\%       \\
                         				& \footnotesize{\bwlcs}      & 59.2\%       & 53.7\%      & 2.6             & 37.9\%       & 60.0\%       \\
                         				& \footnotesize{BoVW}        & \textbf{77.7\%}       & \textbf{69.3\%}      & \textbf{1.6}             & \textbf{60.6\%}       & \textbf{86.7\%}         \\ \midrule
\multirow{3}{*}{\footnotesize{\hspace{-0.1cm}\textbf{Overall}}} & \footnotesize{\bflcs}      & 74.5\%       & 67.2\%      & 2.0             & 60.4\%       & 76.9\%        \\
                         				& \footnotesize{\bwlcs}      & 70.4\%       & 62.2\%      & 2.2             & 55.4\%       & 72.3\%        \\
                         				& \footnotesize{BoVW}        & \textbf{75.3\%}       & \textbf{67.8\%}      & \textbf{1.9}             & \textbf{61.6\%}       & \textbf{78.0\%}      \\ \bottomrule
\end{tabular}
\end{table}

\begin{tcolorbox}[enhanced,skin=enhancedmiddle,borderline={1mm}{0mm}{MidnightBlue}]
\textbf{Answer for \ref{tango:rq:lcs_performance}}: Combining  ordered visual information (via f-LCS and w-LCS) with the orderless BoVW improves the results for four of the six apps. However, across all apps, BoVW performs more consistently.
\end{tcolorbox}

\subsection{RQ3: Combining Visual and Textual Information}
\label{tango:results_combined}

\begin{table}[t]
\centering
\small
\caption{Effectiveness of the best \tangoc, \tangov, and \tangot.}
\vspace{-0.5em}
\label{tab:combined_best}
\begin{tabular}{c|c|ccc|cc}
\toprule
\textbf{App}             & \textbf{Config.} & \textbf{mRR} & \textbf{mAP} & \textbf{\footnotesize{$\mu$Rk}} & \textbf{\footnotesize{HIT@1}} & \textbf{\footnotesize{HIT@2}} \\ \midrule
\multirow{3}{*}{\footnotesize{APOD}}  & \footnotesize{\tangoc} 	   & \textbf{84.4\%}       & \textbf{75.8\%}       & \textbf{1.4}             & \textbf{73.1\%}       & 87.9\%         \\ 
                       				  & \footnotesize{\tangov}       & 80.0\%       & 66.8\%       & 1.7             & 68.1\%       & 82.6\%          \\ 
                      			      & \footnotesize{\tangot}       & 80.8\%       & 75.3\%       & 1.5             & 65.7\%       & \textbf{88.6\%}          \\ \hline
\multirow{3}{*}{\footnotesize{DROID}} & \footnotesize{\tangoc} 	   & \textbf{70.6\%}       & \textbf{66.7\%}       & \textbf{2.2}             & \textbf{55.9\%}       & \textbf{71.0\%}         \\ 
                       				  & \footnotesize{\tangov}       & 64.6\%       & 59.2\%       & 2.6             & 49.5\%       & 61.7\%        \\ 
                       				  & \footnotesize{\tangot}       & 67.9\%       & 64.7\%       & 2.3             & 52.0\%       & 69.8\%          \\ \hline
\multirow{3}{*}{\footnotesize{GNU}}   & \footnotesize{\tangoc} 	   & \textbf{89.5\%}       & \textbf{84.7\%}       & \textbf{1.3}             & \textbf{81.6\%}       & \textbf{94.2\%}         \\ 
                       			      & \footnotesize{\tangov}       & 81.8\%       & 75.1\%       & 1.6             & 70.1\%       & 85.3\%          \\ 
                       				  & \footnotesize{\tangot}       & 84.5\%       & 82.3\%       & 1.4             & 72.2\%       & 92.0\%           \\ \hline
\multirow{3}{*}{\footnotesize{GROW}}  & \footnotesize{\tangoc} 	   & \textbf{81.7\%}       & \textbf{75.4\%}       & \textbf{1.7}             & \textbf{71.4\%}       & \textbf{82.5\%}          \\ 
                       				  & \footnotesize{\tangov}       & 72.7\%       & 68.8\%       & 2.0             & 57.4\%       & 75.6\%          \\ 
                       			 	  & \footnotesize{\tangot}       & 76.8\%       & 69.0\%       & 1.9             & 63.6\%       & 80.1\%          \\ \hline
\multirow{3}{*}{\footnotesize{TIME}}  & \footnotesize{\tangoc} 	   & 59.6\%       & 51.7\%       & 2.8             & 40.2\%       & 58.8\%          \\ 
                      				  & \footnotesize{\tangov}       & \textbf{74.8\%}       & \textbf{67.6\%}       & \textbf{2.3}             & \textbf{63.7\%}       & \textbf{75.9\%}        \\ 
                      				  & \footnotesize{\tangot}       & 47.4\%       & 37.7\%       & 4.0             & 28.3\%       & 44.4\%       \\ \hline
\multirow{3}{*}{\footnotesize{TOK}}   & \footnotesize{\tangoc} 	   & 69.8\%       & 60.8\%       & 2.0             & 50.9\%       & 76.9\%            \\ 
                     				  & \footnotesize{\tangov}       & \textbf{77.7\%}       & \textbf{69.3\%}       & \textbf{1.6}             & \textbf{60.6\%}       & \textbf{86.7\%}          \\ 
                     				  & \footnotesize{\tangot}       & 61.3\%       & 53.3\%       & 2.6             & 42.6\%       & 60.7\%          \\ \midrule
\multirow{3}{*}{\footnotesize{\hspace{-0.1cm}\textbf{Overall}}}   & \footnotesize{\tangoc}   & \textbf{75.9\%}       & \textbf{69.2\%}       & \textbf{1.9}             & \textbf{62.2\%}       & \textbf{78.5\%}           \\ 
                      				  & \footnotesize{\tangov}       & 75.3\%       & 67.8\%       & \textbf{1.9}             & 61.6\%       & 78.0\%            \\ 
                     				  & \footnotesize{\tangot}       & 69.8\%       & 63.7\%       & 2.3             & 54.1\%       & 72.6\%           \\
                       \bottomrule
\end{tabular}
\end{table}

We investigated \tango's effectiveness when combining visual and textual information. We selected the best configurations of \tangov (SimCLR, BoVW,  5  fps,  1k  VW) and \tangot (all-text, 5 fps) from \ref{tango:rq:individual_performance} based on their mRR score and measured its performance overall and per app. We provide the results for the best weight we obtained for \tango's \textit{similarity computation and ranking} which was $w=0.2$, \ie a weight of 0.8 and 0.2 on \tangov and \tangot, respectively. These weights were found by evaluating different $w$ values from zero ($0$) to one ($1$) in increments of $0.1$ and selecting the one leading to the highest overall mRR score. Complete results can be found in our online appendix \cite{tango_appendix}.

Table \ref{tab:combined_best} shows that the overall effectiveness achieved by \tangoc is higher than that achieved by \tangot and \tangov. \tangoc achieves 75.9\%, 69.2\%, 1.9, 62.2\%, and 78.5\% mRR, mAP, avg. rank, HIT@1, and HIT@2, on average. The avg. improvement margin of \tangoc is substantially higher for \tangot (6.2\%/5.5\% mRR/mAP) than for \tangov  (0.7\%/1.4\% mRR/mAP). 

Our analysis of the per-app results explains these differences. Table \ref{tab:combined_best} reveals that combining visual and textual information substantially increases the performance over just using one of the information types alone, except for the TIME and TOK apps. This is because \tangot's effectiveness is substantially lower for these apps, compared to the visual version (see Table \ref{tab:all_best_indiv}), due to the aforementioned vocabulary agreement. Thus, incorporating the textual information significantly harms the performance of \tangoc.

\subsubsection{A Better Combination of Visual and Textual Information}
  
The results indicate that combining visual and textual information is beneficial for most of our studied apps but harmful for a subset (TIME and TOK). This is because the textual information used alone, for TIME and TOK, leads to low performance. The analysis we made for \tangot in \ref{tango:rq:individual_performance}, revealed that the reason for the low performance of \tangot lies in the similar amount of vocabulary overlap between duplicate and non-duplicate videos. Fortunately, based on this amount of vocabulary, we can predict the performance of \tangot for new video-based bug reports as follows \cite{Chaparro2016a}. In practice, the issue tracker will contain reports marked as duplicates (reporting the same bugs) from previous submissions of bug reports as well as non-duplicates (reporting unique bugs). This information can be used to compute the vocabulary agreement between duplicates and non-duplicates, which can be used to predict how well \tangot would perform for new reports.

Based on this, we defined a new approach for \tango, which is based on the vocabulary agreement metric from \cite{Chaparro2016a} applied on existing duplicate and non-duplicate reports. This approach dictates that if the difference of vocabulary agreement between existing duplicates and non-duplicates is greater than a certain threshold, then \tango should combine visual and textual information. Otherwise, \tango should only use the visual information because it is likely that the combination would not be better than using the visual information alone.

From the vocabulary agreement measurements shown in Table \ref{tab:vocab_agreement}, we infer a proper threshold from the new \tango approach. This threshold may be taken as one value from the interval 8.6\% - 16.9\% (exclusive) because those are the limits that separate the apps for which \tangot obtains low (TIME and TOK) and high performance (APOD, DROID, GNU, and GROW). For practical reasons, we select the threshold to be the middle value: $8.6 + (16.9 - 8.6)/2 = 12.8\%$. In future work, we plan to further evaluate this threshold on other apps.

We implemented this approach for \tango, using 0.2 as weight, and measured its effectiveness. This approach resulted in a mRR, mAP, avg. rank, HIT@1 and HIT@2 of 79.8\%, 73.2\%, 1.7, 67.7\%, and 83\%, respectively. The approach leads to a substantial improvement (\ie 3.9\% / 4.1\% higher mRR / mAP) over \tangoc shown in Table~\ref{tab:combined_best}.

The results mean that the best version of \tango is able to suggest correct duplicate video-based bug reports in the first or second position of the returned candidate list for 83\% of the duplicate detection tasks.

\begin{tcolorbox}[enhanced,skin=enhancedmiddle,borderline={1mm}{0mm}{MidnightBlue}]
\textbf{Answer for \ref{tango:rq:combination_performance}}: Combining visual and textual information significantly improves results for 4 of 6 apps. However, due to the vocabulary agreement issue, across all apps, this approach is similar in effectiveness to using visual information alone. Accounting for this vocabulary overlap issue through a selective combination of visual and textual information via a threshold, \tango achieves the highest effectiveness: an mRR, mAP, avg. rank, HIT@1, and HIT@2 of 79.8\%, 73.2\%, 1.7, 67.7\%, and 83\%, respectively.
\end{tcolorbox}

\subsection{RQ4: Time Saved Discovering Duplicates}
\label{tango:results_user_study}

As expected, the participants were successful in finding the duplicate videos for all 20 tasks. In only one task, one participant incorrectly flagged a video as duplicate because it was highly similar to the query. Participants found the first duplicate video in 96.4 seconds and watched 4.3 videos on avg. across all tasks to find it. Participants also found all the duplicates in 263.8 seconds on avg. by watching the entire corpus of videos. This means they spent 20.3 seconds in watching one video on average.

We compared these results with the measurements taken from \tango's best version (\ie selective \tango) on the tasks the participants completed. \tango achieved a 1.5 avg. rank, which means that, by using \tango, they would only have to watch  one or two videos on avg. to find the first duplicate. This would have resulted in $(4.3-1.5) / 4.3 =65.1\%$ of the time saved. In other words, instead of spending $20.3 \times 4 = 81.2$ seconds (on avg.) finding a duplicate for a given task, the participants could have spent $20.3 \times 1.5 = 30.5$ seconds. These results indicate the potential of \tango to help developers save time when finding duplicates.

\begin{tcolorbox}[enhanced,skin=enhancedmiddle,borderline={1mm}{0mm}{MidnightBlue}]
\textbf{Answer for \ref{tango:rq:time}}: On average, \tango's best-performing configuration can save 65.1\% of the time participants spend finding duplicate videos.
\end{tcolorbox}

\section{\tango Limitations \& Threats to Validity}\label{tango:limits}

\textbf{Limitations.} \tango has three main limitations that motivate future work. 
 
The first one stems from the finding that textual information may not be beneficial for some apps. The best \tango version implements an approach for detecting this situation, based on a threshold for the difference in vocabulary overlap between duplicate and non-duplicate videos, which is used for selectively combining visual or textual information.
This threshold is based on the collected data and may not generalize to other apps. Second, 
the visual TF-IDF representation for the videos is based on the mobile app images from the RICO dataset, rather than on the videos found in the tasks' corpus, as it is typically done in text retrieval. Additionally, we considered single images as documents rather than groups of frames that  make up a video. These decisions were made to improve the generalization of \tango's visual features and to support projects that have limited training data. Third, differences in themes and languages across video-based bug reports for an application could have an impact in the performance of \tango. We believe that different themes (\ie dark vs. light modes) will not significantly impact \tango since the SimCLR model is trained to account for such color differences by performing color jittering and gray-scaling augmentations. However, additional experiments are needed to validate this conjecture. For different languages, \tango currently assumes the text in an application to be English when performing OCR and textual similarity. Therefore, its detection effectiveness where the bug reports display different languages (\eg English vs. French) could be negatively impacted. We will investigate this aspect in our future work.

\textbf{Internal \& Construct Validity.} 
Most of the mobile app bugs in our dataset were introduced by MutAPK \cite{Escobar-Velasquez:ASE'19}, and hence potentially may not resemble real bugs. However, MutAPK's mutation operators were empirically derived from a large-scale study of real Android faults, and prior research lends credence of the ability of mutants to resemble real faults~\cite{Andrews:ICSE05}. We intentionally selected generated mutants from a range of operators to increase the diversity of our set of bugs and mitigate potential biases. Another potential threat is related to using real bugs from issue trackers that cannot be reproduced or that do not manifest graphically. We mitigated this threat by using a small, carefully vetted subset of real bugs that were analyzed by multiple authors before being used in our dataset. 
We did not observe major differences in the results between mutants and real bugs.

Another threat to validity is that our approach to construct the duplicate detection tasks does not take into account bug report timestamps, which would be typical in a realistic scenario~\cite{Rakha:TSE'18}, and timestamps could be used as a baseline ordering of videos for comparing against the ranking given by \tango. The lack of timestamps stems from the fact that we were not able to collect the video-based bug reports from existing mobile projects. We mitigated this threat in our user study by randomizing the ordering of the corpus videos given to the participants. We consider this as a reasonable baseline for evaluating our approach considering that, to the best of our knowledge, (1) no existing datasets, with timestamps, are available for conducting research on video-based duplicate detection, and (2) no existing duplicate detectors work exclusively on video-based bug reports, as \tango does.

\textbf{External Validity.} We selected a diverse set of apps that have different functionality, screens, and visual designs, in an attempt to mitigate threats to external validity. Additionally, our selection of bugs also attempted to select diverse bug types (crashes and non-crashes), and the duplicate videos were recorded by different participants. As previously discussed, there is the potential that \tango's different parameters \& thresholds may not generalize to video data from other apps.




\section{Bibliographical Notes}

The paper supporting the content described in this Chapter was written in collaboration with other members of the SEMERU group at William \& Mary and a researcher from George Mason University. I have received permission from the publisher and co-authors to reprint sections of this work:

\textbf{Cooper, N.}, Bernal-C\'{a}rdenas, C., Chaparro, O., Moran, K., \& Poshyvanyk, D. (2021, May). It takes two to tango: Combining visual and textual information for detecting duplicate video-based bug reports. In 2021 IEEE/ACM 43rd International Conference on Software Engineering (ICSE) (pp. 957-969). IEEE.

\chapter{Impact Analysis}
\label{sec:athena}
Modern software systems are long-lived, with extensive development and maintenance histories. 

Projects experience churn in the developers or teams working on them, and can consist of millions of lines of code. Even understanding the potential cascading impacts of seemingly simple code changes can be a difficult proposition. As such, the premise of impact analysis (IA) is that a given change may result in \textit{undesirable side effects}, such as a fault that leads to an erroneous program state, caused by unintended interactions between the changes and other parts of a software system~\cite{kagdi2013integrating}. Thus, the task of IA involves estimating an impact set of entities, usually classes or methods of a software system, from a given change to an entity, also usually a class or a method~\cite{Bohner&Arnold1996b}. This process can be cognitively challenging for developers, as reasoning about complex interactions of a software system requires careful comprehension of large volumes of code. Given that many important engineering and maintenance tasks -- such as bug fixing and refactoring -- require code changes, they necessarily require IA as well.  This process is typically performed \textit{manually} by developers, but given its complexity, researchers have proposed a range of approaches for automating it.

The most recent approaches which aim to automate IA utilize coupling metrics between parts of a software system as a proxy for how likely a change in one part will affect another~\cite{briand1999using}. Popular coupling metrics include evolutionary/logical coupling~\cite{gall2003cvs, zimmermann2005mining} where change histories are used to inform future changes or conceptual coupling~\cite{poshyvanyk2009using, revelle2011using} where underlying textual similarities among code elements are used to estimate impact sets. This formulation has also been found to be better aligned with how developers perceive coupling~\cite{bavota2013empirical}. However, making use of evolutionary information requires careful mining of change histories, and past work that measures textual similarity across code entities make use of textual representations that are limited in their ability to accurately assess the semantic similarity of terms and identifiers used in source code \cite{Scalabrino2006,ReadabilityJSME}. Other techniques have made use of static~\cite{tip1994survey, gallagher1990using, zhao2002change, korpi2007supporting, vidacs2007macro, santelices2010probabilistic} or dynamic~\cite{arisholm2004dynamic, hassoun2004dynamic} analyses to aid in impact set estimation. However, static analysis techniques tend to suffer from trade-offs in the soundness and completeness of their analyses~\cite{Bonett:USENIX18,osti_10334413,9402554} and running programs to collect dynamic information can be prohibitively expensive, particularly for large projects.

Recently, Deep Learning has seen great success in code generation and understanding tasks such as code search \cite{husain2019codesearchnet, feng2020codebert, guo2020graphcodebert, guo2022unixcoder}, code completion \cite{ciniselli2021empirical, chen2021evaluating, austin2021program, fried2022incoder}, bug fixing \cite{gupta2017deepfix, gupta2019deep, tufano2019learning, yin2018learning}, clone detection \cite{White2016, Liu2018c, buch2019learning, perez:msr19, Saini2018, Tufano2018a, yu2019neural, zhang2019novel, zhao2018deepsim}, \etc. Generally, these approaches learn code semantics from unimodal code-only data or bimodal (comment, code pair) data and map code snippets to high dimensional vector representations that can be used to automate downstream code-related tasks. However, despite their success, none of these approaches have been applied to the task of IA. Therefore, in this dissertation, we aim to explore the potential of adapting three representative transformer-based \cite{vaswani2017attention} models \cite{feng2020codebert, guo2020graphcodebert, guo2022unixcoder} which utilize their corresponding semantically rich code representations to advance automated approaches for IA.

However, there are at least two issues that complicate the application of neural models for source code to the task of impact analysis. First, we currently lack extremely large-scale, vetted datasets or benchmarks that would allow a neural model to directly learn the task of impact analysis. This is due to the fact that deriving an IA dataset requires substantial human effort, as impact sets cannot be easily mined from software repositories without manual validation. Despite this, neural embeddings could be used to calculate semantic similarities between code entities to estimate conceptual coupling. However, while such an approach would already represent an advancement over past work, it ignores the context the code finds itself in, \ie how the code is used within a software system. Recent work has shown this information to be important to developers when determining relevance of retrieved code snippets for code search \cite{husain2019codesearchnet}, a similar task to IA.

To overcome these limitations, and advance the task of automated impact analysis, we introduce \athena, a DL-based approach to IA that integrates \textit{both} rich neural representations of code snippets and structural information about a given code snippet's relationships with other entities in a project using call graph information. Specifically, \athena builds a software system's call graph, where nodes represent methods and edges represent the call dependencies between methods. This call graph information is then used to aggregate the neural representations of the methods using an Embedding Propagation strategy inspired by work on Graph Convolutional Networks (GCNs) \cite{kipf2016semi} that requires no additional training. In this way, \athena not only utilizes the local code semantics within the methods, but also utilizes the information of global call dependencies for the task of IA.  More importantly, our technique does not require a specialized training procedure, and makes use of existing neural language models for code that can be trained in an unsupervised manner on massive datasets of code mined at scale.



Evaluating our proposed approach effectively also presents challenges. Existing IA benchmarks have been found to contain tangled commits resulting in inaccurate impact sets~\cite{herbold2022fine}. Additionally, they tend to be small, \ie less than ten software systems. Therefore, to evaluate \athena for the task of IA, we created a large-scale IA benchmark, called {\sc Alexandria}, that leverages an existing dataset of fine-grained, manually untangled commit information from bug-fixes~\cite{herbold2022fine}. The benchmark consists of \imcommits commits across \imrepos open-source Java projects, which we use to construct 4,405 IA tasks -- where each task consists of a query method and a set of impacted methods. Using standard information retrieval metrics of mRR, mAP, and HIT@10, we find \athena significantly (based on statistical tests) improve over the neural semantic-only baseline without call graph information by 4.58\%-5.32\%, 3.85\%-4.62\%, and 5.67\%-7.51\%, respectively. 

\noindent In aggregate, we make the following contributions:

\begin{itemize}[leftmargin=1.2em]
    \item A new large-scale evaluation benchmark for impact analysis, called {\sc Alexandria}, composed of 4,405 IA tasks from 910 commits of \imrepos open source software systems;
    \item \athena, a novel approach to automated impact analysis that utilizes call graph information as well as neural code semantics to estimate impact sets; 
    \item A thorough set of ablations and qualitative analyses showing the improved performance is attributable to integrating call-graph information;
    \item{A comprehensive online appendix ~\cite{athena-tool} that contains the code for \athena, our IA benchmark {\sc Alexandria} and our experimental infrastructure to allow for the replication of this work to help foster future work that aims to advance automated IA.}
\end{itemize}

\section{Athena's Approach}\label{athena:approach}

\begin{figure}[t]
	\centering %
	\caption{\normalsize{Overview of the Workflow of the Athena Impact Analysis Approach}}
	\includegraphics[width=\textwidth]{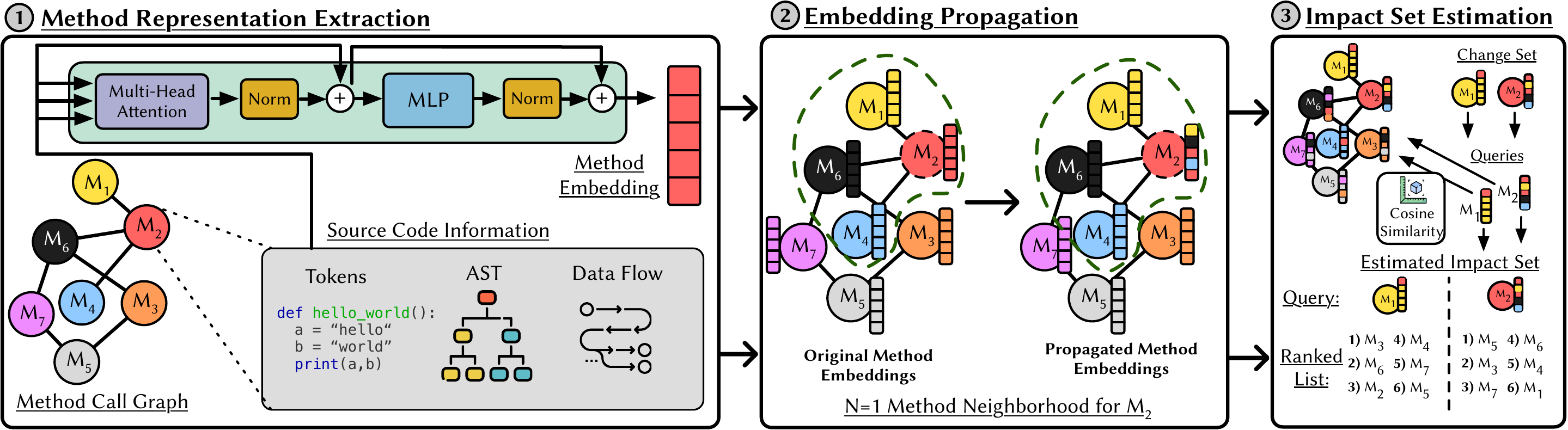}
	\label{fig:athena}
\end{figure}

We formulate impact analysis as an information retrieval task where if a developer intends to modify a method (\ie query method) in a software system, \athena will return a ranked list of other methods being potentially impacted in descending order of likelihood. All methods but the query are used as the search corpus. Formally, for a software system $S$ containing a set of methods $S = \{ m_1, m_2, ..., m_n \}$, a change to one of the methods $m_i \in S$ triggers \athena to rank all other methods thus estimating the impact set.

Figure~\ref{fig:athena} provides an overview of the \athena approach. \athena first builds a call graph generator to aid in capturing the call dependencies among all methods across an entire software system in which the nodes represent the methods, and the edges represent the call dependencies between methods. All methods are then converted into method tokens which are processed by a state-of-the-art neural-based models, \eg CodeBERT, GraphCodeBERT and UniXcoder, fine-tuned on the code search task, to extract the method representations. Next, \athena analyzes the global call dependencies and propagates information from the neural embeddings of "neighbor" nodes in a method call graph to a given target method. Therefore, the initial representation vectors are updated based on a propagation strategy inspired by Graph Convolutional Networks (GCNs)~\cite{kipf2016semi}  so that each method node incorporates the contextual information from its neighboring methods. The cosine similarity between the updated representations of a given query method and each method in the corpus is computed to obtain a final ranked list. We will now discuss each step of \athena in detail.

\subsection{Software System Call Graph Generator}

The first step of \athena is to build a call graph to capture method caller-callee dependencies across a software system. Call graphs can be generated either statically or dynamically. Given that dynamic approaches may result in incomplete information~\cite{xie2002empirical} as they usually analyze only a small set of execution paths (and also require extensive test suites and/or manual construction of scenarios for execution trace collection), \athena uses static analysis to build more complete call-graphs in a more efficient manner. Since existing tools such as DOOP ~\cite{bravenboer2009strictly} and WALA ~\cite{fink2012wala} only analyze the call dependencies within a given package but ignore the dependencies across packages, we created our own tool for generating call graphs that covers all call dependencies, even across packages, without requiring tests and only using the code files of a system as input.

Algorithm \ref{alg:callgraph} provides an overview of our call graph generation process. For each \texttt{\small .java} production source file of the repository, \athena identifies all methods, imported packages, and all method invocation statements by using the Tree-Sitter \cite{tree_sitter} library, which builds a concrete syntax tree for each file and supports searching for various patterns in the tree. The type of the return values and the method arguments are not parsed to save time and ensure scalability of the approach. Next, we resolve each method invocation statement in each file to obtain the index of the caller method and the information needed for finding the callee by analyzing the file in a bottom-up manner using the Tree-sitter query syntax. To handle the inheritance relationship between classes, if the callee is not found in the class based on the given file path with the class name, the method will be further searched in its extended class. We use both the method name and \# arguments (rather than the complete signature) to identify each method. Thus, some overloaded methods cannot be uniquely identified. Therefore, the edges are added between the caller and each of the overloaded callee methods if they have the same method name and \# arguments. For nested method calls, only the outermost call is resolved and the others are discarded as return value types are not resolved.

\begin{algorithm}[h]
    \footnotesize
    \SetAlgoLined 
    \KwIn{A software repository $R$} 
    \KwOut{A call graph representing method invocations in the project} 
    $V \leftarrow \varnothing, I \leftarrow \varnothing, C \leftarrow \varnothing, E \leftarrow \varnothing $  
    
    \ForEach{$file \in R$}{ 
        \If{$file \notin \textit{test files}$}{ 
            \tcc{Identify all methods, imported classes/packages, and calls in the file} 
            
            $methods, imports, calls = fileContents(file)$
            
            $V = V \cup methods, I = I \cup imports, C = C \cup calls$ 
        } 
    } 

    \ForEach{$file \in R$}{ 
        \ForEach{$call \in calls$}{ 
            \tcc{Identify the caller index, and the import module, class name, method name, \# arguments of callee} 
            
            $caller_{idx}, name_{cls}, name_{mthd}, n_{args}, file_{path} = callInfo(M, I, call, file)$             
            \tcc{Find the callee in the software based on the file path, class name, method name, \# arguments}  
            $called_{idx} = callInfo(path_{file}, name_{cls}, name_{mthd}, n_{args}, M)$
            
            $E = E \cup (caller_{idx}, callee_{idx})$  
        } 
    } 
    \Return{$G = (V, E)$} 
    \caption{Call Graph Generation from Repository} 
    \label{alg:callgraph} 
\end{algorithm} 

By using our tool, a static directed call graph $G = (V, E)$ is constructed, where $V$ is the set of method nodes identified by the method index and its content and $E$ is the set of edges representing the method invocation relationships. Each edge in $E$ is a pair of (caller, callee) indices. The method content is directly attached to each method node instead of using only partial information, such as method signatures as in previous techniques, to help facilitate the process of method representation extraction.  

\subsection{Method Representation Extraction}

To extract method representations, all the methods in the software system are first pre-processed to be converted into method tokens. Since the three neural-based models we evaluate treat the comment and code in separate ways by using a special $[SEP]$ token, we first remove all the docstrings and comments from a method to obtain only code. The methods are then parsed into an AST using the Tree-Sitter library ~\cite{tree_sitter} to obtain the ordered method tokens, and composite identifier names are split into subtokens based on the preprocessing pipeline from CodeSearchNet~\cite{husain2019codesearchnet}.

Next, the method tokens are passed through a neural-based code model, \eg CodeBERT \cite{feng2020codebert}, GraphCodeBERT \cite{guo2020graphcodebert} or UniXcoder \cite{guo2022unixcoder}, to generate representation vectors, as shown in Figure~\ref{fig:athena}-\circled{1}. We chose CodeBERT as one of our models since it was one of the first bi-modal pre-trained models which captures the semantic connections between natural language (\ie comment) and programming languages (\ie code) so that the information from the natural language comments can enhance the model's code understanding. CodeBERT is a representative model that makes full use of the sequence information existing in the comment and code. 

In Contrast to CodeBERT, GraphCodeBERT further extracts the DFG in the code to learn the inherent semantic-level structure information by incorporating the relationship of “where-the-value-comes-from” between variables. Finally, UniXcoder is one of the latest open-source code representation models that achieved state-of-the-art performance in code understanding and generation tasks by utilizing the AST instead of the DFG to learn rich syntactic information from code. Both the AST and DFG are mapped into sequence structures to be easily learned by transformers.

CodeBERT, GraphCodeBERT and UniXcoder are pretrained on 2.3 million (comment, code) pairs across six programming languages from the CodeSearchNet dataset. Code representations can be directly obtained from the pre-trained models, but the pre-training objectives (\ie masked language modeling, replaced token detection, code fragment representation learning, \etc) are quite different from IA and these representations are too general to represent one specific Java programming language. Although these models have been further fine-tuned for downstream tasks, none of them have been fine-tuned or evaluated for IA. In the absence of large available training dataset for IA, we fine-tuned the three models based on the code search task. Code search aims to retrieve relevant code snippets given a natural language query, and we choose this task as a proxy for IA because the models implicitly learn the functional goal of each code snippet during the fine-tuning, and it is highly likely that methods with similar functional goals get changed together when performing IA, as shown in the conceptual coupling metric introduced by Poshyvanyk \etal \cite{poshyvanyk2009using}. 

Specifically, we fine-tuned the three models on all $164,923$ (comment, code) pairs of the CodeSearchNet Java split based on the Siamese framework according to similar fine-tuning pipelines shared by Ranasinghe~\etal~\cite{ranasinghe2019semantic}. Each code snippet in the paired data is a complete method from a software repository. During the fine-tuning process, the comment and method tokens are first converted into token IDs based on the tokenizer from the three models and then passed into the comment encoder and method encoder, respectively, to obtain the comment and method embeddings. Both encoders share the same configurations with the same model parameters and weights. Then, for each batch of data, the distance between the comment and its corresponding method is minimized in the embedding space by using the standard cross-entropy loss function. We use the AdamW \cite{kingma2014adam} optimizer and the same hyperparameters recommended by the three respective models (\eg \# epochs, learning rate, batch size \etc) for fine-tuning, and the whole process was performed on an Ubuntu 20.04 server with an NVIDIA A100 40GB GPU. For GraphCodeBERT, we use the same approach of including the DFG information of the method during the fine-tuning process. CodeBERT and UniXcoder models only use the method information without any additional content. These trained models are then used to generate the embedding representations of the methods for the IA task, \ie we perform zero-shot learning \cite{thakur2021beir} without any further training for the IA task.

\subsection{Embedding Propagation}

By using any one of the three neural-based models, we obtain the initial method embeddings of all methods in a software system call graph $G = (V,E),$ where $|V| = N$. 

However, these embeddings are only equipped with the local method semantics but ignore the global dependencies between methods. Therefore, we utilize an embedding propagation strategy to update each of them based on the embeddings of its neighbor methods to combine the local semantics and structural information, \ie DFG or AST, with global dependency information from the call graph. We visualize this process in Figure~\ref{fig:athena}-\circled{2}. Formally, this is represented as the following:

\begin{equation}
    m_i = f(m_i, m^{nebr}_1, m^{nebr}_2, ..., m^{nebr}_k),
\end{equation}

\noindent where $m_i$ is the method being updated through the embedding propagation strategy $f$ with its neighbors $m^{nebr}_j (1 \le j \le k)$. Since a change in the callee method can still require a change in the caller method and this dependency is not capture in a directed call graph, we treat all edges as undirected for the propagation. Specifically, our embedding propagation strategy is inspired from the Graph Convolutional Network \cite{kipf2016semi} which adopts layer-wise propagation on the neural networks motivated by a localized first-order approximation of spectral graph convolutions:

\begin{equation}
    M' = \sigma( \tilde{D} ^ {-\frac{1}{2}} \tilde{A} \tilde{D} ^ {-\frac{1}{2}} M W),
\end{equation}

\noindent where $\sigma$ represents an activation function and $W$ is a trainable weight matrix. $\tilde{A} = A + I_N$ denotes the adjacency matrix of $G$ with self-connections. $I_N$ is the identity matrix and $\tilde{D}_{ii} = \sum_j \tilde{A}_{ij}$. This propagation strategy has been modified using a renormalization method ~\cite{kipf2016semi} in order to mitigate the effects of numerical instabilities and exploding/vanishing gradients when matrix multiplication operators are repeated during the training of the deep neural network. Since we do not train the call graph in this phase, our embedding propagation strategy is directly derived from the first-order approximation of localized spectral filters on graphs, which can be summarized as follows:

\begin{equation}
    M' = (I_N + w D ^ {-\frac{1}{2}} A D ^ {-\frac{1}{2}}) M.
\end{equation}

$M \in \mathbb{R}^{N \times F}$ represents the matrix of all the method embeddings from $G$ and $M'$ represents the updated matrix by incorporating the information from the neighbor methods. $F$ denotes the dimension of each method embedding (\ie $768$). $A$ is the adjacency matrix of $G$ without self-connections and $D$ is the degree matrix of $A$ so that the adjacency matrix is normalized by $D$ with respect to both the row and the column. $w$ is a constant to balance the information from the original method with contextual information from the neighbor methods. Moreover, in order to evaluate the effect of the distance of neighbor methods used for embedding propagation, neighbor methods in other orders (distances) are also utilized in addition to the direct neighbors: 

\begin{equation}
    M' = (I_N + w \sum_i D_i ^ {-\frac{1}{2}} A_i D_i ^ {-\frac{1}{2}}) M,
\end{equation}

\noindent where $1 \le i \le 3$ since we at most take into account the neighbor methods within three orders due to computational constraints. After the Embedding propagation strategy has completed, all of the identified methods in a given project will have an augmented embedding calculated by propagating the \textit{original} method embedding from neighbors (\ie as generated by one of our three neural models) to the target method, as illustrated at the top of Figure~\ref{fig:athena}-\circled{3}.

\subsection{Impact Set Estimation}

Finally, as illustrated in Figure~\ref{fig:athena}-\circled{3}, \athena computes the cosine similarity between the augmented embeddings of a given query method and the augmented embeddings for each method in the search corpus. Based on the cosine similarity scores, our approach returns a ranked list in descending order to help developers find other methods that are possibly affected and likely to be modified.

\section{Evaluation}\label{athena:eval}
To evaluate \athena's effectiveness in the IA task, we investigate the following research questions (RQs):

\begin{enumerate}[label=\textbf{RQ$_\arabic*$:}, ref=\textbf{RQ$_\arabic*$}, wide, labelindent=5pt]\setlength{\itemsep}{0.2em}
    \item \label{rq:baseline}{\textit{How well do traditional information retrieval and neural-based  techniques perform on the task of impact analysis?}}
    \item \label{rq:athena}{\textit{Does augmenting baseline techniques with call-graph information improve the effectiveness on impact analysis?}}
    \item \label{rq:ablation}{\textit{What leads to the difference in effectiveness (or lack thereof) between baseline techniques and techniques augmented with call-graph information?}}
    \item \label{rq:impact}{\textit{How do properties of different impact analysis tasks affect our studied techniques?}}
\end{enumerate}

\begin{table}[t]
\centering
\small
\caption{Dataset statistics of our evaluation benchmark\vspace{-0.3cm}}
\label{tabs:dataset}
\begin{tabular}{l|c|c|c|c}
\toprule
Settings  &  \# queries   &  \# commits    & ground-truth set   & corpus  \\  \hline
1 - whole    & 4,405  & 910    & 15.14 &   3,346    \\ 
2 - inner    & 3,379  & 734    & 3.47  &   31    \\  
3 - outer    & 2,999  & 444    & 17.21  &   3,440    \\  \hline

\end{tabular}
\end{table}

\subsection{Impact Analysis Benchmark: Alexandria}

As one of the most common and important types of changes to a software system, we chose to focus our efforts on bug fixes. Many of the existing datasets, such as the one by Tufano~\etal~\cite{tufano2019empirical}, are quite large, but unfortunately do not contain manually vetted data to be useful for IA. In contrast, existing IA datasets tend to be quite small only containing fewer than ten software projects~\cite{liu2018code,wang2018integrated}. Further, there is the potential for \textit{tangling} in software repository commits, wherein a commit which claims to be fixing a bug, both fixes the bug and may include additional unrelated changes, such as refactorings. As a result, the ground-truth impact sets from previous benchmarks may carry with them inaccuracies in their impact sets, as some methods may not actually be affected by a given change.

Recently, Herbold \etal \cite{herbold2022fine} introduced a large dataset consisting of 3,498 commits (\ie changes) from 28 Java projects, with the purpose of studying the tangling that occurs in bug fixing commits. In this dataset, each changed line was annotated with its type of change, whether it was modified to fix a bug, or was a change to tests, whitespace, a refactoring, or a documentation change. The data were annotated by four participants, and consensus was obtained if at least three participants agreed on the annotation to ensure accuracy. This paper also illustrated that many of the changes in the bug-fixing commits were changes to non-production artifacts, such as tests or documentation, rather than bug fixes. Therefore, we constructed our evaluation benchmark at method-level granularity based on this dataset which manually untangles the commits so that we know exactly which methods are changed for addressing one single concern, namely fixing one bug. 

\textbf {Impact Set Construction.}
To construct the impact set, we systematically mined the dataset from Herbold \etal \cite{herbold2022fine}. Specifically, for each changed line in the production code file labeled as \textit{“contributes to the bug fix”}, we added the corresponding method to our benchmark by recording the information of GitHub Diff URL, repository name, commit ID, parent commit ID, file path, method name, line numbers where the method starts and ends corresponding to both current commit and parent commit. Since Herbold \etal \cite{herbold2022fine} does not provide the method-related information, such as method name and line numbers where the method starts and ends, we checked-out the source code of the repository for the parent commit. We then used the srcML library \cite{collard2013srcml} to locate the changed methods based on the labeled changed line number. We use the snapshot of the software system that corresponds to the parent commit of a given as that is the state in which the change will be applied. Then, for each parent commit, we formulate the changed method set based on the concurrently changed methods. Since there is no clear single query method, \ie which method was changed ``first'' in the commit, we treat each method in the changed method set as a potential query, whereas the others are the ground truth impact set. Each query to impact set pair is considered a task. From developers' point of view, they usually at least know where the change starts, and intend to know which other methods need to be modified. We further post-process the dataset by removing commits that contain only one changed method. This process of formulating co-changing methods into impact analysis tasks has been widely used by past work to assess IA approaches~\cite{lehnert2011taxonomy,kagdi2013integrating,Gethers2010,gethers:icse12}.

\textbf{Task Definition and Settings.} For each changed method set $M = \{m_1, m_2, ..., m_n\}, n \geq 2$, we perform the impact analysis task with the query being $ \forall~m_i \in M $ and the ground-truth impact set being $M - {m_i}$. The search corpus is all methods except the query in the production files from the corresponding repository (Setting 1 - whole). In practice, the developer would determine whether a method should be modified by inspecting the corpus of methods in the order given by the specific approach. After inspecting our dataset, we found that methods in the same file could potentially be changed together. To mitigate the effect of possible shortcuts taken by approaches which pay more attention to all methods from the same file as the query, we formulate two more specific task settings: (i) The methods in both the ground-truth impact set and the search corpus are from the same file as the query (Setting 2 - inner). (ii) The methods in both the ground-truth impact set and the search corpus are from different files than the query (Setting 3 - outer). For setting 2, we focus on the methods that need to be modified in the same file as the query, while for setting 3, we focus on the methods that need to be modified in other files. The data corresponding to these two settings are further filtered by discarding the impact sets whose size is less than two.

\textbf{Dataset Statistics.}  Two software projects in \cite{herbold2022fine} are no longer accessible (\textit{santuario-java} and \textit{wss4j}) and for the software project \textit{eagle}, we could not build any valid impact set,\ie the size of the impact set less than two. As a result, our benchmark contains 25 open-source software systems and \# tasks/queries per software project is shown in Table ~\ref{tabs:repo_results}. Moreover, for each of the three settings, Table \ref{tabs:dataset} shows \# tasks/queries, \# commits, the average number of methods in the ground-truth impact set and in the search corpus respectively. 

Compared to Setting 2 (inner) which retrieves three or four affected methods from 31 methods, the Setting 3 (outer) is far more challenging which retrieves 17 or 18 methods from the corpus with 3,440 methods on average.	 

\subsection{Evaluation Metrics} 

We use standard information retrieval metrics to measure the effectiveness of the proposed approaches, namely mRR (mean Reciprocal Rank), mAP (mean Average Precision) and Hit@k. Specifically, for each task, the ranked list obtained from the proposed approach is compared with the expected impact set from the ground truth. Given one query method, we computed the \textit{rank} of the first actually affected method found in the ranked list, which indicates how many methods developers have to check to find the first one that needs to be modified. Then, we computed the \textit{reciprocal rank} for each task and averaged it across all the tasks to obtain the final mRR score. The mRR score measures the ability of the approach in helping developers find at least one method that needs to be modified. Correspondingly, the AP score for each task is calculated and averaged across tasks to obtain the final mAP score. AP is the average of the precision values which are computed after each ground-truth method in the impact set is retrieved to approximate the area under the uninterpolated \textit{Precision-Recall} curve. mAP scores measure the ability of the approach in helping developers find all the methods that need to be modified. Moreover, we use HIT@k to measure the proportion of successful tasks for the cut point k. A successful task means finding at least one affected method among the top-k (10 in this dissertation) results returned by the approach.

Most prior IA techniques ~\cite{liu2018code} ~\cite{cai2015comprehensive} use \textit{Precision, Recall} and \textit{F-measure} to evaluate their approaches since they consider the IA as a binary classification task by finding the possible affected methods based on dependencies instead of analyzing all methods in the repository. Therefore, what their method produces is not a ranked list, but an unordered estimated impact set, which is then directly compared with the ground truth impact set to obtain an F-score (\ie the harmonic mean of the \textit {Precision} and \textit {Recall} values). However, Poshyvanyk \etal \cite{poshyvanyk2009using} previously formulated impact analysis as an information retrieval task, but adapted prior metrics to the IR setting. We argue that IR metrics provide a more realistic representation of the potential benefit that an automated IA approach may actually provide to a developer in a recommender system setting. Furthermore, mAP score is more accurate than F-measure because it analyzes \textit{Precision-Recall} relationship globally rather than just based on the mean value calculation.

\subsection{ATHENA Configurations} 

Three models are generated using our proposed approach, namely \athena (CodeBERT), \athena (GraphCodeBERT) and \athena (UniXcoder), whose initial method representations are extracted based on each of the three neural-based code understanding models and we set $w = 0.5$ for information balancing. To compare with our approach, we also conducted the experiments based on these models without call graph information, using only the initial method representations without embedding propagation to compute the cosine similarity which serve as the corresponding baselines, including CodeBERT, GraphCodeBERT and UniXcoder.  

Moreover, we use the traditional \textit{bag-of-words} model (\ie TF-IDF) to obtain method representations where the term frequency (TF) is computed as the number of times a given code token appears in all tokens of its corresponding method, and the inverse document frequency (IDF) is calculated as the number of occurrences of a given code token in all code tokens built from the search corpus which includes all the production methods in a repository. Since TF-IDF representations are not real meaningful embeddings but frequency numbers without fixed vector length, we directly use a simple similarity weighting strategy for the \athena (TF-IDF), in which the cosine distance between the query and each of its neighbor methods is reduced by 50\% to obtain the final ranked list.  

In addition, instead of using call graphs, we build class graphs to further validate the effectiveness of call graphs when applied to IA, where the edges are added between each pair of methods in the same class resulting in many small strongly connected graphs for a repository. Due to the magnitude of \# edges, we still use the similarity weighting strategy to obtain class-graph-based CodeBERT, GraphCodeBERT, and UniXcoder respectively. Note that due to the nature of strongly connected graphs, there is no difference when neighbor methods of different orders are taken into account.

\section{Results}\label{athena:results}

\begin{table}[]
\centering
\small
\caption{Effectiveness of the baseline models and their \athena versions with the call graph information\vspace{-0.3cm}}
\label{tabs:results}
\begin{tabular}{c|c|c|ccc}
\hline
\textbf{Type}              & \textbf{Models}                & \textbf{Setting} & \textbf{mAP}   & \textbf{mRR}   & \textbf{HIT@10} \\ \hline
\multirow{12}{*}{Baseline} & \multirow{3}{*}{TF-IDF}        & Whole            & 22.89          & 45.36          & 63.91           \\
                           &                                & Inner            & 61.59          & 71.12          & 92.51           \\
                           &                                & Outer            & 14.97          & 32.01          & 46.62           \\ \cline{2-6} 
                           & \multirow{3}{*}{CodeBERT}      & Whole            & 23.77          & 46.25          & 67.20            \\
                           &                                & Inner            & 62.50          & 73.00          & 94.58           \\
                           &                                & Outer            & 19.16          & 37.81          & 53.55           \\ \cline{2-6} 
                           & \multirow{3}{*}{GraphCodeBERT} & Whole            & 24.42          & 46.70           & 68.79           \\
                           &                                & Inner            & 63.97          & 73.04          & 94.32           \\
                           &                                & Outer            & 19.85          & 38.13          & 54.09           \\ \cline{2-6} 
                           & \multirow{3}{*}{UniXcoder}     & Whole            & 23.65          & 45.96          & 66.95           \\
                           &                                & Inner            & 61.95          & 72.07          & 93.58           \\
                           &                                & Outer            & 19.19          & 37.33          & 52.38           \\ \hline
\multirow{12}{*}{\athena}   & \multirow{3}{*}{TF-IDF}        & Whole            & 23.96          & 47.41          & 70.60           \\
                           &                                & Inner            & 60.69          & 69.42          & 93.40           \\
                           &                                & Outer            & 15.79          & 33.48          & 50.78           \\ \cline{2-6}  
                           & \multirow{3}{*}{CodeBERT}      & Whole            & 27.38          & 50.96          & 73.24           \\
                           &                                & Inner            & 63.28          & 73.32          & \textbf{95.59}  \\
                           &                                & Outer            & 21.60          & \textbf{41.43} & \textbf{59.49}           \\ \cline{2-6} 
                           & \multirow{3}{*}{GraphCodeBERT} & Whole            & \textbf{28.27} & \textbf{51.28} & \textbf{74.46}  \\
                           &                                & Inner            & \textbf{63.78} & \textbf{73.73} & 95.12           \\
                           &                                & Outer            & \textbf{22.12} & 41.40          & 59.29  \\ \cline{2-6} 
                           & \multirow{3}{*}{UniXcoder}     & Whole            & 27.28          & 49.91          & 72.83           \\
                           &                                & Inner            & 62.75          & 72.40          & 94.58           \\
                           &                                & Outer            & 21.35          & 40.20          & 57.55           \\ \hline

\end{tabular}
\end{table}

\subsection{\ref{rq:baseline}: Baseline Performance on IA}

Table \ref{tabs:results} gives an overview of our different models both without (Baseline) and with (Athena) call graph information, which we will discuss in the next subsection. The first observation is that for setting 1 (whole), the traditional TF-IDF and neural-based models CodeBERT, GraphCodeBERT, and UnixCoder all achieve similar mAP scores, 22.89, 23.77, 24.25, and 23.65, respectively. 

The second observation is that GraphCodeBERT, regardless of setting or metric, is the best performing model. This is in opposition of work from Guo \etal \cite{guo2022unixcoder}, showing UnixCoder to outperform GraphCodeBERT. This might be due to GraphCodeBERT being the only model using additional information through Dataflow Graphs (DFGs) during the fine-tuning phase, thereby being able to fully utilize the semantic and structural information within the method.

When deconstructing setting 1 (whole) into its constituents, setting 2 (inner) and setting 3 (outer), we observe that these models perform best on the setting 2 case, \ie when only considering changes within the same class. For example, GraphCodeBERT's mAP score for setting 2 is $63.97$, yet when considering only changes outside of the query method's class it achieves a score of $19.85$. This result is intuitive, as there are many more detractors in the setting 3 than setting 2 when ranking methods for estimating the impact~set.

One of the surprising observations is that TF-IDF performs quite strong in setting 1 and 2 compared to neural-based approaches to code representation. However, for setting 3, it suffers significantly achieving an mAP score of $14.97$ compared to the next worst CodeBERT performance of $19.16$.  The reason behind this is that TF-IDF is particular good at keyword matching ~\cite{husain2019codesearchnet} rather than the understanding of underlying code semantics, and keyword overlapping is more common for the methods within the same file as the query as compared to those in different files, so TF-IDF ranked all these methods higher than others. Meanwhile, the methods within the same file are more likely to be actually affected based on the ratio of the size of ground-truth impact set to the corpus size from Table \ref{tabs:dataset}. Therefore, the performance of TF-IDF is comparable to neural-based models in setting 1. However, when based on keyword matching only, the relative positions of these methods hardly change, thus having little impact on the accuracy of settings 2 and 3. More details about this phenomenon are explained in \ref{rq:ablation}. In contrast, neural-based models focus more on the underlying semantics understanding, which affects the relative positions for the methods in the same file as the query and the methods in other files, contributing to higher mRR and mAP in both setting 2 and 3.

\subsection{\ref{rq:athena}: \athena Performance on IA}

Table \ref{tabs:results} also shows the results of each of the models with call graph information integrated. As observed, each neural-based \athena models improves significantly (Wilcoxon’s paired test $p < 0.05$) compared to its corresponding baseline in setting 1 (whole). Specifically, \athena achieves an improvement of $3.61$ mAP for CodeBERT, $3.85$ mAP for GraphCodeBERT, and $3.63$ mAP for UnixCoder, but only obtains an improvement of $1.07$ mAP for TF-IDF because of the employed simpler similarity weighting strategy.

Similar to the previous analysis, when looking at the constituents of setting 1, we see that setting 2 (inner) is not greatly improved. The improvements largely come from setting 3 (outer), which is intuitive since the integration of the call graph information adds contextual information about the rest of the software system through connections that extend outside of the query method's class. For example, for the best performing overall neural-based model GraphCodeBERT, when using \athena to integrate the call graph information improves by $2.27$ mAP for setting 3 (outer), yet sees a slight decrease for setting 2 (inner).

\begin{table}[]
\centering
\small
\caption{Effectiveness of \athena for different configurations\vspace{-0.3cm}}
\label{tabs:nebrs}
\begin{tabular}{r|ccc}
\hline
\multicolumn{1}{c|}{\textbf{Configuration}} & \textbf{mAP}         & \textbf{mRR}         & \textbf{HIT@10}      \\ \hline
\textbf{CodeBERT}                           & 23.77                & 46.25                & 67.20                 \\
\textbf{ATHENA (CodeBERT)}                  & \textbf{27.38}       & \textbf{50.96}       & \textbf{73.24}       \\
\textbf{+1 Neighbor}                        & 26.01                & 49.06                & 72.05                \\
\textbf{+3 Neighbor}                        & 27.23                & 50.67                & 73.49                \\ \hline
\textbf{GraphCodeBERT}                      & 24.42                & 46.70                & 68.79                \\
\textbf{ATHENA (GraphCodeBERT)}             & \textbf{28.27}       & \textbf{51.28}       & 74.46                \\
\textbf{+1 Neighbor}                        & 26.75                & 49.33                & 73.39                \\
\textbf{+3 Neighbor}                        & 28.22                & 51.18                & \textbf{74.53}       \\ \hline
\textbf{UnixCoder}                          & 23.65                & 45.96                & 66.95                \\
\textbf{ATHENA (UnixCoder)}                 & 27.28                & 49.91                & 72.83                \\
\textbf{+1 Neighbor}                        & 25.68                & 48.10                & 70.87                \\
\textbf{+3 Neighbor}                        & \textbf{27.63}       & \textbf{50.27}       & \textbf{73.14}       \\ \hline       
\end{tabular}
\end{table}

The results from Table \ref{tabs:results} of the \athena versions of the models utilize the best performing integration of neighborhood information, namely using second order neighbors. However, we are also interested in how the number of neighbor orders, \ie neighbors of neighbors \etc, impact performance. Table \ref{tabs:nebrs} shows the neural-based models under different number of neighbor orders, with the one marked \athena being two order neighbors used in Table \ref{tabs:results}. As shown, even with one order neighbors, the neural-based models are significantly improved over their baselines with an improvement of $2.24$ mAP for CodeBERT, $2.33$ mAP for GraphCodeBERT, and $2.03$ mAP for UnixCoder on setting 1 (whole). Further increasing the number of orders to two also see another significant improvement across models of $1.22$ mAP for CodeBERT, $1.52$ mAP for GraphCodeBERT, and $1.6$ mAP for UnixCoder on setting 1. However, increasing the order neighbors indefinitely does not continually lead to large gains as shown with three orders \emph{vs} two: there is very little change in terms of mAP and for CodeBERT and GraphCodeBERT, we see a decrease.

We conducted additional experiments by concatenating the tokens of the method with its corresponding comment for CodeBERT and \athena (CodeBERT), but we did not observe an obvious benefit (the results are available in our online replication package ~\cite{athena-tool}). Therefore, all the presented results are based on code only, without comment information.

\subsection{\ref{rq:ablation}: In-Depth Analysis of the Improvement}

\begin{figure}[t]
	\centering
	\caption{\normalsize{Three qualitative examples for illustrating the effectiveness of \athena.}}
	\includegraphics[width=\textwidth]{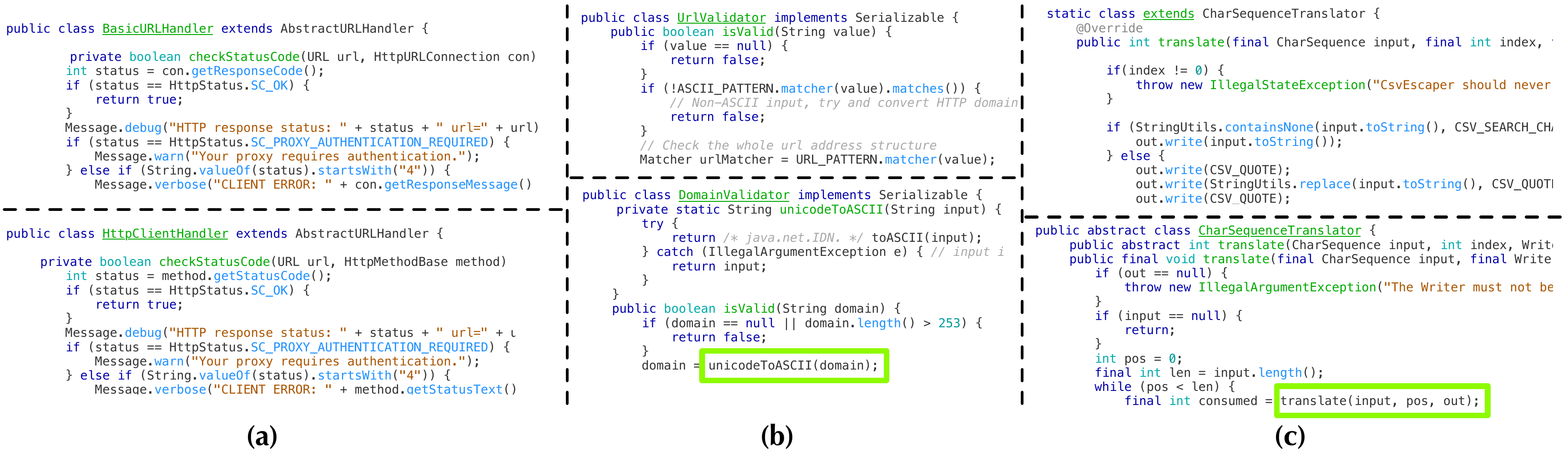}
	\label{fig:example}
\end{figure}

\begin{table}[]
\centering
\small
\caption{Effectiveness of three neural-based models by using class graphs\vspace{-0.3cm}}
\label{tabs:class_graph}
\begin{tabular}{c|c|c|ccc}
\hline
\textbf{Type}              & \textbf{Models}                & \textbf{Setting} & \textbf{mAP}   & \textbf{mRR}   & \textbf{HIT@10} \\ \hline
\multirow{9}{*}{Class Graph}& \multirow{3}{*}{CodeBERT}     & Whole            & 32.94          & 56.15          & 80.32            \\
                           &                                & Inner            & 62.50          & 73.00          & 94.58           \\
                           &                                & Outer            & 19.16          & 37.81          & 53.55           \\ \cline{2-6} 
                           & \multirow{3}{*}{GraphCodeBERT} & Whole            & 33.59          & 56.46          & 80.89           \\
                           &                                & Inner            & 63.97          & 73.04          & 94.32           \\
                           &                                & Outer            & 19.85          & 38.13          & 54.09           \\ \cline{2-6} 
                           & \multirow{3}{*}{UnixCoder}     & Whole            & 32.74          & 55.19          & 79.36           \\
                           &                                & Inner            & 61.95          & 72.07          & 93.58           \\
                           &                                & Outer            & 19.19          & 37.33          & 52.38           \\ \hline

\end{tabular}
\end{table}

In Table \ref{tabs:results}, the improvement produced by \athena compared to the baseline in setting 1 (whole) is distributed between the improvements in setting 2 (inner) and in setting 3 (outer), with setting 3 accounting for the major part of the improvement. 

Specifically, in setting 3, \athena (CodeBERT) outperforms CodeBERT by 3.62\% on mRR, 2.44\% on mAP and 5.94\% on HIT@10, \athena (GraphCodeBERT) is better than GraphCodeBERT by 3.27\% on mRR, 2.27\% on mAP and 5.20\% on HiT@10, and \athena (UniXCoder) outperforms UniXCoder by 2.87\% on mRR, 2.16\% on mAP and 5.17\% on HIT@10, which accounts for 76.65\% \& 67.59\%, 71.40\% \& 58.96\%, and 72.66\% \& 59.50\% of the improvement on mRR \& mAP in setting 1 respectively. Since setting 3 is a more challenging setting than setting 2 which finds the actual affected method in other files, \athena exhibits a better ability to reason about change impact sets across file boundaries -- which may be a more cognitively challenging task. Moreover, the \athena version of each neural model still performs \textit{significantly} better than \athena (TF-IDF) in each of the three settings based on the Wilcoxon’s paired test, and the improvement also comes mainly from the setting 3 (+7.92\% on mRR, +6.33\% on mAP, and 8.51\% on HIT@10 when compared to \athena (GraphCodeBERT)).

To further validate the effectiveness of call graphs combined with code semantics when applied to IA, the performance of class graph-based CodeBERT, GraphCodeBERT, and UniXcoder is shown in Table \ref{tabs:class_graph}. The class graph-based models improve their corresponding baselines by larger margins than our call graph-based models in setting 1, but as expected, they perform the same as their baselines in setting 2 and 3. Since all methods in the same file as the query are drawn closer to the query based on the similarity weighting strategy, they are all ranked higher than other methods. Given that the methods within the same file are more likely to be  affected, we obtain a larger improvement in setting 1 as mentioned in \ref{rq:baseline}.  However, their relative positions in the ranked list do not change, so there is no impact on the performance in setting 2 and 3. This also explains why TF-IDF is comparable to neural-based models in setting 1, but worse than them in setting 2 and 3. In contrast, the improvement produced by our call graph-based models in setting 1 effectively contributes to the improvement in setting 2 and 3.

\subsection{\ref{rq:impact}: Qualitative Analyses on IA Tasks}

\begin{table}[]
\centering
\footnotesize
\caption{Effectiveness of \athena for each software system\vspace{-0.3cm}}
\label{tabs:repo_results}
\begin{tabular}{l|c|cc|cc}
\hline
\multicolumn{1}{c|}{\multirow{2}{*}{\textbf{Repo Name}}} & \multirow{2}{*}{\textbf{Queries}} & \multicolumn{2}{c|}{\textbf{$Baseline_{GCB}$}} & \multicolumn{2}{c}{\textbf{$Athena_{GCB}$}} \\ \cline{3-6} 
\multicolumn{1}{c|}{}                                    &                                   & \textbf{mAP}           & \textbf{mRR}           & \textbf{mAP}          & \textbf{mRR}         \\ \hline
ant-ivy                                                  & 785                               & 19.14                  & 29.59                  & \textbf{22.78}        & \textbf{34.83}       \\
archiva                                                  & 43                                & 14.21                  & 41.07                  & \textbf{21.67}        & \textbf{48.51}       \\
commons-bcel                                             & 138                               & 25.93                  & 45.45                  & \textbf{30.31}        & \textbf{48.28}       \\
commons-beanutils                                        & 42                                & 36.96                  & 51.65                  & \textbf{37.22}        & \textbf{53.49}       \\
commons-codec                                            & 41                                & 36.79                  & 47.93                  & \textbf{41.55}        & \textbf{48.87}       \\
commons-collections                                      & 73                                & 24.70                   & 30.25                  & \textbf{24.88}        & \textbf{30.52}       \\
commons-compress                                         & 260                               & 19.86                  & 30.06                  & \textbf{27.8}         & \textbf{40.91}       \\
commons-configuration                                    & 253                               & 21.95                  & 30.43                  & \textbf{26.56}        & \textbf{35.5}        \\
commons-dbcp                                             & 91                                & 48.45                  & 60.00                  & \textbf{50.32}        & \textbf{62.31}       \\
commons-digester                                         & 22                                & \textbf{14.40}          & \textbf{17.59}         & 11.64                 & 15.83                \\
commons-io                                               & 58                                & 49.40                   & 57.11                  & \textbf{54.48}        & \textbf{63.33}       \\
commons-jcs                                              & 221                               & \textbf{24.07}         & \textbf{40.11}         & 23.16                 & 39.12                \\
commons-lang                                             & 115                               & 42.45                  & 51.28                  & \textbf{54.54}        & \textbf{60.64}       \\
commons-math                                             & 589                               & 31.57                  & 42.04                  & \textbf{38.03}        & \textbf{48.64}       \\
commons-net                                              & 171                               & 28.56                  & 40.00                  & \textbf{30.27}        & \textbf{40.84}       \\
commons-scxml                                            & 114                               & 17.31                  & 27.01                  & \textbf{25.48}        & \textbf{38.25}       \\
commons-validator                                        & 35                                & 41.87                  & 45.71                  & \textbf{43.02}        & \textbf{46.29}       \\
commons-vfs                                              & 166                               & 22.29                  & 30.27                  & \textbf{27.85}        & \textbf{37.75}       \\
deltaspike                                               & 5                                 & 34.08                  & 41.86                  & \textbf{50.28}        & \textbf{50.28}       \\
giraph                                                   & 527                               & 27.43                  & 44.48                  & \textbf{34.29}        & \textbf{51.16}       \\
gora                                                     & 174                               & \textbf{23.76}         & \textbf{36.95}         & 17.21                 & 27.76                \\
jspwiki                                                  & 12                                & 13.24                  & \textbf{52.12}         & \textbf{15.54}        & 51.20                 \\
opennlp                                                  & 141                               & 23.66                  & 34.40                   & \textbf{27.26}        & \textbf{37.44}       \\
parquet                                                  & 324                               & 19.97                  & 37.24                  & \textbf{21.53}        & \textbf{39.52}       \\
systemml                                                 & 5                                 & 46.19                  & \textbf{50.35}         & \textbf{47.32}        & 50.10     \\ \hline           
\end{tabular}
\end{table}

We begin our analysis of impact tasks by looking at the performance of our studied tehcniques across our different studied software projects. Table \ref{tabs:repo_results} provides a finer grained picture of the improvements per repository the best performing \athena model achieves over its corresponding baseline. 

As shown, \athena improves performance on 22 of 25 repositories in terms of mAP and 20 of 25 in terms of mRR. We investigated the reasons for the failure of \athena in those repositories, especially for the project \textit{commons-digester} and \textit{gora} due to the relatively larger performance difference between \athena (GraphCodeBERT) and GraphCodeBERT. For \textit{commons-digester}, this is mainly due to the number of method calls in the method, \eg there are more than 10 called methods in the query, while there is only one or two called methods in the method from the impact set. As for the project \textit{gora}, we found that there are many semantically similar pairs in the changed method set, such as (serialize, deserialize), (encodeInt, decodeInt) \etc, but since their goals are opposite, the methods they called are quite different. Therefore, \athena does not improve in identifying the actual affected method after incorporating the call graph information into the original semantics.

Now that we have examined the performance of \athena across IA tasks at a repository level, we will now discuss some exemplars from our benchmark that showcase how incorporating both structural information and semantic information can benefit the task of automated impact analysis.

\noindent{\textbf{Example 1: The Importance of Semantics.}} Figure \ref{fig:example} (a) shows two methods from different classes. The top method \texttt{\small checkStatusCode (URL, HttpURLConnection)} from class \texttt{\small BasicURLHandler} is the query method in this exemplar and the bottom method \texttt{\small checkStatusCode (URL, HttpMethodBase)} is one of the impact set methods. They share no call graph information that a structural IA technique could use to determine these methods are coupled. Yet, a change in one requires a change in the other. This is representative of conceptual coupling \cite{poshyvanyk2009using}, where the concepts of the two methods, \ie both performing a check on a status code, couples them together making it more likely that change in one would result in a change in the other. Utilizing the semantic information between the methods, \ie their keywords, either through a traditional TF-IDF or a neural based approach is necessary to determine that these two methods are highly coupled.

\noindent{\textbf{Example 2: The Importance of Combining Call Graph and Semantics.}} Figure \ref{fig:example} (b) shows a different example where The \texttt{\small isValid} method from the class \texttt{\small UrlValidator} is the query method, and the \texttt{\small unicodeToASCII} method from the class \texttt{\small DomainValidator} is part of the impact set. Note that the other \texttt{\small isValid} method from the class \texttt{\small DomainValidator} (which shares a name with our query method) is not part of the impact set. When examining the traditional TF-IDF and best performing neural-based GraphCodeBERT approaches, they both fail to rank the impacted method \texttt{\small unicodeToASCII} high, ranking it at 176 and 120, respectively. However, our \athena version of GraphCodeBERT achieved a ranking of 31, so we aimed to understand why this occured. We found that the method \texttt{\small isValid} in the \texttt{\small DomainValidator} calls the ground truth \texttt{\small unicodeToASCII} method. Our \athena (GraphCodeBERT), obtains a rank of 31 for the ground truth associated with the query even though there is no direct method invocation between the query and \texttt{\small unicodeToASCII} impacted method. Therefore, utilizing the embedding propagation strategy of our approach, \texttt{\small unicodeToASCII} was updated with the information from the method \texttt{\small isValid} (in the \texttt{\small DomainValidator} class) that is more semantically similar to the query, thus helping improve the ground truth rank.

\noindent{\textbf{Example 3: Call Graphs Improve IA with more code semantic overlap.}} Fig. ~\ref{fig:example} (c) presents an example with query being the method \texttt{\small int\_translate} from the class \texttt{\small CsvEscaper} and the ground truth being the method \texttt{\small void\_translate} from the class \texttt{\small CharSequenceTranslator}. Owing to their similar code semantics, the original GraphCodeBERT ranks the ground truth 8th. In \athena (GraphCodeBERT), the information of the abstract method \texttt{\small int\_translate} from the same class \texttt{\small CharSequenceTranslator} is incorporated into the method \texttt{\small void\_translate} because of a call dependence between them, which thus improves the similarity between the query and the ground truth since the abstract method \texttt{\small int\_translate} is more semantically similar to the query than \texttt{\small void\_translate} in terms of the method signature. Therefore, \athena (GraphCodeBERT) achieves a higher rank of 6 for the target \texttt{\small void\_translate} method.

As can be observed from these examples, there are clear benefits when the call graph information is combined with the local code semantics, and we saw this pattern hold after investigating additional cases where the \athena outperforms its corresponding baseline. The contextual information obtained from the global call dependencies among methods enriches the original semantics of the methods, which indeed helps to identify the impact set associated with the given query.

\section{Threats to Validity}\label{athena:threats}

\noindent \textbf{Threats to Internal Validity:} To reduce potential issues from internal threats to validity, we studied three different DL models and a non-DL model when validating our proposed method of incorporating call graph information to improve IA. Additionally, we constructed our benchmark from commits that have been manually annotated and had the changes made to fix bugs untangled from other changes such as ones to documentation to ensure our ground truth labels are high quality. Additionally, we ensured there were no overlapping repositories between our training on the CodeSearchNet corpus and testing on our impact analysis dataset.

\noindent \textbf{Threats to External Validity:} To lessen the potential for threats to external validity, we used a significantly larger set of projects, 25 compared to previous work that used around five, and tested our method across different DL and non-DL models to show generalizability. One potential issue with generality is that we only evaluated our approach on Java and Apache projects, therefore, our approach may not generalize to other programming languages such as Python or to different types of projects. However, the DL models we used have shown success across multiple languages and so most likely the same would apply to our approach.

\noindent \textbf{Limitations:} We leave many of our approaches' limitations to future work. Particularly, utilizing more specialized DL architectures that might improve results. One type that we explored initially was Graph Neural Networks, namely GraphSage \cite{hamilton2017inductive} and Graph Attention Network (GAT) \cite{velivckovic2017graph}. However, their performance was subpar so we refocused on the Transformer architecture, but additional investigation of such architectures is warranted. Another future direction is related to incorporating other software engineering artifacts into our approach such as change requests, test cases, architecture diagrams, \etc. Similarly, integrating other types of software specific information besides call graph is another potentially fruitful area. Another area that would cause for future work is how well this improves developer productivity when performing code changes. Lastly, due to the flexibility of our approach, future work could look into applying \athena to other software engineering tasks such as code search, code comprehension, or other maintenance tasks, such as commit message generation \cite{cortes-coy_automatically_2014}.

\section{Bibliographical Notes}

The paper supporting the content described in this Chapter was written in collaboration with other members of the SEMERU group at William \& Mary and researchers from Universit\'{a} della Svizzera Italiana. It is currently under review for publication.

Yan, Y., \textbf{Cooper, N.}, Moran, K., Poshyvanyk, D., \& Bavota, G. (2023, March). Combining Call Graphs and Neural Code Semantics to Improve Automated Impact Analysis. Under Review.
\chapter{On the Generalizability of Transformer Models for Code Completion}
\label{sec:completeformer}

Large Language Models (LLMs) for code have achieved state-of-the-art results across a variety of software engineering tasks such as code completion \cite{White2015a, ciniselli2021empirical}, code reviews \cite {TufanoICSE22,9402025}, clone detection \cite{White2016}, program repair \cite{tufano2019empirical, Chen:2019}, testing \cite{WatsonAsserts,TufanoMutation,9270362}, and others \cite{watson2022systematic,MastropaoloT5,tufano2019learning,9825825,9284034}. The Transformer \cite{vaswani2017attention} has been at the center of these improvements due to its attention mechanism and ability to be highly parallelized, allowing for more efficient training compared to previous models such as Recurrent Neural Networks \cite{rumelhart1985learning}.

When applied to code completion, Transformers take as input an incomplete code component and try to predict the missing code tokens (\eg \cite{ciniselli2021empirical, chen2021evaluating, austin2021program, fried2022incoder}). As already happened in the field of Natural Language Processing \cite{press2020shortformer}, there has been a recent push in increasing the maximum length of the sequences (incomplete code components) on which Transformers are trained and tested. This is due to the fact that longer sequences (i) allow to provide the model with additional contextual information which can help with improving the prediction performance; and (ii) can help in simulating more variegate code completion scenarios. This, however, has a substantial cost to pay in terms of training time \cite{press2022alibi}.

It comes without surprise that efforts have been made in the NLP literature to address this issue (\eg \cite{dai2019transformer,raffel2020exploring,press2022alibi,sun2022length}): The most recent work targets the generalization of Transformers to longer sequences than those they have been originally trained for \cite{press2022alibi,sun2022length}. This allows to efficiently train a model on short sequences and, then, perform the inference on longer sequences without a significant performance degradation. This is, for instance, the goal of the ALiBi (Attention with Linear Biases) attention mechanism for Transformers \cite{press2022alibi}, which has been successfully used in NLP.

If solutions such as ALiBi properly work on source code as well, they could substantially help reduce the training cost of code completion models such as the popular GitHub Copilot \cite{github}. This is the focus of our work. We aim to investigate the extent to which solutions proposed in the NLP literature can support the generalization of Transformers on source code. We focus on three state-of-the-art solutions and one baseline. \add{The first (baseline), Sinusoidal \cite{vaswani2017attention}, uses Absolution Positional Encodings (APEs) by defining sine and cosine functions to generate positional embeddings that the authors original hypothesized would help the model to generalize. The second, xPOS \cite{su2021roformer,sun2022length}, is a hybrid between APEs and Relative Positional Encodings (RPEs) and applies rotations to the sine and cosine positional embedding to incorporate relative position information along with a attention resolution metric to improve generalization. The third, ALiBi \cite{press2022alibi}, offers a simple solution of modifying the attention mechanism to weight positions far away as less important than ones closer. The last, T5 \cite{raffel2020exploring}, similarly modifies the attention mechanism by adding a learned bias term that influences the attention given to a token.}\eject

We want to assess whether models trained on sequences of a specific length are able to generalize, \ie not incur a significant performance degradation, on sequences being longer (or shorter) than the training ones. To accomplish this, we built four datasets (\emph{short}, \emph{medium}, \emph{long}, and \emph{mix}) featuring incomplete Python and Java functions having different lengths. Then, we train 32 Transformers, namely four models (Sinusoidal, xPOS, ALiBi, and T5) for each of the four datasets and two programming languages. The performance of the 32 models has been evaluated on a series of test sets of previously unseen Python and Java functions of various different lengths, studying the generalization of their predictions. For example, we verified if the models trained on \emph{short} datasets are able to work on instances in the test set having a length inline with the examples in the \emph{long} dataset.

Overall, we make the following contributions:
\begin{enumerate}
    \item \add{A large systematic benchmark for evaluating the generalization of LLMs for code completion of different lengths across two programming languages;}
    \item \add{An empirical study on whether current generalization approaches extend to encoder-decoder architectures for the task of code completion;}
    \item \add{A set of results and implications that can be leveraged by researchers and developers of these models for navigating trade-offs.}
\end{enumerate}

\section{Background}\label{completeformer:back}

\begin{figure}[t]
    \begin{center}
		\includegraphics[width=0.50\linewidth]{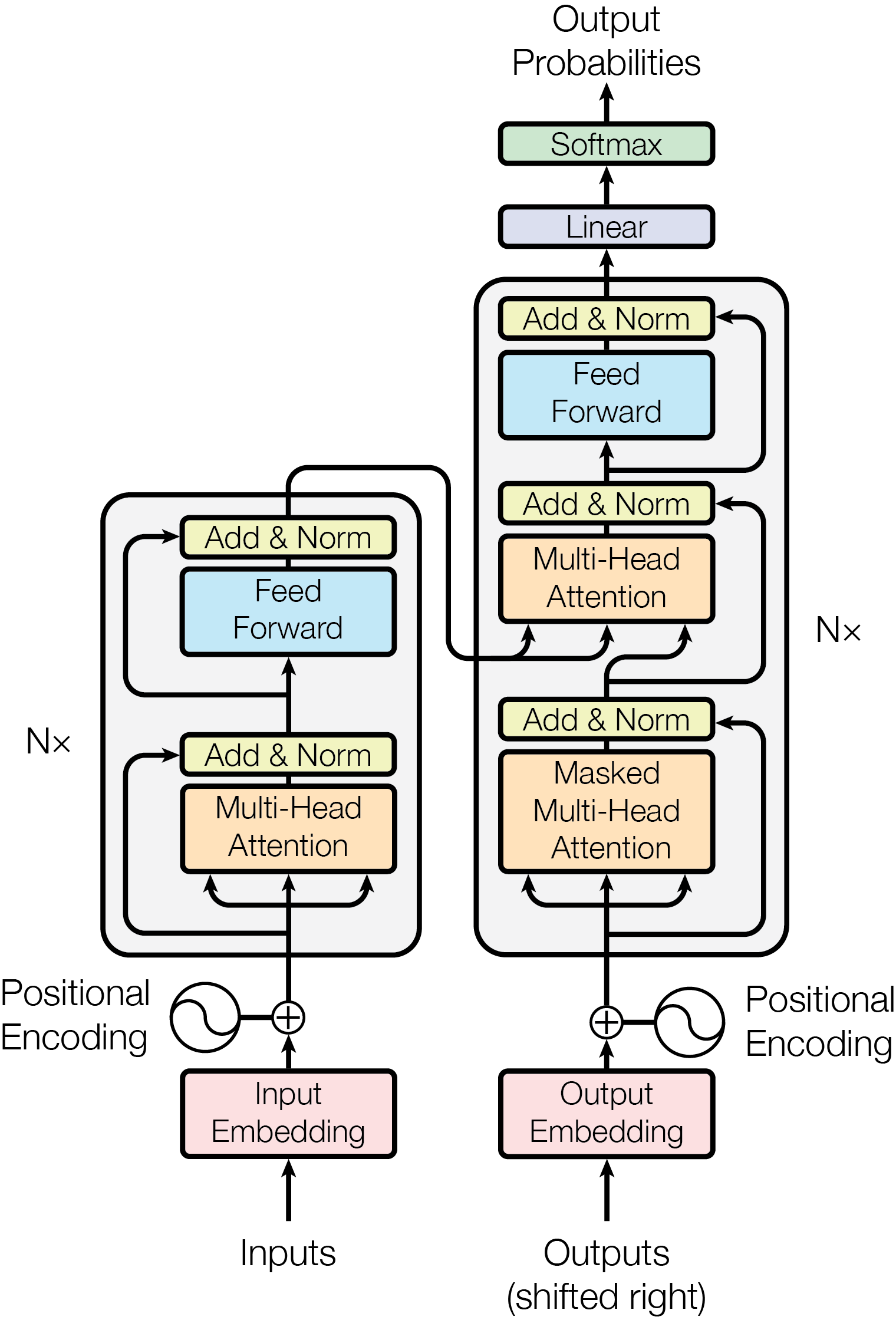}
        \caption{Sequence to Sequence Transformer Overview from the original paper \cite{vaswani2017attention}. The left part is the encoder and the right part is the decoder.}
        \label{fig:transformer}
    \end{center}
\end{figure}

\begin{figure}[t]
    \begin{center}
		\includegraphics[width=0.9\linewidth]{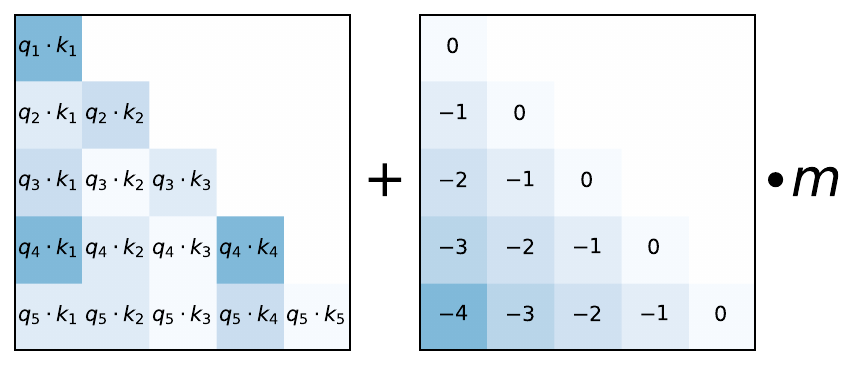}
        \caption{ALiBi Overview from the original paper \cite{press2022alibi}}
        \label{fig:alibi}
    \end{center}
    \vspace{-0.3cm}
\end{figure}

We introduce some mathematical background to understand the specifics of the position encoding schemes, namely Sinusoidal, xPOS, ALiBi, and Relative, and how they apply to our study.

This dissertation focuses on the Sequence to Sequence Transformer \cite{vaswani2017attention} (see Fig. \ref{fig:transformer}). It takes as input a sequence of tokens $C'={x_1,x_2,...,x_n}$ (in our case, the code to be completed), and outputs a target sequence, $M$, that is similarly decomposed into tokens $M={y_1,y_2,...,y_m}$. $M$ represents in our context the missing piece of code to be predicted. Thus, the combination $C = C' + M$ equals the complete code snippet. Note that the $+$ operator does not imply an append, but rather a combination regardless of where the missing part $M$ is placed in $C'$. The decomposition of both $C'$ and $M$ is done through tokenization using a trained Byte Pair Encoding tokenizer \cite{sennrich2015neural} such that $C' \text{ or } M \rightarrow s_0,s_1,...s_T$, where $s_i \in \mathbb{V}$ and $\mathbb{V}$ is the vocabulary of the tokenizer. We use the same vocabulary for both $C'$ and $M$. These tokens are passed through an Embedding layer, which is shared across the encoder and decoder, to get the token's vector representation, which will be updated progressive through various Attention and Feed Forward layers, we will explain further.

The output $M$ is done in an autoregressive manner, where the output of one time step from the model is fed back into the decoder portion of the Transformer for generating the probability distribution of the next token. Formally, the probability distribution of token $y_i$ is conditioned on the output of the encoder $Z$ and the $y_{<i}$ previously generated tokens: $p(M)=\prod^m_i{p(y_i|y_{<i}, Z)}$.

The Transformer architecture itself has no way of modeling sequential information. Therefore, sequential information is injected either at the bottom of the network (Fig. \ref{fig:transformer}) or at each Transformer block of the network depending on the positional scheme. For our four positional schemes, only one injects the positional information at the bottom of the network, namely sinusoidal, the rest of the schemes inject the information at each Transformer block.

The Transformer block's composition depends on if it is part of the encoder or decoder. In the encoder portion, the attention is computed by transforming a given sequence, $s=(s_1,s_2,...,s_n)$ where $s_i \in \mathbb{R}^{d_s}$ into a new sequence of the same size, $z=(z_1,z_2,...,z_n)$ where $z_i \in \mathbb{R}^{d_z}$ via a weighted sum where the weight can be intuitively thought as the amount of attention to pay to the value of $s_j$ in the sequence. $\mathbb{R}^{d_s}$ represents the embedding space of the Embedding layer that transforms the discrete token into a continuous vector of size $d_s$ and similarly $\mathbb{R}^{d_z}$ represents the vector space that is composed of $d_z$ dimensions where $d_z$ can be the same or a different size than $d_s$. The weighted sum can be represented as follows:

\begin{equation}
    z_i=\sum^n_{j=1}a_{ij}(s_jW^V)
\end{equation}

With $W^V$ being a learned value weight matrix and $a_{ij}$ being calculated with the following softmax formula:

\begin{equation}
    a_{ij}=\frac{\text{exp }e_{ij}}{\sum^n_{k=1}\text{exp }e_{ik}}
\end{equation}

And $e_{ij}$ being calculated by taking the corresponding $s_i$ and $s_j$ tokens and multiplying them by a learned query and key weight matrix, $W^Q$ and $W^K$, to get the corresponding query and key vectors. These vectors are then put through a compatibility function, namely the scaled dot product:

\begin{equation}
    e_{ij} = \frac{(s_i W^Q)(s_j W^K)^T}{\sqrt{d_k}}
\end{equation}

As it can be seen, no positional information is present in these operations. So, the original Transformer injected positional information into the beginning of the network by adding a position vector to the token vector to encode this information for the rest of the network. Let us now discuss each positional encoding scheme.\smallskip

\add{\textbf{Sinusoidal:} Sinusoidal is the original scheme proposed in the Transformer paper \cite{vaswani2017attention}. The information is added directly to the token embeddings at the beginning of the network. Concretely, the same dimension of the token embeddings is used for the position embedding and the sinusoidals switch between sine and cosine:}

\begin{equation}
    PE_{pos, 2i} = sin(pos/10,000^{2_i / d_{model}})
\end{equation}

\begin{equation}
    PE_{pos, 2i + 1} = cos(pos/10,000^{2_i / d_{model}})
\end{equation}

\add{Where $pos$ refers to the position in the sequence and $i$ refers to the specific dimension of the position embedding. The authors chose this scheme as they believed it would allow for the model to learn to use relative positions. However, this approach is generally considered an Absolute Positional Encoding (APE) scheme.}\smallskip

\add{\textbf{xPOS:} Sun \etal \cite{sun2022length} extended work from Su \etal \cite{su2021roformer} which  made the observation that the dot product between queries and keys are where information is shared between different tokens. Therefore, position information can also be added of the relative position between the different tokens. Specifically, Su \etal \cite{su2021roformer} wanted to find an operation that satisfies the following:}

\begin{equation}
    (f_q(x_m, m), f_k(x_n, n)) = g(x_m, x_n, m - n)
\end{equation}

\add{Where functions $f_q$ and $f_k$ add this relative information to the token embeddings $x_m$ and $x_n$. respectively. To accomplish this, the authors introduced Rotary, which uses rotations of the token embeddings based on their position so that the relative position of the tokens are preserved through this dot product. Namely, they define the function $g$ that satisfies this to be}

\begin{equation}
    g(x_m, x_n, m - n) = Re( (W_q \cdot x_m) (W_k \cdot x_n) * e^{i(m - n) \theta } )
\end{equation}

where the function $Re$ takes only the real part of the complex number. This is specifically for the 2D case, however, this is generalizable to any dimension that the token embeddings belong to. 

For a complete discussion of the equations and their generalization, we refer readers to the original paper \cite{su2021roformer}.

\add{Sun \etal \cite{sun2022length} extended this approach to have similar extrapolation abilities of ALiBi, discussed below, while still having better performance. To accomplish this, the authors introduced the idea of \textit{attention resolution} where a model's attention should monotonically decrease as the pair wise distance between tokens increases, similar to ALiBi. To integrate this into the rotation matrix, they apply an exponential decay that adds this property. They show that this gives a good trade-off between the Rotary performance and ALiBi's ability to extrapolate to longer than seen during training sequences. For our experiments with xPOS, we use the original implementation from Su \etal \cite{su2021roformer} for the encoder and the extension, xPOS, by Sun \etal \cite{sun2022length} for the decoder since the extension is unable to be applied directly to an encoder.}\smallskip

\textbf{ALiBi:} In ALiBi, the positional information is injected by modifying the equation above by adding a static bias:

\begin{equation}
    e_{ij} = \frac{(s_i W^Q)(s_j W^K)^T}{\sqrt{d_k}}+m^h(j-i)
\end{equation}

Where $m$ is a head-specific scalar that is selected before training. We use the same geometric series for initializing these $m$ values per head as in the original ALiBi paper, namely starting from $2^\frac{-8}{n_{heads}}$ and using the same value as the ratio. Intuitively, this static bias penalizes query and key vectors that are far away from each other. Figure \ref{fig:alibi} visualizes this process. Specifically, it shows how queries and keys corresponding to the same token do not receive any reduction whereas mismatched queries and keys receive a reduction proportional to their relative distance.

This process was originally designed for decoder-only Transformer models. However, since we use an encoder-decoder Transformer model, we additionally use the bidirectional version for the encoder portion as outlined in a post by the original author\footnote{\url{https://github.com/ofirpress/attention_with_linear_biases/issues/5}}.\smallskip

\textbf{T5:} Similar to ALiBi, T5 introduces a bias inside the softmax equation that is based on distance. Specifically, this bias is a learned scalar that is added to the query and key dot product $(s_i W^Q)(s_j W^K)^T+b^h_{ij}$ where each attention head, $h$, has a different learned bias. T5 introduces this idea of buckets, which are the different learned biases, where the different $ij$ pairs logarithmically map up to a relative position of 128 beyond which the same $ij$ pairs are mapped to the same bucket.\smallskip

After each attention operation in the encoder Transformer block follows a normalized residual connection, a simple Feed Forward Multi-Layer Perceptron, and another normalized residual connection. A similar process happens in the decoder block. However, there is an additional attention mechanism that happens after the normal self-attention is applied to the transformed output token embeddings, which considers the encoder's representation of the input, $Z$. This is known as cross-attention and follows the same process as self-attention except the keys and values are constructed from the representation $Z$ while the query is built from the output representation. Intuitively, this can be thought of as the output tokens requesting specific information from the input tokens.

\section{Study Design}\label{completeformer:eval}
The \emph{goal} of our study is to determine whether popular positional encoding schemes for length generalization work for the task of code completion. Namely, we seek to answer the following question:

\begin{enumerate}[label=\textbf{RQ$_{\arabic*}$:}, ref=\textbf{RQ$_{\arabic*}$}, wide, labelindent=5pt]\setlength{\itemsep}{0.2em}
      \item \label{rq:extrapolate}\textit{To what extent can different positional encoding schemes generalize to different code lengths for the task of code completion?}
\end{enumerate}

In the context of code completion this means studying whether models trained on completing code sequences having a specific length generalize when used to complete shorter/longer sequences. This is analogous to whether models are able to utilize different amounts of information, in terms of the input tokens, as compared to what they have seen during training. Naturally, shorter instances require shorter training time. Yet, it is unclear if a model trained on short code completions can generalize to also work on longer ones and \textit{vice versa}.

While answering RQ$_1$ we also check whether our findings generalize to multiple programming languages. In particular, we contextualize our study to Python and Java, as languages  often adopted in code completion studies (see \eg \cite{ciniselli2021empirical,svyatkovskiy2021fast}). Also, we consider two different code completion tasks recently used in the code completion study by Ciniselli \etal \cite{ciniselli2021empirical}: \emph{statement-level} and \emph{block-level} completion. The former is the classic code completion task in which the last $n$ tokens of a single code statement are masked, with the model in charge of predicting them. The latter, instead, possibly extends the completion to multiple statements, masking the last $n$ tokens in a block of code (\eg the body of an \texttt{if} statement) and asking the model to guess them.

In the following we detail the procedure used to (i) collect the training/testing datasets employed in our study (Section \ref{subsec:data}), and (ii) perform the required data collection and analysis (Section \ref{subsec:experiments}).

\subsection{Dataset Construction}\label{subsec:data}

To build the datasets needed for our study, we mined data from GitHub open source projects. In particular we: (i) used the GitHub search tool by Dabi\'c \etal \cite{Dabic:msr2021data} to identify all Python and Java GitHub projects having at least 100 commits, 10 contributors, 10 stars and 10 issues (to exclude toy projects); (ii) sorted the projects by number of stars; (iii) cloned the top 3k and extracted from each of them the functions in the master branch (to only rely on functions likely to be syntactically correct); (iv) removed all functions containing non-ascii characters (to avoid problems when reading data); (vi) removed all duplicates (to avoid leaking of information between the training, validation, and test sets we created out of this dataset); (vii) removed from the dataset all instances consisting of more than 1,024 tokens. The latter is a procedure usually adopted in applications of DL4SE (\eg \cite{leclair-mcmillan-2019-recommendations,Tufano:testGen,Chen:2019}). Indeed, too long instances make the training of DL models too expensive, also motivating our investigation into the length generalizability. 

Such a process resulted in the collection of $\sim$4M Python functions and $\sim$4.5M Java functions which we further processed to create datasets aimed at answering our research question.

\subsubsection{Java dataset: statement-level code completion task}
The Java dataset will cover the statement-level code completion task. The goal is to build three datasets featuring instances (\ie functions)  having different lengths, namely \emph{short}, \emph{medium}, and \emph{long} instances. Basically, with ``length'' we refer to the number of input tokens provided to the model. In all three datasets we keep constant the complexity of the prediction to generate (\ie the number of masked tokens). Indeed, only in this way we can ``isolate'' the impact on performance of changing the input length. 

Since the code completion task, which we are focusing on, requires the masking of code statements, we start by removing for the set of collected Java functions all of those that do not contain any statements. We also removed all Javadoc comments, since we are focusing on completion tasks within the body of a function. We then compute the number of Java tokens within each of the remaining 3,833,445 functions: this results in a distribution of functions' \emph{length}. 

We then compute a second distribution representing the number of Java tokens within the code statements in the subject functions, observing a median of 11 tokens per statement. The idea is to mask in each instance of the three datasets we will create (\ie \emph{short}, \emph{medium}, \emph{long}) the exact same number of tokens (11). As previously mentioned, this is done to keep constant the ``complexity'' of the completion task and better isolate the impact of the sequence length on the observed performance.

Given such a constraint, we remove all the functions not containing any valid statements to mask, \ie a statements consisting of at least 11 Java tokens. We sort the remaining 1,855,578 functions by their length and split them into three sets of the same size, obtaining: (i) a first set of functions with lengths ranging from 6 to 96 (\emph{short} dataset); (ii) a second set of functions with lengths ranging from 97 to 180 (\emph{medium}); and (iii) a third set of functions with lengths ranging from 181 to 1024 (\emph{long}).

For each function $F$ in each of the three datasets, we created $n$ training instances, where $n$ is the number of valid statements to mask $F$ contains (\ie the statements having at least 11 tokens). This may end up in generating duplicates due to different functions from which we masked the only part being different, thus resulting in duplicates. For this reason, we perform a second deduplication round on all the datasets.

The final set of (masked) functions is then flattened to obtain the model input by replacing all new line characters (\ie `\texttt{\textbackslash n}’) with a special tag, ⟨NEW\_LINE⟩, and remove all tabs (\ie `\texttt{\textbackslash t}’) and white spaces used to indent the code. We randomly split each set (\emph{short}, \emph{medium}, \emph{long}) into training (80\%), validation (10\%) and test (10\%) by making sure that all the instances obtained from the same function fall into the same set.

Finally, for each of the three datasets (\emph{short}, \emph{medium}, \emph{long}), we limit the number of training instances to 280k and, proportionally, those of the test and evaluation set to 35k. This is done to reduce the training cost of our study, as explained later, required to train and test 32 DL models. These numbers (\ie 280k and 35k) are inherited from the (smaller) Python dataset that we will describe in the next section. In other words, we aligned the size of the Java and of the Python datasets towards the smallest one (Python), due to our computational resources.

To summarize, at the end of this process we have three datasets (\emph{short}, \emph{medium}, and \emph{long}) each split into training, validation, and test, all containing the same number of instances having, however, different lengths.

We also built a fourth dataset, named \emph{mixed}, consisting of a mix of the three lengths: 1/3 of instances comes from the \emph{short} dataset, 1/3 from the \emph{medium}, and 1/3 from the \emph{long}. In this case we only built the training and the validation sets, since we will test the model trained on the mixed dataset on the \emph{short}, \emph{medium} and \emph{long} test sets.

\subsubsection{Python dataset: block-level code completion task}
Similarly to what we discussed for Java, we build three Python datasets (\emph{short}, \emph{medium}, and \emph{long}) featuring instances (functions) having different lengths but characterized by the same task complexity (\ie same number of masked tokens to predict). The main difference between the Java and the Python dataset is that the latter simulates block-level completion, thus possibly featuring completions spanning across multiple statements. 

The process used to build the Python datasets resembles the one we presented for Java. Thus, we only briefly summarize it here. We removed all functions not containing any code block (2,833,017). For consistency with the Java datasets, we decided to keep the same task complexity, meaning that we target the masking of the last 11 Python tokens within a given block. Thus, we remove from the dataset all functions not containing any valid block to mask (\ie a block consisting of at least 11 tokens).

We then sort the remaining functions by their length and split them into three sets of the same size: \emph{short} (featuring functions with length from 30 to 150), \emph{medium} (from 151 to 309), and \emph{long} (from 310 to 1024). For each function $F$ we created $n$ instances, each one having a different block featuring its last 11 tokens masked. For the same reasons previously explained, we remove any duplicates created at this stage.
 
We replace all characters used to indent the code (\ie `\texttt{\textbackslash n}’, `\texttt{\textbackslash t}’ and extra white spaces) with a special tag: \texttt{⟨TAB⟩}. This allows to flatten each function without losing information about the indentation, which is fundamental for the Python syntax. 

Finally, we split each dataset (\emph{short}, \emph{medium}, \emph{long}) into training, validation, and test using the same procedure described for Java. For each dataset, the training contains 280k instances, while the evaluation and test contain 35k instances. These numbers have been dictated by the smallest dataset involved in our study, being the \emph{short} Python dataset. Aligning the size of all datasets removes another possible confounding factor.

\subsection{Data Collection \& Analysis}\label{subsec:experiments}

\begin{table}[]
\centering
\caption{Hyperparameters used and searched.}
\label{tab:hyperparams}
\begin{tabular}{l|r}
\hline
\textbf{Hyperparameter} & \textbf{Values} \\ \hline
Learning Rate           & 1e-4            \\
Batch Size              & 256             \\
Inner Dimension         & 512             \\
Encoder Max Length      & 1,024            \\
Encoder Layers          & 6               \\
Encoder Heads           & 8               \\
Decoder Max Length      & 128             \\
Decoder Layers          & 8               \\
Decoder Heads           & 6               \\ \hline
\end{tabular}
\end{table}

To answer our research question, we train eight models (four per each of the subject languages, namely Python and Java) for each of the four experimented position encoding schemes (\ie Sinusoidal, xPOS, ALiBi, and T5). This leads to a total of 32 trained models. The four models for each language have been trained on datasets featuring code completions having inputs (\ie the Java or Python function to complete) characterized by different lengths (\ie \emph{short}, \emph{medium}, \emph{long}, and \emph{mix}). Then, each of these models have been used to generate predictions on three test sets featuring code completions of different lengths (\ie \emph{short}, \emph{medium}, and \emph{long}). This allows us to verify if, for example, a model trained on \emph{short} code completions can generalize to a test set containing \emph{long} instances. Also, we can verify whether a model trained on code completions having a mixture of lengths (\ie featuring short, medium, and long sequences) can achieve on each of the three test sets (\ie \emph{short}, \emph{medium}, and \emph{long}) results competitive with those of models specialized (\ie trained only) on instances having a specific length. For example, we can check whether the model trained on the mixture of lengths achieves on the \emph{short} test set performance comparable to those of the model trained on the \emph{short} training set. Remember that the amount of instances in each training set is fixed. Thus, observed differences should be due to the length of the employed training instances.

To reduce confounding factors, we used the same hyperparameters amongst all 32 models. The adopted hyperparameters are those suggested in the paper originally proposing the Transformer architecture \cite{vaswani2017attention} and are reported in Table \ref{tab:hyperparams}. This design decision also avoided the need for an expensive hyperparameters tuning involving 32 different models.

\eject

All models have been trained with the Adam optimizer \cite{kingma2014adam} with a cosine learning rate scheduler using a warmup of 2,000 steps. We used a vocabulary size of 50k for the tokenizer, which was shared across all the models.

For implementing and training the Transformers, we used x-transformers \cite{xtransformers} and Pytorch Lightning \cite{Falcon_PyTorch_Lightning_2019}. Additionally, when generating samples, we used Nucleus Sampling \cite{holtzman2019curious} with a $top_p = 0.95$ and stopped generations once the \texttt{⟨EOS⟩} token was produced or the maximum number of tokens, $128$, were produced. We trained all models for a maximum of five epochs and used the best performing checkpoint based on validation loss, which happened to always be the models trained on all five epochs.

To assess the performance of the models on the test sets, we collect the predictions they generate and measure the percentage of Exact Match (EM) with respect to the expected target. An EM indicates that the code generated by the model for the completion instance is identical to the target (\ie the one we masked). We also compute metrics usually adopted in the assessment of generative models, namely the BLEU \cite{Papineni2002}, ChrF \cite{popovic2015chrf}, Levenshtein Distance \cite{levenshtein1966binary}, ROUGE \cite{lin2004rouge}, and METEOR \cite{banerjee2005meteor} score with respect to the target.

BLEU  is a popular automatic metric for machine translation tasks due to the high correlation to human judgement. It has become a standard metric in code completion tasks \cite{lu2021codexglue} since it measures the overlap of a predicted sequence and a set of reference sequences in terms of n-grams. ChrF is a character level metric which averages the F-score of 1 to 6-grams of characters. Levenshtein Distance is a measure of the minimal edit operations (\ie insert, modify, and remove), that would be needed to convert the predicted sequence into the target one, and it has been used in assessing the models' performance in previous code completion studies \cite{ciniselli2021empirical}. ROUGE, and specifically RougeL, is a metric that measures the longest common subsequence between the predicted and ground truth sequences. Lastly, Meteor is also an F-score, where the recall is weighted nine times more than the precision. Additionally, predictions are penalized for not having adjacent unigrams that exist in the ground truth. We also measure the Cross-Entropy of the generated predictions (\ie a measure of the surprise of the model when predicting the ground truth sequence).

While we computed all the above-described metrics, we only discuss the results achieved in terms of EM, ChrF, and RougeL. The former (EM) is an easy-to-interpret proxy of the model's performance. ChrF and RougeL, instead, have been found to be best at measuring performance compared to human evaluation and allow to claim significance (95\% confidence) if the difference between two models on code generation tasks is greater than two points \cite{evtikhiev2022out}. Our full analysis can be found in our replication package \cite{replication}. For implementing these metrics, we used the Huggingface's datasets library \cite{lhoest-etal-2021-datasets}, which contains a large selection of automated metrics for the evaluation of generative models.

\section{Results and Discussion}\label{completeformer:results}
\setlength\tabcolsep{3pt}
\begin{table}[]
\centering
\caption{Exact Match Score ($\uparrow$) achieved by the different position encoding schemes.\vspace{-0.3cm}}

\footnotesize
\centering
\label{tab:exact_results}
\begin{tabular}{cc|cccc|cccc}
\hline
\multirow{2}{*}{\textbf{Test Set}} & \multirow{2}{*}{\textbf{Encoding}}        & \multicolumn{4}{c|}{\textbf{Java}}                                  & \multicolumn{4}{c}{\textbf{Python}}    \\
\multicolumn{2}{c|}{}                                  & \textbf{Short} & \textbf{Medium} & \textbf{Long} & \textbf{Avg. $\Delta$}                      & \textbf{Short} & \textbf{Medium} & \textbf{Long} & \textbf{Avg. $\Delta$}   \\ \hline
\multirow{4}{*}{\textbf{Short}}  & \textbf{Sinusoidal} & \textbf{10.81\%}        & 2.91\%          & 0.50\%        & 84.23\%                    & \textbf{1.57\%}         & 0.26\%          & 0.03\%        & 90.76\%  \\
                                 & \textbf{xPOS}     & \textbf{12.49\%}        & 4.87\%          & 0.85\%        & \textcolor{red}{77.10\%}   & \textbf{2.33\%}         & 0.64\%          & 0.07\%        & 84.76\%  \\
                                 & \textbf{ALiBi}      & \textbf{10.97\%}        & 3.09\%          & 0.36\%        & 84.28\%                    & \textbf{1.48\%}         & 0.26\%          & 0.03\%        & 90.20\%  \\
                                 & \textbf{T5}         & \cellcolor{black}\textbf{\textcolor{white}{18.39\%}} 
                                                                                 & 6.66\%          & 1.38\%        & 78.14\%                    & \cellcolor{black}\textbf{\textcolor{white}{5.77\%}} 
                                                                                                                                                                          & 3.19\%          & 1.51\%        & \textcolor{red}{59.27\%} \\ \hline
                                 
\multirow{4}{*}{\textbf{Medium}} & \textbf{Sinusoidal} & 1.61\%   & \textbf{6.67\%}   & 3.32\%        & 63.04\%            & 0.57\%         & \textbf{1.33\%}          & 0.72\%        & 51.50\%  \\
                                 & \textbf{xPOS}     & 2.85\%   & \textbf{9.36\%}   & 8.02\%        & 41.93\%            & 1.53\%         & \textbf{2.84\%}          & 1.73\%        & 42.61\%  \\
                                 & \textbf{ALiBi}      & 1.56\%   & \textbf{6.61\%}   & 3.68\%        & 60.36\%            & 0.49\%         & \textbf{1.36\%}          & 0.64\%        & 58.46\%  \\
                                 & \textbf{T5}         & 5.29\%   & \cellcolor{black}\textbf{\textcolor{white}{14.06\%}}
                                                                                     & 11.33\%  & \textcolor{red}{40.90\%} & 3.16\%         & \cellcolor{black}\textbf{\textcolor{white}{5.71\%}}
                                                                                                                                                                            & 5.99\%        & \textcolor{red}{19.88\%} \\ \hline
                                 
\multirow{4}{*}{\textbf{Long}}   & \textbf{Sinusoidal} & 0.60\%  & 2.85\%  & \textbf{8.27\%}   & 79.14\%           & 0.05\%         & 0.61\%   & \textbf{8.81\%}        & 96.25\%          \\
                                 & \textbf{xPOS}     & 1.57\%  & 5.47\%  & \textbf{11.33\%}  & 68.93\%           & 0.48\%         & 1.80\%   & \textbf{11.34\%}       & 89.95\%          \\
                                 & \textbf{ALiBi}      & 0.57\%  & 2.79\%  & \textbf{8.29\%}   & 79.73\%           & 0.08\%         & 0.50\%   & \textbf{8.84\%}        & 96.72\%          \\
                                 & \textbf{T5}         & 3.62\%  & 9.59\%  & \cellcolor{black}\textbf{\textcolor{white}{17.03\%}}       
                                                                                     & \textcolor{red}{61.22\%}    & 3.10\%   & 7.19\%   & \cellcolor{black}\textbf{\textcolor{white}{20.73\%}} & \textcolor{red}{75.18\%}
\end{tabular}
\end{table}
\begin{table}[]
\caption{ChrF Score ($\uparrow$) achieved by the different position encoding schemes.\vspace{-0.3cm}}
\footnotesize
\centering
\label{tab:chrf_results}
\begin{tabular}{cc|cccc|cccc}
\hline
\multirow{2}{*}{\textbf{Test Set}} & \multirow{2}{*}{\textbf{Encoding}}  & \multicolumn{4}{c|}{\textbf{Java}}                          & \multicolumn{4}{c}{\textbf{Python}}                     \\
\multicolumn{2}{c|}{}                                    & \textbf{Short}   & \textbf{Medium}  & \textbf{Long}    & \textbf{Avg. $\Delta$}            & \textbf{Short}   & \textbf{Medium}  & \textbf{Long}    & \textbf{Avg. $\Delta$}\\ \hline
\multirow{4}{*}{\textbf{Short}}   & \textbf{Sinusoidal}  & \textbf{30.72\%} & 17.70\% & 13.19\% & 49.72\%                                     & \textbf{37.48\%} & 32.56\% & 30.00\% & 16.54\%  \\
                         & \textbf{xPOS}               & \textbf{34.57\%} & 23.36\% & 16.89\% & \textcolor{red}{41.78\%}                                   & \textbf{41.83\%} & 37.51\% & 33.06\% & 15.65\%  \\
                         & \textbf{ALiBi}                & \textbf{30.95\%} & 17.46\% & 13.02\% & 50.76\%                                     & \textbf{37.50\%} & 33.10\% & 30.31\% & 15.45\%  \\
                         & \textbf{T5} & \cellcolor{black}\textbf{\textcolor{white}{42.97\%}} & 27.43\%  & 17.37\% & 47.87\% & \cellcolor{black}\textbf{\textcolor{white}{50.86\%}} & 48.41\% & 43.30\% & \textcolor{red}{9.84\%}\\ \hline

\multirow{4}{*}{\textbf{Medium}}  & \textbf{Sinusoidal}  & 16.57\% & \textbf{25.01\%} & 19.94\% & 27.01\%                                     & 35.99\% & \textbf{37.17\%} & 35.21\% & \textcolor{red}{4.22\%}  \\
                         & \textbf{xPOS}               & 21.41\% & \textbf{30.96\%} & 27.68\% & 20.72\%                                     & 40.84\% & \textbf{43.95\%} & 39.30\% & 8.83\%  \\
                         & \textbf{ALiBi}                & 16.39\% & \textbf{24.93\%} & 20.29\% & 26.43\%                                     & 35.65\% & \textbf{36.88\%} & 34.67\% & 4.66\%   \\
                         & \textbf{T5} & 29.18\% & \cellcolor{black}\textbf{\textcolor{white}{39.67\%}} & 35.95\%  & \textcolor{red}{17.91\%} & 47.45\% & \cellcolor{black}\textbf{\textcolor{white}{52.61\%}} & 51.36\% & 6.09\% \\ \hline

\multirow{4}{*}{\textbf{Long}}    & \textbf{Sinusoidal}  & 13.93\% & 17.19\% & \textbf{25.03\%} & 37.83\%                                     & 32.79\% & 34.72\% & \textbf{41.71\%} & \textcolor{red}{19.07\%}          \\
                         & \textbf{xPOS}               & 18.30\% & 25.18\% & \textbf{32.38\%} & 32.86\%                                     & 38.36\% & 42.00\% & \textbf{50.58\%} & 20.56\%          \\
                         & \textbf{ALiBi}                & 13.47\% & 17.24\% & \textbf{24.91\%} & 38.36\%                                     & 32.36\% & 34.50\% & \textbf{41.66\%} & 19.76\%           \\
                         & \textbf{T5} & 27.14\% & 36.42\% & \cellcolor{black}\textbf{\textcolor{white}{44.93\%}} & \textcolor{red}{29.27\%}  & 48.14\% & 54.86\% & \cellcolor{black}\textbf{\textcolor{white}{64.96\%}} & 20.72\%
\end{tabular}
\end{table}
\begin{table*}[]
\centering
\caption{ROUGE-L Score ($\uparrow$) achieved by the different position encoding schemes.\vspace{-0.3cm}}

\footnotesize
\centering
\label{tab:rougel_results}
\begin{tabular}{cc|cccc|cccc}
\hline

\multirow{3}{*}{\textbf{Test Set}} & \multirow{3}{*}{\textbf{Encoding}}      & \multicolumn{4}{c|}{\textbf{Java}}                 & \multicolumn{4}{c}{\textbf{Python}}\\
\multicolumn{2}{c|}{}                      & \textbf{Short} & \textbf{Medium} & \textbf{Long} & \textbf{Avg. $\Delta$}                    & \textbf{Short}     & \textbf{Medium} & \textbf{Long} & \textbf{Avg. $\Delta$}\\ \hline
\multirow{4}{*}{\textbf{Short}}  & \textbf{Sinusoidal}  & \textbf{38.80\%}   & 24.56\%     & 17.68\%   & 45.57\%                  & \textbf{39.39\%}   & 33.61\%         & 30.16\%       & 19.05\% \\
                                 & \textbf{xPOS}     & \textbf{45.61\%}   & 35.23\%     & 25.66\%   & \textcolor{red}{33.25\%}               & \textbf{45.01\%}   & 40.10\%         & 32.96\%       & 18.84\% \\
                                 & \textbf{ALiBi}      & \textbf{39.24\%}   & 24.71\%     & 17.85\%   & 45.77\%                                              & \textbf{39.41\%}   & 33.81\%         & 30.62\%       & 18.26\% \\
                                 & \textbf{T5}         & \cellcolor{black}\textbf{\textcolor{white}{52.47\%}}  
                                                                            & 38.99\%     & 25.92\%   & 38.15\%                   & \cellcolor{black}\textbf{\textcolor{white}{52.84\%}}   & 50.46\%   & 44.92\%  & \textcolor{red}{9.75\%}\\ \hline

\multirow{4}{*}{\textbf{Medium}} & \textbf{Sinusoidal} & 22.08\%    & \textbf{29.63\%}    & 23.63\%    & 22.87\%                   & 36.26\%        & \textbf{37.15\%}         & 33.82\%       & 5.68\% \\
                                 & \textbf{xPOS}     & 30.98\%    & \textbf{41.28\%}    & 36.71\%    & 18.01\%                   & 43.14\%        & \textbf{46.43\%}         & 41.28\%       & 9.09\% \\
                                 & \textbf{ALiBi}      & 21.39\%    & \textbf{29.57\%}    & 24.18\%    & 22.95\%                   & 35.98\%        & \textbf{36.74\%}         & 33.81\%       & \textcolor{red}{5.02\%}\\
                                 & \textbf{T5}         & 39.12\%    & \cellcolor{black}\textbf{\textcolor{white}{49.36\%}}    
                                                                                          & 45.66\%    & \textcolor{red}{14.12\%}                            & 49.04\%        & \cellcolor{black}\textbf{\textcolor{white}{54.49\%}} & 53.14\% & 6.24\% \\ \hline

\multirow{4}{*}{\textbf{Long}}   & \textbf{Sinusoidal} & 19.05\%    & 22.27\%     & \textbf{26.60\%}   & 22.33\%                   & 32.80\%        & 34.07\%         & \textbf{39.83\%}       & \textcolor{red}{16.06\%}\\
                                 & \textbf{xPOS}     & 26.79\%    & 36.05\%     & \textbf{41.01\%}   & 23.38\%                   & 41.17\%        & 44.39\%         & \textbf{51.65\%}       & 17.17\% \\
                                 & \textbf{ALiBi}      & 18.69\%    & 22.33\%     & \textbf{26.66\%}   & 23.07\%                   & 32.70\%        & 33.90\%         & \textbf{39.80\%}       & 16.33\% \\
                                 & \textbf{T5}         & 36.20\%    & 46.35\%     & \cellcolor{black}\textbf{\textcolor{white}{52.84\%}}   
                                                                                                       & \textcolor{red}{21.89\%}  & 49.71\%        & 56.35\%         & \cellcolor{black}\textbf{\textcolor{white}{65.52\%}}  & 19.06\%
\end{tabular}
\end{table*}

Tables \ref{tab:exact_results}, \ref{tab:chrf_results}, and \ref{tab:rougel_results} show the results in terms of EM, ChrF, and RougeL, respectively, achieved by the four positional encoding schemes when trained on datasets featuring code completions of different lengths (columns) and tested on the \emph{short}, \emph{medium}, and \emph{long} test sets (rows). 

The results are reported for both Java and Python. To provide a concrete example, let us consider the EM results reported in the Table \ref{tab:exact_results}. Here, the Sinusoidal schema trained on \emph{short} Java completions generated 10.81\% EM predictions when tested on \emph{short} instances (\ie those resembling the training set instances). 

Instead, when the training is performed on instances having a \emph{medium} length, the percentage of EM predictions drops, on the same \emph{short} test set, to 2.91\%, finally falling at 0.50\% when the training was performed on \emph{long} instances. 

Similar results are observed for Python in which, however, the percentage of EM predictions is substantially lower, moving from 1.57\% achieved on the \emph{short} test set when training on \emph{short} instances down to the 0.03\% when tested on the \emph{long} ones.

Tables \ref{tab:exact_results}, \ref{tab:chrf_results}, and \ref{tab:rougel_results} also contain two ``Avg. $\Delta$'' columns (one per language). Given a row in one of the tables (\eg the first row in Table \ref{tab:exact_results} reporting the performance of the Sinusoidal schema when run on the \emph{short} test set), the Avg. $\Delta$ indicates the relative change in performance observed, on average, for the models trained on different lengths (in our example, those trained on \emph{medium} and \emph{long} completions) when compared the one specialized on lengths related to that row (\ie \emph{short}). Indeed, the 84.22\% shown as average $\Delta$ in the subject row is the result of:

$$
\frac{\frac{10.81\% - 2.91\%}{10.81\%} + \frac{10.81\% - 0.50\%}{10.81}}{2} = 84.22\%
$$

where 10.81\% is the percentage of EM predictions for Java generated by the Sinusoidal schema when trained and tested on code completions of \emph{short} length, while 2.91\% and 0.50\% are the EM scores achieved by the Sinusoidal schema when trained on \emph{medium} and \emph{long} instances, respectively, while still being tested on \emph{short} instances.

Finally, Tables \ref{tab:exact_results}, \ref{tab:chrf_results}, and \ref{tab:rougel_results} adopt three styles to highlight findings in the context of a specific test set length. Let us focus on the Java results achieved on the \emph{short} test set in terms of EM (Table \ref{tab:exact_results}). The black box shows the best-performing combination of $\langle$\emph{encoding schema}, \emph{training length}$\rangle$ for such a test set (\ie T5 trained on \emph{short} completions). The bold values highlight, for each encoding schema, the best-performing training length for such a test set (\ie in all cases, training on \emph{short} instances works better when testing on \emph{short} instances). The red value in each ``Avg. $\Delta$'' column highlights, instead, the encoding schema manifesting the lowest relative drop in performances when moving from a training length matching the instances in the test set (\eg training on \emph{short}, testing on \emph{short}) to the other training lengths. In this case, the lowest relative drop in terms of EM predictions is exhibited by xPOS. Note that a lowest relative drop indicates a better ability of the encoding schema to generalize to unseen lengths. 

The first observation that can be made from the three tables is that T5's positional encoding schema performs better than all other approaches. Such a finding is consistently captured by all metrics, including ChfR and RougeL for which the difference is always substantially higher than two points, indicating a statistically significant difference at 95\% confidence \cite{evtikhiev2022out}. Being the best performing one, however, does not save T5 from a strong general observation that can be made across the board for all training schemes: They all suffer from a major degradation in performance when applied on code completions having a length different from the one they have been trained on. Interestingly, the degradation is not only observed when the models are tested on instances being longer (likely more complex to handle) than those they have been trained on, but also in the opposite direction. This can be easily seen in Tables \ref{tab:exact_results}, \ref{tab:chrf_results}, and \ref{tab:rougel_results} by the fact that (i) the average $\Delta$ values are always positive, and (ii) the bold values in a given test set length are always associated with the same length in the training set.\eject

The second observation concerns the encoding schema reporting the lowest average drop in performance (red values in the ``Avg. $\Delta$'' column in the three tables). Overall, also from this perspective, T5 seems to be the best choice. There are a few exceptions to this trend, depending on the test set under analysis and on the metric used as proxy for performance. 

For example, on the \emph{short} Java test set, T5 is the second best in class in terms of EM and ChfR score, while confirming its leadership when looking at the RougeL score. By considering all 18 combinations of test set length (3), language (2), and evaluation metrics (3), T5 is the one exhibiting the lowest relative drop in 11 (61\%) of cases, and the second-best in additional 3 cases (17\%). Still, as observed, T5 also exhibits major drops in performance when working on sequence lengths unseen during training. For example, there is an absolute drop of 13.41\% in terms of EM predictions when testing T5 on the \emph{long} dataset when trained on \emph{short} sequences as compared to the one trained on \emph{long} sequences (17.03\% \emph{vs} 3.62\%). The trend is confirmed when looking at the ChfR and the RougeL scores.

xPOS is the second best performing schema, both in terms of absolute performance and generalization to different lengths. ALiBi and Sonusoidal follow, exhibiting similar performances from both perspectives. 

\vspace{0.2cm}
\begin{resultbox}
\textbf{Take Away \#1:} T5's positional encoding scheme achieves the best overall performance across metrics, lengths, and languages. Also, it is also better at generalizing to unseen lengths. In general, however, all encoding schemes suffer generalization issues for unseen lengths.  
\end{resultbox}

\textbf{Differences across languages.} Overall, our main findings hold on both languages. These include: (i) the lack of generalizability to unseen lengths of any of the experimented encoding schemes; and (ii) the superiority of T5 both in terms of absolute performance and relative drop when dealing with unseen lengths. 

We do not compare the absolute performance achieved on the two languages since (i) the test sets are different, (ii) the code completion tasks are different (statement-level \emph{vs.} block-level), and (iii) the syntaxt of the two languages make the prediction tasks quite different, since Python requires the generation of the \texttt{<TAB>} indentation tokens while Java does not.

\vspace{0.2cm}
\begin{resultbox}
\textbf{Take Away \#2:} On both languages, all encoding schemes struggle to generalize to unseen lengths. T5 confirms its superiority on both Java and Python code.
\end{resultbox}

\begin{table*}[]
\caption{Exact Match Mix ($\uparrow$) achieved by the different position encoding schemes.\vspace{-0.3cm}}

\centering
\label{tab:exact_mix}
\begin{tabular}{c|rrr|rrr}
\hline
\multirow{2}{*}{}   & \multicolumn{3}{c|}{\textbf{Java - Test Set}}               & \multicolumn{3}{c}{\textbf{Python - Test Set}}              \\
                    & \textbf{Short} & \textbf{Medium} & \textbf{Long} & \textbf{Short} & \textbf{Medium} & \textbf{Long} \\ \hline
\textbf{Sinusoidal} & 9.85\%         & 7.05\%          & 8.81\%        & 1.03\%         & 1.04\%          & 7.50\%        \\
\textbf{$\Delta$}      & -8.88\%         & +5.70\%         & +6.53\%       & -34.39\%         &-21.80\%          & -14.87\%        \\ \hline
\textbf{xPOS}     & 11.99\%        & 9.73\%          & 11.42\%       & 2.49\%         & 2.76\%          & 10.67\%       \\
\textbf{$\Delta$}      & -4.00\%         & +3.95\%         & +0.79\%       & +6.87\%        & -2.82\%          & -5.91\%        \\ \hline
\textbf{ALiBi}      & 10.12\%        & 6.85\%          & 8.53\%        & 1.00\%         & 0.98\%          & 7.34\%        \\
\textbf{$\Delta$}      & -7.75\%         & +3.63\%         & +2.90\%       & -32.43\%         & -27.94\%          & -16.97\%        \\ \hline
\textbf{T5}         & 19.11\%        & 17.57\%         & 18.63\%       & 3.19\%         & 3.69\%          & 12.22\%       \\
\textbf{$\Delta$}      & +3.92\%        & +24.96\%         & +9.40\%       & -44.71\%         & -35.38\%          & -41.05\%       
\end{tabular}
\end{table*}

\begin{table*}[]
\caption{ChrF Mix ($\uparrow$) achieved by the different position encoding schemes.\vspace{-0.3cm}}
\centering
\label{tab:chrf_mix}
\begin{tabular}{c|rrr|rrr}
\hline
\multirow{2}{*}{}   & \multicolumn{3}{c|}{\textbf{Java - Test Set}}               & \multicolumn{3}{c}{\textbf{Python - Test Set}}              \\
                    & \textbf{Short} & \textbf{Medium} & \textbf{Long} & \textbf{Short} & \textbf{Medium} & \textbf{Long} \\ \hline
\textbf{Sinusoidal} & 30.49\%        & 27.04\%           & 26.60\%         & 36.48\%    & 36.24\%         & 40.91\%   \\
\textbf{$\Delta$}   & -0.75\%        & +8.12\%           & +6.27\%         & -2.67\%    & -2.50\%         & -1.92\%   \\ \hline
\textbf{xPOS}     & 34.97\%        & 33.20\%           & 33.34\%         & 43.83\%    & 45.07\%         & 51.33\%   \\
\textbf{$\Delta$}   & +1.16\%        & +7.24\%           & +2.96\%         & +4.78\%    & +2.55\%         & +1.48\%   \\ \hline
\textbf{ALiBi}      & 30.42\%        & 26.72\%           & 26.28\%         & 36.44\%    & 36.14\%         & 40.93\%   \\
\textbf{$\Delta$}   & -1.71\%        & +7.18\%           & +5.50\%         & -2.83\%    & -2.01\%         & -1.75\%   \\ \hline
\textbf{T5}         & 45.47\%        & 46.11\%           & 47.73\%         & 45.94\%    & 48.33\%         & 55.00\%   \\
\textbf{$\Delta$}   & +5.82\%        & +16.23\%          & +6.23\%         & -9.67\%    & -8.14\%         & -15.33\%         
\end{tabular}
\end{table*}
\begin{table*}[]
\caption{ROUGE-L Mix ($\uparrow$) achieved by the different position encoding schemes.\vspace{-0.3cm}}

\centering
\label{tab:rouge_mix}
\begin{tabular}{c|rrr|rrr}
\hline
\multirow{2}{*}{}   & \multicolumn{3}{c|}{\textbf{Java - Test Set}}               & \multicolumn{3}{c}{\textbf{Python - Test Set}}              \\
                    & \textbf{Short} & \textbf{Medium} & \textbf{Long} & \textbf{Short} & \textbf{Medium} & \textbf{Long} \\ \hline
\textbf{Sinusoidal} & 36.85\%        & 31.54\%         & 29.37\%       & 37.85\%        & 36.18\%         & 39.35\%    \\
\textbf{$\Delta$}      & -5.03\%        & +6.45\%         & +10.41\%      & -3.91\%        & -2.61\%         & -1.21\%    \\ \hline
\textbf{xPOS}     & 45.26\%        & 43.36\%         & 42.60\%       & 46.27\%        & 47.33\%         & 52.25\%    \\
\textbf{$\Delta$}      & -0.77\%        & +5.04\%         & +3.88\%       & +1.94\%        & +1.94\%         & +1.16\%    \\ \hline
\textbf{ALiBi}      & 36.83\%        & 31.03\%         & 28.82\%       & 37.80\%        & 36.10\%         & 39.28\%    \\
\textbf{$\Delta$}      & -6.14\%        & +4.94\%         & +8.10\%       & -1.74\%        & -1.74\%         & -1.31\%    \\ \hline
\textbf{T5}         & 52.60\%        & 53.96\%         & 55.18\%       & 48.20\%        & 50.47\%         & 55.94\%    \\
\textbf{$\Delta$}      & +0.25\%        & +9.32\%         & +4.43\%       & -8.78\%        & -7.38\%         & -14.62\%      
\end{tabular}
\end{table*}

\textbf{Impact of training diversity.} Tables \ref{tab:exact_mix} (EM), \ref{tab:chrf_mix} (ChrF), and \ref{tab:rouge_mix} (RougeL) report the results achieved by the four encoding schemas (rows) when trained on the \emph{mix} dataset (\ie the one featuring a mixture of instances taken from the \emph{short}, \emph{medium}, and \emph{long} datasets)  and tested on the three datasets featuring sequences of different length (columns). 

Note that the \emph{mix} training dataset has exactly the same number of instances of the other length-specific datasets, thus do not introducing a confounding variable related to the training size. 

The $\Delta$ associated to each combination of encoding schema and test set is the relative change in performance with respect to the same schema exclusively trained for sequences of the corresponding length. For example, in terms of ChrF score (Table \ref{tab:chrf_mix}), T5 trained on \emph{short} Java sequences achieves a 42.97\% ChrF score when tested on the \emph{short} dataset. Such a score grows to 45.47\% when T5 is trained on the \emph{mix} dataset, with a relative improvement equal to (45.47\% - 42.97\%)/42.97\% = +5.82\%. The achieved findings confirm the superiority of T5 in this scenario as well. 

Most importantly, we found that relying on a mixture of lengths during training is generally sufficient to achieve results approaching, and in some cases improving, than those achieved by specifically training the model for the target sequence length. 

Indeed, by comparing the relative $\Delta$ reported, for example, in Table \ref{tab:chrf_mix} (ChfR scores when training on the \emph{mix} dataset) to those reported in Table \ref{tab:chrf_results} (ChfR scores when training on datasets featuring functions having different lengths), it is possible to observe a major difference in terms of magnitude of the deltas, with those in \ref{tab:chrf_mix} being substantially smaller. This indicates that, while in some cases training on a specific length range $l$ could help in achieving better performances on test instances fitting $l$, training on a mixture of lengths is a safe choice, since it would not result in dramatic lost of performances as those observed in Table \ref{tab:chrf_results}.

\vspace{0.2cm}
\begin{resultbox}
\textbf{Take Away \#3:} Training on a mixture of lengths being representative of those that will be seen during testing but also including other types of lengths might be the safest choice in most of cases. Only in scenarios in which even minor increases in performance are considered valuable, experimenting with a combination of models specialized on different lengths might be worthwhile, to then decide the best strategy to adopt.
\end{resultbox}

\section{Threats to Validity and Limitations}\label{completeformer:threats}

\noindent\textbf{Threats to Internal Validity.} In order to control for various levels of bias that can creep into our evaluation, we ensured to hold as many variables as possible in our datasets and models constant. This involved ensuring that there were no duplicates across the different training splits, both in the input and target \cite{allamanis2019adverse}. Also, we held constant the hyperparameters across our different models and only changed the type of length they were trained on. However, despite these thorough mitigation strategies, bias can still be present in our empirical study. 


\noindent\textbf{Threats to Construct Validity.} To mitigate threats to construct validity, we calculated a range of different metrics that have been commonly used in code completion literature. Additionally, we focus our discussion either on metrics that  have been shown to correlate with human preference and that are more statistically stable \cite{evtikhiev2022out} (\ie ChfR and RougeL) or that allow for a simple interpretation such as EM.

While there have been new recent metrics that are specific to code data for code completion, namely CodeBLEU \cite{ren2020codebleu} and functional-correctness \cite{hendrycks2021measuring, chen2021evaluating}, we did not compute these for the following reasons. CodeBLEU has been shown to be not as stable as ChrF and RougeL  \cite{evtikhiev2022out}. Unfortunately, functional-correctness was not even an option for our evaluation due to the lack of unit tests for our test examples. 

Besides the metric used, the type of code completion performed can result in bias as there has been some studies showing that synthetic benchmarks of code completion where the completions are randomized do not necessarily reflect the performance of real-world code completions \cite{hellendoorn2019code}. 

\noindent\textbf{Threats to External Validity.} We investigated two popular Transformer architectures to mitigate the threats to external validity of our results. Additionally, we measured multiple types of metrics and constructed our datasets in such a way as to hopefully mimic realistic code completion scenarios. Additionally, we used two popular programming languages, namely Java and Python, to better ensure our results generalize across languages.

\section{Bibliographical Notes}

The paper supporting the content described in this Chapter was written in collaboration with other members of the SEMERU group at William \& Mary and researchers from Universit\'{a} della Svizzera italiana. It is currently under review for publication.

\textbf{Cooper, N.}, Tufano, R., Bavota, G., \& Poshyvanyk, D. (2023, March). On the Generalizability of Transformer Models for Code Completion. Under Review.

\chapter{Conclusions \& Future Research}
\label{sec:conclusion}

In this dissertation, I have discussed exploration to answer my original research question: \textit{In what ways can the software development process be improved through leveraging Deep Learning techniques on the vast amounts of unstructured software engineering artifacts?} Specifically, I discussed our work in first conducting a literature review and then using the findings to guide the development of intelligent software development and maintenance tools to assist software engineers in a variety of tasks. To give a succinct overview, I discuss each projects' conclusions and future work below.


\section{DL4SE Literature Review}
\label{dlslr:conclusion}

In Chapter \ref{sec:dlslr}, we presented a systematic literature review on the primary studies related to DL4SE from the top software engineering research venues. Our work heavily relied on the guidelines laid out by Kitchenham \etal for performing systematic literature reviews in software engineering. We began by establishing a set of research questions that we wanted to answer pertaining to applications of DL models on SE tasks. We then empirically developed a search string to extract the relevant primary studies to the research questions we wanted to answer. We supplemented our searching process with snowballing and manual additions of papers that were not captured by our systematic approach but were relevant to our study. We then classified the relevant pieces of work using a set of agreed upon inclusion and exclusion criteria. After distilling out a set of relevant papers, we extracted the necessary information from those papers to answer our research questions. Through the extraction process and the nature of our research questions, we inherently generated a taxonomy which pertains to different aspects of applying a DL-based approach to a SE task.

\textbf{Future work.} Our hope is that this SLR provides future SE researchers with the necessary information and intuitions for applying DL in new and interesting ways within the field of SE. The concepts described in this review should aid researchers and developers in understanding where DL can be applied and the necessary considerations for applying these complex models to automate SE tasks.

\section{Video-Based Bug Reporting}\label{tango:conclusion}

Chapter \ref{sec:tango} presented \tango, an approach that combines visual and textual information to help developers find duplicate video-based bug reports. Our empirical evaluation, conducted on $4,680$ duplicate detection tasks created from 180 video-based bug reports from six mobile apps, illustrates that \tango is able to effectively identify duplicate reports and save developer effort. Specifically, \tango correctly suggests duplicate video-based bug reports within the top-2 candidate videos for 83\% of the tasks, and saves  65.1\% of the time that humans spend finding duplicate videos.

\textbf{Future work.} Future work can focus on addressing \tango's limitations and extending \tango's evaluation. Specifically, research can (1) explore additional ways to address the vocabulary overlap problem, (2) investigate the resilience of \tango to different app characteristics such as the use of different themes, languages, and screen sizes, (3) extend \tango for detecting duplicate bug reports that contain multimedia information (text, images, and videos), (4) evaluate \tango using data from additional apps, and (5) assess the usefulness of \tango in industrial settings.

\section{Impact Analysis with Deep Learning and Call Graphs}\label{athena:conclusion}

\new{In Chapter \ref{sec:athena}, we proposed \athena, an information retrieval technique for the task of impact analysis that combines neural code semantic embedding information with structural software call graph information. Additionally, we constructed a large benchmark for impact analysis that is constructed from manually verified bug fixing commits to prevent tangling. On our new benchmark, \athena outperforms techniques without contextual call graph information by a large margin (+4.58\% mRR, +3.85\% mAP, and +5.67\% HIT@10) and is robust across software systems with 22 out of 25 systems seeing improvement. Additionally, through our analysis, we found \athena's performance increase is on its ability to better find impacted methods to a change when the methods are outside of the query method's class.}

\new{\textbf{Future work.} New research can explore using \athena for other tasks in software engineering such as code search, clone detection, traceability, \etc. Additionally, looking at different ways of integrating the two types of information such as through the usage of Graph Neural Networks. Lastly, we believe exploring \athenas ability to work with other languages to see if the approach generalizes is also a fruitful endeavor.}

\section{Generalization of Code Completion Models}\label{completeformer:conclusion}

\new{Lastly, in Chapter \ref{sec:completeformer}, we explored the generalization ability of popular decoder-only Transformer position encoding schemes that have shown success in Natural Language Processing that can be extended to the encoder-decoder Transformer and code completion task. Specifically, we investigated four different positional encoding schemes, namely Sinusoidal \cite{vaswani2017attention}, xPOS \cite{su2021roformer, sun2022length}, ALiBi \cite{press2022alibi}, and T5 \cite{raffel2020exploring}, which have been proposed as a way to boost this generalization ability and represent the most popular position encoding types, \ie Absolute Positional Encoding and Relative Positional Encoding.}

\new{Overall, our results demonstrate that none of the studied positional encoding schemes has the ability to generalize to unseen lengths. While these findings suggest that there are currently no ``shortcuts'' for researchers or developers of tools utilizing these models to efficiently train on short lengths (which are less expensive to process) and generalize to longer lengths, it is interesting to note that training on a mixture of lengths should represent a safe compromise in most of cases. Still, the possible drop in performance this may result in should be considered and assessed case by case, depending on the context in which the models must be used, the targeted programming language, and the actual focus on performance.}

\new{Moreover, it is worth considering that our conclusions are only based on performance proxies we adopted (\ie EM, ChrF, and RougeL). Different trade-offs come into play if best performance is not the only constraint. For example, while T5 is the best performing positional encoding, it is also the slowest and most memory intensive both for training and inference. Therefore, in performance-critical settings it might be more beneficial to use xPOS, which achieves less performance, but is more efficient. Similar observations can be made for the Sinusoidal and ALiBi schemes for which, however, the cost to pay in terms of performance as compared to T5 is noticeably higher.}

\new{\textbf{Future work.} Given our findings and the potential impact that the generalizability of code completions models can have on the software engineering community in terms of training efficiency, we believe future research should explicitly target this problem with research focused outside of different positional encoding schemes and possibly involving additional architectural changes to the Transformer or even proposing completely novel architectures.}



\makeThesisBib{main}

\begin{thebibliography}{100}

\bibitem{tesseract-ocr}
Tesseract ocr library \url{https://github.com/tesseract-ocr/tesseract/wiki}.

\bibitem{acm-artifact-review}
Acm artifact review policies
  \url{https://www.acm.org/publications/policies/artifact-review-badging},
  2019.

\bibitem{antennapod}
Antennapod \url{https://github.com/AntennaPod/AntennaPod}, 2019.

\bibitem{droidweight}
Droidweight \url{https://github.com/sspieser/droidweight}, 2019.

\bibitem{gnucash}
Gnucash \url{https://github.com/codinguser/gnucash-android}, 2019.

\bibitem{growtracker}
Growtracker \url{https://tinyurl.com/yy9oezom}, 2019.

\bibitem{time-tracker}
Time tracker \url{https://github.com/netmackan/ATimeTracker}, 2019.

\bibitem{token}
Token \url{https://github.com/markmcavoy/androidtoken}, 2019.

\bibitem{birdeatsbugs}
Bird eats bugs \url{https://birdeatsbug.com/}, 2020.

\bibitem{bugsquasher}
Bug squasher \url{https://thebugsquasher.com/}, 2020.

\bibitem{bugclipper}
Bugclipper \url{http://bugclipper.com}, 2020.

\bibitem{bugreplay}
Bugreplay \url{https://www.bugreplay.com/ }, 2020.

\bibitem{bugsee}
Bugsee \url{https://www.bugsee.com/}, 2020.

\bibitem{Instabug}
Instabug \url{https://instabug.com/screen-recording}, 2020.

\bibitem{lucene-tfidfsimilarity}
Lucene's tfidfsimilarity javadoc - \url{https://tinyurl.com/ybhqqrqm}, 2020.

\bibitem{outklip}
Outklip \url{https://outklip.com/}, 2020.

\bibitem{pytesseract}
Python tesseract \url{https://github.com/madmaze/pytesseract}, 2020.

\bibitem{snaffu}
Snaffu \url{https://snaffu.squarespace.com/}, 2020.

\bibitem{testfairy}
Testfairy \url{https://testfairy.com}, 2020.

\bibitem{ubertesters}
Ubertesters \url{https://ubertesters.com/bug-reporting-tools/}, 2020.

\bibitem{core}
Welcome to core \url{https://www.core.edu.au/home}.
\newblock 2023.

\bibitem{abu-mastafa}
{\sc Yaser~S. Abu-Mostafa, Malik Magdon-Ismail, and Hsuan-Tien Lin}.
\newblock {\em Learning from data: a short course}.
\newblock AMLbook.com, 2012.

\bibitem{Alahmadi:EMSE20}
{\sc Mohammad Alahmadi, Abdulkarim Khormi, Biswas Parajuli, Jonathan Hassel,
  Sonia Haiduc, and Piyush Kumar}.
\newblock Code localization in programming screencasts.
\newblock {\em EMSE'20}, 25(2):1536--1572, 2020.

\bibitem{allal2023santacoder}
{\sc Loubna~Ben Allal, Raymond Li, Denis Kocetkov, Chenghao Mou, Christopher
  Akiki, Carlos~Munoz Ferrandis, Niklas Muennighoff, Mayank Mishra, Alex Gu,
  Manan Dey, et~al.}
\newblock Santacoder: don't reach for the stars!
\newblock {\em arXiv preprint arXiv:2301.03988}, 2023.

\bibitem{allamanis2019adverse}
{\sc Miltiadis Allamanis}.
\newblock The adverse effects of code duplication in machine learning models of
  code.
\newblock In {\em Proceedings of the 2019 ACM SIGPLAN International Symposium
  on New Ideas, New Paradigms, and Reflections on Programming and Software},
  pages 143--153, 2019.

\bibitem{Allamanis2015}
{\sc Miltiadis Allamanis, Earl~T. Barr, Christian Bird, and Charles Sutton}.
\newblock Suggesting accurate method and class names.
\newblock In {\em Proceedings of the 2015 10th Joint Meeting on Foundations of
  Software Engineering}, ESEC/FSE 2015, pages 38--49, New York, NY, USA, 2015.
  ACM.

\bibitem{allamanis2018learning}
{\sc Miltiadis Allamanis, Marc Brockschmidt, and Mahmoud Khademi}.
\newblock Learning to represent programs with graphs.
\newblock In {\em 6th International Conference on Learning Representations,
  {ICLR} 2018, Vancouver, BC, Canada, April 30 - May 3, 2018, Conference Track
  Proceedings}. OpenReview.net, 2018.

\bibitem{Allamanis2016}
{\sc Miltiadis Allamanis, Hao Peng, and Charles Sutton}.
\newblock A convolutional attention network for extreme summarization of source
  code.
\newblock In {\em Proceedings of the 33nd International Conference on Machine
  Learning, {ICML} 2016, New York City, NY, USA, June 19-24, 2016},
  Maria{-}Florina Balcan and Kilian~Q. Weinberger, editors, volume~48 of {\em
  {JMLR} Workshop and Conference Proceedings}, pages 2091--2100. JMLR.org,
  2016.

\bibitem{alon2019code2vec}
{\sc Uri Alon, Meital Zilberstein, Omer Levy, and Eran Yahav}.
\newblock code2vec: Learning distributed representations of code.
\newblock {\em Proceedings of the ACM on Programming Languages}, 3(POPL):1--29,
  2019.

\bibitem{ami2022crypto}
{\sc Amit~Seal Ami, Nathan Cooper, Kaushal Kafle, Kevin Moran, Denys
  Poshyvanyk, and Adwait Nadkarni}.
\newblock Why crypto-detectors fail: A systematic evaluation of cryptographic
  misuse detection techniques.
\newblock In {\em 2022 IEEE Symposium on Security and Privacy (SP)}, pages
  614--631. IEEE, 2022.

\bibitem{osti_10334413}
{\sc Amit~Seal Ami, Kaushal Kafle, Kevin Moran, Adwait Nadkarni, and Denys
  Poshyvanyk}.
\newblock Systematic mutation-based evaluation of the soundness of
  security-focused android static analysis techniques.
\newblock {\em ACM Transactions on Privacy and Security}, 24(3).

\bibitem{9402554}
{\sc Amit~Seal Ami, Kaushal Kafle, Adwait Nadkarni, Denys Poshyvanyk, and Kevin
  Moran}.
\newblock µse: Mutation-based evaluation of security-focused static analysis
  tools for android.
\newblock In {\em 2021 IEEE/ACM 43rd International Conference on Software
  Engineering: Companion Proceedings (ICSE-Companion)}, pages 53--56, 2021.

\bibitem{Andrews:ICSE05}
{\sc James~H Andrews, Lionel~C Briand, and Yvan Labiche}.
\newblock Is mutation an appropriate tool for testing experiments?
\newblock In {\em ICSE'05}, pages 402--411, 2005.

\bibitem{ANGELOV2020185}
{\sc Plamen Angelov and Eduardo Soares}.
\newblock Towards explainable deep neural networks (xdnn).
\newblock {\em Neural Networks}, 130:185--194, 2020.

\bibitem{antoniol2000identifying}
{\sc Giuliano Antoniol, Gerardo Canfora, Gerardo Casazza, and Andrea De~Lucia}.
\newblock Identifying the starting impact set of a maintenance request: A case
  study.
\newblock In {\em Proceedings of the fourth European conference on software
  maintenance and reengineering}, pages 227--230. IEEE, 2000.

\bibitem{Arabshahi2018}
{\sc Forough Arabshahi, Sameer Singh, and Animashree Anandkumar}.
\newblock Combining symbolic expressions and black-box function evaluations in
  neural programs.
\newblock In {\em 6th International Conference on Learning Representations,
  {ICLR} 2018, Vancouver, BC, Canada, April 30 - May 3, 2018, Conference Track
  Proceedings}. OpenReview.net, 2018.

\bibitem{arisholm2004dynamic}
{\sc Erik Arisholm, Lionel~C Briand, and Audun Foyen}.
\newblock Dynamic coupling measurement for object-oriented software.
\newblock {\em IEEE Transactions on software engineering}, 30(8):491--506,
  2004.

\bibitem{austin2021program}
{\sc Jacob Austin, Augustus Odena, Maxwell Nye, Maarten Bosma, Henryk
  Michalewski, David Dohan, Ellen Jiang, Carrie Cai, Michael Terry, Quoc Le,
  et~al.}
\newblock Program synthesis with large language models.
\newblock {\em arXiv preprint arXiv:2108.07732}, 2021.

\bibitem{aye2021learning}
{\sc Gareth~Ari Aye, Seohyun Kim, and Hongyu Li}.
\newblock Learning autocompletion from real-world datasets.
\newblock In {\em 2021 IEEE/ACM 43rd International Conference on Software
  Engineering: Software Engineering in Practice (ICSE-SEIP)}, pages 131--139.
  IEEE, 2021.

\bibitem{badri2005supporting}
{\sc Linda Badri, Mourad Badri, and Daniel St-Yves}.
\newblock Supporting predictive change impact analysis: a control call graph
  based technique.
\newblock In {\em 12th Asia-Pacific Software Engineering Conference
  (APSEC'05)}, pages 9--pp. IEEE, 2005.

\bibitem{bahdanau2014neural}
{\sc Dzmitry Bahdanau, Kyunghyun Cho, and Yoshua Bengio}.
\newblock Neural machine translation by jointly learning to align and
  translate.
\newblock {\em arXiv preprint arXiv:1409.0473}, 2014.

\bibitem{Balog2016}
{\sc Matej Balog, Alexander~L. Gaunt, Marc Brockschmidt, Sebastian Nowozin, and
  Daniel Tarlow}.
\newblock Deepcoder: Learning to write programs.
\newblock In {\em 5th International Conference on Learning Representations,
  {ICLR} 2017, Toulon, France, April 24-26, 2017, Conference Track
  Proceedings}. OpenReview.net, 2017.

\bibitem{banerjee2005meteor}
{\sc Satanjeev Banerjee and Alon Lavie}.
\newblock Meteor: An automatic metric for mt evaluation with improved
  correlation with human judgments.
\newblock In {\em Proceedings of the acl workshop on intrinsic and extrinsic
  evaluation measures for machine translation and/or summarization}, pages
  65--72, 2005.

\bibitem{Bao:ICSE15}
{\sc Lingfeng Bao, Jing Li, Zhenchang Xing, Xinyu Wang, and Bo~Zhou}.
\newblock scvripper: video scraping tool for modeling developers' behavior
  using interaction data.
\newblock In {\em ICSE'15}, pages 673--676. IEEE Press, 2015.

\bibitem{Bavishi2019}
{\sc Rohan Bavishi, Caroline Lemieux, Roy Fox, Koushik Sen, and Ion Stoica}.
\newblock Autopandas: Neural-backed generators for program synthesis.
\newblock {\em Proc. ACM Program. Lang.}, 3(OOPSLA), October 2019.

\bibitem{bavota2013empirical}
{\sc Gabriele Bavota, Bogdan Dit, Rocco Oliveto, Massimilano Di~Penta, Denys
  Poshyvanyk, and Andrea De~Lucia}.
\newblock An empirical study on the developers' perception of software
  coupling.
\newblock In {\em 2013 35th International Conference on Software Engineering
  (ICSE)}, pages 692--701. IEEE, 2013.

\bibitem{Bay:ECCV'06}
{\sc Herbert Bay, Tinne Tuytelaars, and Luc Van~Gool}.
\newblock Surf: Speeded up robust features.
\newblock volume 3951, pages 404--417, 07 2006.

\bibitem{beltramelli2018pix2code}
{\sc Tony Beltramelli}.
\newblock pix2code: Generating code from a graphical user interface screenshot.
\newblock In {\em Proceedings of the ACM SIGCHI Symposium on Engineering
  Interactive Computing Systems}, pages 1--6, 2018.

\bibitem{Beltramelli:EICS18}
{\sc Tony Beltramelli}.
\newblock pix2code: Generating code from a graphical user interface screenshot.
\newblock In {\em EICS'18}, page~3. ACM, 2018.

\bibitem{BenNun2018}
{\sc Tal Ben{-}Nun, Alice~Shoshana Jakobovits, and Torsten Hoefler}.
\newblock Neural code comprehension: {A} learnable representation of code
  semantics.
\newblock In {\em Advances in Neural Information Processing Systems 31: Annual
  Conference on Neural Information Processing Systems 2018, NeurIPS 2018,
  December 3-8, 2018, Montr{\'{e}}al, Canada}, Samy Bengio, Hanna~M. Wallach,
  Hugo Larochelle, Kristen Grauman, Nicol{\`{o}} Cesa{-}Bianchi, and Roman
  Garnett, editors, pages 3589--3601, 2018.

\bibitem{9839513}
{\sc Carlos Bernal-C\'{a}rdenas, Nathan Cooper, Madeleine Havranek, Kevin
  Moran, Oscar Chaparro, Denys Poshyvanyk, and Andrian Marcus}.
\newblock Translating video recordings of complex mobile app ui gestures into
  replayable scenarios.
\newblock {\em IEEE Transactions on Software Engineering}, pages 1--23, 2022.

\bibitem{3Bettenburg:FSE08}
{\sc Nicolas Bettenburg, Sascha Just, Adrian Schr\"{o}ter, Cathrin Weiss, Rahul
  Premraj, and Thomas Zimmermann}.
\newblock What makes a good bug report?
\newblock In {\em FSE'08}, pages 308--318, New York, NY, USA, 2008. ACM.

\bibitem{Bettenburg:ICSM08}
{\sc Nicolas Bettenburg, R.~Premraj, T.~Zimmermann, and Sunghun Kim}.
\newblock Duplicate bug reports considered harmful... really?
\newblock In {\em ICSM'08}, pages 337--345, Sept 2008.

\bibitem{11Bettenburg:MSR08}
{\sc Nicolas Bettenburg, Rahul Premraj, Thomas Zimmermann, and Sunghun Kim}.
\newblock Extracting structural information from bug reports.
\newblock In {\em MSR'08}, MSR '08, pages 27--30, New York, NY, USA, 2008. ACM.

\bibitem{Bhatia2018}
{\sc Sahil Bhatia, Pushmeet Kohli, and Rishabh Singh}.
\newblock Neuro-symbolic program corrector for introductory programming
  assignments.
\newblock In {\em Proceedings of the 40th International Conference on Software
  Engineering}, ICSE '18, pages 60--70, New York, NY, USA, 2018. ACM.

\bibitem{Bohner&Arnold1996b}
{\sc Shawn~A. Bohner and Robert~S. Arnold}.
\newblock {\em Software Change Impact Analysis}, chapter An Introduction to
  Software Change Impact Analysis, pages 1--26.
\newblock 1996.

\bibitem{Bonett:USENIX18}
{\sc Richard Bonett, Kaushal Kafle, Kevin Moran, Adwait Nadkarni, and Denys
  Poshyvanyk}.
\newblock Discovering flaws in {Security-Focused} static analysis tools for
  android using systematic mutation.
\newblock In {\em 27th USENIX Security Symposium (USENIX Security 18)}, pages
  1263--1280, Baltimore, MD, August 2018. USENIX Association.

\bibitem{Borg2014}
{\sc Markus Borg, Per Runeson, Jens Johansson, and Mika~V. M{\"a}ntyl{\"a}}.
\newblock A {{Replicated Study}} on {{Duplicate Detection}}: {{Using Apache
  Lucene}} to {{Search Among Android Defects}}.
\newblock In {\em ESEM'14}, pages 8:1--8:4, 2014.

\bibitem{bravenboer2009strictly}
{\sc Martin Bravenboer and Yannis Smaragdakis}.
\newblock Strictly declarative specification of sophisticated points-to
  analyses.
\newblock In {\em Proceedings of the 24th ACM SIGPLAN conference on Object
  oriented programming systems languages and applications}, pages 243--262,
  2009.

\bibitem{briand1999using}
{\sc Lionel~C Briand, Jurgen Wust, and Hakim Lounis}.
\newblock Using coupling measurement for impact analysis in object-oriented
  systems.
\newblock In {\em Proceedings IEEE International Conference on Software
  Maintenance-1999 (ICSM'99).'Software Maintenance for Business Change'(Cat.
  No. 99CB36360)}, pages 475--482. IEEE, 1999.

\bibitem{bruch2009learning}
{\sc Marcel Bruch, Martin Monperrus, and Mira Mezini}.
\newblock Learning from examples to improve code completion systems.
\newblock In {\em Proceedings of the 7th joint meeting of the European software
  engineering conference and the ACM SIGSOFT symposium on the foundations of
  software engineering}, pages 213--222, 2009.

\bibitem{tree_sitter}
{\sc Max Brunsfeld, Patrick Thomson, Andrew Hlynskyi, Josh Vera, Phil Turnbull,
  Timothy Clem, Douglas Creager, Andrew Helwer, Rob Rix, Hendrik van Antwerpen,
  Michael Davis, Ika, Tuan-Anh Nguyen, Stafford Brunk, Niranjan Hasabnis,
  bfredl, Mingkai Dong, Vladimir Panteleev, ikrima, Steven Kalt, Kolja Lampe,
  Alex Pinkus, Mark Schmitz, Matthew Krupcale, narpfel, Santos Gallegos, Vicent
  Martí, Edgar, and George Fraser}.
\newblock tree-sitter/tree-sitter: v0.20.7, September 2022.

\bibitem{buch2019learning}
{\sc Lutz B{\"u}ch and Artur Andrzejak}.
\newblock Learning-based recursive aggregation of abstract syntax trees for
  code clone detection.
\newblock In {\em 2019 IEEE 26th International Conference on Software Analysis,
  Evolution and Reengineering (SANER)}, pages 95--104. IEEE, 2019.

\bibitem{DBLP:conf/aaai/BuiJY18}
{\sc Nghi D.~Q. Bui, Lingxiao Jiang, and Yijun Yu}.
\newblock Cross-language learning for program classification using bilateral
  tree-based convolutional neural networks.
\newblock In {\em The Workshops of the The Thirty-Second {AAAI} Conference on
  Artificial Intelligence, New Orleans, Louisiana, USA, February 2-7, 2018},
  volume {WS-18} of {\em {AAAI} Workshops}, pages 758--761. {AAAI} Press, 2018.

\bibitem{bui:saner19}
{\sc Nghi D.~Q. Bui, Yijun Yu, and Lingxiao Jiang}.
\newblock Bilateral dependency neural networks for cross-language algorithm
  classification.
\newblock In {\em 2019 IEEE 26th International Conference on Software Analysis,
  Evolution and Reengineering (SANER)}, pages 422--433, 2019.

\bibitem{Bunel2018}
{\sc Rudy Bunel, Matthew~J. Hausknecht, Jacob Devlin, Rishabh Singh, and
  Pushmeet Kohli}.
\newblock Leveraging grammar and reinforcement learning for neural program
  synthesis.
\newblock In {\em 6th International Conference on Learning Representations,
  {ICLR} 2018, Vancouver, BC, Canada, April 30 - May 3, 2018, Conference Track
  Proceedings}. OpenReview.net, 2018.

\bibitem{buch:saner19}
{\sc Lutz Büch and Artur Andrzejak}.
\newblock Learning-based recursive aggregation of abstract syntax trees for
  code clone detection.
\newblock In {\em 2019 IEEE 26th International Conference on Software Analysis,
  Evolution and Reengineering (SANER)}, pages 95--104, 2019.

\bibitem{Caetano:02}
{\sc Anabela Caetano, Neri Goulart, Manuel Fonseca, and Joaquim Jorge}.
\newblock Javasketchit: Issues in sketching the look of user interfaces.
\newblock In {\em SSS'02}, pages 9--14, 2002.

\bibitem{cai2015comprehensive}
{\sc Haipeng Cai and Raul Santelices}.
\newblock A comprehensive study of the predictive accuracy of dynamic
  change-impact analysis.
\newblock {\em Journal of Systems and Software}, 103:248--265, 2015.

\bibitem{Cai2017}
{\sc Jonathon Cai, Richard Shin, and Dawn Song}.
\newblock Making neural programming architectures generalize via recursion.
\newblock In {\em 5th International Conference on Learning Representations,
  {ICLR} 2017, Toulon, France, April 24-26, 2017, Conference Track
  Proceedings}. OpenReview.net, 2017.

\bibitem{Cai:MM'11}
{\sc Yang Cai, Linjun Yang, Wei Ping, Fei Wang, Tao Mei, Xian-Sheng Hua, and
  Shipeng Li}.
\newblock Million-scale near-duplicate video retrieval system.
\newblock In {\em MM'11}, page 837–838, New York, NY, USA, 2011.

\bibitem{canfora2005impact}
{\sc Gerardo Canfora and Luigi Cerulo}.
\newblock Impact analysis by mining software and change request repositories.
\newblock In {\em 11th IEEE International Software Metrics Symposium
  (METRICS'05)}, pages 9--pp. IEEE, 2005.

\bibitem{canfora2006fine}
{\sc Gerardo Canfora and Luigi Cerulo}.
\newblock Fine grained indexing of software repositories to support impact
  analysis.
\newblock In {\em Proceedings of the 2006 international workshop on Mining
  software repositories}, pages 105--111, 2006.

\bibitem{Carreira:17}
{\sc Joao Carreira and Andrew Zisserman}.
\newblock Quo vadis, action recognition? a new model and the kinetics dataset.
\newblock In {\em CVPR'17}, 2017.

\bibitem{Chang:UIST11}
{\sc Tsung-Hsiang Chang, Tom Yeh, and Rob Miller}.
\newblock Associating the visual representation of user interfaces with their
  internal structures and metadata.
\newblock In {\em Proceedings of the 24th Annual ACM Symposium on User
  Interface Software and Technology}, UIST '11, pages 245--256, New York, NY,
  USA, 2011. ACM.

\bibitem{Chaparro:FSE'19}
{\sc Oscar Chaparro, Carlos Bernal-C\'{a}rdenas, Jing Lu, Kevin Moran, Andrian
  Marcus, Massimiliano Di~Penta, Denys Poshyvanyk, and Vincent Ng}.
\newblock Assessing the quality of the steps to reproduce in bug reports.
\newblock In {\em ESEC/FSE'19}, Bergamo, Italy, Ausgust 2019 2019.

\bibitem{Chaparro2016a}
{\sc Oscar Chaparro, Juan~Manuel Florez, and Andrian Marcus}.
\newblock On the vocabulary agreement in software issue descriptions.
\newblock In {\em ICSME'16}, pages 448--452, 2016.

\bibitem{Chaparro2019}
{\sc Oscar Chaparro, Juan~Manuel Florez, Unnati Singh, and Andrian Marcus}.
\newblock Reformulating queries for duplicate bug report detection.
\newblock In {\em SANER'19}, pages 218--229, 2019.

\bibitem{Charmaz:groundedtheory}
{\sc K.~Charmaz}.
\newblock {\em Constructing Grounded Theory}.
\newblock SAGE Publications Inc., 2006.

\bibitem{Chechik:JMLR'10}
{\sc Gal Chechik, Varun Sharma, Uri Shalit, and Samy Bengio}.
\newblock Large scale online learning of image similarity through ranking.
\newblock {\em JMLR'10}, 11:1109–1135, March 2010.

\bibitem{Chen2019}
{\sc C.~{Chen}, Z.~{Xing}, Y.~{Liu}, and K.~L.~X. {Ong}}.
\newblock Mining likely analogical apis across third-party libraries via
  large-scale unsupervised api semantics embedding.
\newblock {\em IEEE Transactions on Software Engineering}, pages 1--1, 2019.

\bibitem{Chen2018d}
{\sc Chao Chen, Wenrui Diao, Yingpei Zeng, Shanqing Guo, and Chengyu Hu}.
\newblock Drlgencert: Deep learning-based automated testing of certificate
  verification in {SSL/TLS} implementations.
\newblock In {\em 2018 {IEEE} International Conference on Software Maintenance
  and Evolution, {ICSME} 2018, Madrid, Spain, September 23-29, 2018}, pages
  48--58. {IEEE} Computer Society, 2018.

\bibitem{Chen2018}
{\sc Chunyang Chen, Ting Su, Guozhu Meng, Zhenchang Xing, and Yang Liu}.
\newblock From ui design image to gui skeleton: A neural machine translator to
  bootstrap mobile gui implementation.
\newblock In {\em Proceedings of the 40th International Conference on Software
  Engineering}, ICSE '18, pages 665--676, New York, NY, USA, 2018. ACM.

\bibitem{Chen2016}
{\sc G.~{Chen}, C.~{Chen}, Z.~{Xing}, and B.~{Xu}}.
\newblock Learning a dual-language vector space for domain-specific
  cross-lingual question retrieval.
\newblock In {\em 2016 31st IEEE/ACM International Conference on Automated
  Software Engineering (ASE)}, pages 744--755, Sep. 2016.

\bibitem{Chen:FSE20}
{\sc Jieshan Chen, Mulong Xie, Zhenchang Xing, Chunyang Chen, Xiwei Xu, and
  Liming Zhu}.
\newblock Object detection for graphical user interface: Old fashioned or deep
  learning or a combination?
\newblock In {\em ESEC/FSE’20}, page to appear, 2020.

\bibitem{chen2021evaluating}
{\sc Mark Chen, Jerry Tworek, Heewoo Jun, Qiming Yuan, Henrique Ponde
  de~Oliveira Pinto, Jared Kaplan, Harri Edwards, Yuri Burda, Nicholas Joseph,
  Greg Brockman, et~al.}
\newblock Evaluating large language models trained on code.
\newblock {\em arXiv preprint arXiv:2107.03374}, 2021.

\bibitem{Chen2018a}
{\sc Qingying Chen and Minghui Zhou}.
\newblock A neural framework for retrieval and summarization of source code.
\newblock In {\em Proceedings of the 33rd ACM/IEEE International Conference on
  Automated Software Engineering}, ASE 2018, pages 826--831, New York, NY, USA,
  2018. ACM.

\bibitem{Chen:SimCLR'20}
{\sc Ting Chen, Simon Kornblith, Mohammad Norouzi, and Geoffrey Hinton}.
\newblock A simple framework for contrastive learning of visual
  representations.
\newblock In {\em ICML'20}, pages 1597--1607, 2020.

\bibitem{Chen:ICDM'06}
{\sc X.~{Chen} and C.~{Zhang}}.
\newblock An interactive semantic video mining and retrieval
  platform--application in transportation surveillance video for incident
  detection.
\newblock In {\em ICDM'06}, pages 129--138, 2006.

\bibitem{DBLP:conf/iclr/ChenLS18}
{\sc Xinyun Chen, Chang Liu, and Dawn Song}.
\newblock Towards synthesizing complex programs from input-output examples.
\newblock In {\em 6th International Conference on Learning Representations,
  {ICLR} 2018, Vancouver, BC, Canada, April 30 - May 3, 2018, Conference Track
  Proceedings}. OpenReview.net, 2018.

\bibitem{Chen2018e}
{\sc Xinyun Chen, Chang Liu, and Dawn Song}.
\newblock Tree-to-tree neural networks for program translation.
\newblock In {\em Advances in Neural Information Processing Systems 31},
  S.~Bengio, H.~Wallach, H.~Larochelle, K.~Grauman, N.~Cesa-Bianchi, and
  R.~Garnett, editors, pages 2547--2557. Curran Associates, Inc., 2018.

\bibitem{Chen:2019}
{\sc Zimin Chen, Steve Kommrusch, Michele Tufano, Louis{-}No{\"{e}}l Pouchet,
  Denys Poshyvanyk, and Martin Monperrus}.
\newblock Sequencer: Sequence-to-sequence learning for end-to-end program
  repair.
\newblock {\em CoRR}, abs/1901.01808, 2019.

\bibitem{Choetkiertikul2018}
{\sc M.~{Choetkiertikul}, H.~K. {Dam}, T.~{Tran}, A.~{Ghose}, and J.~{Grundy}}.
\newblock Predicting delivery capability in iterative software development.
\newblock {\em IEEE Transactions on Software Engineering}, 44(6):551--573, June
  2018.

\bibitem{Choetkiertikul2019}
{\sc M.~{Choetkiertikul}, H.~K. {Dam}, T.~{Tran}, T.~{Pham}, A.~{Ghose}, and
  T.~{Menzies}}.
\newblock A deep learning model for estimating story points.
\newblock {\em IEEE Transactions on Software Engineering}, 45(7):637--656, July
  2019.

\bibitem{Choetkiertikul2017}
{\sc Morakot Choetkiertikul, Hoa~Khanh Dam, Truyen Tran, and Aditya Ghose}.
\newblock Predicting the delay of issues with due dates in software projects.
\newblock {\em Empirical Software Engineering}, 22(3):1223--1263, Jun 2017.

\bibitem{ciniselli2021empirical}
{\sc Matteo Ciniselli, Nathan Cooper, Luca Pascarella, Antonio Mastropaolo,
  Emad Aghajani, Denys Poshyvanyk, Massimiliano Di~Penta, and Gabriele Bavota}.
\newblock An empirical study on the usage of transformer models for code
  completion.
\newblock {\em IEEE Transactions on Software Engineering}, 2021.

\bibitem{9463129}
{\sc Matteo Ciniselli, Nathan Cooper, Luca Pascarella, Denys Poshyvanyk,
  Massimiliano Di~Penta, and Gabriele Bavota}.
\newblock An empirical study on the usage of bert models for code completion.
\newblock In {\em 2021 IEEE/ACM 18th International Conference on Mining
  Software Repositories (MSR)}, pages 108--119, 2021.

\bibitem{clark2020electra}
{\sc Kevin Clark, Minh-Thang Luong, Quoc~V Le, and Christopher~D Manning}.
\newblock Electra: Pre-training text encoders as discriminators rather than
  generators.
\newblock {\em arXiv preprint arXiv:2003.10555}, 2020.

\bibitem{collard2013srcml}
{\sc Michael~L Collard, Michael~John Decker, and Jonathan~I Maletic}.
\newblock srcml: An infrastructure for the exploration, analysis, and
  manipulation of source code: A tool demonstration.
\newblock In {\em 2013 IEEE International conference on software maintenance},
  pages 516--519. IEEE, 2013.

\bibitem{connor2022can}
{\sc Aidan Connor, Aaron Harris, Nathan Cooper, and Denys Poshyvanyk}.
\newblock Can we automatically fix bugs by learning edit operations?
\newblock In {\em 2022 IEEE International Conference on Software Analysis,
  Evolution and Reengineering (SANER)}, pages 782--792. IEEE, 2022.

\bibitem{replication}
{\sc N.~Cooper, R.~Tufano, G.~Bavota, and D.~\&~Poshyvanyk}.
\newblock Completeformer replication package
  \url{https://github.com/WM-SEMERU/completeformer}.
\newblock {\em GitHub}, 2023.

\bibitem{tango_appendix}
{\sc Nathan Cooper, Carlos Bernal-Cárdenas, Oscar Chaparro, Kevin Moran, and
  Denys Poshyvanyk}.
\newblock Tango's online appendix \url{https://github.com/ncoop57/tango}, 2020.

\bibitem{Corley2015}
{\sc C.~S. Corley, K.~Damevski, and N.~A. Kraft}.
\newblock Exploring the use of deep learning for feature location.
\newblock In {\em 2015 {{IEEE International Conference}} on {{Software
  Maintenance}} and {{Evolution}} ({{ICSME}})}, ICSME'15, pages 556--560,
  September 2015.
\newblock ISSN:.

\bibitem{cortes-coy_automatically_2014}
{\sc Luis~Fernando Cortes-Coy, Mario Linares-Vasquez, Jairo Aponte, and Denys
  Poshyvanyk}.
\newblock On {Automatically} {Generating} {Commit} {Messages} via
  {Summarization} of {Source} {Code} {Changes}.
\newblock In {\em 2014 {IEEE} 14th {International} {Working} {Conference} on
  {Source} {Code} {Analysis} and {Manipulation}}, pages 275--284, Victoria, BC,
  Canada, September 2014. IEEE.

\bibitem{Cummins2018}
{\sc Chris Cummins, Pavlos Petoumenos, Alastair Murray, and Hugh Leather}.
\newblock Compiler fuzzing through deep learning.
\newblock In {\em Proceedings of the 27th ACM SIGSOFT International Symposium
  on Software Testing and Analysis}, ISSTA 2018, pages 95--105, New York, NY,
  USA, 2018. ACM.

\bibitem{cvitkovic:icml19}
{\sc Milan Cvitkovic, Badal Singh, and Animashree Anandkumar}.
\newblock Open vocabulary learning on source code with a graph-structured
  cache.
\newblock In {\em Proceedings of the 36th International Conference on Machine
  Learning}, Kamalika Chaudhuri and Ruslan Salakhutdinov, editors, volume~97 of
  {\em Proceedings of Machine Learning Research}, pages 1475--1485. PMLR,
  09--15 Jun 2019.

\bibitem{Dabic:msr2021data}
{\sc Ozren Dabic, Emad Aghajani, and Gabriele Bavota}.
\newblock Sampling projects in github for {MSR} studies.
\newblock In {\em 18th {IEEE/ACM} International Conference on Mining Software
  Repositories, {MSR} 2021}, pages 560--564. {IEEE}, 2021.

\bibitem{dai2019transformer}
{\sc Zihang Dai, Zhilin Yang, Yiming Yang, Jaime Carbonell, Quoc~V Le, and
  Ruslan Salakhutdinov}.
\newblock Transformer-xl: Attentive language models beyond a fixed-length
  context.
\newblock {\em arXiv preprint arXiv:1901.02860}, 2019.

\bibitem{Dam2018}
{\sc H.~K. {Dam}, T.~{Tran}, T.~T.~M. {Pham}, S.~W. {Ng}, J.~{Grundy}, and
  A.~{Ghose}}.
\newblock Automatic feature learning for predicting vulnerable software
  components.
\newblock {\em IEEE Transactions on Software Engineering}, pages 1--1, 2018.

\bibitem{dam:msr19}
{\sc Hoa~Khanh Dam, Trang Pham, Shien~Wee Ng, Truyen Tran, John Grundy, Aditya
  Ghose, Taeksu Kim, and Chul-Joo Kim}.
\newblock Lessons learned from using a deep tree-based model for software
  defect prediction in practice.
\newblock In {\em 2019 IEEE/ACM 16th International Conference on Mining
  Software Repositories (MSR)}, pages 46--57, 2019.

\bibitem{deerwester1990indexing}
{\sc Scott Deerwester, Susan~T Dumais, George~W Furnas, Thomas~K Landauer, and
  Richard Harshman}.
\newblock Indexing by latent semantic analysis.
\newblock {\em Journal of the American society for information science},
  41(6):391--407, 1990.

\bibitem{Deka:UIST'17}
{\sc Biplab Deka, Zifeng Huang, Chad Franzen, Joshua Hibschman, Daniel Afergan,
  Yang Li, Jeffrey Nichols, and Ranjitha Kumar}.
\newblock Rico: A mobile app dataset for building data-driven design
  applications.
\newblock In {\em UIST'17}, 2017.

\bibitem{10.5555/3305381.3305483}
{\sc Yuntian Deng, Anssi Kanervisto, Jeffrey Ling, and Alexander~M. Rush}.
\newblock Image-to-markup generation with coarse-to-fine attention.
\newblock In {\em Proceedings of the 34th International Conference on Machine
  Learning - Volume 70}, ICML’17, page 980–989. JMLR.org, 2017.

\bibitem{Deshmukh2017}
{\sc J.~{Deshmukh}, A.~K. {M}, S.~{Podder}, S.~{Sengupta}, and N.~{Dubash}}.
\newblock Towards accurate duplicate bug retrieval using deep learning
  techniques.
\newblock In {\em 2017 IEEE International Conference on Software Maintenance
  and Evolution (ICSME)}, pages 115--124, Sep. 2017.

\bibitem{devanbu_deep_2020}
{\sc Prem Devanbu, Matthew Dwyer, Sebastian Elbaum, Michael Lowry, Kevin Moran,
  Denys Poshyvanyk, Baishakhi Ray, Rishabh Singh, and Xiangyu Zhang}.
\newblock Deep {Learning} \& {Software} {Engineering}: {State} of {Research}
  and {Future} {Directions}, September 2020.
\newblock arXiv:2009.08525 [cs].

\bibitem{Devlin2017}
{\sc Jacob Devlin, Rudy Bunel, Rishabh Singh, Matthew~J. Hausknecht, and
  Pushmeet Kohli}.
\newblock Neural program meta-induction.
\newblock In {\em Advances in Neural Information Processing Systems 30: Annual
  Conference on Neural Information Processing Systems 2017, December 4-9, 2017,
  Long Beach, CA, {USA}}, Isabelle Guyon, Ulrike von Luxburg, Samy Bengio,
  Hanna~M. Wallach, Rob Fergus, S.~V.~N. Vishwanathan, and Roman Garnett,
  editors, pages 2080--2088, 2017.

\bibitem{devlin2018bert}
{\sc Jacob Devlin, Ming-Wei Chang, Kenton Lee, and Kristina Toutanova}.
\newblock Bert: Pre-training of deep bidirectional transformers for language
  understanding.
\newblock {\em arXiv preprint arXiv:1810.04805}, 2018.

\bibitem{DBLP:conf/icml/DevlinUBSMK17}
{\sc Jacob Devlin, Jonathan Uesato, Surya Bhupatiraju, Rishabh Singh,
  Abdel{-}rahman Mohamed, and Pushmeet Kohli}.
\newblock Robustfill: Neural program learning under noisy {I/O}.
\newblock In {\em Proceedings of the 34th International Conference on Machine
  Learning, {ICML} 2017, Sydney, NSW, Australia, 6-11 August 2017}, Doina
  Precup and Yee~Whye Teh, editors, volume~70 of {\em Proceedings of Machine
  Learning Research}, pages 990--998. {PMLR}, 2017.

\bibitem{Dixon:CHI11}
{\sc Morgan Dixon, Daniel Leventhal, and James Fogarty}.
\newblock Content and hierarchy in pixel-based methods for reverse engineering
  interface structure.
\newblock In {\em Proceedings of the SIGCHI Conference on Human Factors in
  Computing Systems}, CHI '11, pages 969--978, New York, NY, USA, 2011. ACM.

\bibitem{10.5555/3000292.3000299}
{\sc Pedro Domingos}.
\newblock Occam’s two razors: The sharp and the blunt.
\newblock In {\em Proceedings of the Fourth International Conference on
  Knowledge Discovery and Data Mining}, KDD’98, page 37–43. AAAI Press,
  1998.

\bibitem{Douze:TM'10}
{\sc M.~{Douze}, H.~{Jegou}, and C.~{Schmid}}.
\newblock An image-based approach to video copy detection with spatio-temporal
  post-filtering.
\newblock {\em TMM'10}, 12(4):257--266, 2010.

\bibitem{dubois2020location}
{\sc Yann Dubois, Gautier Dagan, Dieuwke Hupkes, and Elia Bruni}.
\newblock {L}ocation {A}ttention for {E}xtrapolation to {L}onger {S}equences.
\newblock In {\em Proceedings of the 58th Annual Meeting of the Association for
  Computational Linguistics}, pages 403--413, Online, July 2020. Association
  for Computational Linguistics.

\bibitem{Duchi2011}
{\sc John Duchi, Elad Hazan, and Yoram Singer}.
\newblock Adaptive subgradient methods for online learning and stochastic
  optimization.
\newblock {\em J. Mach. Learn. Res.}, 12:2121--2159, July 2011.

\bibitem{Ellis2018}
{\sc Kevin Ellis, Lucas Morales, Mathias Sabl\'{e}-Meyer, Armando Solar-Lezama,
  and Josh Tenenbaum}.
\newblock Learning libraries of subroutines for neurally\textendash guided
  bayesian program induction.
\newblock In {\em Advances in Neural Information Processing Systems 31},
  S.~Bengio, H.~Wallach, H.~Larochelle, K.~Grauman, N.~Cesa-Bianchi, and
  R.~Garnett, editors, pages 7805--7815. Curran Associates, Inc., 2018.

\bibitem{Ellis2018a}
{\sc Kevin Ellis, Daniel Ritchie, Armando Solar-Lezama, and Josh Tenenbaum}.
\newblock Learning to infer graphics programs from hand-drawn images.
\newblock In {\em Advances in Neural Information Processing Systems 31},
  S.~Bengio, H.~Wallach, H.~Larochelle, K.~Grauman, N.~Cesa-Bianchi, and
  R.~Garnett, editors, pages 6059--6068. Curran Associates, Inc., 2018.

\bibitem{escobar-velasquez_enabling_2022}
{\sc Camilo Escobar-Velasquez, Mario Linares-Vasquez, Gabriele Bavota, Michele
  Tufano, Kevin Moran, Massimiliano Di~Penta, Christopher Vendome, Carlos
  Bernal-C\'{a}rdenas, and Denys Poshyvanyk}.
\newblock Enabling {Mutant} {Generation} for {Open}- and {Closed}-{Source}
  {Android} {Apps}.
\newblock {\em IEEE Transactions on Software Engineering}, 48(1):186--208,
  January 2022.

\bibitem{Escobar-Velasquez:ASE'19}
{\sc C.~{Escobar-Velásquez}, M.~{Osorio-Riaño}, and M.~{Linares-Vásquez}}.
\newblock Mutapk: Source-codeless mutant generation for android apps.
\newblock In {\em ASE'19}, pages 1090--1093, 2019.

\bibitem{evtikhiev2022out}
{\sc Mikhail Evtikhiev, Egor Bogomolov, Yaroslav Sokolov, and Timofey Bryksin}.
\newblock Out of the bleu: how should we assess quality of the code generation
  models?
\newblock {\em arXiv preprint arXiv:2208.03133}, 2022.

\bibitem{fakhoury:saner19}
{\sc Sarah Fakhoury, Venera Arnaoudova, Cedric Noiseux, Foutse Khomh, and
  Giuliano Antoniol}.
\newblock Keep it simple: Is deep learning good for linguistic smell detection?
\newblock In {\em 2018 IEEE 25th International Conference on Software Analysis,
  Evolution and Reengineering (SANER)}, pages 602--611, 2018.

\bibitem{Falcon_PyTorch_Lightning_2019}
{\sc William Falcon and {The PyTorch Lightning team}}.
\newblock {PyTorch Lightning}, 3 2019.

\bibitem{8812083}
{\sc Ming Fan, Xiapu Luo, Jun Liu, Meng Wang, Chunyin Nong, Qinghua Zheng, and
  Ting Liu}.
\newblock Graph embedding based familial analysis of android malware using
  unsupervised learning.
\newblock In {\em 2019 IEEE/ACM 41st International Conference on Software
  Engineering (ICSE)}, pages 771--782, 2019.

\bibitem{Fazzini2023}
{\sc Mattia Fazzini, Kevin Moran, Carlos Bernal-Cárdenas, Tyler Wendland,
  Alessandro Orso, and Denys Poshyvanyk}.
\newblock Enhancing mobile app bug reporting via real-time understanding of
  reproduction steps.
\newblock {\em IEEE Transactions on Software Engineering}, 49(3):1246--1272,
  2023.

\bibitem{feng2020codebert}
{\sc Zhangyin Feng, Daya Guo, Duyu Tang, Nan Duan, Xiaocheng Feng, Ming Gong,
  Linjun Shou, Bing Qin, Ting Liu, Daxin Jiang, et~al.}
\newblock Codebert: A pre-trained model for programming and natural languages.
\newblock {\em arXiv preprint arXiv:2002.08155}, 2020.

\bibitem{fink2012wala}
{\sc Stephen Fink and Julian Dolby}.
\newblock Wala--the tj watson libraries for analysis, 2012.

\bibitem{Freedman:07}
{\sc David Freedman, Robert Pisani, and Roger Purves}.
\newblock {\em Statistics (4th edn.)}.
\newblock W. W. Norton \& Company, 2007.

\bibitem{fried2022incoder}
{\sc Daniel Fried, Armen Aghajanyan, Jessy Lin, Sida Wang, Eric Wallace, Freda
  Shi, Ruiqi Zhong, Wen-tau Yih, Luke Zettlemoyer, and Mike Lewis}.
\newblock Incoder: A generative model for code infilling and synthesis.
\newblock {\em arXiv preprint arXiv:2204.05999}, 2022.

\bibitem{Frisson:CHI16}
{\sc Christian Frisson, Sylvain Malacria, Gilles Bailly, and Thierry Dutoit}.
\newblock Inspectorwidget: A system to analyze users behaviors in their
  applications.
\newblock In {\em CHI'16}, pages 1548--1554. ACM, 2016.

\bibitem{Fu:FSE17}
{\sc Wei Fu and Tim Menzies}.
\newblock Easy over hard: A case study on deep learning.
\newblock In {\em Proceedings of the 2017 11th Joint Meeting on Foundations of
  Software Engineering}, ESEC/FSE'17, pages 49--60, New York, NY, USA, 2017.
  ACM.

\bibitem{gall2003cvs}
{\sc Harald Gall, Mehdi Jazayeri, and Jacek Krajewski}.
\newblock Cvs release history data for detecting logical couplings.
\newblock In {\em Sixth International Workshop on Principles of Software
  Evolution, 2003. Proceedings.}, pages 13--23. IEEE, 2003.

\bibitem{gallagher1990using}
{\sc Keith~Brian Gallagher}.
\newblock {\em Using program slicing in software maintenance}.
\newblock University of Maryland, Baltimore County, 1990.

\bibitem{Gao2018}
{\sc Jian Gao, Xin Yang, Ying Fu, Yu~Jiang, and Jiaguang Sun}.
\newblock Vulseeker: A semantic learning based vulnerability seeker for
  cross-platform binary.
\newblock In {\em Proceedings of the 33rd ACM/IEEE International Conference on
  Automated Software Engineering}, ASE 2018, pages 896--899, New York, NY, USA,
  2018. ACM.

\bibitem{gao:saner19}
{\sc Sa~Gao, Chunyang Chen, Zhenchang Xing, Yukun Ma, Wen Song, and Shang-Wei
  Lin}.
\newblock A neural model for method name generation from functional
  description.
\newblock In {\em 2019 IEEE 26th International Conference on Software Analysis,
  Evolution and Reengineering (SANER)}, pages 414--421, 2019.

\bibitem{Gaunt2017}
{\sc Alexander~L. Gaunt, Marc Brockschmidt, Nate Kushman, and Daniel Tarlow}.
\newblock Differentiable programs with neural libraries.
\newblock In {\em Proceedings of the 34th International Conference on Machine
  Learning}, Doina Precup and Yee~Whye Teh, editors, volume~70 of {\em
  Proceedings of Machine Learning Research}, pages 1213--1222, International
  Convention Centre, Sydney, Australia, 06--11 Aug 2017. PMLR.

\bibitem{gethers:icse12}
{\sc Malcom Gethers, Bogdan Dit, Huzefa Kagdi, and Denys Poshyvanyk}.
\newblock Integrated impact analysis for managing software changes.
\newblock In {\em 2012 34th International Conference on Software Engineering
  (ICSE)}, pages 430--440, 2012.

\bibitem{Gethers2010}
{\sc Malcom Gethers and Denys Poshyvanyk}.
\newblock Using relational topic models to capture coupling among classes in
  object-oriented software systems.
\newblock In {\em 2010 {IEEE} International Conference on Software
  Maintenance}. {IEEE}, September 2010.

\bibitem{gilpin2019explaining}
{\sc Leilani~H. Gilpin, David Bau, Ben~Z. Yuan, Ayesha Bajwa, Michael Specter,
  and Lalana Kagal}.
\newblock Explaining explanations: An overview of interpretability of machine
  learning, 2019.

\bibitem{github}
{\sc GitHub}.
\newblock Github copilot your ai pair programmer, 2023.

\bibitem{Godefroid2017}
{\sc Patrice Godefroid, Hila Peleg, and Rishabh Singh}.
\newblock Learn\&\#38;fuzz: Machine learning for input fuzzing.
\newblock In {\em Proceedings of the 32Nd IEEE/ACM International Conference on
  Automated Software Engineering}, ASE 2017, pages 50--59, Piscataway, NJ, USA,
  2017. IEEE Press.

\bibitem{Goodfellow2016}
{\sc Ian Goodfellow, Yoshua Bengio, and Aaron Courville}.
\newblock {\em Deep Learning}.
\newblock The MIT Press, 2016.

\bibitem{Gu2018}
{\sc Xiaodong Gu, Hongyu Zhang, and Sunghun Kim}.
\newblock Deep code search.
\newblock In {\em Proceedings of the 40th International Conference on Software
  Engineering}, ICSE '18, pages 933--944, New York, NY, USA, 2018. ACM.

\bibitem{Gu2016}
{\sc Xiaodong Gu, Hongyu Zhang, Dongmei Zhang, and Sunghun Kim}.
\newblock Deep api learning.
\newblock In {\em Proceedings of the 2016 24th ACM SIGSOFT International
  Symposium on Foundations of Software Engineering}, FSE 2016, pages 631--642,
  New York, NY, USA, 2016. ACM.

\bibitem{guo:saner19}
{\sc C.~Guo, W.~Wang, Y.~Wu, N.~Dong, Q.~Ye, J.~Xu, and S.~Zhang}.
\newblock Systematic comprehension for developer reply in mobile system forum.
\newblock In {\em 2019 IEEE 26th International Conference on Software Analysis,
  Evolution and Reengineering (SANER)}, pages 242--252, Los Alamitos, CA, USA,
  feb 2019. IEEE Computer Society.

\bibitem{guo2022unixcoder}
{\sc Daya Guo, Shuai Lu, Nan Duan, Yanlin Wang, Ming Zhou, and Jian Yin}.
\newblock Unixcoder: Unified cross-modal pre-training for code representation.
\newblock {\em arXiv preprint arXiv:2203.03850}, 2022.

\bibitem{guo2020graphcodebert}
{\sc Daya Guo, Shuo Ren, Shuai Lu, Zhangyin Feng, Duyu Tang, Shujie Liu, Long
  Zhou, Nan Duan, Alexey Svyatkovskiy, Shengyu Fu, et~al.}
\newblock Graphcodebert: Pre-training code representations with data flow.
\newblock {\em arXiv preprint arXiv:2009.08366}, 2020.

\bibitem{Guo2017}
{\sc Jin Guo, Jinghui Cheng, and Jane Cleland-Huang}.
\newblock Semantically enhanced software traceability using deep learning
  techniques.
\newblock In {\em Proceedings of the 39th International Conference on Software
  Engineering}, ICSE '17, pages 3--14, Piscataway, NJ, USA, 2017. IEEE Press.

\bibitem{gupta2019}
{\sc Rahul Gupta, Aditya Kanade, and Shirish Shevade}.
\newblock Deep reinforcement learning for syntactic error repair in student
  programs.
\newblock {\em Proceedings of the AAAI Conference on Artificial Intelligence},
  33:930--937, 07 2019.

\bibitem{gupta2019deep}
{\sc Rahul Gupta, Aditya Kanade, and Shirish Shevade}.
\newblock Deep reinforcement learning for syntactic error repair in student
  programs.
\newblock In {\em Proceedings of the AAAI Conference on Artificial
  Intelligence}, volume~33, pages 930--937, 2019.

\bibitem{gupta2017deepfix}
{\sc Rahul Gupta, Soham Pal, Aditya Kanade, and Shirish Shevade}.
\newblock Deepfix: Fixing common c language errors by deep learning.
\newblock In {\em Proceedings of the aaai conference on artificial
  intelligence}, volume~31, 2017.

\bibitem{Gusfield:LCS'97}
{\sc Dan Gusfield}.
\newblock Algorithms on stings, trees, and sequences: Computer science and
  computational biology.
\newblock {\em Acm Sigact News}, 28(4):41--60, 1997.

\bibitem{8811988}
{\sc Huong Ha and Hongyu Zhang}.
\newblock Deepperf: Performance prediction for configurable software with deep
  sparse neural network.
\newblock In {\em 2019 IEEE/ACM 41st International Conference on Software
  Engineering (ICSE)}, pages 1095--1106, 2019.

\bibitem{hamilton2017inductive}
{\sc Will Hamilton, Zhitao Ying, and Jure Leskovec}.
\newblock Inductive representation learning on large graphs.
\newblock {\em Advances in neural information processing systems}, 30, 2017.

\bibitem{Han2017}
{\sc Z.~{Han}, X.~{Li}, Z.~{Xing}, H.~{Liu}, and Z.~{Feng}}.
\newblock Learning to predict severity of software vulnerability using only
  vulnerability description.
\newblock In {\em 2017 IEEE International Conference on Software Maintenance
  and Evolution (ICSME)}, pages 125--136, Sep. 2017.

\bibitem{Hao:TM'17}
{\sc Y.~{Hao}, T.~{Mu}, R.~{Hong}, M.~{Wang}, N.~{An}, and J.~Y. {Goulermas}}.
\newblock Stochastic multiview hashing for large-scale near-duplicate video
  retrieval.
\newblock {\em TMM'17}, 19(1):1--14, 2017.

\bibitem{Harer2018}
{\sc Jacob Harer, Onur Ozdemir, Tomo Lazovich, Christopher Reale, Rebecca
  Russell, Louis Kim, and peter chin}.
\newblock Learning to repair software vulnerabilities with generative
  adversarial networks.
\newblock In {\em Advances in Neural Information Processing Systems 31},
  S.~Bengio, H.~Wallach, H.~Larochelle, K.~Grauman, N.~Cesa-Bianchi, and
  R.~Garnett, editors, pages 7933--7943. Curran Associates, Inc., 2018.

\bibitem{hassoun2004dynamic}
{\sc Youssef Hassoun, Roger Johnson, and Steve Counsell}.
\newblock A dynamic runtime coupling metric for meta-level architectures.
\newblock In {\em Eighth European Conference on Software Maintenance and
  Reengineering, 2004. CSMR 2004. Proceedings.}, pages 339--346. IEEE, 2004.

\bibitem{Hatcher2004}
{\sc Erik Hatcher and Otis Gospodnetic}.
\newblock {\em Lucene in Action}.
\newblock Manning Publications, 2004.

\bibitem{havranek_v2s_2021}
{\sc Madeleine Havranek, Carlos Bernal-C\'{a}rdenas, Nathan Cooper, Oscar
  Chaparro, Denys Poshyvanyk, and Kevin Moran}.
\newblock {V2S}: {A} {Tool} for {Translating} {Video} {Recordings} of {Mobile}
  {App} {Usages} into {Replayable} {Scenarios}.
\newblock In {\em 2021 {IEEE}/{ACM} 43rd {International} {Conference} on
  {Software} {Engineering}: {Companion} {Proceedings} ({ICSE}-{Companion})},
  pages 65--68, Madrid, ES, May 2021. IEEE.

\bibitem{He2020}
{\sc Jianjun He, Ling Xu, Meng Yan, Xin Xia, and Yan Lei}.
\newblock Duplicate bug report detection using dual-channel convolutional
  neural networks.
\newblock In {\em ICPC'20}, page to appear, 2020.

\bibitem{He:CVPR'16}
{\sc K.~{He}, X.~{Zhang}, S.~{Ren}, and J.~{Sun}}.
\newblock Deep residual learning for image recognition.
\newblock In {\em CVPR'16}, pages 770--778, 2016.

\bibitem{Hellendoorn2018}
{\sc Vincent~J. Hellendoorn, Christian Bird, Earl~T. Barr, and Miltiadis
  Allamanis}.
\newblock Deep learning type inference.
\newblock In {\em Proceedings of the 2018 26th ACM Joint Meeting on European
  Software Engineering Conference and Symposium on the Foundations of Software
  Engineering}, ESEC/FSE 2018, pages 152--162, New York, NY, USA, 2018. ACM.

\bibitem{Hellendoorn2017}
{\sc Vincent~J. Hellendoorn and Premkumar Devanbu}.
\newblock Are deep neural networks the best choice for modeling source code?
\newblock In {\em Proceedings of the 2017 11th Joint Meeting on Foundations of
  Software Engineering}, ESEC/FSE 2017, pages 763--773, New York, NY, USA,
  2017. ACM.

\bibitem{Hellendoorn2018a}
{\sc Vincent~J. Hellendoorn, Premkumar~T. Devanbu, and Mohammad~Amin Alipour}.
\newblock On the naturalness of proofs.
\newblock In {\em Proceedings of the 2018 26th ACM Joint Meeting on European
  Software Engineering Conference and Symposium on the Foundations of Software
  Engineering}, ESEC/FSE 2018, pages 724--728, New York, NY, USA, 2018. ACM.

\bibitem{hellendoorn2019code}
{\sc Vincent~J Hellendoorn, Sebastian Proksch, Harald~C Gall, and Alberto
  Bacchelli}.
\newblock When code completion fails: A case study on real-world completions.
\newblock In {\em 2019 IEEE/ACM 41st International Conference on Software
  Engineering (ICSE)}, pages 960--970. IEEE, 2019.

\bibitem{hendrycks2021measuring}
{\sc Dan Hendrycks, Steven Basart, Saurav Kadavath, Mantas Mazeika, Akul Arora,
  Ethan Guo, Collin Burns, Samir Puranik, Horace He, Dawn Song, et~al.}
\newblock Measuring coding challenge competence with apps.
\newblock {\em arXiv preprint arXiv:2105.09938}, 2021.

\bibitem{herbold2022fine}
{\sc Steffen Herbold, Alexander Trautsch, Benjamin Ledel, Alireza
  Aghamohammadi, Taher~A Ghaleb, Kuljit~Kaur Chahal, Tim Bossenmaier, Bhaveet
  Nagaria, Philip Makedonski, Matin~Nili Ahmadabadi, et~al.}
\newblock A fine-grained data set and analysis of tangling in bug fixing
  commits.
\newblock {\em Empirical Software Engineering}, 27(6):125, 2022.

\bibitem{Hindle2016}
{\sc Abram Hindle, Anahita Alipour, and Eleni Stroulia}.
\newblock A contextual approach towards more accurate duplicate bug report
  detection and ranking.
\newblock {\em EMSE'16}, 21(2):368--410, 2016.

\bibitem{hindle2016naturalness}
{\sc Abram Hindle, Earl~T Barr, Mark Gabel, Zhendong Su, and Premkumar
  Devanbu}.
\newblock On the naturalness of software.
\newblock {\em Communications of the ACM}, 59(5):122--131, 2016.

\bibitem{Hindle2018}
{\sc Abram Hindle and Curtis Onuczko}.
\newblock Preventing duplicate bug reports by continuously querying bug
  reports.
\newblock {\em EMSE'18}, pages 1--35, 2018.

\bibitem{hoang:msr19}
{\sc Thong Hoang, Hoa Khanh~Dam, Yasutaka Kamei, David Lo, and Naoyasu
  Ubayashi}.
\newblock Deepjit: An end-to-end deep learning framework for just-in-time
  defect prediction.
\newblock In {\em 2019 IEEE/ACM 16th International Conference on Mining
  Software Repositories (MSR)}, pages 34--45, 2019.

\bibitem{hochreiter1997long}
{\sc Sepp Hochreiter and J{\"u}rgen Schmidhuber}.
\newblock Long short-term memory.
\newblock {\em Neural computation}, 9(8):1735--1780, 1997.

\bibitem{holtzman2019curious}
{\sc Ari Holtzman, Jan Buys, Li~Du, Maxwell Forbes, and Yejin Choi}.
\newblock The curious case of neural text degeneration.
\newblock {\em arXiv preprint arXiv:1904.09751}, 2019.

\bibitem{hou2010towards}
{\sc Daqing Hou and David~M Pletcher}.
\newblock Towards a better code completion system by api grouping, filtering,
  and popularity-based ranking.
\newblock In {\em Proceedings of the 2nd International Workshop on
  Recommendation Systems for Software Engineering}, pages 26--30, 2010.

\bibitem{Hu:FSE18}
{\sc Gang Hu, Linjie Zhu, and Junfeng Yang}.
\newblock {AppFlow}: using machine learning to synthesize robust, reusable {UI}
  tests.
\newblock In {\em ESEC/FSE'18}, pages 269--282. {ACM} Press.

\bibitem{10.1145/3196321.3196334}
{\sc Xing Hu, Ge~Li, Xin Xia, David Lo, and Zhi Jin}.
\newblock Deep code comment generation.
\newblock In {\em Proceedings of the 26th Conference on Program Comprehension},
  ICPC '18, page 200–210, New York, NY, USA, 2018. Association for Computing
  Machinery.

\bibitem{Huang2018}
{\sc Q.~{Huang}, X.~{Xia}, D.~{Lo}, and G.~C. {Murphy}}.
\newblock Automating intention mining.
\newblock {\em IEEE Transactions on Software Engineering}, pages 1--1, 2018.

\bibitem{huang2014probabilistic}
{\sc Yuan Huang, Xiangping Chen, Qiwen Zou, and Xiaonan Luo}.
\newblock A probabilistic neural network-based approach for related software
  changes detection.
\newblock In {\em 2014 21st Asia-Pacific Software Engineering Conference},
  volume~1, pages 279--286. IEEE, 2014.

\bibitem{huo:tse19}
{\sc Xuan Huo, Ferdian Thung, Ming Li, David Lo, and Shu-Ting Shi}.
\newblock Deep transfer bug localization.
\newblock {\em IEEE Transactions on Software Engineering}, 47(7):1368--1380,
  2021.

\bibitem{hupkes2020compositionality}
{\sc Dieuwke Hupkes, Verna Dankers, Mathijs Mul, and Elia Bruni}.
\newblock Compositionality decomposed: How do neural networks generalise?
\newblock {\em Journal of Artificial Intelligence Research}, 67:757--795, 2020.

\bibitem{husain2019codesearchnet}
{\sc Hamel Husain, Ho-Hsiang Wu, Tiferet Gazit, Miltiadis Allamanis, and Marc
  Brockschmidt}.
\newblock Codesearchnet challenge: Evaluating the state of semantic code
  search.
\newblock {\em arXiv preprint arXiv:1909.09436}, 2019.

\bibitem{DBLP:journals/corr/IoffeS15}
{\sc Sergey Ioffe and Christian Szegedy}.
\newblock Batch normalization: Accelerating deep network training by reducing
  internal covariate shift.
\newblock {\em CoRR}, abs/1502.03167, 2015.

\bibitem{izadi2022codefill}
{\sc Maliheh Izadi, Roberta Gismondi, and Georgios Gousios}.
\newblock Codefill: Multi-token code completion by jointly learning from
  structure and naming sequences.
\newblock {\em arXiv preprint arXiv:2202.06689}, 2022.

\bibitem{Jiang2017}
{\sc He~Jiang, Najam Nazar, Jingxuan Zhang, Tao Zhang, and Zhilei Ren}.
\newblock {{PRST}}: {{A PageRank}}-{{Based Summarization Technique}} for
  {{Summarizing Bug Reports}} with {{Duplicates}}.
\newblock {\em International Journal of Software Engineering and Knowledge
  Engineering}, 27(06):869--896, 2017.

\bibitem{Jiang:16}
{\sc Y.~{Jiang} and J.~{Wang}}.
\newblock Partial copy detection in videos: A benchmark and an evaluation of
  popular methods.
\newblock {\em TBD'16}, 2(1):32--42, 2016.

\bibitem{Jiang:IVR'07}
{\sc Yu-Gang Jiang, Chong-Wah Ngo, and Jun Yang}.
\newblock Towards optimal bag-of-features for object categorization and
  semantic video retrieval.
\newblock In {\em IVR'07}, pages 494--501, 07 2007.

\bibitem{jin2018hidden}
{\sc Xianhao Jin and Francisco Servant}.
\newblock The hidden cost of code completion: Understanding the impact of the
  recommendation-list length on its efficiency.
\newblock In {\em Proceedings of the 15th International Conference on Mining
  Software Repositories}, pages 70--73, 2018.

\bibitem{Jing:19}
{\sc Weizhen Jing, Xiushan Nie, Chaoran Cui, Xiaoming Xi, Gongping Yang, and
  Yilong Yin}.
\newblock Global-view hashing: harnessing global relations in near-duplicate
  video retrieval.
\newblock {\em WWW'19}, 22(2):771--789, 2019.

\bibitem{Huang:CVPR'97}
{\sc {Jing Huang}, S.~R. {Kumar}, M.~{Mitra}, {Wei-Jing Zhu}, and R.~{Zabih}}.
\newblock Image indexing using color correlograms.
\newblock In {\em CVPR'97}, pages 762--768, 1997.

\bibitem{Jones:2014}
{\sc N.~Jones}.
\newblock Seven best practices for optimizing mobile testing efforts.
\newblock Technical Report G00248240, Gartner.

\bibitem{Jegou:CVPR'10}
{\sc H.~{Jégou}, M.~{Douze}, C.~{Schmid}, and P.~{Pérez}}.
\newblock Aggregating local descriptors into a compact image representation.
\newblock In {\em CVPR'10}, pages 3304--3311, 2010.

\bibitem{kagdi2013integrating}
{\sc Huzefa Kagdi, Malcom Gethers, and Denys Poshyvanyk}.
\newblock Integrating conceptual and logical couplings for change impact
  analysis in software.
\newblock {\em Empirical Software Engineering}, 18:933--969, 2013.

\bibitem{Vijayakumar2018}
{\sc Ashwin Kalyan, Abhishek Mohta, Oleksandr Polozov, Dhruv Batra, Prateek
  Jain, and Sumit Gulwani}.
\newblock Neural-guided deductive search for real-time program synthesis from
  examples.
\newblock In {\em 6th International Conference on Learning Representations,
  {ICLR} 2018, Vancouver, BC, Canada, April 30 - May 3, 2018, Conference Track
  Proceedings}. OpenReview.net, 2018.

\bibitem{Kang2017}
{\sc Li~Kang}.
\newblock Automated {{Duplicate Bug Reports Detection}} - {{An Experiment}} at
  {{Axis Communication AB}}.
\newblock Master's thesis, 2017.

\bibitem{Karampatsis2019}
{\sc Rafael{-}Michael Karampatsis and Charles Sutton}.
\newblock Maybe deep neural networks are the best choice for modeling source
  code.
\newblock {\em CoRR}, abs/1903.05734, 2019.

\bibitem{katz:saner19}
{\sc Deborah~S. Katz, Jason Ruchti, and Eric Schulte}.
\newblock Using recurrent neural networks for decompilation.
\newblock In {\em 2018 IEEE 25th International Conference on Software Analysis,
  Evolution and Reengineering (SANER)}, pages 346--356, 2018.

\bibitem{kingma2014adam}
{\sc Diederik Kingma and Jimmy Ba}.
\newblock Adam: A method for stochastic optimization.
\newblock {\em International Conference on Learning Representations}, 12 2014.

\bibitem{Kingma2014}
{\sc Diederik~P. Kingma and Jimmy Ba}.
\newblock Adam: A method for stochastic optimization, 2014.

\bibitem{kipf2016semi}
{\sc Thomas~N Kipf and Max Welling}.
\newblock Semi-supervised classification with graph convolutional networks.
\newblock {\em arXiv preprint arXiv:1609.02907}, 2016.

\bibitem{Kitchenham2007}
{\sc B.~Kitchenham and S~Charters}.
\newblock Guidelines for performing systematic literature reviews in software
  engineering, 2007.

\bibitem{kiyono2021shape}
{\sc Shun Kiyono, Sosuke Kobayashi, Jun Suzuki, and Kentaro Inui}.
\newblock Shape: Shifted absolute position embedding for transformers.
\newblock {\em arXiv preprint arXiv:2109.05644}, 2021.

\bibitem{Klein2014}
{\sc Nathan Klein, Christopher~S. Corley, and Nicholas~A. Kraft}.
\newblock New {{Features}} for {{Duplicate Bug Detection}}.
\newblock In {\em MSR'14}, pages 324--327. {ACM}, 2014.

\bibitem{Kordopatis-Zilos:TM'19}
{\sc G.~{Kordopatis-Zilos}, S.~{Papadopoulos}, I.~{Patras}, and
  I.~{Kompatsiaris}}.
\newblock Fivr: Fine-grained incident video retrieval.
\newblock {\em TMM'19}, 21(10):2638--2652, 2019.

\bibitem{Kordopatis-Zilos:ICCVW'17}
{\sc G.~{Kordopatis-Zilos}, S.~{Papadopoulos}, I.~{Patras}, and
  Y.~{Kompatsiaris}}.
\newblock Near-duplicate video retrieval with deep metric learning.
\newblock In {\em ICCVW'17}, pages 347--356, 2017.

\bibitem{Kordopatis-Zilos:17}
{\sc Giorgos Kordopatis-Zilos, Symeon Papadopoulos, Ioannis Patras, and Ioannis
  Kompatsiaris}.
\newblock Near-duplicate video retrieval by aggregating intermediate cnn
  layers.
\newblock In {\em MMM'17}, volume 10132, pages 251--263, 01 2017.

\bibitem{korpi2007supporting}
{\sc Jaakko Korpi and Jussi Koskinen}.
\newblock Supporting impact analysis by program dependence graph based forward
  slicing.
\newblock In {\em Advances and innovations in systems, computing sciences and
  software engineering}, pages 197--202. Springer, 2007.

\bibitem{Kraaij:11}
{\sc W.~Kraaij and G.~Awad}.
\newblock Trecvid 2011 content-based copy detection: Task overview.
\newblock in Online Proc. TRECVid, 2010,2011.

\bibitem{NIPS2012_4824}
{\sc Alex Krizhevsky, Ilya Sutskever, and Geoffrey~E Hinton}.
\newblock Imagenet classification with deep convolutional neural networks.
\newblock In {\em Advances in Neural Information Processing Systems 25},
  F.~Pereira, C.~J.~C. Burges, L.~Bottou, and K.~Q. Weinberger, editors, pages
  1097--1105. Curran Associates, Inc., 2012.

\bibitem{lake2018generalization}
{\sc Brenden Lake and Marco Baroni}.
\newblock Generalization without systematicity: On the compositional skills of
  sequence-to-sequence recurrent networks.
\newblock In {\em International conference on machine learning}, pages
  2873--2882. PMLR, 2018.

\bibitem{Lam2015}
{\sc A.~N. {Lam}, A.~T. {Nguyen}, H.~A. {Nguyen}, and T.~N. {Nguyen}}.
\newblock Combining deep learning with information retrieval to localize buggy
  files for bug reports (n).
\newblock In {\em 2015 30th IEEE/ACM International Conference on Automated
  Software Engineering (ASE)}, pages 476--481, Nov 2015.

\bibitem{Lasecki:ACM'15}
{\sc Walter~S. Lasecki, Juho Kim, Nick Rafter, Onkur Sen, Jeffrey~P. Bigham,
  and Michael~S. Bernstein}.
\newblock Apparition: Crowdsourced user interfaces that come to life as you
  sketch them.
\newblock In {\em CHI'15}, page 1925–1934, New York, NY, USA, 2015.
  Association for Computing Machinery.

\bibitem{Lazar2014}
{\sc Alina Lazar, Sarah Ritchey, and Bonita Sharif}.
\newblock Improving the {{Accuracy}} of {{Duplicate Bug Report Detection Using
  Textual Similarity Measures}}.
\newblock In {\em MSR'14}, pages 308--311, 2014.

\bibitem{le2014distributed}
{\sc Quoc Le and Tomas Mikolov}.
\newblock Distributed representations of sentences and documents.
\newblock In {\em International conference on machine learning}, pages
  1188--1196. PMLR, 2014.

\bibitem{Le2018a}
{\sc Tien-Duy~B. Le, Lingfeng Bao, and David Lo}.
\newblock Dsm: A specification mining tool using recurrent neural network based
  language model.
\newblock In {\em Proceedings of the 2018 26th ACM Joint Meeting on European
  Software Engineering Conference and Symposium on the Foundations of Software
  Engineering}, ESEC/FSE 2018, pages 896--899, New York, NY, USA, 2018. ACM.

\bibitem{10.1145/3213846.3213876}
{\sc Tien-Duy~B. Le and David Lo}.
\newblock Deep specification mining.
\newblock In {\em Proceedings of the 27th ACM SIGSOFT International Symposium
  on Software Testing and Analysis}, ISSTA 2018, page 106–117, New York, NY,
  USA, 2018. Association for Computing Machinery.

\bibitem{10.1109/ICSE.2019.00087}
{\sc Alexander LeClair, Siyuan Jiang, and Collin McMillan}.
\newblock A neural model for generating natural language summaries of program
  subroutines.
\newblock In {\em Proceedings of the 41st International Conference on Software
  Engineering}, ICSE '19, page 795–806. IEEE Press, 2019.

\bibitem{leclair-mcmillan-2019-recommendations}
{\sc Alexander LeClair and Collin McMillan}.
\newblock Recommendations for datasets for source code summarization.
\newblock In {\em Proceedings of the 2019 Conference of the North {A}merican
  Chapter of the Association for Computational Linguistics: Human Language
  Technologies, Volume 1 (Long and Short Papers)}, pages 3931--3937, June 2019.

\bibitem{Lecun:CNN'98}
{\sc Y.~{Lecun}, L.~{Bottou}, Y.~{Bengio}, and P.~{Haffner}}.
\newblock Gradient-based learning applied to document recognition.
\newblock {\em Proceedings of the IEEE}, 86(11):2278--2324, 1998.

\bibitem{Lee2017}
{\sc Sun-Ro Lee, Min-Jae Heo, Chan-Gun Lee, Milhan Kim, and Gaeul Jeong}.
\newblock Applying deep learning based automatic bug triager to industrial
  projects.
\newblock In {\em Proceedings of the 2017 11th Joint Meeting on Foundations of
  Software Engineering}, ESEC/FSE 2017, pages 926--931, New York, NY, USA,
  2017. ACM.

\bibitem{lehnert2011taxonomy}
{\sc Steffen Lehnert}.
\newblock A taxonomy for software change impact analysis.
\newblock In {\em Proceedings of the 12th International Workshop on Principles
  of Software Evolution and the 7th annual ERCIM Workshop on Software
  Evolution}, pages 41--50, 2011.

\bibitem{Lerch2013a}
{\sc J.~Lerch and M.~Mezini}.
\newblock Finding {{Duplicates}} of {{Your Yet Unwritten Bug Report}}.
\newblock In {\em CSMR'13}, pages 69--78, 2013.

\bibitem{levenshtein1966binary}
{\sc Vladimir~I Levenshtein et~al.}
\newblock Binary codes capable of correcting deletions, insertions, and
  reversals.
\newblock In {\em Soviet physics doklady}, volume~10, pages 707--710. Soviet
  Union, 1966.

\bibitem{Levy2017}
{\sc Dor Levy and Lior Wolf}.
\newblock Learning to align the source code to the compiled object code.
\newblock In {\em Proceedings of the 34th International Conference on Machine
  Learning}, Doina Precup and Yee~Whye Teh, editors, volume~70 of {\em
  Proceedings of Machine Learning Research}, pages 2043--2051, International
  Convention Centre, Sydney, Australia, 06--11 Aug 2017. PMLR.

\bibitem{lhoest-etal-2021-datasets}
{\sc Quentin Lhoest, Albert Villanova~del Moral, Yacine Jernite, Abhishek
  Thakur, Patrick von Platen, Suraj Patil, Julien Chaumond, Mariama Drame,
  Julien Plu, Lewis Tunstall, Joe Davison, Mario {\v{S}}a{\v{s}}ko, Gunjan
  Chhablani, Bhavitvya Malik, Simon Brandeis, Teven Le~Scao, Victor Sanh,
  Canwen Xu, Nicolas Patry, Angelina McMillan-Major, Philipp Schmid, Sylvain
  Gugger, Cl{\'e}ment Delangue, Th{\'e}o Matussi{\`e}re, Lysandre Debut, Stas
  Bekman, Pierric Cistac, Thibault Goehringer, Victor Mustar, Fran{\c{c}}ois
  Lagunas, Alexander Rush, and Thomas Wolf}.
\newblock Datasets: A community library for natural language processing.
\newblock In {\em Proceedings of the 2021 Conference on Empirical Methods in
  Natural Language Processing: System Demonstrations}, pages 175--184, Online
  and Punta Cana, Dominican Republic, November 2021. Association for
  Computational Linguistics.

\bibitem{Li2018}
{\sc D.~{Li}, Z.~{Wang}, and Y.~{Xue}}.
\newblock Fine-grained android malware detection based on deep learning.
\newblock In {\em 2018 IEEE Conference on Communications and Network Security
  (CNS)}, pages 1--2, May 2018.

\bibitem{Li2017}
{\sc Liuqing Li, He~Feng, Wenjie Zhuang, Na~Meng, and Barbara~G. Ryder}.
\newblock Cclearner: A deep learning-based clone detection approach.
\newblock {\em 2017 IEEE International Conference on Software Maintenance and
  Evolution (ICSME)}, pages 249--260, 2017.

\bibitem{Li2019}
{\sc Mingyang Li, Lin Shi, and Qing Wang}.
\newblock Are all duplicates value-neutral? an empirical analysis of duplicate
  issue reports.
\newblock In {\em QRS'19}, pages 272--279. IEEE, 2019.

\bibitem{10.1145/3360588}
{\sc Yi~Li, Shaohua Wang, Tien~N. Nguyen, and Son Van~Nguyen}.
\newblock Improving bug detection via context-based code representation
  learning and attention-based neural networks.
\newblock {\em Proc. ACM Program. Lang.}, 3(OOPSLA), October 2019.

\bibitem{Liang2018}
{\sc Chen Liang, Mohammad Norouzi, Jonathan Berant, Quoc~V Le, and Ni~Lao}.
\newblock Memory augmented policy optimization for program synthesis and
  semantic parsing.
\newblock In {\em Advances in Neural Information Processing Systems 31},
  S.~Bengio, H.~Wallach, H.~Larochelle, K.~Grauman, N.~Cesa-Bianchi, and
  R.~Garnett, editors, pages 9994--10006. Curran Associates, Inc., 2018.

\bibitem{likhomanenko2021cape}
{\sc Tatiana Likhomanenko, Qiantong Xu, Gabriel Synnaeve, Ronan Collobert, and
  Alex Rogozhnikov}.
\newblock Cape: Encoding relative positions with continuous augmented
  positional embeddings.
\newblock {\em Advances in Neural Information Processing Systems},
  34:16079--16092, 2021.

\bibitem{Lin2018}
{\sc B.~{Lin}, F.~{Zampetti}, G.~{Bavota}, M.~{Di Penta}, M.~{Lanza}, and
  R.~{Oliveto}}.
\newblock Sentiment analysis for software engineering: How far can we go?
\newblock In {\em 2018 IEEE/ACM 40th International Conference on Software
  Engineering (ICSE)}, pages 94--104, May 2018.

\bibitem{lin2004rouge}
{\sc Chin-Yew Lin}.
\newblock Rouge: A package for automatic evaluation of summaries.
\newblock In {\em Text summarization branches out}, pages 74--81, 2004.

\bibitem{SO2004}
{\sc Mario Linares-V\'{a}squez, Gabriele Bavota, Massimiliano Di~Penta, Rocco
  Oliveto, and Denys Poshyvanyk}.
\newblock How do api changes trigger stack overflow discussions? a study on the
  android sdk.
\newblock In {\em Proceedings of the 22nd International Conference on Program
  Comprehension}, ICPC 2014, page 83–94, New York, NY, USA, 2014. Association
  for Computing Machinery.

\bibitem{10.5555/2820518.2820534}
{\sc Mario Linares-V\'{a}squez, Martin White, Carlos Bernal-C\'{a}rdenas, Kevin
  Moran, and Denys Poshyvanyk}.
\newblock Mining android app usages for generating actionable gui-based
  execution scenarios.
\newblock In {\em Proceedings of the 12th Working Conference on Mining Software
  Repositories}, MSR '15, page 111–122. IEEE Press, 2015.

\bibitem{8094467}
{\sc Mario Linares-Vásquez, Cárlos Bernal-C\'{a}rdenas, Kevin Moran, and
  Denys Poshyvanyk}.
\newblock How do developers test android applications?
\newblock In {\em 2017 IEEE International Conference on Software Maintenance
  and Evolution (ICSME)}, pages 613--622, 2017.

\bibitem{8094439}
{\sc Mario Linares-Vásquez, Kevin Moran, and Denys Poshyvanyk}.
\newblock Continuous, evolutionary and large-scale: A new perspective for
  automated mobile app testing.
\newblock In {\em 2017 IEEE International Conference on Software Maintenance
  and Evolution (ICSME)}, pages 399--410, 2017.

\bibitem{Liu2018c}
{\sc Bingchang Liu, Wei Huo, Chao Zhang, Wenchao Li, Feng Li, Aihua Piao, and
  Wei Zou}.
\newblock {$\alpha$}diff: Cross-version binary code similarity detection with
  dnn.
\newblock In {\em Proceedings of the 33rd ACM/IEEE International Conference on
  Automated Software Engineering}, ASE 2018, pages 667--678, New York, NY, USA,
  2018. ACM.

\bibitem{Liu2016}
{\sc Chang Liu, Xinyun Chen, Richard Shin, Mingcheng Chen, and Dawn Song}.
\newblock Latent attention for if-then program synthesis.
\newblock In {\em Advances in Neural Information Processing Systems 29}, D.~D.
  Lee, M.~Sugiyama, U.~V. Luxburg, I.~Guyon, and R.~Garnett, editors, pages
  4574--4582. Curran Associates, Inc., 2016.

\bibitem{Liu2018}
{\sc Hui Liu, Zhifeng Xu, and Yanzhen Zou}.
\newblock Deep learning based feature envy detection.
\newblock In {\em Proceedings of the 33rd ACM/IEEE International Conference on
  Automated Software Engineering}, ASE 2018, pages 385--396, New York, NY, USA,
  2018. ACM.

\bibitem{Liu2018d}
{\sc K.~{Liu}, D.~{Kim}, T.~F. {Bissyande}, S.~{Yoo}, and Y.~{Le Traon}}.
\newblock Mining fix patterns for findbugs violations.
\newblock {\em IEEE Transactions on Software Engineering}, pages 1--1, 2018.

\bibitem{Liu2013}
{\sc K.~Liu, H.~Beng~Kuan Tan, and H.~Zhang}.
\newblock Has this bug been reported?
\newblock In {\em WCRE'13}, pages 82--91, 2013.

\bibitem{8812134}
{\sc Kui Liu, Dongsun Kim, Tegawendé~F. Bissyandé, Taeyoung Kim, Kisub Kim,
  Anil Koyuncu, Suntae Kim, and Yves Le~Traon}.
\newblock Learning to spot and refactor inconsistent method names.
\newblock In {\em 2019 IEEE/ACM 41st International Conference on Software
  Engineering (ICSE)}, pages 1--12, 2019.

\bibitem{Liu2017}
{\sc P.~{Liu}, X.~{Zhang}, M.~{Pistoia}, Y.~{Zheng}, M.~{Marques}, and
  L.~{Zeng}}.
\newblock Automatic text input generation for mobile testing.
\newblock In {\em 2017 IEEE/ACM 39th International Conference on Software
  Engineering (ICSE)}, pages 643--653, May 2017.

\bibitem{DBLP:conf/aaai/LiuLPW19}
{\sc Xiao Liu, Xiaoting Li, Rupesh Prajapati, and Dinghao Wu}.
\newblock Deepfuzz: Automatic generation of syntax valid {C} programs for fuzz
  testing.
\newblock In {\em The Thirty-Third {AAAI} Conference on Artificial
  Intelligence, {AAAI} 2019, The Thirty-First Innovative Applications of
  Artificial Intelligence Conference, {IAAI} 2019, The Ninth {AAAI} Symposium
  on Educational Advances in Artificial Intelligence, {EAAI} 2019, Honolulu,
  Hawaii, USA, January 27 - February 1, 2019}, pages 1044--1051. {AAAI} Press,
  2019.

\bibitem{liu2018code}
{\sc Xiaoyu Liu, LiGuo Huang, Alexander Egyed, and Jidong Ge}.
\newblock Do code data sharing dependencies support an early prediction of
  software actual change impact set?
\newblock {\em Journal of Software: Evolution and Process}, 30(11):e1960, 2018.

\bibitem{liu:saner19}
{\sc Yibin Liu, Yanhui Li, Jianbo Guo, Yuming Zhou, and Baowen Xu}.
\newblock Connecting software metrics across versions to predict defects.
\newblock In {\em 2018 IEEE 25th International Conference on Software Analysis,
  Evolution and Reengineering (SANER)}, pages 232--243, 2018.

\bibitem{Lowe:JCV'04}
{\sc David Lowe}.
\newblock Distinctive image features from scale-invariant keypoints.
\newblock {\em JCV'04}, 60:91--, 11 2004.

\bibitem{lu2021codexglue}
{\sc Shuai Lu, Daya Guo, Shuo Ren, Junjie Huang, Alexey Svyatkovskiy, Ambrosio
  Blanco, Colin Clement, Dawn Drain, Daxin Jiang, Duyu Tang, Ge~Li, Lidong
  Zhou, Linjun Shou, Long Zhou, Michele Tufano, Ming Gong, Ming Zhou, Nan Duan,
  Neel Sundaresan, Shao~Kun Deng, Shengyu Fu, and Shujie Liu}.
\newblock Codexglue: A machine learning benchmark dataset for code
  understanding and generation, 2021.

\bibitem{Ma2018}
{\sc Lei Ma, Felix Juefei-Xu, Fuyuan Zhang, Jiyuan Sun, Minhui Xue, Bo~Li,
  Chunyang Chen, Ting Su, Li~Li, Yang Liu, Jianjun Zhao, and Yadong Wang}.
\newblock Deepgauge: Multi-granularity testing criteria for deep learning
  systems.
\newblock In {\em Proceedings of the 33rd ACM/IEEE International Conference on
  Automated Software Engineering}, ASE 2018, pages 120--131, New York, NY, USA,
  2018. ACM.

\bibitem{MacLeod:ICPC'15}
{\sc L.~{MacLeod}, M.~{Storey}, and A.~{Bergen}}.
\newblock Code, camera, action: How software developers document and share
  program knowledge using youtube.
\newblock In {\em ICPC'15}, pages 104--114, 2015.

\bibitem{8811893}
{\sc Rabee~Sohail Malik, Jibesh Patra, and Michael Pradel}.
\newblock Nl2type: Inferring javascript function types from natural language
  information.
\newblock In {\em 2019 IEEE/ACM 41st International Conference on Software
  Engineering (ICSE)}, pages 304--315, 2019.

\bibitem{mandelin2005jungloid}
{\sc David Mandelin, Lin Xu, Rastislav Bod{\'\i}k, and Doug Kimelman}.
\newblock Jungloid mining: helping to navigate the api jungle.
\newblock {\em ACM Sigplan Notices}, 40(6):48--61, 2005.

\bibitem{Mao:ASE17}
{\sc Ke~Mao, Mark Harman, and Yue Jia}.
\newblock Crowd intelligence enhances automated mobile testing.
\newblock In {\em ASE'17}, pages 16--26, Piscataway, NJ, USA, 2017. IEEE Press.

\bibitem{MastropaoloT5}
{\sc Antonio Mastropaolo, Nathan Cooper, David~Nader Palacio, Simone
  Scalabrino, Denys Poshyvanyk, Rocco Oliveto, and Gabriele Bavota}.
\newblock Using transfer learning for code-related tasks.
\newblock {\em IEEE Transactions on Software Engineering}, pages 1--20, 2022.

\bibitem{mastropaolo2021studying}
{\sc Antonio Mastropaolo, Simone Scalabrino, Nathan Cooper, David~Nader
  Palacio, Denys Poshyvanyk, Rocco Oliveto, and Gabriele Bavota}.
\newblock Studying the usage of text-to-text transfer transformer to support
  code-related tasks.
\newblock In {\em 2021 IEEE/ACM 43rd International Conference on Software
  Engineering (ICSE)}, pages 336--347. IEEE, 2021.

\bibitem{mikolov2013distributed}
{\sc Tomas Mikolov, Ilya Sutskever, Kai Chen, Greg~S Corrado, and Jeff Dean}.
\newblock Distributed representations of words and phrases and their
  compositionality.
\newblock {\em Advances in neural information processing systems}, 26, 2013.

\bibitem{10.1109/ICSE.2019.00084}
{\sc Facundo Molina, Renzo Degiovanni, Pablo Ponzio, Germ\'{a}n Regis, Nazareno
  Aguirre, and Marcelo Frias}.
\newblock Training binary classifiers as data structure invariants.
\newblock In {\em Proceedings of the 41st International Conference on Software
  Engineering}, ICSE '19, page 759–770. IEEE Press, 2019.

\bibitem{9284034}
{\sc K.~Moran, D.~N. Palacio, C.~Bernal-C\'{a}rdenas, D.~McCrystal,
  D.~Poshyvanyk, C.~Shenefiel, and J.~Johnson}.
\newblock Improving the effectiveness of traceability link recovery using
  hierarchical bayesian networks.
\newblock In {\em 2020 IEEE/ACM 42nd International Conference on Software
  Engineering (ICSE)}, pages 873--885, Los Alamitos, CA, USA, oct 2020. IEEE
  Computer Society.

\bibitem{Moran2018}
{\sc K.~P. Moran, C.~Bernal-Cárdenas, M.~Curcio, R.~Bonett, and
  D.~Poshyvanyk}.
\newblock Machine learning-based prototyping of graphical user interfaces for
  mobile apps.
\newblock {\em IEEE Transactions on Software Engineering}, pages 1--1, 2018.

\bibitem{moran_automated_2018}
{\sc Kevin Moran, Boyang Li, Carlos Bernal-Cardenas, Dan Jelf, and Denys
  Poshyvanyk}.
\newblock Automated reporting of {GUI} design violations for mobile apps.
\newblock In {\em Proceedings of the 40th {International} {Conference} on
  {Software} {Engineering}}, pages 165--175, Gothenburg Sweden, May 2018. ACM.

\bibitem{Moran:FSE15}
{\sc Kevin Moran, Mario Linares-V\'{a}squez, Carlos Bernal-C\'{a}rdenas, and
  Denys Poshyvanyk}.
\newblock Auto-completing bug reports for android applications.
\newblock Bergamo, Italy, August-September 2015.

\bibitem{Moran:ICST16}
{\sc Kevin Moran, Mario Linares-V{\'a}squez, Carlos Bernal-C{\'a}rdenas,
  Christopher Vendome, and Denys Poshyvanyk}.
\newblock Automatically discovering, reporting and reproducing android
  application crashes.
\newblock In {\em ICST'16}, pages 33--44. IEEE, 2016.

\bibitem{7965246}
{\sc Kevin Moran, Mario Linares-Vasquez, Carlos Bernal-C\'{a}rdenas,
  Christopher Vendome, and Denys Poshyvanyk}.
\newblock Crashscope: A practical tool for automated testing of android
  applications.
\newblock In {\em 2017 IEEE/ACM 39th International Conference on Software
  Engineering Companion (ICSE-C)}, pages 15--18, 2017.

\bibitem{10.1145/3238147.3238203}
{\sc Kevin Moran, Cody Watson, John Hoskins, George Purnell, and Denys
  Poshyvanyk}.
\newblock Detecting and summarizing gui changes in evolving mobile apps.
\newblock In {\em Proceedings of the 33rd ACM/IEEE International Conference on
  Automated Software Engineering}, ASE '18, page 543–553, New York, NY, USA,
  2018. Association for Computing Machinery.

\bibitem{9825825}
{\sc Kevin Moran, Ali Yachnes, George Purnell, Junayed Mahmud, Michele Tufano,
  Carlos~Bernal C\'{a}rdenas, Denys Poshyvanyk, and Zach H'Doubler}.
\newblock An empirical investigation into the use of image captioning for
  automated software documentation.
\newblock In {\em 2022 IEEE International Conference on Software Analysis,
  Evolution and Reengineering (SANER)}, pages 514--525, 2022.

\bibitem{10.5555/3015812.3016002}
{\sc Lili Mou, Ge~Li, Lu~Zhang, Tao Wang, and Zhi Jin}.
\newblock Convolutional neural networks over tree structures for programming
  language processing.
\newblock In {\em Proceedings of the Thirtieth AAAI Conference on Artificial
  Intelligence}, AAAI'16, page 1287–1293. AAAI Press, 2016.

\bibitem{Murali2017}
{\sc Vijayaraghavan Murali, Swarat Chaudhuri, and Chris Jermaine}.
\newblock Bayesian specification learning for finding api usage errors.
\newblock In {\em Proceedings of the 2017 11th Joint Meeting on Foundations of
  Software Engineering}, ESEC/FSE 2017, pages 151--162, New York, NY, USA,
  2017. ACM.

\bibitem{DBLP:conf/iclr/MuraliQCJ18}
{\sc Vijayaraghavan Murali, Letao Qi, Swarat Chaudhuri, and Chris Jermaine}.
\newblock Neural sketch learning for conditional program generation.
\newblock In {\em 6th International Conference on Learning Representations,
  {ICLR} 2018, Vancouver, BC, Canada, April 30 - May 3, 2018, Conference Track
  Proceedings}. OpenReview.net, 2018.

\bibitem{Nayebi2020}
{\sc Maleknaz Nayebi}.
\newblock Eye of the mind: Image processing for social coding.
\newblock In {\em ICSE'20}, page 49–52, 2020.

\bibitem{neishi2019relation}
{\sc Masato Neishi and Naoki Yoshinaga}.
\newblock On the relation between position information and sentence length in
  neural machine translation.
\newblock In {\em Proceedings of the 23rd Conference on Computational Natural
  Language Learning (CoNLL)}, pages 328--338, 2019.

\bibitem{newman2020eos}
{\sc Benjamin Newman, John Hewitt, Percy Liang, and Christopher~D Manning}.
\newblock The eos decision and length extrapolation.
\newblock {\em arXiv preprint arXiv:2010.07174}, 2020.

\bibitem{nguyen:saner19}
{\sc Anh~Tuan Nguyen, Trong~Duc Nguyen, Hung~Dang Phan, and Tien~N. Nguyen}.
\newblock A deep neural network language model with contexts for source code.
\newblock In {\em 2018 IEEE 25th International Conference on Software Analysis,
  Evolution and Reengineering (SANER)}, pages 323--334, 2018.

\bibitem{Nguyen2012}
{\sc Anh~Tuan Nguyen, Tung~Thanh Nguyen, Tien~N. Nguyen, David Lo, and
  Chengnian Sun}.
\newblock Duplicate {{Bug Report Detection}} with a {{Combination}} of
  {{Information Retrieval}} and {{Topic Modeling}}.
\newblock In {\em ASE'12}, pages 70--79, 2012.

\bibitem{nguyen2019combining}
{\sc Son Nguyen, Tien Nguyen, Yi~Li, and Shaohua Wang}.
\newblock Combining program analysis and statistical language model for code
  statement completion.
\newblock In {\em 2019 34th IEEE/ACM International Conference on Automated
  Software Engineering (ASE)}, pages 710--721. IEEE, 2019.

\bibitem{Nguyen:ASE15}
{\sc Tuan~Anh Nguyen and Christoph Csallner}.
\newblock Reverse engineering mobile application user interfaces with remaui.
\newblock In {\em ASE'15}, pages 248--259, Washington, DC, USA, 2015.

\bibitem{nijkamp2022codegen}
{\sc Erik Nijkamp, Bo~Pang, Hiroaki Hayashi, Lifu Tu, Huan Wang, Yingbo Zhou,
  Silvio Savarese, and Caiming Xiong}.
\newblock Codegen: An open large language model for code with multi-turn
  program synthesis.
\newblock {\em arXiv preprint arXiv:2203.13474}, 2022.

\bibitem{nijkamp2023codegen}
{\sc Erik Nijkamp, Bo~Pang, Hiroaki Hayashi, Lifu Tu, Huan Wang, Yingbo Zhou,
  Silvio Savarese, and Caiming Xiong}.
\newblock Codegen: An open large language model for code with multi-turn
  program synthesis.
\newblock In {\em The Eleventh International Conference on Learning
  Representations}, 2023.

\bibitem{Ott2018}
{\sc J.~{Ott}, A.~{Atchison}, P.~{Harnack}, A.~{Bergh}, and E.~{Linstead}}.
\newblock A deep learning approach to identifying source code in images and
  video.
\newblock In {\em 2018 IEEE/ACM 15th International Conference on Mining
  Software Repositories (MSR)}, pages 376--386, May 2018.

\bibitem{palacio2023theory}
{\sc David~N. Palacio, Nathan Cooper, Alvaro Rodriguez, Kevin Moran, and Denys
  Poshyvanyk}.
\newblock Toward a theory of causation for interpreting neural code models,
  2023.

\bibitem{8919098}
{\sc David~N. Palacio, Daniel McCrystal, Kevin Moran, Carlos Bernal-Cárdenas,
  Denys Poshyvanyk, and Chris Shenefiel}.
\newblock Learning to identify security-related issues using convolutional
  neural networks.
\newblock In {\em 2019 IEEE International Conference on Software Maintenance
  and Evolution (ICSME)}, pages 140--144, 2019.

\bibitem{PALOMBA2018143}
{\sc Fabio Palomba, Mario Linares-V{\'a}squez, Gabriele Bavota, Rocco Oliveto,
  Massimiliano~Di Penta, Denys Poshyvanyk, and Andrea~De Lucia}.
\newblock Crowdsourcing user reviews to support the evolution of mobile apps.
\newblock {\em Journal of Systems and Software}, 137:143--162, 2018.

\bibitem{7332475}
{\sc Fabio Palomba, Mario Linares-Vásquez, Gabriele Bavota, Rocco Oliveto,
  Massimiliano Di~Penta, Denys Poshyvanyk, and Andrea De~Lucia}.
\newblock User reviews matter! tracking crowdsourced reviews to support
  evolution of successful apps.
\newblock In {\em 2015 IEEE International Conference on Software Maintenance
  and Evolution (ICSME)}, pages 291--300, 2015.

\bibitem{Panichella2019}
{\sc Annibale Panichella}.
\newblock A systematic comparison of search algorithms for topic modelling—a
  study on duplicate bug report identification.
\newblock In {\em SSBSE'19}, pages 11--26. Springer, 2019.

\bibitem{Papineni2002}
{\sc Kishore Papineni, Salim Roukos, Todd Ward, and Wei-Jing Zhu}.
\newblock Bleu: A method for automatic evaluation of machine translation.
\newblock In {\em Proceedings of the 40th Annual Meeting on Association for
  Computational Linguistics}, ACL '02, pages 311--318, Stroudsburg, PA, USA,
  2002. Association for Computational Linguistics.

\bibitem{DBLP:conf/iclr/ParisottoMS0ZK17}
{\sc Emilio Parisotto, Abdel{-}rahman Mohamed, Rishabh Singh, Lihong Li,
  Dengyong Zhou, and Pushmeet Kohli}.
\newblock Neuro-symbolic program synthesis.
\newblock In {\em 5th International Conference on Learning Representations,
  {ICLR} 2017, Toulon, France, April 24-26, 2017, Conference Track
  Proceedings}. OpenReview.net, 2017.

\bibitem{perez:msr19}
{\sc Daniel Perez and Shigeru Chiba}.
\newblock Cross-language clone detection by learning over abstract syntax
  trees.
\newblock In {\em 2019 IEEE/ACM 16th International Conference on Mining
  Software Repositories (MSR)}, pages 518--528, 2019.

\bibitem{Piech2015}
{\sc Chris Piech, Jonathan Huang, Andy Nguyen, Mike Phulsuksombati, Mehran
  Sahami, and Leonidas Guibas}.
\newblock Learning program embeddings to propagate feedback on student code.
\newblock In {\em Proceedings of the 32nd International Conference on Machine
  Learning}, Francis Bach and David Blei, editors, volume~37 of {\em
  Proceedings of Machine Learning Research}, pages 1093--1102, Lille, France,
  07--09 Jul 2015. PMLR.

\bibitem{Ponzanelliz:TSE'19}
{\sc L.~{Ponzanelli}, G.~{Bavota}, A.~{Mocci}, R.~{Oliveto}, M.~D. {Penta},
  S.~{Haiduc}, B.~{Russo}, and M.~{Lanza}}.
\newblock Automatic identification and classification of software development
  video tutorial fragments.
\newblock {\em TSE'19}, 45(5):464--488, 2019.

\bibitem{popovic2015chrf}
{\sc Maja Popovi{\'c}}.
\newblock chrf: character n-gram f-score for automatic mt evaluation.
\newblock In {\em Proceedings of the Tenth Workshop on Statistical Machine
  Translation}, pages 392--395, 2015.

\bibitem{5306298}
{\sc Denys Poshyvanyk}.
\newblock Using information retrieval to support software maintenance tasks.
\newblock In {\em 2009 IEEE International Conference on Software Maintenance},
  pages 453--456, 2009.

\bibitem{poshyvanyk2009using}
{\sc Denys Poshyvanyk, Andrian Marcus, Rudolf Ferenc, and Tibor Gyim{\'o}thy}.
\newblock Using information retrieval based coupling measures for impact
  analysis.
\newblock {\em Empirical software engineering}, 14:5--32, 2009.

\bibitem{10.1145/3276517}
{\sc Michael Pradel and Koushik Sen}.
\newblock Deepbugs: A learning approach to name-based bug detection.
\newblock {\em Proc. ACM Program. Lang.}, 2(OOPSLA), October 2018.

\bibitem{dlse19-report}
{\sc Devanbu Prem, Matthew Dwyer, Sebastian Elbaum, Michael Lowry, Kevin Moran,
  Denys Poshyvanyk, Baishakhi Ray, Rishabh Singh, and Xiangyu Zhang}.
\newblock Deep learning \& software engineering: State of research and future
  directions.
\newblock In {\em Proceedings of the 2019 NSF Workshop on Deep Learning and
  Software Engineering}, 2019.

\bibitem{press2022alibi}
{\sc Ofir Press, Noah Smith, and Mike Lewis}.
\newblock Train short, test long: Attention with linear biases enables input
  length extrapolation.
\newblock In {\em International Conference on Learning Representations}, 2022.

\bibitem{press2020shortformer}
{\sc Ofir Press, Noah~A Smith, and Mike Lewis}.
\newblock Shortformer: Better language modeling using shorter inputs.
\newblock {\em arXiv preprint arXiv:2012.15832}, 2020.

\bibitem{raffel2020exploring}
{\sc Colin Raffel, Noam Shazeer, Adam Roberts, Katherine Lee, Sharan Narang,
  Michael Matena, Yanqi Zhou, Wei Li, Peter~J Liu, et~al.}
\newblock Exploring the limits of transfer learning with a unified text-to-text
  transformer.
\newblock {\em J. Mach. Learn. Res.}, 21(140):1--67, 2020.

\bibitem{Rakha:TSE'18}
{\sc M.~S. {Rakha}, C.~{Bezemer}, and A.~E. {Hassan}}.
\newblock Revisiting the performance evaluation of automated approaches for the
  retrieval of duplicate issue reports.
\newblock {\em TSE'18}, 44(12):1245--1268, 2018.

\bibitem{Rakha2018}
{\sc Mohamed~Sami Rakha, Cor-Paul Bezemer, and Ahmed~E. Hassan}.
\newblock Revisiting the performance of automated approaches for the retrieval
  of duplicate reports in issue tracking systems that perform just-in-time
  duplicate retrieval.
\newblock {\em EMSE}, 23(5):2597--2621, 2018.

\bibitem{ranasinghe2019semantic}
{\sc Tharindu Ranasinghe, Constantin Orǎsan, and Ruslan Mitkov}.
\newblock Semantic textual similarity with siamese neural networks.
\newblock In {\em Proceedings of the International Conference on Recent
  Advances in Natural Language Processing (RANLP 2019)}, pages 1004--1011,
  2019.

\bibitem{NIPS2000_1925}
{\sc Carl~Edward Rasmussen and Zoubin Ghahramani}.
\newblock Occam\textquotesingle s razor.
\newblock In {\em Advances in Neural Information Processing Systems 13}, T.~K.
  Leen, T.~G. Dietterich, and V.~Tresp, editors, pages 294--300. MIT Press,
  2001.

\bibitem{raychev2014code}
{\sc Veselin Raychev, Martin Vechev, and Eran Yahav}.
\newblock Code completion with statistical language models.
\newblock In {\em Proceedings of the 35th ACM SIGPLAN Conference on Programming
  Language Design and Implementation}, pages 419--428, 2014.

\bibitem{10.1145/2594291.2594321}
{\sc Veselin Raychev, Martin Vechev, and Eran Yahav}.
\newblock Code completion with statistical language models.
\newblock In {\em Proceedings of the 35th ACM SIGPLAN Conference on Programming
  Language Design and Implementation}, PLDI '14, page 419–428, New York, NY,
  USA, 2014. Association for Computing Machinery.

\bibitem{DBLP:journals/corr/ReedF15}
{\sc Scott~E. Reed and Nando de~Freitas}.
\newblock Neural programmer-interpreters.
\newblock In {\em 4th International Conference on Learning Representations,
  {ICLR} 2016, San Juan, Puerto Rico, May 2-4, 2016, Conference Track
  Proceedings}, Yoshua Bengio and Yann LeCun, editors, 2016.

\bibitem{ren2020codebleu}
{\sc Shuo Ren, Daya Guo, Shuai Lu, Long Zhou, Shujie Liu, Duyu Tang, Neel
  Sundaresan, Ming Zhou, Ambrosio Blanco, and Shuai Ma}.
\newblock Codebleu: a method for automatic evaluation of code synthesis.
\newblock {\em arXiv preprint arXiv:2009.10297}, 2020.

\bibitem{ren2005chianti}
{\sc Xiaoxia Ren, Barbara~G Ryder, Maximilian Stoerzer, and Frank Tip}.
\newblock Chianti: a change impact analysis tool for java programs.
\newblock In {\em Proceedings of the 27th international conference on Software
  engineering}, pages 664--665, 2005.

\bibitem{Revaud:CVPR'13}
{\sc J.~{Revaud}, M.~{Douze}, C.~{Schmid}, and H.~{Jégou}}.
\newblock Event retrieval in large video collections with circulant temporal
  encoding.
\newblock In {\em CVPR'13}, pages 2459--2466, 2013.

\bibitem{revelle2011using}
{\sc Meghan Revelle, Malcom Gethers, and Denys Poshyvanyk}.
\newblock Using structural and textual information to capture feature coupling
  in object-oriented software.
\newblock {\em Empirical software engineering}, 16:773--811, 2011.

\bibitem{revelle_exploratory_2009}
{\sc Meghan Revelle and Denys Poshyvanyk}.
\newblock An exploratory study on assessing feature location techniques.
\newblock In {\em 2009 {IEEE} 17th {International} {Conference} on {Program}
  {Comprehension}}, pages 218--222, Vancouver, BC, Canada, May 2009. IEEE.

\bibitem{ribeiro:2020}
{\sc Marco~Tulio Ribeiro, Tongshuang Wu, Carlos Guestrin, and Sameer Singh}.
\newblock Beyond accuracy: Behavioral testing of {NLP} models with
  {C}heck{L}ist.
\newblock In {\em Proceedings of the 58th Annual Meeting of the Association for
  Computational Linguistics}, pages 4902--4912, Online, July 2020. Association
  for Computational Linguistics.

\bibitem{robbes2008program}
{\sc Romain Robbes and Michele Lanza}.
\newblock How program history can improve code completion.
\newblock In {\em 2008 23rd IEEE/ACM International Conference on Automated
  Software Engineering}, pages 317--326. IEEE, 2008.

\bibitem{robbes2010improving}
{\sc Romain Robbes and Michele Lanza}.
\newblock Improving code completion with program history.
\newblock {\em Automated Software Engineering}, 17(2):181--212, 2010.

\bibitem{Rodrigues2020}
{\sc Irving~Muller Rodrigues, Daniel Aloise, Eraldo~Rezende Fernandes, and
  Michel Dagenais}.
\newblock A soft alignment model for bug deduplication.
\newblock In {\em MSR’20}, pages 43--53, 2020.

\bibitem{Romansky2017}
{\sc S.~{Romansky}, N.~C. {Borle}, S.~{Chowdhury}, A.~{Hindle}, and
  R.~{Greiner}}.
\newblock Deep green: Modelling time-series of software energy consumption.
\newblock In {\em 2017 IEEE International Conference on Software Maintenance
  and Evolution (ICSME)}, pages 273--283, Sep. 2017.

\bibitem{rosendahl2019analysis}
{\sc Jan Rosendahl, Viet Anh~Khoa Tran, Weiyue Wang, and Hermann Ney}.
\newblock Analysis of positional encodings for neural machine translation.
\newblock In {\em Proceedings of the 16th International Conference on Spoken
  Language Translation}, 2019.

\bibitem{rumelhart1985learning}
{\sc David~E Rumelhart, Geoffrey~E Hinton, and Ronald~J Williams}.
\newblock Learning internal representations by error propagation.
\newblock Technical report, California Univ San Diego La Jolla Inst for
  Cognitive Science, 1985.

\bibitem{Runeson2007}
{\sc P.~Runeson, M.~Alexandersson, and O.~Nyholm}.
\newblock Detection of {{Duplicate Defect Reports Using Natural Language
  Processing}}.
\newblock In {\em ICSE'07}, pages 499--510, 2007.

\bibitem{ryder2001change}
{\sc Barbara~G Ryder and Frank Tip}.
\newblock Change impact analysis for object-oriented programs.
\newblock In {\em Proceedings of the 2001 ACM SIGPLAN-SIGSOFT workshop on
  Program analysis for software tools and engineering}, pages 46--53, 2001.

\bibitem{Saini2018}
{\sc Vaibhav Saini, Farima Farmahinifarahani, Yadong Lu, Pierre Baldi, and
  Cristina~V. Lopes}.
\newblock Oreo: detection of clones in the twilight zone.
\newblock In {\em Proceedings of the 2018 {ACM} Joint Meeting on European
  Software Engineering Conference and Symposium on the Foundations of Software
  Engineering, {ESEC/SIGSOFT} {FSE} 2018, Lake Buena Vista, FL, USA, November
  04-09, 2018}, Gary~T. Leavens, Alessandro Garcia, and Corina~S. Pasareanu,
  editors, pages 354--365. {ACM}, 2018.

\bibitem{Salton:TFIDF86}
{\sc Gerard Salton and Michael~J. McGill}.
\newblock {\em Introduction to Modern Information Retrieval}.
\newblock McGraw-Hill, Inc., USA, 1986.

\bibitem{santelices2010probabilistic}
{\sc Raul Santelices and Mary~Jean Harrold}.
\newblock Probabilistic slicing for predictive impact analysis.
\newblock Technical report, Georgia Institute of Technology, 2010.

\bibitem{Sayyad-Shirabad+Menzies:2005}
{\sc J.~Sayyad~Shirabad and T.J. Menzies}.
\newblock {The {PROMISE} Repository of Software Engineering Databases.}
\newblock School of Information Technology and Engineering, University of
  Ottawa, Canada, 2005.

\bibitem{ReadabilityJSME}
{\sc Simone Scalabrino, Mario Linares-Vásquez, Rocco Oliveto, and Denys
  Poshyvanyk}.
\newblock A comprehensive model for code readability.
\newblock {\em Journal of Software: Evolution and Process}, 30(6):e1958, 2018.
\newblock e1958 smr.1958.

\bibitem{Scalabrino2006}
{\sc Simone Scalabrino, Mario Linares-Vásquez, Denys Poshyvanyk, and Rocco
  Oliveto}.
\newblock Improving code readability models with textual features.
\newblock In {\em 2016 IEEE 24th International Conference on Program
  Comprehension (ICPC)}, pages 1--10, 2016.

\bibitem{Schroeder2017}
{\sc Jan Schroeder, Christian Berger, Alessia Knauss, Harri Preenja, Mohammad
  Ali, Miroslaw Staron, and Thomas Herpel}.
\newblock Predicting and evaluating software model growth in the automotive
  industry.
\newblock In {\em 2017 {IEEE} International Conference on Software Maintenance
  and Evolution, {ICSME} 2017, Shanghai, China, September 17-22, 2017}, pages
  584--593. {IEEE} Computer Society, 2017.

\bibitem{sennrich2015neural}
{\sc Rico Sennrich, Barry Haddow, and Alexandra Birch}.
\newblock Neural machine translation of rare words with subword units.
\newblock {\em arXiv preprint arXiv:1508.07909}, 2015.

\bibitem{codereviewlearn}
{\sc Shu-Ting Shi, Ming Li, David Lo, Ferdian Thung, and Xuan Huo}.
\newblock Automatic code review by learning the revision of source code.
\newblock {\em Proceedings of the AAAI Conference on Artificial Intelligence},
  33:4910--4917, 07 2019.

\bibitem{Shin2018}
{\sc Richard Shin, Illia Polosukhin, and Dawn Song}.
\newblock Improving neural program synthesis with inferred execution traces.
\newblock In {\em Advances in Neural Information Processing Systems 31},
  S.~Bengio, H.~Wallach, H.~Larochelle, K.~Grauman, N.~Cesa-Bianchi, and
  R.~Garnett, editors, pages 8917--8926. Curran Associates, Inc., 2018.

\bibitem{Si2018}
{\sc Xujie Si, Hanjun Dai, Mukund Raghothaman, Mayur Naik, and Le~Song}.
\newblock Learning loop invariants for program verification.
\newblock In {\em Advances in Neural Information Processing Systems 31},
  S.~Bengio, H.~Wallach, H.~Larochelle, K.~Grauman, N.~Cesa-Bianchi, and
  R.~Garnett, editors, pages 7751--7762. Curran Associates, Inc., 2018.

\bibitem{Sivic:CCV'03}
{\sc {Sivic} and {Zisserman}}.
\newblock Video google: a text retrieval approach to object matching in videos.
\newblock In {\em CCV'03}, pages 1470--1477 vol.2, 2003.

\bibitem{10.1145/3540250.3549131}
{\sc Yang Song, Junayed Mahmud, Ying Zhou, Oscar Chaparro, Kevin Moran, Andrian
  Marcus, and Denys Poshyvanyk}.
\newblock Toward interactive bug reporting for (android app) end-users.
\newblock In {\em Proceedings of the 30th ACM Joint European Software
  Engineering Conference and Symposium on the Foundations of Software
  Engineering}, ESEC/FSE 2022, page 344–356, New York, NY, USA, 2022.
  Association for Computing Machinery.

\bibitem{JMLR:v15:srivastava14a}
{\sc Nitish Srivastava, Geoffrey Hinton, Alex Krizhevsky, Ilya Sutskever, and
  Ruslan Salakhutdinov}.
\newblock Dropout: A simple way to prevent neural networks from overfitting.
\newblock {\em Journal of Machine Learning Research}, 15(56):1929--1958, 2014.

\bibitem{su2021roformer}
{\sc Jianlin Su, Yu~Lu, Shengfeng Pan, Bo~Wen, and Yunfeng Liu}.
\newblock Roformer: Enhanced transformer with rotary position embedding.
\newblock {\em arXiv preprint arXiv:2104.09864}, 2021.

\bibitem{Sun2011}
{\sc C.~Sun, D.~Lo, S.~C. Khoo, and J.~Jiang}.
\newblock Towards more accurate retrieval of duplicate bug reports.
\newblock In {\em ASE'11}, pages 253--262, 2011.

\bibitem{Sun2010}
{\sc Chengnian Sun, David Lo, Xiaoyin Wang, Jing Jiang, and Siau-Cheng Khoo}.
\newblock A {{Discriminative Model Approach}} for {{Accurate Duplicate Bug
  Report Retrieval}}.
\newblock In {\em ICSE'10}, pages 45--54, 2010.

\bibitem{Sun2018}
{\sc Shao-Hua Sun, Hyeonwoo Noh, Sriram Somasundaram, and Joseph Lim}.
\newblock Neural program synthesis from diverse demonstration videos.
\newblock In {\em Proceedings of the 35th International Conference on Machine
  Learning}, Jennifer Dy and Andreas Krause, editors, volume~80 of {\em
  Proceedings of Machine Learning Research}, pages 4790--4799,
  Stockholmsmässan, Stockholm Sweden, 10--15 Jul 2018. PMLR.

\bibitem{sun2022length}
{\sc Yutao Sun, Li~Dong, Barun Patra, Shuming Ma, Shaohan Huang, Alon Benhaim,
  Vishrav Chaudhary, Xia Song, and Furu Wei}.
\newblock A length-extrapolatable transformer.
\newblock {\em arXiv preprint arXiv:2212.10554}, 2022.

\bibitem{sun2019}
{\sc Zeyu Sun, Qihao Zhu, Lili Mou, Yingfei Xiong, Ge~Li, and Lu~Zhang}.
\newblock A grammar-based structural cnn decoder for code generation.
\newblock {\em Proceedings of the AAAI Conference on Artificial Intelligence},
  33:7055--7062, 07 2019.

\bibitem{Sureka2010}
{\sc A.~Sureka and P.~Jalote}.
\newblock Detecting {Duplicate} {Bug} {Report} {Using} {Character}
  {N}-{Gram}-{Based} {Features}.
\newblock In {\em ASPEC'10}, pages 366--374, 2010.

\bibitem{6976121}
{\sc J.~{Svajlenko}, J.~F. {Islam}, I.~{Keivanloo}, C.~K. {Roy}, and M.~M.
  {Mia}}.
\newblock Towards a big data curated benchmark of inter-project code clones.
\newblock In {\em 2014 IEEE International Conference on Software Maintenance
  and Evolution}, pages 476--480, 2014.

\bibitem{svyatkovskiy2021fast}
{\sc Alexey Svyatkovskiy, Sebastian Lee, Anna Hadjitofi, Maik Riechert,
  Juliana~Vicente Franco, and Miltiadis Allamanis}.
\newblock Fast and memory-efficient neural code completion.
\newblock In {\em 2021 IEEE/ACM 18th International Conference on Mining
  Software Repositories (MSR)}, pages 329--340. IEEE, 2021.

\bibitem{svyatkovskiy2019pythia}
{\sc Alexey Svyatkovskiy, Ying Zhao, Shengyu Fu, and Neel Sundaresan}.
\newblock Pythia: Ai-assisted code completion system.
\newblock In {\em Proceedings of the 25th ACM SIGKDD International Conference
  on Knowledge Discovery \& Data Mining}, pages 2727--2735, 2019.

\bibitem{Tetko1995NeuralNS}
{\sc I.~Tetko, D.~Livingstone, and A.~I. Luik}.
\newblock Neural network studies, 1. comparison of overfitting and
  overtraining.
\newblock {\em J. Chem. Inf. Comput. Sci.}, 35:826--833, 1995.

\bibitem{thakur2021beir}
{\sc Nandan Thakur, Nils Reimers, Andreas R{\"u}ckl{\'e}, Abhishek Srivastava,
  and Iryna Gurevych}.
\newblock Beir: A heterogenous benchmark for zero-shot evaluation of
  information retrieval models.
\newblock {\em arXiv preprint arXiv:2104.08663}, 2021.

\bibitem{thaller:saner19}
{\sc Hannes Thaller, Lukas Linsbauer, and Alexander Egyed}.
\newblock Feature maps: A comprehensible software representation for design
  pattern detection.
\newblock In {\em 2019 IEEE 26th International Conference on Software Analysis,
  Evolution and Reengineering (SANER)}, pages 207--217, 2019.

\bibitem{Thung2014}
{\sc Ferdian Thung, Pavneet~Singh Kochhar, and David Lo}.
\newblock {{DupFinder}}: {{Integrated Tool Support}} for {{Duplicate Bug Report
  Detection}}.
\newblock In {\em ASE'14}, pages 871--874, 2014.

\bibitem{Tian2012}
{\sc Y.~Tian, C.~Sun, and D.~Lo}.
\newblock Improved {{Duplicate Bug Report Identification}}.
\newblock In {\em CSMR'12}, pages 385--390, 2012.

\bibitem{Tian2018}
{\sc Yuchi Tian, Kexin Pei, Suman Jana, and Baishakhi Ray}.
\newblock Deeptest: Automated testing of deep-neural-network-driven autonomous
  cars.
\newblock In {\em Proceedings of the 40th International Conference on Software
  Engineering}, ICSE '18, pages 303--314, New York, NY, USA, 2018. ACM.

\bibitem{tip1994survey}
{\sc Frank Tip}.
\newblock {\em A survey of program slicing techniques}.
\newblock Centrum voor Wiskunde en Informatica Amsterdam, 1994.

\bibitem{Tran:ICCV'15}
{\sc D.~{Tran}, L.~{Bourdev}, R.~{Fergus}, L.~{Torresani}, and M.~{Paluri}}.
\newblock Learning spatiotemporal features with 3d convolutional networks.
\newblock In {\em ICCV'15}, pages 4489--4497, 2015.

\bibitem{tu2014localness}
{\sc Zhaopeng Tu, Zhendong Su, and Premkumar Devanbu}.
\newblock On the localness of software.
\newblock In {\em Proceedings of the 22nd ACM SIGSOFT International Symposium
  on Foundations of Software Engineering}, pages 269--280, 2014.

\bibitem{9270362}
{\sc M.~Tufano, J.~Kimko, S.~Wang, C.~Watson, G.~Bavota, M.~Di Penta, and
  D.~Poshyvanyk}.
\newblock Deepmutation: A neural mutation tool.
\newblock In {\em 2020 IEEE/ACM 42nd International Conference on Software
  Engineering: Companion Proceedings (ICSE-Companion)}, pages 29--33, Los
  Alamitos, CA, USA, oct 2020. IEEE Computer Society.

\bibitem{Tufano:testGen}
{\sc Michele Tufano, Dawn Drain, Alexey Svyatkovskiy, Shao~Kun Deng, and Neel
  Sundaresan}.
\newblock Unit test case generation with transformers.
\newblock {\em CoRR}, abs/2009.05617, 2020.

\bibitem{tufano2019learning}
{\sc Michele Tufano, Jevgenija Pantiuchina, Cody Watson, Gabriele Bavota, and
  Denys Poshyvanyk}.
\newblock On learning meaningful code changes via neural machine translation.
\newblock In {\em 2019 IEEE/ACM 41st International Conference on Software
  Engineering (ICSE)}, pages 25--36. IEEE, 2019.

\bibitem{Tufano2018a}
{\sc Michele Tufano, Cody Watson, Gabriele Bavota, Massimiliano Di~Penta,
  Martin White, and Denys Poshyvanyk}.
\newblock Deep learning similarities from different representations of source
  code.
\newblock In {\em Proceedings of the 15th International Conference on Mining
  Software Repositories}, MSR '18, pages 542--553, New York, NY, USA, 2018.
  ACM.

\bibitem{Tufano2018}
{\sc Michele Tufano, Cody Watson, Gabriele Bavota, Massimiliano Di~Penta,
  Martin White, and Denys Poshyvanyk}.
\newblock An empirical investigation into learning bug-fixing patches in the
  wild via neural machine translation.
\newblock In {\em Proceedings of the 33rd ACM/IEEE International Conference on
  Automated Software Engineering}, ASE 2018, pages 832--837, New York, NY, USA,
  2018. ACM.

\bibitem{TufanoMutation}
{\sc Michele Tufano, Cody Watson, Gabriele Bavota, Massimiliano Di~Penta,
  Martin White, and Denys Poshyvanyk}.
\newblock Learning how to mutate source code from bug-fixes.
\newblock In {\em 2019 IEEE International Conference on Software Maintenance
  and Evolution (ICSME)}, pages 301--312, 2019.

\bibitem{tufano2019empirical}
{\sc Michele Tufano, Cody Watson, Gabriele Bavota, Massimiliano~Di Penta,
  Martin White, and Denys Poshyvanyk}.
\newblock An empirical study on learning bug-fixing patches in the wild via
  neural machine translation.
\newblock {\em ACM Transactions on Software Engineering and Methodology
  (TOSEM)}, 28(4):1--29, 2019.

\bibitem{TufanoICSE22}
{\sc Rosalia Tufano, Simone Masiero, Antonio Mastropaolo, Luca Pascarella,
  Denys Poshyvanyk, and Gabriele Bavota}.
\newblock Using pre-trained models to boost code review automation.
\newblock In {\em 2022 IEEE/ACM 44th International Conference on Software
  Engineering (ICSE)}, pages 2291--2302, 2022.

\bibitem{9402025}
{\sc Rosalia Tufano, Luca Pascarella, Michele Tufano, Denys Poshyvanyk, and
  Gabriele Bavota}.
\newblock Towards automating code review activities.
\newblock In {\em 2021 IEEE/ACM 43rd International Conference on Software
  Engineering (ICSE)}, pages 163--174, 2021.

\bibitem{vaswani2017attention}
{\sc Ashish Vaswani, Noam Shazeer, Niki Parmar, Jakob Uszkoreit, Llion Jones,
  Aidan~N Gomez, {\L}ukasz Kaiser, and Illia Polosukhin}.
\newblock Attention is all you need.
\newblock {\em Advances in neural information processing systems}, 30, 2017.

\bibitem{vaucher2008discovering}
{\sc St{\'e}phane Vaucher, Houari Sahraoui, and Jean Vaucher}.
\newblock Discovering new change patterns in object-oriented systems.
\newblock In {\em 2008 15th Working Conference on Reverse Engineering}, pages
  37--41. IEEE, 2008.

\bibitem{velivckovic2017graph}
{\sc Petar Veli{\v{c}}kovi{\'c}, Guillem Cucurull, Arantxa Casanova, Adriana
  Romero, Pietro Lio, and Yoshua Bengio}.
\newblock Graph attention networks.
\newblock {\em arXiv preprint arXiv:1710.10903}, 2017.

\bibitem{vidacs2007macro}
{\sc L{\'a}szl{\'o} Vid{\'a}cs, {\'A}rp{\'a}d Besz{\'e}des, and Rudolf Ferenc}.
\newblock Macro impact analysis using macro slicing.
\newblock 2007.

\bibitem{Wan2018}
{\sc Yao Wan, Zhou Zhao, Min Yang, Guandong Xu, Haochao Ying, Jian Wu, and
  Philip~S. Yu}.
\newblock Improving automatic source code summarization via deep reinforcement
  learning.
\newblock In {\em Proceedings of the 33rd ACM/IEEE International Conference on
  Automated Software Engineering}, ASE 2018, pages 397--407, New York, NY, USA,
  2018. ACM.

\bibitem{Wang2019}
{\sc Junjie Wang, Mingyang Li, Song Wang, Tim Menzies, and Qing Wang}.
\newblock Images don’t lie: Duplicate crowdtesting reports detection with
  screenshot information.
\newblock {\em IST}, 110:139--155, 2019.

\bibitem{Wang2020}
{\sc Junjie Wang, Ye~Yang, Tim Menzies, and Qing Wang}.
\newblock isense2. 0: Improving completion-aware crowdtesting management with
  duplicate tagger and sanity checker.
\newblock {\em TOSEM}, 29(4):1--27, 2020.

\bibitem{Wang2017}
{\sc Ke~Wang, Rishabh Singh, and Zhendong Su}.
\newblock Dynamic neural program embeddings for program repair.
\newblock In {\em 6th International Conference on Learning Representations,
  {ICLR} 2018, Vancouver, BC, Canada, April 30 - May 3, 2018, Conference Track
  Proceedings}. OpenReview.net, 2018.

\bibitem{xtransformers}
{\sc Phil Wang}.
\newblock X-transformers.
\newblock {\em GitHub}, 2023.

\bibitem{wang:msr19}
{\sc Shaohua Wang, NhatHai Phan, Yan Wang, and Yong Zhao}.
\newblock Extracting api tips from developer question and answer websites.
\newblock In {\em 2019 IEEE/ACM 16th International Conference on Mining
  Software Repositories (MSR)}, pages 321--332, 2019.

\bibitem{Wang2016}
{\sc Shaowei Wang and David Lo}.
\newblock Amalgam+: Composing rich information sources for accurate bug
  localization.
\newblock {\em Journal of Software: Evolution and Process}, 28(10):921--942,
  2016.

\bibitem{8502853}
{\sc Song Wang, Taiyue Liu, Jaechang Nam, and Lin Tan}.
\newblock Deep semantic feature learning for software defect prediction.
\newblock {\em IEEE Transactions on Software Engineering}, 46(12):1267--1293,
  2020.

\bibitem{wang2018integrated}
{\sc Wei Wang, Yun He, Tong Li, Jiajun Zhu, and Jinzhuo Liu}.
\newblock An integrated model for information retrieval based change impact
  analysis.
\newblock {\em Scientific Programming}, 2018:1--13, 2018.

\bibitem{Wang2008}
{\sc Xiaoyin Wang, Lu~Zhang, Tao Xie, John Anvik, and Jiasu Sun}.
\newblock An {{Approach}} to {{Detecting Duplicate Bug Reports Using Natural
  Language}} and {{Execution Information}}.
\newblock In {\em ICSE'08}, pages 461--470, 2008.

\bibitem{watson2022systematic}
{\sc Cody Watson, Nathan Cooper, David~Nader Palacio, Kevin Moran, and Denys
  Poshyvanyk}.
\newblock A systematic literature review on the use of deep learning in
  software engineering research.
\newblock {\em ACM Transactions on Software Engineering and Methodology
  (TOSEM)}, 31(2):1--58, 2022.

\bibitem{watson_palacio_cooper_moran_poshyvanyk}
{\sc Cody Watson, David Palacio, Nathan Cooper, Kevin Moran, and Denys
  Poshyvanyk}.
\newblock Data analysis for the systematic literature review of dl4se.

\bibitem{WatsonAsserts}
{\sc Cody Watson, Michele Tufano, Kevin Moran, Gabriele Bavota, and Denys
  Poshyvanyk}.
\newblock On learning meaningful assert statements for unit test cases.
\newblock In {\em 2020 IEEE/ACM 42nd International Conference on Software
  Engineering (ICSE)}, pages 1398--1409, 2020.

\bibitem{Wen2018}
{\sc M.~{Wen}, R.~{Wu}, and S.~C. {Cheung}}.
\newblock How well do change sequences predict defects? sequence learning from
  software changes.
\newblock {\em IEEE Transactions on Software Engineering}, pages 1--1, 2018.

\bibitem{White2016}
{\sc M.~White, M.~Tufano, C.~Vendome, and D.~Poshyvanyk}.
\newblock Deep learning code fragments for code clone detection.
\newblock In {\em 2016 31st {{IEEE}}/{{ACM International Conference}} on
  {{Automated Software Engineering}} ({{ASE}})}, ASE'16, pages 87--98,
  September 2016.
\newblock ISSN:.

\bibitem{7181432}
{\sc Martin White, Mario Linares-Vásquez, Peter Johnson, Carlos
  Bernal-Cárdenas, and Denys Poshyvanyk}.
\newblock Generating reproducible and replayable bug reports from android
  application crashes.
\newblock In {\em 2015 IEEE 23rd International Conference on Program
  Comprehension}, pages 48--59, 2015.

\bibitem{white:saner19}
{\sc Martin White, Michele Tufano, Matías Martínez, Martin Monperrus, and
  Denys Poshyvanyk}.
\newblock Sorting and transforming program repair ingredients via deep learning
  code similarities.
\newblock In {\em 2019 IEEE 26th International Conference on Software Analysis,
  Evolution and Reengineering (SANER)}, pages 479--490, 2019.

\bibitem{White2015a}
{\sc Martin White, Christopher Vendome, Mario Linares-V\'{a}squez, and Denys
  Poshyvanyk}.
\newblock Toward deep learning software repositories.
\newblock In {\em Proceedings of the 12th Working Conference on Mining Software
  Repositories}, MSR '15, pages 334--345, Piscataway, NJ, USA, 2015. IEEE
  Press.

\bibitem{Wu:MM'07}
{\sc Xiao Wu, Alexander~G. Hauptmann, and Chong-Wah Ngo}.
\newblock Practical elimination of near-duplicates from web video search.
\newblock In {\em MM'07}, page 218–227, New York, NY, USA, 2007. Association
  for Computing Machinery.

\bibitem{wu2022memorizing}
{\sc Yuhuai Wu, Markus~N Rabe, DeLesley Hutchins, and Christian Szegedy}.
\newblock Memorizing transformers.
\newblock {\em arXiv preprint arXiv:2203.08913}, 2022.

\bibitem{xia2004change}
{\sc Franck Xia}.
\newblock A change impact dependency measure for predicting the maintainability
  of source code.
\newblock In {\em Proceedings of the 28th Annual International Computer
  Software and Applications Conference, 2004. COMPSAC 2004.}, volume~2, pages
  22--23. IEEE, 2004.

\bibitem{xie2020explainable}
{\sc Ning Xie, Gabrielle Ras, Marcel van Gerven, and Derek Doran}.
\newblock Explainable deep learning: A field guide for the uninitiated, 2020.

\bibitem{xie:saner19}
{\sc Rui Xie, Long Chen, Wei Ye, Zhiyu Li, Tianxiang Hu, Dongdong Du, and
  Shikun Zhang}.
\newblock Deeplink: A code knowledge graph based deep learning approach for
  issue-commit link recovery.
\newblock In {\em 2019 IEEE 26th International Conference on Software Analysis,
  Evolution and Reengineering (SANER)}, pages 434--444, 2019.

\bibitem{xie2002empirical}
{\sc Tao Xie and David Notkin}.
\newblock An empirical study of java dynamic call graph extractors.
\newblock {\em University of Washington CSE Technical Report}, pages 02--12,
  2002.

\bibitem{xing2004data}
{\sc Zhenchang Xing and Eleni Stroulia}.
\newblock Data-mining in support of detecting class co-evolution.
\newblock In {\em SEKE}, volume~4, pages 123--128. Citeseer, 2004.

\bibitem{xing2004understanding}
{\sc Zhenchang Xing and Eleni Stroulia}.
\newblock Understanding class evolution in object-oriented software.
\newblock In {\em Proceedings. 12th IEEE International Workshop on Program
  Comprehension, 2004.}, pages 34--43. IEEE, 2004.

\bibitem{Xu2016}
{\sc B.~Xu, D.~Ye, Z.~Xing, X.~Xia, G.~Chen, and S.~Li}.
\newblock Predicting semantically linkable knowledge in developer online forums
  via convolutional neural network.
\newblock In {\em 2016 31st {{IEEE}}/{{ACM International Conference}} on
  {{Automated Software Engineering}} ({{ASE}})}, ASE'16, pages 51--62,
  September 2016.
\newblock ISSN:.

\bibitem{Yadid:2016}
{\sc Shir Yadid and Eran Yahav}.
\newblock Extracting code from programming tutorial videos.
\newblock In {\em Onward!'16}, page 98–111, New York, NY, USA, 2016. ACM.

\bibitem{athena-tool}
{\sc Y.~Yan, N.~Cooper, K.~Moran, D.~Poshyvanyk, and G.~Bavota}.
\newblock Athena online appendix \url{https://github.com/WM-SEMERU/athena},
  2023.

\bibitem{yang2017language}
{\sc Yixiao Yang, Yu~Jiang, Ming Gu, Jiaguang Sun, Jian Gao, and Han Liu}.
\newblock A language model for statements of software code.
\newblock In {\em 2017 32nd IEEE/ACM International Conference on Automated
  Software Engineering (ASE)}, pages 682--687. IEEE, 2017.

\bibitem{yang2019enhancing}
{\sc Zebin Yang, Aijun Zhang, and Agus Sudjianto}.
\newblock Enhancing explainability of neural networks through architecture
  constraints, 2019.

\bibitem{10.1145/3196398.3196408}
{\sc Pengcheng Yin, Bowen Deng, Edgar Chen, Bogdan Vasilescu, and Graham
  Neubig}.
\newblock Learning to mine aligned code and natural language pairs from stack
  overflow.
\newblock In {\em Proceedings of the 15th International Conference on Mining
  Software Repositories}, MSR '18, page 476–486, New York, NY, USA, 2018.
  Association for Computing Machinery.

\bibitem{yin2018learning}
{\sc Pengcheng Yin, Graham Neubig, Miltiadis Allamanis, Marc Brockschmidt, and
  Alexander~L Gaunt}.
\newblock Learning to represent edits.
\newblock {\em arXiv preprint arXiv:1810.13337}, 2018.

\bibitem{DBLP:conf/iclr/YinNABG19}
{\sc Pengcheng Yin, Graham Neubig, Miltiadis Allamanis, Marc Brockschmidt, and
  Alexander~L. Gaunt}.
\newblock Learning to represent edits.
\newblock In {\em 7th International Conference on Learning Representations,
  {ICLR} 2019, New Orleans, LA, USA, May 6-9, 2019}. OpenReview.net, 2019.

\bibitem{ying2004predicting}
{\sc Annie~TT Ying, Gail~C Murphy, Raymond Ng, and Mark~C Chu-Carroll}.
\newblock Predicting source code changes by mining change history.
\newblock {\em IEEE transactions on Software Engineering}, 30(9):574--586,
  2004.

\bibitem{yu2019neural}
{\sc Hao Yu, Wing Lam, Long Chen, Ge~Li, Tao Xie, and Qianxiang Wang}.
\newblock Neural detection of semantic code clones via tree-based convolution.
\newblock In {\em 2019 IEEE/ACM 27th International Conference on Program
  Comprehension (ICPC)}, pages 70--80. IEEE, 2019.

\bibitem{yuan2021explainability}
{\sc Hao Yuan, Haiyang Yu, Shurui Gui, and Shuiwang Ji}.
\newblock Explainability in graph neural networks: A taxonomic survey, 2021.

\bibitem{Zeiler2012}
{\sc Matthew~D. Zeiler}.
\newblock Adadelta: An adaptive learning rate method, 2012.

\bibitem{zhang2019novel}
{\sc Jian Zhang, Xu~Wang, Hongyu Zhang, Hailong Sun, Kaixuan Wang, and Xudong
  Liu}.
\newblock A novel neural source code representation based on abstract syntax
  tree.
\newblock In {\em 2019 IEEE/ACM 41st International Conference on Software
  Engineering (ICSE)}, pages 783--794. IEEE, 2019.

\bibitem{Zhang2018}
{\sc Lisa Zhang, Gregory Rosenblatt, Ethan Fetaya, Renjie Liao, William Byrd,
  Matthew Might, Raquel Urtasun, and Richard Zemel}.
\newblock Neural guided constraint logic programming for program synthesis.
\newblock In {\em Advances in Neural Information Processing Systems 31},
  S.~Bengio, H.~Wallach, H.~Larochelle, K.~Grauman, N.~Cesa-Bianchi, and
  R.~Garnett, editors, pages 1737--1746. Curran Associates, Inc., 2018.

\bibitem{Zhang2018a}
{\sc Mengshi Zhang, Yuqun Zhang, Lingming Zhang, Cong Liu, and Sarfraz
  Khurshid}.
\newblock Deeproad: Gan-based metamorphic testing and input validation
  framework for autonomous driving systems.
\newblock In {\em Proceedings of the 33rd ACM/IEEE International Conference on
  Automated Software Engineering}, ASE 2018, pages 132--142, New York, NY, USA,
  2018. ACM.

\bibitem{zhang:saner19}
{\sc Zhuo Zhang, Yan Lei, Xiaoguang Mao, and Panpan Li}.
\newblock Cnn-fl: An effective approach for localizing faults using
  convolutional neural networks.
\newblock In {\em 2019 IEEE 26th International Conference on Software Analysis,
  Evolution and Reengineering (SANER)}, pages 445--455, 2019.

\bibitem{Zhao:ICSE19}
{\sc Dehai Zhao, Zhenchang Xing, Chunyang Chen, Xin Xia, and Guoqiang Li}.
\newblock Actionnet: Vision-based workflow action recognition from programming
  screencasts.
\newblock In {\em 2019 IEEE/ACM 41st International Conference on Software
  Engineering (ICSE)}, pages 350--361, 2019.

\bibitem{Zhao:ICSE20}
{\sc Dehai Zhao, Zhenchang Xing, Chunyang Chen, Xiwei Xu, Liming Zhu, Guoqiang
  Li, and Jinshui Wang}.
\newblock Seenomaly: Vision-based linting of gui animation effects against
  design-don’t guidelines.
\newblock In {\em ICSE’20}, 2020.

\bibitem{zhao2018deepsim}
{\sc Gang Zhao and Jeff Huang}.
\newblock Deepsim: deep learning code functional similarity.
\newblock In {\em Proceedings of the 2018 26th ACM Joint Meeting on European
  Software Engineering Conference and Symposium on the Foundations of Software
  Engineering}, pages 141--151, 2018.

\bibitem{Zhao:PAMI'07}
{\sc Guoying Zhao and Matti Pietikäinen}.
\newblock Dynamic texture recognition using local binary patterns with an
  application to facial expressions.
\newblock {\em PAMI'07}, 29:915--28, 07 2007.

\bibitem{8730177}
{\sc Hui Zhao, Zhihui Li, Hansheng Wei, Jianqi Shi, and Yanhong Huang}.
\newblock Seqfuzzer: An industrial protocol fuzzing framework from a deep
  learning perspective.
\newblock In {\em 2019 12th IEEE Conference on Software Testing, Validation and
  Verification (ICST)}, pages 59--67, 2019.

\bibitem{zhao2002change}
{\sc Jianjun Zhao, Hongji Yang, Liming Xiang, and Baowen Xu}.
\newblock Change impact analysis to support architectural evolution.
\newblock {\em Journal of software maintenance and evolution: research and
  practice}, 14(5):317--333, 2002.

\bibitem{Zhao2018}
{\sc Jinman Zhao, Aws Albarghouthi, Vaibhav Rastogi, Somesh Jha, and Damien
  Octeau}.
\newblock Neural-augmented static analysis of android communication.
\newblock In {\em Proceedings of the 2018 {ACM} Joint Meeting on European
  Software Engineering Conference and Symposium on the Foundations of Software
  Engineering, {ESEC/SIGSOFT} {FSE} 2018, Lake Buena Vista, FL, USA, November
  04-09, 2018}, Gary~T. Leavens, Alessandro Garcia, and Corina~S. Pasareanu,
  editors, pages 342--353. {ACM}, 2018.

\bibitem{21Zhou:CIKM12}
{\sc Jian Zhou and Hongyu Zhang}.
\newblock Learning to rank duplicate bug reports.
\newblock In {\em CIKM '12}, pages 852--861, New York, NY, USA, 2012. ACM.

\bibitem{zimmermann2005mining}
{\sc Thomas Zimmermann, Andreas Zeller, Peter Weissgerber, and Stephan Diehl}.
\newblock Mining version histories to guide software changes.
\newblock {\em IEEE Transactions on Software Engineering}, 31(6):429--445,
  2005.

\bibitem{Zohar2018}
{\sc Amit Zohar and Lior Wolf}.
\newblock Automatic program synthesis of long programs with a learned garbage
  collector.
\newblock In {\em Advances in Neural Information Processing Systems 31},
  S.~Bengio, H.~Wallach, H.~Larochelle, K.~Grauman, N.~Cesa-Bianchi, and
  R.~Garnett, editors, pages 2094--2103. Curran Associates, Inc., 2018.

\end{thebibliography}
\end{document}